\documentclass{amsart}

\usepackage{latexsym,amsfonts,amsmath,amssymb,mathrsfs,pifont,epsfig,extarrows,xy}
\usepackage{graphics}
\usepackage{indentfirst}
\usepackage{footmisc}
\usepackage{MnSymbol}

\usepackage{ifpdf}
\ifpdf
\usepackage{epstopdf}
\usepackage{hyperref}
\else
\usepackage[hypertex]{hyperref}
\fi

\newtheorem{Thm}{Theorem}
\newtheorem{Prop}[Thm]{Proposition}

\newtheorem{Cor}[Thm]{Corollary}
\theoremstyle{definition}
\newtheorem{Def}[Thm]{Definition}
\newtheorem{Rem}[Thm]{Remark}
\newtheorem{Eg}[Thm]{Example}

\numberwithin{equation}{section} \numberwithin{Thm}{section}

\newcommand{\brem}{\begin{Rem}}
\newcommand{\erem}{\end{Rem}\begin{center}*\hspace{2cm}*\hspace{2cm}*\end{center}}
\newcommand{\beg}{\begin{Eg}}
\newcommand{\eeg}{\end{Eg}}
\newcommand{\bedef}{\begin{Def}}
\newcommand{\exdef}{\begin{flushright}$\checkmark$\end{flushright}\end{Def}}
\newcommand{\berop}{\begin{Prop}}
\newcommand{\eerop}{\end{Prop}}
\newcommand{\bethe}{\begin{Thm}}
\newcommand{\ethe}{\end{Thm}}
\newcommand{\becor}{\begin{Cor}}
\newcommand{\ecor}{\end{Cor}}
\newcommand{\beroof}{\noindent\begin{proof}}
\newcommand{\eroof}{\qed\end{proof}}

\def\cA{\mathcal{A}}
\def\cB{\mathcal{B}}
\def\cC{\mathcal{C}}
\def\cD{\mathcal{D}}

\def\cF{\mathcal{F}}
\def\cG{\mathcal{G}}
\def\ceH{\mathcal{H}}

\def\cJ{\mathcal{J}}

\def\ceL{\mathcal{L}}
\def\cM{\mathcal{M}}

\def\cO{\mathcal{O}}

\def\cS{\mathcal{S}}
\def\cT{\mathcal{T}}
\def\cU{\mathcal{U}}

\def\xcA{\mathscr{A}}
\def\xcB{\mathscr{B}}
\def\xcC{\mathscr{C}}
\def\xcD{\mathscr{D}}

\def\xcF{\mathscr{F}}

\def\xcI{\mathscr{I}}

\def\xcL{\mathscr{L}}
\def\xcM{\mathscr{M}}

\def\xcO{\mathscr{O}}
\def\xcP{\mathscr{P}}
\def\xcQ{\mathscr{Q}}

\def\xcV{\mathscr{V}}
\def\xcW{\mathscr{W}}
\def\xcX{\mathscr{X}}

\def\bC{{\mathbb{C}}}

\def\bH{{\mathbb{H}}}

\def\bN{{\mathbb{N}}}

\def\bR{{\mathbb{R}}}
\def\bS{{\mathbb{S}}}
\def\bT{{\mathbb{T}}}
\def\bZ{{\mathbb{Z}}}

\def\a{\alpha}
\def\b{\beta}
\def\g{\gamma}
\def\G{\Gamma}
\def\d{\delta}
\def\D{\Delta}
\def\ep{\epsilon}
\def\vep{\varepsilon}

\def\tht{\theta}
\def\Th{\Theta}

\def\la{\lambda}

\def\om{\omega}
\def\Om{\Omega}

\def\si{\sigma}
\def\Si{\Sigma}
\def\vsi{\varsigma}

\def\Bgt{\gt{B}}

\def\Dgt{\gt{D}}

\def\Egt{\gt{E}}

\def\ggt{\gt{g}}

\def\igt{\gt{i}}
\def\Igt{\gt{I}}
\def\jgt{\gt{j}}
\def\Jgt{\gt{J}}
\def\kgt{\gt{k}}

\def\Pgt{\gt{P}}

\def\tgt{\gt{t}}
\def\Tgt{\gt{T}}

\def\Vgt{\gt{V}}

\newcommand{\sfd}{{\mathsf d}}

\newcommand{\sfi}{{\mathsf i}}
\newcommand{\sfI}{{\mathsf I}}

\newcommand{\sfJ}{{\mathsf J}}
\newcommand{\sfk}{{\mathsf k}}

\newcommand{\sfL}{{\mathsf L}}

\newcommand{\sfN}{{\mathsf N}}

\newcommand{\sfp}{{\mathsf p}}
\newcommand{\sfP}{{\mathsf P}}

\newcommand{\sfT}{{\mathsf T}}

\newcommand{\txB}{{\rm B}}

\newcommand{\txE}{{\rm E}}

\newcommand{\txF}{{\rm F}}
\newcommand{\txg}{{\rm g}}
\newcommand{\txG}{{\rm G}}

\newcommand{\txH}{{\rm H}}

\newcommand{\txK}{{\rm K}}

\newcommand{\txm}{{\rm m}}
\newcommand{\txM}{{\rm M}}

\newcommand{\txp}{{\rm p}}

\def\bd{{\dot \b}}

\def\x{\times}
\def\ox{\otimes}

\def\lx{{\hspace{-0.04cm}\ltimes\hspace{-0.05cm}}}

\def\lact{\vartriangleright}

\def\emb{\hookrightarrow}

\def\p{\partial}

\def\id{{\rm id}}

\def\dim{{\rm dim}}

\def\ker{{\rm ker}}

\def\curv{{\rm curv}}
\def\Hol{{\rm Hol}}

\def\det{{\rm det}}
\def\diag{\textrm{diag}}

\def\tr{{\rm tr}}

\def\Vol{{\rm Vol}}

\newcommand{\faff}[1]{P^{\sfk}_{+}(#1)}

\newtheorem{defn}{Definition}[section]

\newtheorem{exer}{Exercise}[section]
\newtheorem{soln}{Solution}[section]

\newtheorem{rmq}{Remark}[section]

\newcommand{\beq}{\begin{equation}}
\newcommand{\eeq}{\end{equation}}
\newcommand{\beqa}{\begin{eqnarray}}
\newcommand{\eeqa}{\end{eqnarray}}
\newcommand{\begt}{\begin{gather}}
\newcommand{\bal}{\begin{align}}
\newcommand{\eal}{\end{align}}

\newcommand{\barr}{\begin{array}}
\newcommand{\earr}{\end{array}}
\newcommand{\ben}{\begin{enumerate}}
\newcommand{\een}{\end{enumerate}}
\newcommand{\bit}{\begin{itemize}}
\newcommand{\eit}{\end{itemize}}
\newcommand{\bdef}{\begin{defn}}
\newcommand{\eedef}{\end{defn}}
\newcommand{\bermq}{\begin{rmq}}
\newcommand{\eermq}{\end{rmq}}
\newcommand{\bex}{\begin{exer}}
\newcommand{\eex}{\end{exer}}
\newcommand{\besol}{\begin{soln}}
\newcommand{\eesol}{\end{soln}}
\newcommand{\tx}[1]{\textrm{#1}}

\newcommand{\gt}[1]{\mathfrak{#1}}

\def\1{\mathbb{I}}
\def\2{[2]_q}

\newcommand{\ovl}[1]{\overline{#1}}

\newcommand{\nn}{\nonumber}

\newcommand{\qq}{\begin{eqnarray}}

\newcommand{\ee}{{\rm e}}
\newcommand{\qqq}{\end{eqnarray}}

\newcommand{\wrt}{with respect to }

\def\qU2{\cU_q (\mathfrak{su}(2))}

\def\resu2{\textrm{REA}_q(\mathfrak{su}(2))}

\def\ad{{\rm ad}}
\def\Ad{{\rm Ad}}

\def\vC{\check{C}}

\def\vH{\check{H}}

\newcommand{\unl}[1]{\underline{#1}}

\def\vd{\check{\d}}

\def\con{\righthalfcup}

\newcommand{\Reqref}[1]{Eq.\,\eqref{#1}}
\newcommand{\Rcite}[1]{Ref.\,\cite{#1}}
\newcommand{\Rxcite}[2]{Ref.\,\cite[#1]{#2}}
\newcommand{\Rfig}[1]{Fig.\,\ref{#1}}

\def\obj{{\rm Ob}}

\def\morf{{\rm Mor}}

\def\Diff{{\xcD iff}}

\def\ev{{\rm ev}}

\def\bd1{{\boldsymbol{1}}}
\def\brd0{{\boldsymbol{0}}}

\def\bgrb{\gt{BGrb}}

\def\ggtk{\widehat{\gt{g}}_\sfk}

\newcommand{\cGk}{\cG_\sfk}
\newcommand{\epk}{\ep_\sfk}

\newcommand{\pr}{{\rm pr}}

\newcommand{\exd}{{\rm d}}
\newcommand{\pLie}[1]{\,{-\hspace{-12pt}\xcL\hspace{-4pt}}_{#1}}

\newcommand{\uj}{{\rm U}(1)}
\newcommand{\sug}{{\rm SU}(2)}

\newcommand{\Mup}{{}^{\tx{\tiny $M$}}\hspace{-2pt}}

\newcommand{\xcMup}{{}^{\tx{\tiny $\xcM$}}\hspace{-2pt}}

\newcommand{\Qup}{{}^{\tx{\tiny $Q$}}\hspace{-2pt}}

\newcommand{\Tnup}{{}^{\tx{\tiny $T_n$}}\hspace{-2pt}}

\newcommand{\Gup}{{}^{\tx{\tiny $\txG$}}\hspace{-2pt}}

\newcommand{\alxydim}[2]{\begin{aligned}\xymatrix#1{#2}\end{aligned}}

\newcommand\void[1]{}

\xyoption{all}

\begin{document}

\begin{flushright}
ZMP-HH/10-2\\
Hamburger Beitr\"age zur Mathematik Nr.\,360\\
\end{flushright}
\vskip 5.0em

\title{\mbox{Defects, dualities and the geometry of strings via
gerbes} \\ \mbox{I. Dualities and state fusion through
defects${}^\dagger$}}

\author{Rafa\l ~R.~Suszek${}^*$}
\address{\emph{Address:}
Katedra Metod Matematycznych Fizyki, Wydzia\l ~Fizyki Uniwersytetu
Warszawskiego, ul.\,Ho\.za 74, PL-00-682 Warszawa, Poland}
\email{suszek@fuw.edu.pl}
\thanks{${}^*$ The author's work was done partly under the EPSRC
First Grant EP/E005047/1, the PPARC rolling grant PP/C507145/1 and
the Marie Curie network `Superstring Theory' (MRTN-CT-2004-512194).
He was also funded by the Collaborative Research Centre 676
``Particles, Strings and the Early Universe -- the Structure of
Matter and Space-Time''.\\
${}^\dagger$ The contents of the paper are an extended version of
lectures given by the author at the Albert-Einstein-Institut in
Potsdam in April 2009.}


\begin{abstract}
This is the first of a series of papers discussing canonical aspects
of the two-dimensional non-linear sigma model in the presence of
conformal defects on the world-sheet in the framework of gerbe
theory. In the paper, the basic tools of the state-space analysis
are introduced, such as the symplectic structure on the state space
of the sigma model and the pre-quantum bundle over it, and a
relation between the defects and dualities of the sigma model is
established. Also, a state-space description of the
splitting-joining interaction of the string across the defect is
presented, leading to an interpretation of the geometric data
associated to junctions of defect lines in terms of intertwiners
between the incoming and outgoing sectors of the theory in
interaction.
\end{abstract}

\keywords{Sigma models, dualities, defects; Gerbes}

\maketitle

\tableofcontents

\section{Introduction}\label{sec:intro}

The study of symmetries of a physical model and of dualities that
relate it to other points in the relevant moduli space has often
proved an indispensable source of knowledge on the structure of the
physical system of interest -- suffice it to invoke the use of
Ward--Takahashi identities in the derivation of correlation
functions of a quantum field theory, the `duality net' of consistent
(super)string backgrounds, or the AdS/CFT duality that has gained us
insights into a strongly coupled QCD-type field theory through the
study of a weakly coupled string theory. This by now well-ingrained
and widely exploited constatation forms the base of a series of
papers, of which the present one is the first, discussing the r\^ole
played by world-sheet defects in the description of symmetries and
dualities of a large class of two-dimensional field theories with
conformal symmetry known as non-linear $\si$-models, first
considered in all generality in \Rcite{Friedan:1985phd}. These are
-- in the simplest setting -- theories of harmonic embeddings
$\,X:\Si\to M\,$ of an oriented two-dimensional lorentzian manifold
$\,\Si$,\ termed the world-sheet, in a metric manifold
$\,(M,\txg)$,\ called the target space and generically equipped with
additional cohomological structure. In the purely bosonic setting,
and in the absence of the dilaton field on $\,M$,\ this additional
structure comes from the 2-category $\,\bgrb^\nabla(M)\,$ of bundle
gerbes (with connection) over $\,M$,\ introduced in
\Rcite{Stevenson:2000wj} and further elaborated in
\Rcite{Waldorf:2007mm}. Its objects, termed bundle gerbes with
connection, or gerbes for short, were first considered, in rather
abstract terms, in \Rcite{Giraud:1971}. The more intuitive
(hyper)cohomological description was given only in
\Rcite{Brylinski:1993ab}, and an intrinsic geometric definition
followed in Refs.\,\cite{Murray:1994db,Murray:1999ew}. Gerbes are
geometric structures representing classes in the second real Deligne
(hyper)cohomology group of $\,M$.\ As such, they serve to give a
rigorous \emph{global} definition of the topological term in the
action functional of the $\si$-model on a closed world-sheet. The
term is determined by a specific Cheeger-Simons differential
character, to wit, the so-called surface holonomy of the gerbe over
$\,\Si\,$ and it is locally expressed as the integral of the
pullback, along $\,X$,\ of a primitive of the closed curvature
3-form of the gerbe to the world-sheet. The function of the Deligne
cohomology, first brought into the picture in
Refs.\,\cite{Alvarez:1984es,Gawedzki:1987ak}, transcends the
definition of the classical action functional -- indeed, it gives
rise to a natural classification scheme for $\si$-models on a given
target space in terms of equivalence classes of gerbes, which -- for
a given curvature -- span the sheaf-cohomology group
$\,H^2\bigl(M,\uj\bigr)$,\ and it canonically defines the
pre-quantum bundle of the theory, thus establishing the basis of the
geometric quantisation scheme. This was realised already in
\Rcite{Gawedzki:1987ak} and subsequently employed in
\Rcite{Felder:1988sd} in the setting of the Wess--Zumino--Witten
(WZW) $\si$-model with a compact simple connected Lie group as the
target space. Over and above these, gerbe theory provides us with
concrete tools for constructing new theories from the existing ones
by way of gauging subgroups $\,\txK\,$ of the isometry group of
$\,(M,\txg)\,$ and through orientifolding, both procedures being
founded on the notion of an equivariant (resp.\ twisted-equivariant,
cp.\ Refs.\,\cite{Schreiber:2005mi,Gawedzki:2008um}) structure on
the gerbe. The structure uses distinguished 1- and 2-cells of the
2-category of bundle gerbes over the nerve of the action groupoid
$\,\txK\lx M$,\ and it descends to a gerbe structure on the quotient
space $\,M/\txK\,$ under suitable conditions. Since, furthermore,
all gerbes on $\,M/\txK\,$ can be obtained in this manner, cp.\
\Rcite{Gawedzki:2010rn}, it gives us a classification of
$\si$-models on the quotient space (resp.\ that of $\si$-models for
unorientable world-sheets in the case of orientifolding). The
existence and uniqueness theorems established in
Refs.\,\cite{Gawedzki:2002se,Gawedzki:2003pm} and
\Rcite{Gawedzki:2007uz} for the orbifolded resp.\ orientifolded
variants of the WZW model, for which there exists an explicit
construction of the gerbe (worked out, in steps, in
Refs.\,\cite{Gawedzki:1987ak,Chatterjee:1998,Meinrenken:2002}), are
in perfect agreement with known results of the structure-heavy
Conformal Field-Theory (CFT) analyses of modular invariants from
Refs.\,\cite{Felder:1988sd,Kreuzer:1994,Brunner:2002em} and those of
the categorial quantisation of the $\si$-model, reported in
\Rcite{Fuchs:2004dz}. Analogous statements from
\Rcite{Gawedzki:2010rn} pertaining to the case of continuous group
actions, the latter presenting an additional complication due to the
coupling between the gerbe and the principal $\txK$-connection on a
(generically non-trivial) principal $\txK$-bundle over $\,\Si\,$
which may fail because of global gauge anomalies, go far beyond the
long-established results of both the geometric and algebraic
discussion of
Refs.\,\cite{Hull:1989,Hull:1990ms,Figueroa:1994ns,Figueroa:1994dj},
and the (conformal) field-theoretic analysis of
Refs.\,\cite{Goddard:1984vk,Gawedzki:1988nj,Hori:1994nc,Fuchs:1995tq}.
\medskip

The more general physical meaning of the full-blown 2-categorial
structure associated with bundle gerbes and its naturality in the
context of the two-dimensional field theory have been brought to the
fore by the construction of the multi-phase $\si$-model in
\Rcite{Runkel:2008gr}. In this construction, the two-dimensional
spacetime $\,\Si\,$ (or its euclidean version) is split into a
collection of domains $\,\Si_i,\ i\in\ovl{1,N}\subset\bN$,\ each
carrying its own phase of the full theory (i.e.\ a choice $\,M_i\,$
of a connected component of the target space with the attendant
structures of the metric and the gerbe) and separated from adjacent
domains by lines of discontinuity of the embedding field $\,X$,\
termed defect lines and mapped by $\,X\,$ into a correspondence
space $\,Q\,$ called the bi-brane world-volume. Defect lines, in
turn, intersect at the so-called defect junctions, sent by $\,X\,$
into another correspondence space $\,T$,\ dubbed the inter-bi-brane
world-volume. The constitutive elements of the construction of
\Rcite{Runkel:2008gr} and the ensuing definition of the action
functional of the $\si$-model are recalled in Section
\ref{sec:lagr}.

An important source of inspiration for the construction, with its
assignment of distinguished 1-cells of $\,\bgrb^\nabla(Q)\,$ to
defect lines and 2-cells of $\,\bgrb^\nabla(T)\,$ to defect
junctions, were the earlier findings of
Refs.\,\cite{Freed:1999vc,Kapustin:1999di,Carey:2002,Gawedzki:2002se,Gawedzki:2004tu}
and \Rcite{Fuchs:2007fw} in which the relevant cohomological
structures had been identified over the submanifolds of $\,M\,$ and
$\,M_i\x M_j$,\ respectively, defining the codomain (the so-called
D-brane or $\cG$-brane world-volume) of the restriction of $\,X\,$
to connected components of the boundary of $\,\Si\,$ in the former
case, and determining the discontinuity of $\,X\,$ along a defect
line homeomorphic to $\,\bS^1\,$ in the latter case. These
cohomological structures, corresponding to vector bundles twisted by
the gerbe in a well-defined manner, contribute their own part to the
classification scheme of consistent $\si$-models on world-sheets
with defects and straightforwardly accommodate a variant of the
orientifolding and gauging constructions set up for the bulk gerbes,
as demonstrated in
Refs.\,\cite{Gawedzki:2002se,Gawedzki:2004tu,Gawedzki:2008um,Gawedzki:2010ggd}.
It well deserves to be pointed out that the ensuing explicit
constructions of orbifold and orientifold $\cG$-branes in the
controlled setting of the WZW model, presented in
Refs.\,\cite{Gawedzki:2004tu,Gawedzki:2010G}, indicate the presence
of an essentially new species of $\cG$-brane, dubbed the non-abelian
brane in the original \Rcite{Gawedzki:2004tu}, over those conjugacy
classes in the target Lie group which are invariant under the action
of non-cyclic components of $\,\txK$.\ These $\cG$-branes have
properties suggestive of an interpretation in terms of irresoluble
stacks of fixed-point fractional branes of
\Rcite{Diaconescu:1997br}. Their existence, peculiar to the maximal
orbifold $\,{\rm Spin}(4n)/(\bZ_2\x\bZ_2)\,$ and ubiquitous on
proper orientifolds of group manifolds, had not been predicted by
the standard CFT methods, and so they provide a tangible example of
a novel string-theoretic insight gained by purely gerbe-theoretic
methods. The intriguing internal (open-string) dynamics of these
branes still awaits an in-depth treatment.

A piece of motivation that is more immediately related to the
subject matter of the present paper, and also of a more
field-theoretic flavour (as seen from the two-dimensional
perspective), comes from the CFT studies of conformal interfaces
(i.e.\ defect lines transmissive to that half of the conformal
symmetries of either of the two phases of the theory supported over
the two domains of the world-sheet separated by the defect line
which preserve that line). The concept originated from the
condensed-matter considerations of \Rcite{Oshikawa:1996dj} and was
later transplanted into the string-theoretic domain in
\Rcite{Petkova:2000ip} (in a purely operator-algebraic language) and
in \Rcite{Bachas:2001vj} (in a more geometric world-sheet terms).
Subsequent studies have diverged into a variety of specialised
directions, including the classificatory analysis and specific
constructions of
Refs.\,\cite{Fuchs:2002cm,Quella:2002ct,Fuchs:2007tx,Fuchs:2007fw,Bachas:2009mc,Gawedzki:2010ggd}
for various distinguished classes of CFT (such as, e.g., the free
boson, the (gauged) WZW model and, more generally, an arbitrary
rational CFT), the discussion of the fusion of conformal interfaces
in the quantum r\'egime in
Refs.\,\cite{Petkova:2000ip,Fuchs:2002cm,Bachas:2007td}, alongside a
description of perturbed defect CFTs and Renormalisation-Group (RG)
flows in their presence advanced in
Refs.\,\cite{Bachas:2004sy,Alekseev:2007in,Runkel:2007wd,Kormos:2009sk,Bachas:2009mc},
and related to certain integrable structures of CFT in
Refs.\,\cite{Runkel:2007wd,Manolopoulos:2009np}. Analogous results
have also been obtained in the context of supersymmetric
two-dimensional field theories, cf., e.g.,
Refs.\,\cite{Brunner:2007qu,Brunner:2008fa,Brunner:2009zt,Brunner:2010xm}.
The studies carried out to date, and in particular those reported in
Refs.\,\cite{Frohlich:2004ef,Frohlich:2006ch,Schweigert:2007wd,Runkel:2008gr,Sarkissian:2008dq,Bachas:2008jd},
bear ample evidence of a prominent r\^ole played by conformal
interfaces in establishing correspondences between phases of CFT, in
encoding order-disorder dualities among various CFTs, and in mapping
into one another their RG flows as well as UV and IR fixed points of
the latter. Finally, they can be associated with the so-called
spectrum-generating symmetries of string theory, relating -- via
fusion with boundary states (a process that has not been fully
understood up to now) -- the D-brane categories of a dual pair of
CFTs. All this leads to a natural question as to a state-space
interpretation of the conformal world-sheet defects of
\Rcite{Runkel:2008gr} and the attendant cohomological structures on
the codomain of $\si$-model fields, encompassing the data carried by
both the defect lines and their junctions. This question is at the
core of the present paper, and in our search for an answer, we shall
be guided by insights inferred from the detailed treatment of the
maximally symmetric WZW defects in
Refs.\,\cite{Fuchs:2007fw,Runkel:2008gr,Runkel:2009su,Runkel:2010}.
The latter provide an excellent setting in which to look, in
particular, for conditions necessary and sufficient for the
correspondence between the phases of the $\si$-model determined by
the defect to be compatible with the module structure on the
respective state spaces with respect to the action of an extended
current symmetry algebra. This issue is put in a wider
generalised-geometric context and subsequently elaborated at great
length in \Rcite{Suszek:2010b}, forming the second part of the
series opened by the present article.\medskip

A natural framework for establishing the sought-after state-space
interpretation of the conformal defects is provided by the canonical
description of the $\si$-model. The description can be derived in
the so-called covariant (or first-order) formalism of
Refs.\,\cite{Gawedzki:1972ms,Kijowski:1973gi,Kijowski:1974mp,Kijowski:1976ze,Szczyrba:1976,Kijowski:1979dj}
which leads to a systematic reconstruction of the symplectic
structure on the state space of the two-dimensional field theory of
interest. The basic tools of the formalism are introduced in Section
\ref{sub:can-gen}. These are subsequently applied, in the remainder
of Section \ref{sec:can}, to the two qualitatively distinct sectors
of string theory present on a generic world-sheet with an embedded
defect, that is the untwisted sector, composed of strings
represented by smooth loops embedded in the respective connected
components of the target space, and the twisted sector, with states
represented by piecewise smooth maps from the unit circle into the
target space, with point-like discontinuities which can be
understood as resulting from transversal intersections with defect
lines. A prototypical example of the latter sector is provided by
the twisted sector of string theory on an orbifold of a smooth
target space, first discussed in
Refs.\,\cite{Dixon:1985jw,Dixon:1986jc}, in which case the
discontinuities are determined by elements of the orbifold group.
The upshot of the analysis carried out in Section \ref{sec:can} is a
full-fledged canonical description of the classical (bosonic) string
with a multi-phase world-sheet.

The key advantage of working with the global geometric structures
from $\,\bgrb^\nabla(M\sqcup Q\sqcup T)\,$ in the classical setting
is that they actually afford inroads into the quantum r\'egime of
the theory. As mentioned already in the opening paragraph of the
present section, this fact has been known ever since the
introduction of the hypercohomological language into rigorous
studies of the two-dimensional $\si$-model in
\Rcite{Gawedzki:1987ak}, which is where the transgression map was
defined. The latter is a cohomology map canonically assigning to the
1-isomorphism class of the gerbe of the $\si$-model with target
space $\,M\,$ the isomorphism class of a circle bundle over its
configuration space $\,\sfL M\equiv C^\infty(\bS^1,M)\,$ (the
free-loop space of $\,M$) with a connection whose curvature yields,
upon pullback to the state space $\,\sfP_{\si,\emptyset}\cong\sfT^*
\sfL M\,$ of the theory and correction by a canonical (and
topologically trivial) term, the symplectic form of the
(defect-free) $\si$-model. In other words, the gerbe determines a
pre-quantum bundle $\,\ceL_{\si,\emptyset}\to\sfP_{\si,\emptyset}\,$
of the closed string, a prerequisite of its geometric quantisation,
cp., e.g., \Rcite{Woodhouse:1992de}. The important novel result for
strings with multi-phase world-sheets, anticipated by the findings
of Refs.\,\cite{Gawedzki:2002se,Gawedzki:2004tu} and derived in
Section \ref{sec:can} in analogy with the original result for the
untwisted sector, is the existence of a straightforward
generalisation of the transgression map to the twisted sector of the
string. This generalised cohomology map canonically assigns the
isomorphism class of a pre-quantum bundle over the space of twisted
states to the equivalence class of a coupled pair consisting of the
$\si$-model gerbe and the associated bi-brane, a fact following
directly from Theorem \ref{thm:trans-tw}.\medskip

\begin{figure}[hbt]~\\[5pt]

$$
 \raisebox{-50pt}{\begin{picture}(50,50)
  \put(-79,-4){\scalebox{0.25}{\includegraphics{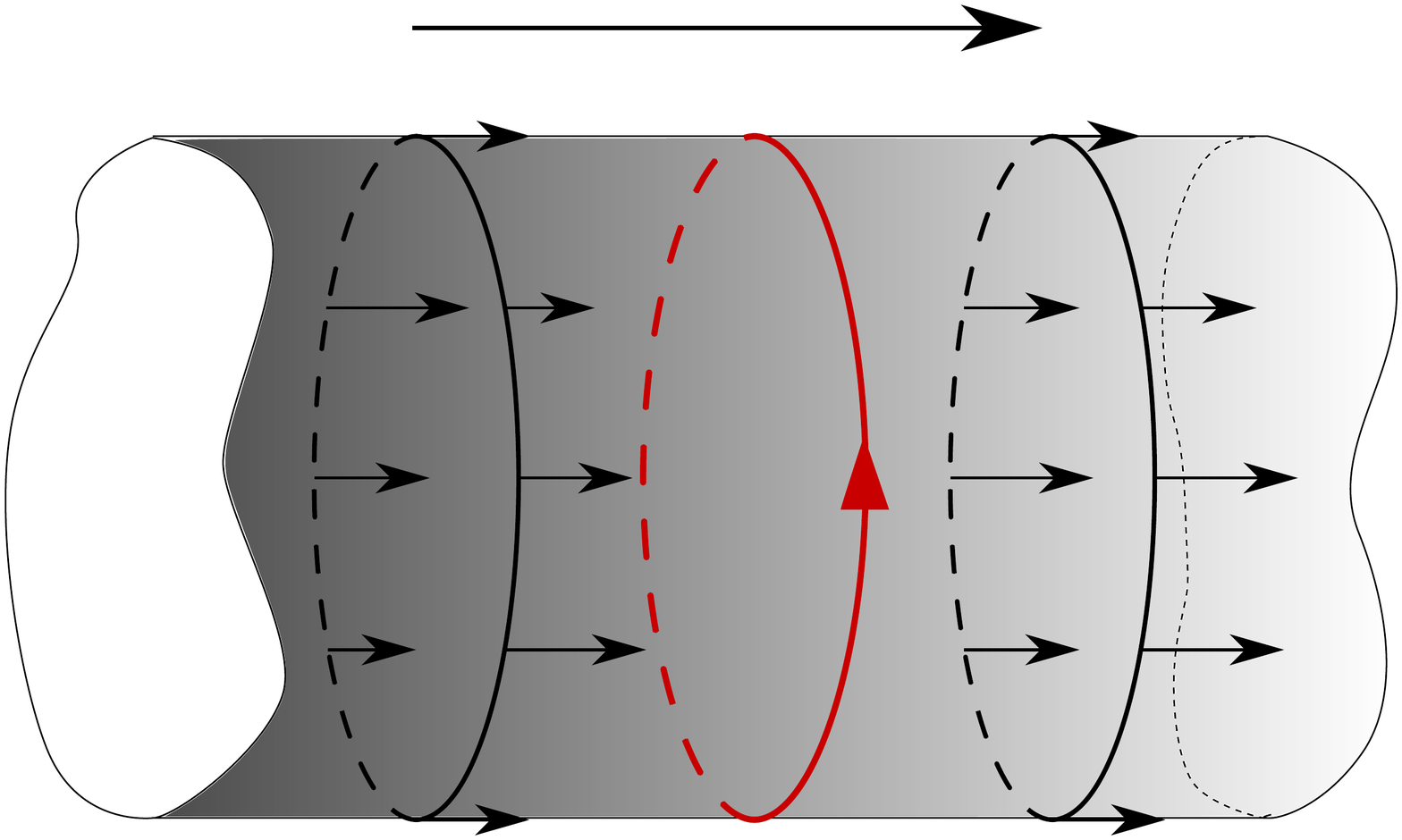}}}
  \end{picture}
  \put(0,0){
     \setlength{\unitlength}{.60pt}\put(-28,-16){
     \put(-10,40)     { $\ell_{1,2}$  }
     \put(-82,40)     { $\psi_1$ }
     \put(58,40)      { $\psi_2$ }
     \put(-160,133)   { ${\rm CFT}_1$   }
     \put(135,133)    { ${\rm CFT}_2$   }
     \put(-10,250)    { $t$ }
            }\setlength{\unitlength}{1pt}}}
$$

\caption{The correspondence between states mediated by the defect
line: the state $\,\psi_1\,$ from the phase $\,{\rm CFT}_1\,$ is
transferred to the state $\,\psi_2\,$ from the phase $\,{\rm
CFT}_2\,$ across the defect line $\,\ell_{1,2}$.\ The arrow above
the picture represents the world-sheet time direction.}
\label{fig:defect-corr}
\end{figure}

The canonical formalism thus reconstructed constitutes an excellent
basis for phrasing the question about the r\^ole of defects in a
rigorous manner. Guided by the simple geometric intuition conveyed
by the world-sheet picture of the cross-defect identification of
states effected by the propagation of the closed string, as
illustrated in Figure \ref{fig:defect-corr},
we are led to investigate conditions under which the data of the
defect (including the gluing condition to be imposed on the
lagrangean fields of the model) define a duality between the phases
of the $\si$-model separated by the defect. The point of departure
for these investigations is the identification of a (pre-quantum)
duality with an isomorphism
\qq\nn
\pr_1^*\ceL_{\si,\emptyset}\vert_{\Igt_\si}\cong\pr_2^*\ceL_{\si,
\emptyset}\vert_{\Igt_\si}
\qqq
of the pullbacks of the pre-quantum bundle along the canonical
projections $\,\pr_\a:\sfP_{\si,\emptyset}\x\sfP_{\si,\emptyset}\to
\sfP_{\si,\emptyset},\ \a\in\{1,2\}\,$ over the graph $\,\Igt_\si\,$
of a symplectomorphism that preserves the hamiltonian density of the
$\si$-model. The relevant conditions are stated in Theorem
\ref{thm:def-dual}, and it is worth underlining that they single out
bi-brane world-volumes $\,Q\,$ that are surjectively submersed onto
the target space $\,M$,\ in keeping with the results of
\Rcite{Fuchs:2009si}. The reverse question as to the circumstances
under which a duality gives rise to consistent defect data is
subsequently examined for an important class of dualities in the
remainder of Section \ref{sec:def-as-iso}, culminating in Theorems
\ref{thm:duali-T-bib} and \ref{thm:duali-N-bib}. Altogether, the
findings of Section \ref{sec:def-as-iso} establish a rather strong
and general correspondence between the so-called topological defects
and dualities of string theory, the latter including, in particular,
symmetries of a single phase induced from distinguished isometries
of the target space and the proper (T-)duality between string models
on topologically non-equivalent principal torus bundles, of the kind
originally discussed in
Refs.\,\cite{Buscher:1987qj,Buscher:1987sk}.\medskip

\begin{figure}[hbt]~\\[5pt]

$$
 \raisebox{-50pt}{\begin{picture}(50,50)
  \put(-79,-4){\scalebox{0.25}{\includegraphics{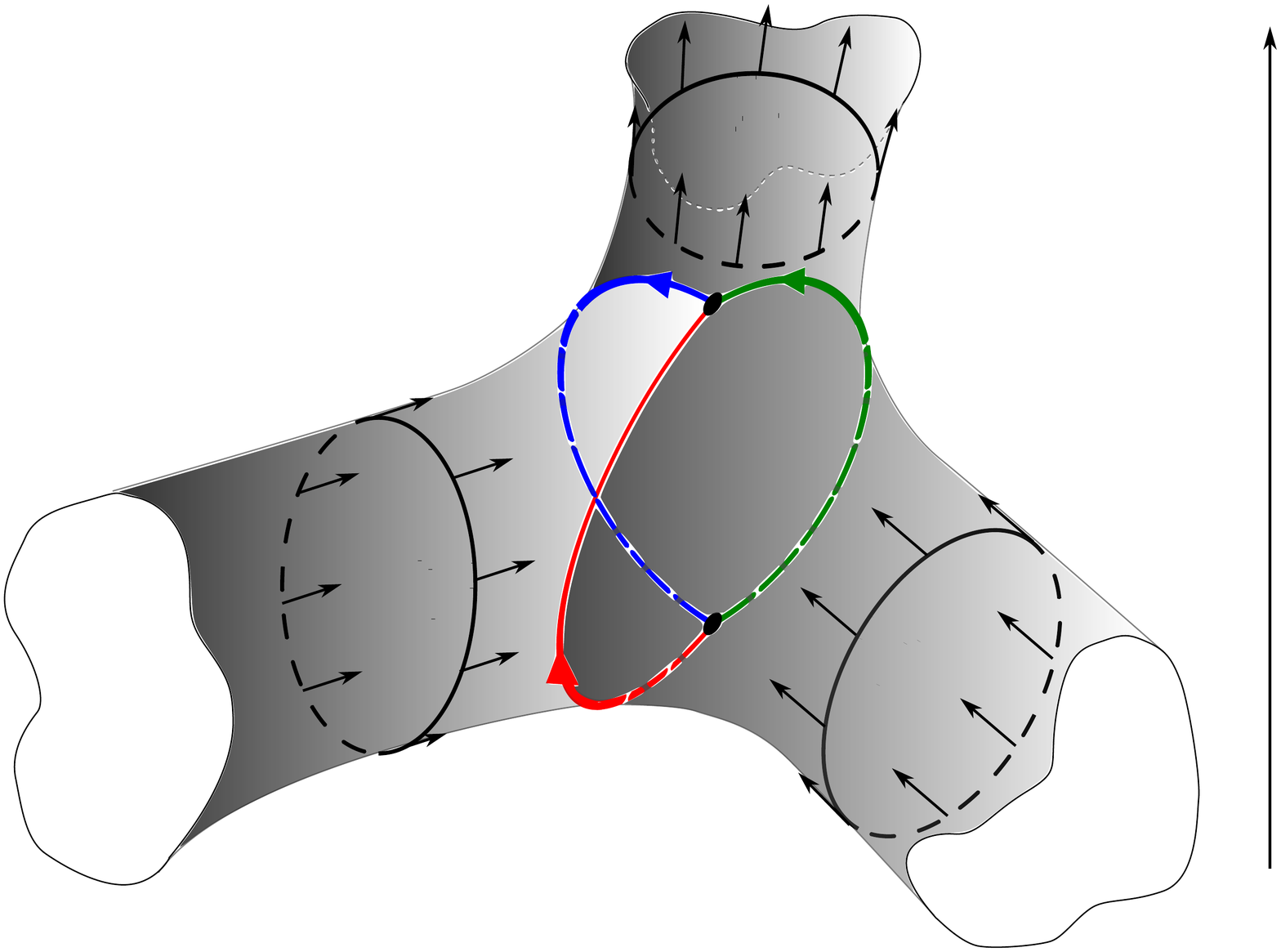}}}
  \end{picture}
  \put(0,0){
     \setlength{\unitlength}{.60pt}\put(-28,-16){
     \put(-35,57)     { $\ell_{1,2}$  }
     \put(48,165)     { $\ell_{2,3}$  }
     \put(-64,177)    { $\ell_{3,1}$  }
     \put(-3,162)     { $\jmath_{1,2,3}$  }
     \put(-35,96)     { $\jmath^\vee_{1,2,3}$  }
     \put(-86,47)     { $\psi_1$ }
     \put(25,30)      { $\psi_2$ }
     \put(-45,210)    { $\psi_3$ }
     \put(-163,72)    { ${\rm CFT}_1$   }
     \put(76,40)      { ${\rm CFT}_2$   }
     \put(-5,260)     { ${\rm CFT}_3$   }
     \put(160,140)    { $t$ }
            }\setlength{\unitlength}{1pt}}}
$$

\caption{The basic $\,2\to 1\,$ splitting-joining interaction across
the three-valent defect junction. The arrow to the right of the
picture represents the time direction.} \label{fig:defect-junct-int}
\end{figure}

Just as defect-line data constrain the `tunelling' of the closed
string between images (with respect to the embedding map) of the
supports of adjacent phases of the theory, those carried by defect
junctions are of relevance to the stringy interaction processes,
represented by regions in the world-sheet homeomorphic to a sphere
with (at least) three punctures and an embedded defect subgraph. The
simple world-sheet intuition behind this statement is depicted in
Figure \ref{fig:defect-junct-int}.
It can readily be formalised in the canonical language, in which one
is led to expect the emergence of symplectomorphic identifications
among multi-string states in interaction, lifting to isomorphisms of
the associated pullback pre-quantum bundles over an interaction
subspace spanned by these states. Theorems
\ref{thm:cross-def-int-untw} and \ref{thm:cross-def-int-tw} confirm
these expectations independently for each of the two sectors of the
state space of the $\si$-model, that is for the untwisted sector and
the twisted sector, altogether giving rise to a canonical picture of
the cross-defect splitting-joining interaction of the string. In
this picture, the data of the 2-isomorphism associated with a defect
junction are shown to transgress, in a canonical manner, to local
data of the expected isomorphism of the pullback pre-quantum
bundles. This result can be viewed as a logical completion of the
transgression scheme for $\,\bgrb^\nabla(M\sqcup Q\sqcup T)\,$
anticipated by the findings of \Rcite{Gawedzki:1987ak}. A natural,
if also merely implicit in the present treatment, consequence of the
existence of cross-defect identifications among multi-string states
is an interpretation of the defect-junction data in terms of
intertwiners between representations of (current) symmetry algebras
furnished by the multi-string state spaces that enter the definition
of the interaction subspace. This interpretation is substantiated in
\Rcite{Suszek:2010b}.\medskip

The clear-cut canonical interpretation of world-sheet defects, and
-- in particular -- the relation between the latter and dualities of
string theory, in conjunction with the by now rich knowledge --
gathered in
Refs.\,\cite{Gawedzki:2003pm,Gawedzki:2004tu,Schreiber:2005mi,Gawedzki:2007uz,Gawedzki:2008um}
and further enhanced in \Rcite{Gawedzki:2010rn} through the study of
continuous group actions -- on the gerbe-theoretic structure of
multi-phase world-sheets and on their r\^ole in defining string
theory on quotient target spaces suggest a natural generalisation of
the notion of a smooth (pseudo-)riemannian manifold with extra
(cohomological) structure. Inspired by the pioneering work
\cite{Hull:2004in}, but also taking into account the findings of
\Rcite{Jureit:2006yf}, where multi-phase world-sheets were
considered from the point of view of a consistent interaction of the
closed string on the orbifolded target space, we may conceive a
situation in which the string world-sheet is mapped into a target
space modelled on a (pseudo-)riemannian manifold with a gerbe over
it only \emph{locally}, with geometric data (those of the metric and
of the gerbe) from local charts glued together by means of local
data of the bi-brane implementing \emph{bona fide} $\si$-model
dualities. This ultimately leads to the concept of a non-geometric
background with the structure of a duality `quotient' (provided that
the latter can be defined in a meaningful manner), or a
D(uality)-fold, generalising the idea of a T-fold based on the
T-duality group. The correspondence between gerbes on $\txK$-spaces
and those on $\txK$-quotients of the latter is a strong indication
that string theory on a would-be D-fold prerequires a
self-consistent hierarchy of cohomological structures, to wit, the
bulk gerbe, the duality bi-brane and the basic inter-bi-brane (for
three-valent defect junctions) from which all components of the
inter-bi-brane associated with defect junctions of valence higher
that 3 could be induced in a well-defined and physically intuitive
manner. The relevant intuition derives from the observation that an
insertion of a local defect field for a defect junction of valence
$\,n>3\,$ can be generated in a suitably regularised limiting
procedure of bringing together a number of insertions of local
defect fields for defect junctions of valence 3. The existence of
such a hierarchy was first discussed (and illustrated with the
explicit example of the central-jump WZW defect) in
\cite{Runkel:2008gr} under the name of defect-junction data with
induction. In Remark \ref{rem:duality-scheme}, the original
discussion is extended and rephrased in the language of simplicial
objects in the category of differentiable manifolds, inspired by the
study of equivariant structures and purely physical considerations,
whereupon the notion of a simplicial string background is
introduced. This is then conjectured to be the point of departure in
any consistent construction of a D-fold, to which we are hoping to
return in the future.\medskip

Prior to concluding this introductory section, let us add a few more
comments on the structure of the present paper. First of all, we
have decided to organise the discourse into a collection of
definitions, propositions and theorems, interspersed with examples
and occasional remarks of a looser nature. Secondly, the more
technical proofs have been relegated to the appendices. Finally,
several open questions have been collected in the closing Section
\ref{sec:out}, alongside a brief recapitulation of the results.
\bigskip

\noindent{\bf Acknowledgements:} The author is much beholden to
S.~Fredenhagen, I.~Runkel and, in particular, to K.~Gaw{\c{e}}dzki
for discussions and their sustained interest in the project reported
in the present paper. He also gratefully acknowledges the kind
hospitality of Laboratoire de Physique de l'\'Ecole Normale
Sup\'erieure de Lyon, Albert-Einstein-Institut in Potsdam, Bereich
Algebra und Zahlentheorie des Departments Mathematik an der
Universit\"at Hamburg and Matematyczne Centrum
Konferencyjno-Badawcze IM PAN in B{\c{e}}dlewo, where parts of this
work was carried out.

\section{Defects in the lagrangean picture}\label{sec:lagr}

In the present paper, we shall be concerned with a theory of bosonic
fields on an oriented two-dimensional space-time in the presence of
domain walls that split the space-time into domains supporting the
respective phases of the field theory. Fields of the theory take
values in differentiable manifolds with additional geometric
structure, captured neatly by gerbe theory, and a working knowledge
of the rudiments thereof is assumed throughout this paper. The
reader unfamiliar with the theory is referred to, e.g., the
literature cited in the Introduction. Thus, let us begin with
\bedef\label{def:bckgrnd}
A \textbf{string background} is a triple $\,\Bgt=(\cM,\cB,\cJ )\,$
composed of the following geometric structures:
\bit
\item the \textbf{target} $\,\cM=(M,\txg,\cG)\,$ consisting of
a manifold $\,M$,\ termed the \textbf{target space}, with a metric
$\,\txg$,\ a closed 3-form $\,\txH\,$ and an abelian gerbe $\,\cG\,$
(with connection) of curvature $\,\txH$;
\item the \textbf{$\cG$-bi-brane} $\,\cB=\bigl(Q,\iota_\a,\om,\Phi\
\vert\ \a\in\{1,2\}\bigr)\,$ consisting of a manifold $\,Q$,\ termed
the \textbf{$\cG$-bi-brane world-volume}, with a 2-form $\,\om$,\
termed the \textbf{$\cG$-bi-brane curvature}, and a pair of smooth
maps $\,\iota_\a:Q\to M,\ \a\in\{1,2\}$,\ and of a gerbe
1-isomorphism (a $(\iota_1^*\cG,\iota_2^*\cG)$-bi-module)
\qq\nn
\Phi\ :\ \iota_1^*\cG\xrightarrow{\cong}\iota_2^*\cG\ox I_\om\,,
\qqq
written in terms of a trivial gerbe $\, I_\om\,$ with curving
$\,\om$,\ obeying the identity
\qq\nn
\D_Q\txH=-\sfd\om\,,\qquad\D_Q:=\iota_2^*-\iota_1^*\,;
\qqq
\item the \textbf{$(\cG,\cB)$-inter-bi-brane} $\,\cJ=\bigl(T_n,
\bigl(\vep^{k,k+1}_n,\pi^{k,k+1}_n \ \vert\ k\in\ovl{1,n}\bigr),
\varphi_n\ \vert\ n\in\bN_{\geq 3}\bigr)$,\ with $\,\ovl{1,n}=\{\ \
k\in\bZ \quad\vert\quad 1\leq k\leq n \}$,\ consisting of a disjoint
sum of manifolds $\,\bigsqcup_{n\in\bN_{\geq 3}}\,T_n=:T$,\ termed
the \textbf{$(\cG, \cB)$-inter-bi-brane world-volume}, with a
collection of orientation maps $\,\vep^{k,k+ 1}_n:T_n\to\{-1,+1\}\,$
and smooth maps $\,\pi^{k,k+1}_n:T_n\to Q\,$ subject to the
constraints
\qq\label{eq:proto-simpl}
\iota_2^{\vep_n^{k-1,k}}\circ\pi_n^{k-1,k}=\iota_1^{\vep_n^{k,k+1}}
\circ\pi_n^{k,k+1}\,,\qquad k\in\ovl{1,n}\,,
\qqq
with $\,(\iota_1^{+1},\iota_2^{+1})=(\iota_1,\iota_2)\,$ and
$\,(\iota_1^{-1},\iota_2^{-1})=(\iota_2,\iota_1)$,\ and of
distinguished gerbe 2-isomorphisms
\qq\label{diag:2iso}
\xy (50,0)*{\bullet}="G3"+(5,4)*{\cG_n^3\ox I_{\om_n^{1,2}+\om_n^{2,
3}}}; (25,-20)*{\bullet}="G2"+(-10,0)*{\cG_n^2\ox I_{\om_n^{1,2}}};
(75,-20)*{\ \vdots}="dots"; (85,-20)*{\,;};
(35,-40)*{\bullet}="G1"+(0,-4)*{\cG_n^1};
(65,-40)*{\bullet}="G1add"+(10.5,-4)*{\cG_n^1\ox I_{\om_n^{1,2}+
\om_n^{2,3}+\ldots+\om_n^{n,1}}}; (50,-40)*{}="id";
\ar@{->}|{\Phi_n^{2,3}\ox\id} "G2";"G3"
\ar@{->}|{\Phi_n^{3,4}\ox\id} "G3";"dots" \ar@{->}|{\Phi_n^{1,2}}
"G1";"G2" \ar@{->}|{\Phi_n^{n,1}\ox\id} "dots";"G1add"
\ar@{=}|{\id_{\cG_n^1}} "G1"+(2,0);"G1add"+(-2,0)
\ar@{=>}|{\varphi_n} "G3"+(0,-3);"id"+(0,+3)
\endxy
\qqq
written in terms of 1-isomorphisms $\,\Phi_n^{k,k+1}=\pi_n^{k,k+1\,
*}\Phi^{\vep_n^{k,k+1}}$,\ with $\,\Phi^{+1}=\Phi\,$ and $\,\Phi^{-
1}=\Phi^\vee\,$ (the dual 1-isomorphism), between gerbes $\,\cG_n^k=
(\iota_1^{\vep_n^{k,k+1}}\circ\pi_n^{k,k+1})^*\cG$,\ and the trivial
gerbes with global curvings $\,\om_n^{k,k+1}=\vep_n^{k,k+1}
\,\pi_n^{k,k+1\,*}\om$.\ The latter satisfy the Defect-Junction
Identity (DJI)
\qq\nn
\D_{T_n}\om=0\,,\qquad\D_{T_n}:=\sum_{k=1}^n\,\vep_n^{k,k+1}\,
\pi_n^{k,k+1\,*}\,.
\qqq
\eit
\exdef \noindent In the subsequent sections, we shall oftentimes
have a need for a more explicit description of the gerbe-theoretic
concepts invoked in Definition \ref{def:bckgrnd}. For this reason,
we recall
\bedef\label{def:loco}
Let $\,\xcM\,$ be a differentiable manifold, and let $\,\cS^q_\xcM,\
q\in\ovl{0,\dim\,\xcM}\,$ be the following sheaves over $\,\xcM$:
\bit
\item $\cS^0_\xcM:=\unl\uj_\xcM$,\ the sheaf of locally smooth
$\uj$-valued maps on $\,\xcM$;
\item $\cS^q_\xcM:=\unl\Om^q(\xcM),\ q>0$,\ the sheaf of locally
smooth (real) $q$-forms on $\,\xcM$.
\eit
Given the differential Deligne complex
\qq\nn
\cD(n)_\xcM^\bullet\ :\ \cS^0_\xcM\xrightarrow{\ \sfd^{(0)}:=
\frac{1}{\sfi}\sfd\log\ }\cS^1_\xcM \xrightarrow{\ \sfd^{(1)}:=\sfd\
}\cS^2_\xcM \xrightarrow{\ \sfd^{(2)}:=\sfd\ }\cS^3_\xcM
\xrightarrow{\ \sfd^{(3)}:=\sfd\ }\cdots\xrightarrow{\ \sfd^{(n-1)}:
=\sfd\ }\cS^n_\xcM\,,
\qqq
denote by $\,\cA^{n,\bullet}( \cO_\xcM)\,$ the diagonal sub-complex
of the \v Cech--Deligne double complex $\,\vC^\bullet\bigl(\cO_\xcM,
\cD(n)_\xcM^\bullet\bigr)\,$ obtained, for a given choice
$\,\cO_\xcM=\{\cO^\xcM_i\}_{i\in\xcI}\,$ of a good open cover of
$\,\xcM\,$ (with non-empty multiple intersections of its elements
denoted as $\,\cO^\xcM_{i_1}\cap\cO^\xcM_{i_2}\cap\cdots\cap
\cO^\xcM_{i_n}=: \cO^\xcM_{i_1 i_1\ldots i_n}\,$ and assumed
contractible),\ by extending $\,\cD(n)_\xcM^\bullet\,$ through the
\v Cech complexes
\qq\nn
\vC^0(\cO_\xcM,\cS_\xcM^q)\xrightarrow{\ \vd^{(0)}\ }\vC^1(\cO_\xcM
,\cS_\xcM^q)\xrightarrow{\ \vd^{(1)}\ }\vC^2(\cO_\xcM,\cS_\xcM^q)
\xrightarrow{\ \vd^{(2)}\ }\cdots
\qqq
associated to $\,\cO_\xcM$,\ and with the standard \v Cech
coboundary operators
\qq\nn
\vd^{(p)}\ &:&\ \vC^p(\cO_\xcM,\cS_\xcM^q)\to\vC^{p+1}(\cO_\xcM,
\cS_\xcM^q)\cr\cr
&:&\ (s_{i_0 i_1 \ldots i_p})\mapsto\bigl((\vd^{(p)}s)_{i_0 i_1
\ldots i_{p+1}}):=\left(\sum_{k=0}^{p+1}\,(-1)^k\,s_{i_0 i_1
\underset{\widehat{i_k}}{\ldots} i_{p+1}}\vert_{\cO^\xcM_{i_0 i_1
\ldots i_{p+1}}}\right)\,.
\qqq
The above is written in the additive notation for local sections $\,
s_{i_0 i_1 \ldots i_p}\in\cS_\xcM^q(\cO^\xcM_{i_0 i_1 \ldots i_p})
\,$ of the sheaves $\,\cS_\xcM^q$,\ in which "+" stands for
multiplication of sections if $\,q=0\,$ and for addition of sections
otherwise, and in which multiplication of a section by a real number
$\,c\,$ stands for the raising of the section to the power $\,c\,$
if $\,q=0\,$ and for the multiplying of the section by $\,c\,$
otherwise. This notation shall be used throughout the paper.
Finally, we write as $\,\xcMup D_{(r)}\,$ the Deligne differential
defined component-wise as
\qq\nn
\xcMup D_{(r)}\ :\ \cA^{n,r}(\cO_\xcM)\to\cA^{n,r+1}(\cO_\xcM)\,,
\qquad\qquad\xcMup D_{(r)}\vert_{\vC^p(\cO_\xcM,\cS^q_\xcM)}=
\sfd^{(q)}+(-1)^{q+1}\,\vd^{(p)}\,,
\qqq
with the corresponding Deligne (hyper-)cohomology groups denoted as
$\,\bH^r\bigl(\xcM,\cD(n)^\bullet_\xcM\bigr)$.\ A \textbf{local
presentation of string background $\,\Bgt=(\cM,\cB,\cJ)\,$} consists
of the following data
\bit
\item for the gerbe $\,\cG\,$ over the target space $\,M$,\ a \v
Cech--Deligne cochain
\qq\nn
\cG\xrightarrow{\rm loc.}(B_i,A_{ij},g_{ijk})=:b\in\cA^{3, 2}(\cO_M)
\qqq
with \textbf{curvings} $\,B_i$,\ \textbf{connections} $\,A_{ij}\,$
and \textbf{transition functions} $\,g_{ijk}$,\ satisfying the
cohomological identity
\qq\label{eq:DG-is-H}
\Mup D_{(2)}b=(\txH\vert_{\cO^M_i},0,0,1)\,;
\qqq
the local data $\,b\,$ are determined up to \textbf{gauge
transformations}
\qq\label{eq:gauge-trans-gerbe}
b\mapsto b+\Mup D_{(1)}\pi\,,\qquad\pi:=(\Pi_i,\chi_{ij})\in\cA^{3,
1}(\cO_M)\,;
\qqq
thus, gauge equivalence classes of local data correspond to elements
of $\,\bH^2\bigl(M,\cD(2)^\bullet_M\bigr)$;
\item for the $\cG$-bi-brane 1-isomorphism $\,\Phi$,\ a \v
Cech--Deligne cochain
\qq\nn
\Phi\xrightarrow{\rm loc.}(P_i,K_{ij})=:p\in\cA^{2,1}(\cO_Q)
\qqq
satisfying the cohomological identity
\qq\label{eq:DPhi-is}
\Qup D_{(1)}p=\check{\D}_Q b+\ovl\om\,,
\qqq
in which $\,\ovl\om=(\om\vert_{\cO^Q_i},0,1)\,$ are local data of
the trivial gerbe $\,I_\om$,\ and $\,\check{\D}_Q:=\check{\iota}_2^*
-\check{\iota}_1^*\,$ for the \v Cech-extended $\cG$-bi-brane maps
$\,\check\iota_\a=(\iota_\a,\phi_\a)\,$ with index maps $\,\phi_\a:
\xcI_Q\to\xcI_M\,$ covering the respective manifold maps $\,\iota_\a
:Q\to M\,$ as per
\qq\nn
\iota_\a(\cO^Q_i)\subset\cO^M_{\phi_\a(i)}\,,
\qqq
and in which we use the shorthand notation
\qq\nn
\check{\iota}_\a(B_i,A_{ij},g_{ijk}):=\iota_\a^*(B_{\phi_\a(i)},
A_{\phi_\a(i)\phi_\a(j)},g_{\phi_\a(i)\phi_\a(j)\phi_\a(k)})\,;
\qqq
the local data $\,p\,$ are determined up to \textbf{$\cG$-twisted
gauge transformations}
\qq\label{eq:gauge-trans-bi}
p\mapsto p+\check{\D}_Q\pi-\Qup D_{(0)}w\,,\qquad w:=(W_i)\in\cA^{2
,0}(\cO_Q)\,,
\qqq
with the $\cG$-twist $\,\check{\D}_Q\pi\,$ ensuring that the
defining identity \eqref{eq:DPhi-is} is preserved under a gauge
transformation \eqref{eq:gauge-trans-gerbe};
\item for the $(\cG,\cB)$-inter-bi-brane 2-isomorphisms
$\,\varphi_n$,\ \v Cech--Deligne cochains
\qq\nn
\varphi_n\xrightarrow{\rm loc.}(f_{n,i})=:F_n\in\cA^{1,0}(\cO_{T_n}
)
\qqq
satisfying the cohomological identities
\qq\label{eq:Dphin-is}
\Tnup D_{(0)}F_n=-\check{\D}_{T_n}p\,,
\qqq
in which $\,\check{\D}_{T_n}:=\sum_{k=1}^{n+1}\,\vep_n^{k,k+1}\,
\check{\pi}_n^{k,k+1\,*}\,$ for the \v Cech-extended $(\cG,\cB
)$-inter-bi-brane maps $\,\check\pi_n^{k,k+1}=(\pi_n^{k,k+1},
\psi_n^{k,k+1})\,$ with index maps $\,\psi_n^{k,k+1}:\xcI_{T_n}\to
\xcI_Q\,$ covering the respective manifold maps $\,\pi_n^{k,k+1}:T_n
\to Q\,$ as per
\qq\nn
\pi_n^{k,k+1}(\cO^{T_n}_i)\subset\cO^Q_{\psi_n^{k,k+1}(i)}\,,
\qqq
and in which we use the shorthand notation
\qq\nn
\check{\pi}_n^{k,k+1\,*}(P_i,K_{ij}):=\pi_n^{k,k+1\,*}(P_{\psi_n^{k
,k+1}(i)},K_{\psi_n^{k,k+1}(i)\psi_n^{k,k+1}(j)})\,;
\qqq
the local data $\,F_n\,$ undergo a compensating gauge transformation
\qq\nn
F_n&\mapsto&F_n+\check{\D}_{T_n}w
\qqq
under a $\cG$-twisted gauge transformation
\eqref{eq:gauge-trans-bi}, ensuring that the defining identity
\eqref{eq:Dphin-is} is preserved.
\eit
\exdef \noindent The reader is urged to consult
\Rcite{Brylinski:1993ab} for a thorough introduction to the
cohomological constructs used in the above definition. Here, we
merely point out an important consequence of the cohomological
description of gerbes and 1- and 2-isomorphisms, which provides us
with a natural classification scheme of string backgrounds.
\berop
The set of 1-isomorphism classes of gerbes with a given curvature
over a manifold $\,\xcM\,$ is a torsor under a natural action of the
sheaf-cohomology group $\,H^2\bigl(\xcM,\uj\bigr)$.
\eerop
\berop\label{prop:2iso-class-1iso}
The set of 2-isomorphism classes of 1-isomorphisms between two given
gerbes over a manifold $\,\xcM\,$ is a torsor under a natural action
of the sheaf-cohomology group $\,H^1\bigl( \xcM,\uj\bigr)$.
\eerop
\berop
The set of inequivalent 2-isomorphisms between two given
1-isomorphisms of gerbes over a manifold $\,\xcM\,$ with $\,|\pi_0(
\xcM)|\,$ connected components is a torsor under a natural action of
the sheaf-cohomology group $\,H^0\bigl( \xcM,\uj\bigr)\cong\uj^{|
\pi_0(\xcM)|}$.
\eerop
\noindent All three statements are simple corollaries of the
relation between the Deligne hypercohomology and sheaf cohomology,
taken in conjunction with the contents of Definition \ref{def:loco},
cf., e.g., \cite{Brylinski:1993ab,Gawedzki:2002se,Gomi:2003}.

\begin{figure}[hbt]~\\[5pt]

$$
 \raisebox{-50pt}{\begin{picture}(50,50)
  \put(-79,-4){\scalebox{0.25}{\includegraphics{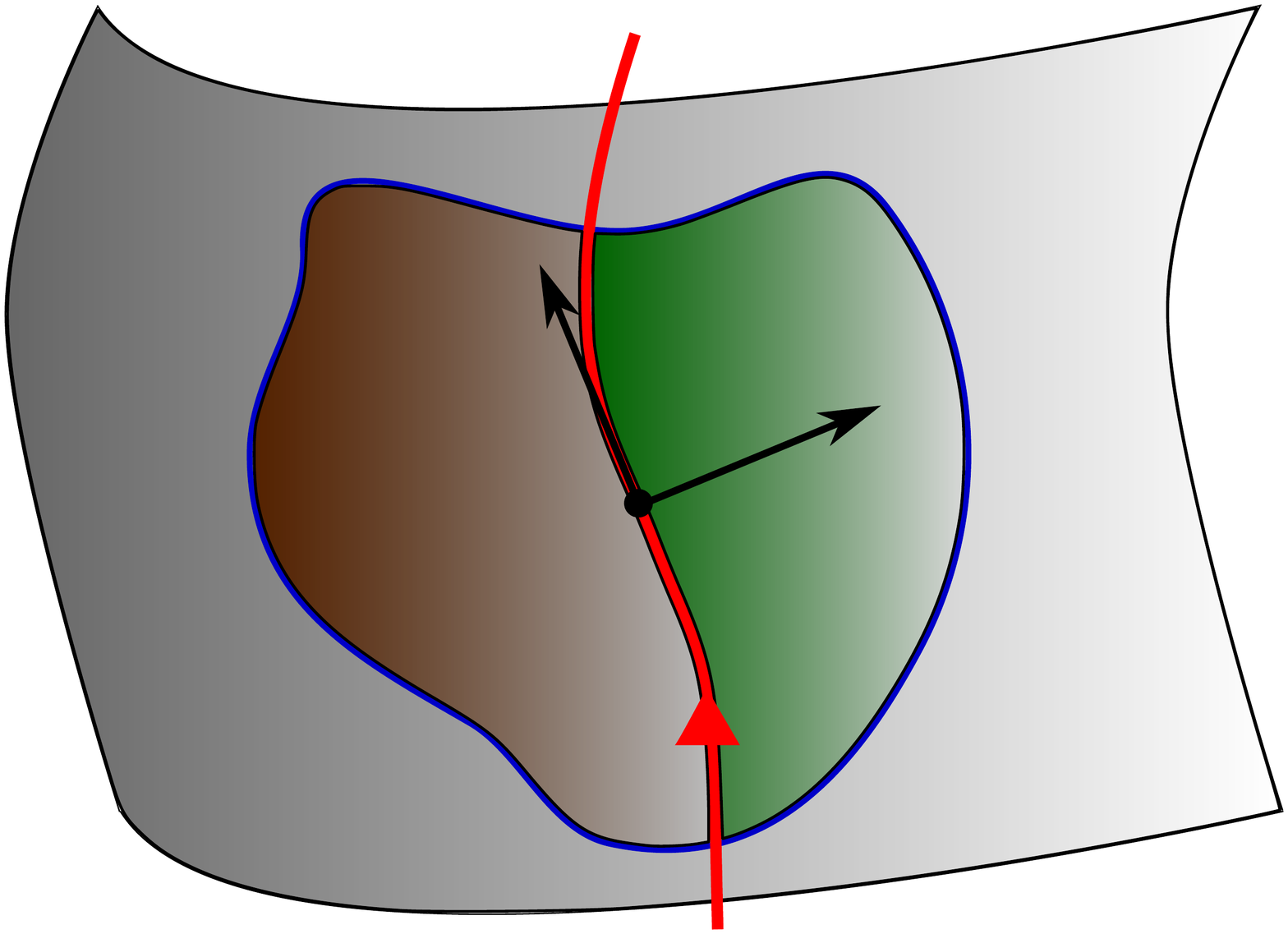}}}
  \end{picture}
  \put(0,0){
     \setlength{\unitlength}{.60pt}\put(-28,-16){
     \put(-45,145)     { $\widehat t$  }
     \put(15,123)     { $\widehat n$ }
     \put(-35,113)      { $p$ }
     \put(-145,195)   { $\Si$   }
     \put(-16,225)    { $\ell$   }
     \put(-80,170)     { $U_1$  }
     \put(20,170)     { $U_2$  }
     \put(-80,63)     { $U$  }
            }\setlength{\unitlength}{1pt}}}
$$

\caption{The right-handed basis $\,(\widehat n,\widehat t)\,$ of the
tangent space $\,\sfT_p\Si\,$ at a point $\,p\,$ on the defect line
$\,\ell\,$ embedded in the world-sheet $\,\Si\,$ and splitting the
neighbourhood $\,U\,$ of $\,p\,$ (inside the blue contour) into
subsets $\,U_1\,$ and $\,U_2$.} \label{fig:nbdry-tan}
\end{figure}

Another auxiliary concept of use in the sequel is introduced in the
following
\bedef\label{def:net-field}
Let $\,\Si\,$ be a closed oriented two-dimensional manifold with an
intrinsic metric $\,\g\,$ of a lorentzian signature\footnote{Note
that -- unlike \Rcite{Runkel:2008gr} -- we are dealing with the
lorentzian version of the world-sheet theory here as we intend to
discuss its canonical structure. It is a classic result in topology,
cf., e.g., \Rxcite{Thm.\,40.10}{Steenrod:1951}, that a global
lorentzian structure can exist on $\,\Si\,$ iff $\,\Si\,$ is
non-compact or $\,\Si\,$ is homeomorphic with a torus or a Klein
bottle. In what follows, we shall mainly be interested in $\,\Si
\cong\bR\x\bS^1\,$ (an infinite cylinder) with a view to a canonical
interpretation of defects. In this case, there are no obstructions
to the existence of a lorentzian metric. In more general situations,
and -- in particular -- in the case of the trinion (also known as
``pair-of-pants'') geometry representing the basic splitting-joining
interaction of strings, we shall disregard the signature problem,
with the implicit understanding that a proper treatment of the
world-sheet metric may require passing to the euclidean version of
the theory, accompanied by the complexification of the field space.
These manipulations are not going to invalidate our conclusions.
\label{foot:mink-vs-eukl}} $\,(-,+)$,\ termed the
\textbf{world-sheet} and split into patches $\,\wp$,\ forming the
patch set $\,\Pgt_\Si$,\ by an embedded oriented graph $\,\G$,\ to
be termed the \textbf{defect quiver}. The graph is composed of a
collection of oriented lines $\,\ell$,\ termed \textbf{defect
lines}, forming the edge set $\,\Egt_\G\,$ of $\,\G\,$ and
intersecting at a number of points $\,\jmath$,\ termed
\textbf{defect junctions} and forming the vertex set $\,\Vgt_\G\,$
of $\,\G$.\ Furthermore, let $\,\Bgt=(\cM,\cB,\cJ)\,$ be a string
background as in Definition \ref{def:bckgrnd}. A
\textbf{network-field configuration $\,(X\,\vert\,\G )\,$ in string
background $\,\Bgt\,$ on world-sheet $\,(\Si,\g)\,$ with defect
quiver $\,\G\,$} is a pair composed of the defect quiver $\,\G\,$
embedded in the world-sheet $\,\Si$,\ together with a map
$\,X:\Si\to M\sqcup Q\sqcup T\,$ such that
\bit
\item $X\,$ restricts to a once differentiable map $\,\Si\setminus\G
\to M$,\ a once differentiable map $\,\G\setminus\Vgt_\G\to Q$,\ and
it sends $\,\Vgt_\G\to T\,$ in such a manner that a defect junction
$\,\jmath\,$ of valence $\,n_\jmath\,$ is mapped to
$\,T_{n_\jmath}$;
\item for every $\,p\in\G\setminus\Vgt_\G\,$ and $\,U\subset\Si\,$ a
small neighbourhood of $\,p\,$ split into subsets $\,U_\a,\ \a\in\{1
,2\}\,$ by $\,\G\,$ so that the vector $\,\widehat n\,$ normal to
$\,\G\,$ at $\,p\,$ and pointing towards $\,U_2\,$ together with the
vector $\,\widehat t\,$ tangent to $\,\G\,$ at $\,p\,$ and
determining its orientation define a right-handed basis $\,(\widehat
n,\widehat t)\,$ of $\,\sfT_p \Si\,$ as in Figure
\ref{fig:nbdry-tan}, the map $\,X\vert_\a\,$ admits a differentiable
extension $\,X_{|\a}:\ovl U_\a\to M\,$ to the closure of $\,U_\a$,\
with the property $\,X_{|\a}(p)=\iota_\a\circ X(p)$;
\item the Defect Gluing Condition (DGC)
\qq\nn
\tx{DGC}_\cB(\psi_{|1},\psi_{|2},X)\equiv\sfp_{|1}\circ\iota_{1\,*}
-\sfp_{|2}\circ\iota_{2\,*}-X_*\widehat t\con\om(X)=0\,,\qquad
\qquad\psi_{|\a}=(X_{|\a},\sfp_{|\a})\,,\\ \label{eq:DGC}
\qqq
is satisfied at each $\,p\in\ell\in\Egt_\G\,$ for a vector
$\,\widehat t\,$ tangent to $\,\G\,$ at $\,p\,$ and determining its
orientation, and for
\qq\nn
\sfp_{|\a}=\txg(X_{|\a})(X_{|\a\,*}\widehat n,\cdot)\,,
\qqq
with a vector $\,\widehat n=\g^{-1}\bigl(\widehat t\con\Vol(\Si,\g),
\cdot\bigr)\,$ written in terms of the metric volume form $\,\Vol(
\Si,\g)\,$ on $\,\Si\,$ and defining $\,X_{|\a\,*}\widehat n\,$ in
terms of a (one-sided) derivative;
\item for $\,\jmath\in\Vgt_\G\,$ an $n_\jmath$-valent defect
junction and $\,\ell_{k,k+1}\,$ a defect line converging at
$\,\jmath$,\ the map $\,X\vert_{\ell_{k,k+1}\setminus\Vgt_\G}\,$
admits a differentiable extension $\,X_{k,k+1}:\ell_{k,k+1}\to Q\,$
such that $\,X_{k,k+1}(\jmath)=\pi_{n_\jmath}^{k,k+1}\circ X(\jmath
)$;
\item for $\,(\jmath,\ell_{k,k+1})\,$ as above, the orientation map
takes the value $\,\vep_{n_\jmath}^{k,k+1}\bigl(X(\jmath)\bigr)=+
1\,$ if $\,\ell_{k,k+1}\,$ is oriented towards $\,\jmath\,$ (an
incoming defect line), and the opposite value $\,\vep_{n_\jmath}^{k,
k+1}\bigl(X(\jmath)\bigr)=-1\,$ otherwise.
\eit
\exdef We may now give a precise description of the main object of
our study.
\bedef\label{def:sigmod}
Let $\,(X\,\vert\,\G)\,$ be a network-field configuration in string
background $\,\Bgt\,$ with field space $\,M\sqcup Q\sqcup T\,$ on
world-sheet $\,(\Si,\g)\,$ with defect quiver $\,\G$,\ and choose a
local presentation of $\,\Bgt\,$ with respect to good open covers
$\,\cO_\xcM,\ \xcM\in\{M,Q,T\}$.\ Furthermore, let $\,\triangle(\Si
)\,$ be a \textbf{triangulation of $\,\Si\,$ subordinate to
$\,\cO_\xcM,\ \xcM\in\{M,Q,T\}\,$ \wrt $\,(X\,\vert\,\G)\,$} in the
following sense:
\bit
\item $\triangle(\Si)\,$ induces a triangulation $\,\triangle(\G)
\subset\triangle(\Si)\,$ of the defect quiver in such a manner that
each defect line $\,\ell\in\Egt_\G\,$ is covered by the edges
$\,e\in\triangle(\G)\,$ and $\,\Vgt_\G\subset\triangle(\G)$;
\item for each plaquette $\,p\in\triangle(\Si)$,\ there exists a
\v Cech index $\,i_p\in\xcI_M\,$ of $\,\cO_M\,$ such that $\,X(p)
\subset\cO^M_{i_p}$,\ which we fix;
\item for each defect edge $\,e\in\triangle(\G)$,\ there exists a
\v Cech index $\,i_e\in\xcI_Q\,$ of $\,\cO_Q\,$ such that $\,X(e)
\subset\cO^Q_{i_e}$,\ which we fix;
\item for each defect vertex $\,\jmath\in\G\,$ of valence
$\,n_\jmath$,\ we fix a \v Cech index $\,i_\jmath\in
\xcI_{T_{n_\jmath}}\,$ of $\,\cO_{T_{n_\jmath}}\,$ such that $\,X(
\jmath)\in\cO^{T_{n_\jmath}}_{i_\jmath}$.
\eit
The \textbf{(two-dimensional) non-linear $\si$-model for
network-field configurations $\,(X\,\vert\,\G)\,$ in string
background $\,\Bgt\,$ on world-sheet $\,(\Si, \g)\,$ with defect
quiver $\,\G\,$} is a theory of continuously differentiable maps
$\,X:\Si \to M\sqcup Q\sqcup T$,\ determined by the principle of
least action applied to the action functional
\qq\label{eq:sigma}
S_\si[(X\,\vert\,\G);\g]=-\tfrac{1}{2}\,\int_\Si\,\txg_X(\sfd X
\overset{\wedge}{,}\star_\g\sfd X)+S_{\rm top}[(X\,\vert\,\G)]\,,
\qqq
in which
\bit
\item $\sfd X=\p_a X^\mu\,\sfd\si^a\ox\p_\mu$,\ in local coordinates
$\,\{\si^a\}^{a\in\{1,2\}}\,$ on $\,\Si\,$ and $\{X^\mu\}^{\mu\in
\ovl{1,\dim\,M}}\,$ on $\,M$,\ and the target-space metric is
assumed to act on the second factor of the tensor product;
\item $\star_\g\,$ is the Hodge operator on $\,\G(\wedge^\bullet
\sfT^*\Si)\,$ determined by the world-sheet metric $\,\g$;
\item the topological term
\qq\nn
S_{\rm top}[(X\,\vert\,\G)]=-\sfi\,\log\Hol_\Bgt(X\,\vert\,\G)
\qqq
is given by the generalised surface holonomy
$\,\Hol_\Bgt(X\,\vert\,\G )\,$ for the network-field configuration
$\,(X\,\vert\,\G)$,\ which, in a triangulation of $\,\Si\,$
subordinate to the good open covers $\,\cO_\xcM,\
\xcM\in\{M,Q,T\}\,$ \wrt $\,(X\,\vert\,\G)\,$ (consisting of
plaquettes $\,p$,\ edges $\,e\,$ and vertices $\,v$) and a local
presentation of $\,\Bgt\,$ as described in Definition
\ref{def:loco}, takes the form
\qq\nn
-\sfi\,\log\Hol_\Bgt(X\,\vert\,\G)&=&\sum_{p\in\triangle(\Si)}
\left[\int_p\,X_p^*B_{i_p}+\sum_{e\subset p}\left(\int_e\,X_e^*
A_{i_p i_e}-\sfi\,\sum_{v\in e}\,\log X^*g_{i_p i_e i_v}^{\vep_{pe
v}}(v)\right)\right]\cr\cr
&&+\sum_{e\in\triangle(\G\setminus\Vgt_\G)}\,\left(\int_e\,X_e^*
P_{i_e}-\sfi\,\sum_{v\in e}\,\log X^*K_{i_e i_v}^{-\vep_{ev}}(v)
\right)\cr\cr
&&-\sfi\,\sum_{\jmath\in\Vgt_\G}\,\log X^*f_{n_\jmath,i_\jmath}(
\jmath)\,.
\qqq
\eit
\exdef An extensive discussion of the various components of the
string background $\,\Bgt\,$ was presented, alongside a derivation
of the local formula for $\,S_{\rm top}[(X\,\vert\,\G)]$,\ in
\Rcite{Runkel:2008gr}, to which we refer the reader for details.
Here, we merely point out the statement of consistency:
\berop\cite[Sec.\,2.7]{Runkel:2008gr}
The topological term $\,S_{\rm top}[(X\,\vert\,\G)]\,$ of the action
functional of the non-linear $\si$-model for network-field
configurations $\,(X\,\vert\,\G)\,$ in string background $\,\Bgt\,$
on world-sheet $\,(\Si,\g)\,$ with defect quiver $\,\G\,$ is
independent of the choice $\,\triangle(\Si)\,$ of the triangulation
and invariant under gauge transformations of the local presentation
of the string background $\,\Bgt$,\ as described in Definition
\ref{def:loco}.
\eerop
\noindent And the statement of symmetry:
\bethe\cite[Sec.\,2.9]{Runkel:2008gr}\label{thm:conf-def}
The non-linear $\si$-model for network-field configurations $\,(X\,
\vert\,\G)\,$ in string background $\,\Bgt\,$ on world-sheet $\,(\Si
,\g)\,$ with defect quiver $\,\G\,$ of Definition \ref{def:sigmod}
is invariant with respect to arbitrary (gauge) transformations
\qq\nn
&X\mapsto X\circ D\,,\qquad\qquad\g\mapsto D^*\g\,,\qquad D\in
\Diff^+_\G(\Si)\,,&\cr\cr
&\g\mapsto\ee^{2w}\cdot\g\,,\qquad\ee^{2w}\in{\rm Weyl}(\g)&
\qqq
from the semidirect product $\,\Diff^+_\G(\Si)\lx{\rm Weyl}(\g)\,$
of the group $\,\Diff^+_\G(\Si)\,$ of those (orientation-preserving)
diffeomorphisms of $\,\Si\,$ that preserve $\,\G$,\ with the group
$\,{\rm Weyl}(\g)\,$ of Weyl rescalings of the metric $\,\g$. \ethe
\brem\label{rem:Weyl-anom} It deserves to be emphasised that a
generic string background $\,\Bgt=(\cM,\cB,\cJ)\,$ does not lead to
a consistent quantum field theory with a non-anomalous realisation
of the gauge symmetries of the classical action functional. Such a
realisation prerequires that the Weyl anomaly of the theory vanish,
which -- in turn -- imposes constraints on the various components of
a consistent string background, cf., e.g.,
Refs.\,\cite{Friedan:1985phd,Braaten:1985is,Gawedzki:1996ias}. \erem

Given the conformal character of the network-field configurations
considered, as stated in Theorem \ref{thm:conf-def}, it is useful to
put the world-sheet metric $\,\g\,$ (locally) in the minkowskian
gauge $\,\g\equiv\eta=\diag(-1,1)$,\ which we impose now for the
remainder of the paper. In this gauge, the theory of maps referred
to in Definition \ref{def:sigmod} is determined by the set of
second-order non-linear differential equations\footnote{The field
equations use the Christoffel symbols
\qq\nn
\bigl\{\begin{smallmatrix} \nu \\ \rho\si\end{smallmatrix}\bigr\}=
\tfrac{1}{2}\,\bigl(\txg^{-1}\bigr)^{\nu\la}\,\bigl(\p_\rho
\txg_{\si\la}+\p_\si\txg_{\rho\la}-\p_\la\txg_{\rho\si}\bigr)
\qqq
of the target-space metric.}
\qq\label{eq:field-eqs}\qquad\qquad
\txg_{\mu\nu}(X)\,\eta^{ab}\,\bigl(\p_a\,\p_b X^\nu+\bigl\{
\begin{smallmatrix} \nu \\ \rho\si \end{smallmatrix}\bigr\}(X)\,\p_a
X^\rho\,\p_b X^\si\bigr)+3\txH_{\mu\rho\si}(X)\,\ep^{ab}\,\p_a
X^\rho\,\p_b X^\si=0\,.
\qqq
In what follows, we shall mainly consider a space-like defect line
$\,\ell\,$ which is the locus of the equation $\,t=0\,$ in the
adapted coordinates $\,(t,\varphi)=(\si_1,\si_2)\,$
\label{page:adapt} in the vicinity of a point $\,p\in\ell$.\ In
these coordinates, $\,( \widehat n,\widehat t)=(\p_t,\p_\varphi)\,$
and our definition of $\,\sfp\,$ coincides with the standard
definition of the kinetic momentum of the $\si$-model field $\,X$,\
whence also the notation. At such a defect line, we have the (local)
differentiable extensions $\,X_{|\a}\,$ of the patch map
$\,X\vert_{\wp_\a}\,$ to the defect edge
\qq\nn
X_{|\a}\vert_{\wp_\a}=X\vert_{\wp_\a}\,,\qquad\qquad X_{|\a}(0,
\varphi)=\iota_\a\circ X(\varphi)\,,
\qqq
with one-sided normal derivatives
\qq\nn
X_{|\a\,*}\widehat n(\varphi)=\lim_{\ep\to 0^+}\frac{X_{|\a}^\mu
\bigl((-1)^\a\,\ep,\varphi\bigr)-X_{|\a}^\mu(0,\varphi)}{(-1)^\a\,
\ep}\,\p_\mu\,.
\qqq

Prior to finishing this introductory section by presenting a couple
of examples of bi-branes, let us add the following preparatory
\brem\label{rem:states} The world-sheet $\,\Si\,$ with an embedded
defect quiver $\,\G\,$ can be understood as a model of multiple
phases, in coexistence and undergoing transitions, of the underlying
CFT, in which the particular phases are represented by the patches
$\,\wp\in\Pgt_\Si$.\ From this point of view, it is natural to set
up the canonical description of the theory in each patch
independently, and only upon completing the task, examine the
relations between the ensuing phase-restricted state spaces imposed
by the defects that separate the phases. In this picture, the defect
lines $\,\ell\in\Egt_\G\,$ appear as space-like domain walls of the
two-dimensional field theory carrying the geometric data that effect
the transition. This is the basic setting in which we shall carry
out our analysis in the next section, phrasing our considerations in
terms of Cauchy data of the dynamical evolution, to be localised on
a space-like Cauchy contour $\,\xcC$.\ In principle, we might attach
the phase (patch) label to the dynamical objects defined over a
given patch but we choose, instead, to shift the dependence on the
phase of the underlying CFT to the definition of the embedding map
$\,X$,\ along the lines of \Rcite{Runkel:2008gr}, which enables us
to develop a unified treatment of all admissible phases at the same
time. The modular invariance of the quantised (euclidean version of
the) CFT leads us to consider a dual of the picture described in
which time-like and space-like contours are swapped. Having set out
with space-like defect lines, we thus end up with time-like ones,
and the obvious question arises as to the nature of the canonical
description of this dual CFT. Motivated by the distinguished example
of boundary defects and the associated $\cG$-branes of string
theory, we should be inclined to formulate our description in terms
of Cauchy data localised on open segments stretched transversally
between defect lines, each contained in a single patch of the
world-sheet. The problem with this description, masked by the
triviality of the patch data for the patch `behind' the defect line
in the boundary case, is that consistency of its formulation in a
single patch prerequires the knowledge of the field configuration
across the defect lines to which the open segment is attached, cf.\
\Reqref{eq:DGC}. Geometrically, this is reflected in the inability
to write down gauge invariant functionals using only the space-time
data assigned to a single patch, augmented by those for the defect
lines bounding it. The unavoidable incompleteness of such a
description prompts us to conceive a formulation in which open
segments stretching between pairs of defect lines are -- instead --
joined, consistently with \Reqref{eq:DGC}, to form a closed contour
$\,C_{\{\ell_k\}}\,$ intersecting transversally a (finite) number
$\,I\in\bN_{>0}\,$ of defect lines $\,\ell_k,\ k\in\ovl{1,I}$,\ with
the dynamical data localised on $\,C_{\{\ell_k\}}$.

Clearly, Cauchy contours which do not intersect the defect quiver
(and the attendant dynamical structures) can be regarded as special
examples of the $\G$-twisted ones, with all defects intersected by
the corresponding Cauchy contour $\,C_\emptyset\equiv C\,$ assumed
trivial. Thus, unless expressly stated otherwise, we shall always
mean both non-trivially $\G$-twisted and untwisted states whenever
we use the notation for the $\G$-twisted ones in the sequel. \erem

\beg\label{eg:triv-def}\textbf{The trivial (inter-)bi-brane.}
\\[-8pt]

Given an arbitrary target $\,\cM=(M,\txg,\cG)$,\ there always exists
the \textbf{trivial $\cG$-bi-brane}
\qq\nn
\cB_{\rm triv}=(M,\id_M,\id_M,0,\id_\cG)\,,
\qqq
and the attendant \textbf{trivial $(\cG,\cB_{\rm triv}
)$-inter-bi-brane}
\qq\nn
\cJ_{\rm triv}=\left(\bigsqcup_{i=1}^{2^n}\,M,\bigl(\vep_n^{k,k+1},
\id_M \ \vert\ k\in\ovl{1,n}\bigr),\id_{\id_\cG}\,\vert\,n\in
\bN_{\geq 3}\right)\,.
\qqq
\eeg\medskip

\beg\label{eg:WZW-def}\textbf{The maximally symmetric WZW defects.}
\\[-8pt]

\noindent \emph{\textbf{The target.}} We consider here a
distinguished class of defects in the Wess--Zumino--Witten (WZW)
$\si$-model of \Rcite{Witten:1983ar}, with target $\,\cM_\sfk=(\txG,
\txg_\sfk,\cGk)$,\ where
\bit
\item[(T.i)] $\txG\,$ is the group manifold of an arbitrary compact
simple 1-connected Lie group, with Lie algebra $\,\ggt\,$ and a
trace $\,\tr_\ggt\,$ on $\,\ggt\,$ normalised such that the equality
\qq\nn
\tr_\ggt(t_A\,t_B)=-\tfrac{1}{2}\,\d_{AB}
\qqq
holds for generators $\,t_A\,$ of $\,\ggt$,\ the latter satisfying
the defining commutation relations
\qq\nn
[t_A,t_B]=f_{ABC}\,t_C\,,
\qqq
with $\,f_{ABC}\,$ the structure constants of $\,\ggt\,$ -- this
prescription yields the standard matrix trace for, e.g., $\,\txG=
\sug$;
\item[(T.ii)] $\txg_\sfk\,$ is the Cartan--Killing metric
\qq\nn
\txg_\sfk=-\tfrac{\sfk}{4\pi}\,\tr_\ggt\bigl(\tht_L\ox\tht_L\bigr)
\,,\qquad\sfk\in\bZ_{>0}\,,
\qqq
written in terms of the standard left-invariant Maurer--Cartan
1-form $\,\tht_L(g)=g^{-1}\,\exd g\in\G(\sfT^*\txG)\ox\ggt\,$ on
$\,\txG$;
\item[(T.iii)] $\cGk=\cG_1^{\ox\sfk}\,$ is the $\sfk$-th power of the
basic gerbe\footnote{Meinrenken's construction of the basic gerbe
for a general (compact simple 1-connected) Lie group was preceded by
that of \Rcite{Gawedzki:1987ak} for $\,\sug\,$ and that of
Refs.\,\cite{Chatterjee:1998,Hitchin:1999} which works for $\,{\rm
SU}(N)$.\ The non-simply connected case was worked out in Refs.\
\cite{Gawedzki:2002se,Gawedzki:2003pm}.} $\,\cG_1\,$ of
\Rcite{Meinrenken:2002}, with curvature equal to the Cartan 3-form
\qq\nn
\txH_\sfk=\tfrac{\sfk}{12\pi}\,\tr_\ggt(\tht_L\wedge\tht_L\wedge
\tht_L)
\qqq
whose cohomology class is the generator of $\,H^3(\txG)\cong\bZ$.
\eit
The action functional for a defect-free world-sheet $\,\Si\,$ is
given by
\qq\nn
S_{{\rm WZW},\sfk}[g]=\tfrac{\sfk}{8\pi}\,\int_\Si\,\tr_\ggt\bigl(
\tht_L(g)\wedge\star_{\rm H}\tht_L(g)\bigr)-\sfi\,\log\Hol_{\cGk}(g)
\,,
\qqq
and the constant $\,\sfk\in\bZ_{>0}\,$ is called the level of the
WZW model. The field equations of the model have the compact form
\qq\nn
\bigl(\eta^{ab}+\ep^{ab}\bigr)\,\p_a\bigl(g^{-1}\,\p_b g\bigr)=0\,,
\qqq
which can further be rewritten, using the light-cone coordinates
$\,\si^\pm=\si^2\pm\si^1\,$ and the attendant derivatives
$\,\p_\pm=\tfrac{\p\ }{\p\si^\pm}$,\ as
\qq\nn
\p_+\bigl(g^{-1}\,\p_-g\bigr)=0\,.
\qqq
A general solution to this equation factorises as
\qq\nn
g(\si)=g_L(\si^+)\cdot g_R(\si^-)^{-1}\,,
\qqq
for independent $\txG$-valued maps $\,g_L\,$ and $\,g_R\,$ on
$\,\bR\,$ with equal monodromies (when viewed as maps on $\,\bR/2\pi
\bZ$), cf.\ \Rcite{Gawedzki:2001rm}. In addition to the standard
conformal symmetry of Theorem \ref{thm:conf-def}, realised, on the
infinitesimal level, by two (chiral) copies of the Lie algebra of
diffeomorphisms of the circle, the bulk theory enjoys a level-$\sfk$
Ka\v c--Moody symmetry, realised on fields by the chiral currents
\qq\nn
J_L(\si)=\tfrac{\sfk}{2\pi}\,g(\si)\,\p_+g(\si)^{-1}\,,\qquad
\qquad J_R(\si)=\tfrac{\sfk}{2\pi}\,g(\si)^{-1}\,\p_-g(\si)
\qqq
that generate the centrally extended current algebra $\,\ggtk^L
\oplus\ggtk^R$.\ Note that the currents become functions of the
respective light-cone coordinates $\,\si^\pm\,$ upon using the field
equations of the $\si$-model. Their (infinitesimal) action
integrates to
\qq\nn
g(\si)\mapsto h_L(\si^+)\cdot g(\si)\cdot h_R(\si^-)^{-1}\,,
\qqq
with independent chiral transformation maps $\,h_L,h_R\in\sfL\txG\,$
from the loop group $\,\sfL\txG\equiv C^\infty(\bS^1,\txG)$.\medskip

\noindent \emph{\textbf{The boundary $\cGk$-bi-brane.}} We shall
first consider defects describing \emph{boundary} maximally
symmetric $\cGk$-bi-branes of the WZW model, or maximally symmetric
$\cGk$-branes for short. To this end, we focus on (the vicinity of)
a connected component $\,\ell\,$ of the edge set $\,\Egt_\G\,$ of
$\,\G$,\ which we take to be an oriented circle embedded in
$\,\Si\,$ at $\,t=0\,$ in a local coordinate system
$\,(t,\varphi)\,$ described on p.\,\pageref{page:adapt}. As argued
in \Rxcite{p.\ 12}{Runkel:2008gr}, the corresponding string
background $\,\Bgt_\sfk^\p=(\cM_\sfk^\p,\cB^\sfk_\p,\cdot)\,$
consists of
\bit
\item[(TD)] the target $\,\cM_\sfk^\p=\cM_\sfk\sqcup\{\bullet\}\,$
given by the disjoint union of the bulk target $\,\cM_\sfk=(\txG,
\txg_\sfk,\cGk)\,$ and a single point $\,\{\bullet\}\,$ with no
structure over it;
\item[(D)] the $\cGk$-bi-brane $\,\cB_\sfk^\p=(Q_\sfk^\p,
\iota_{Q_\sfk^\p},\bullet,\om_\sfk^\p,\Phi_\sfk^\p)$,\ with
\bit
\item[(D.i)] the world-volume
\qq\nn
Q^\p_\sfk=\bigsqcup_{\la\in\faff{\ggt}}\,\xcC_\la\,,\qquad\qquad
\xcC_\la=\bigl\{\ \Ad_x\ee_\la \quad\vert\quad x\in\txG\ \bigr\}\,,
\qqq
given by the disjoint sum of the conjugacy classes of Cartan
elements $\,\ee_\la=\ee^{\frac{2\pi\sfi\,\la}{\sfk}}\in\txG\,$
labelled by weights $\,\la\,$ from the fundamental affine Weyl
alcove\footnote{In what follows, we shall always identify $\,\ggt\,$
with its dual $\,\ggt^*\,$ using the Cartan--Killing metric.} at
level $\,\sfk$,
\qq\label{eq:faffggt}
P^\sfk_+(\ggt)=\sfk\,\xcA_W(\ggt)\cap P(\ggt)\,,
\qqq
the latter being the intersection of the weight lattice $\,P(\ggt
)\,$ of $\,\ggt\,$ with its $\sfk$-inflated Weyl alcove $\,\sfk\,
\xcA_{\rm W}(\ggt)$,\ i.e.\ a subset
\qq\nn
\xcA_{\rm W}(\ggt)=\bigl\{\ \la\in\tgt \quad\big\vert\quad \tr_\ggt(
\la\cdot\tht)\leq 1\quad\land\quad\tr_\ggt(\la\cdot\a_i)\geq 0\,,\ i
\in\ovl{1,{\rm rank}\,\ggt} \ \bigr\}
\qqq
of the Cartan subalgebra $\,\tgt\subset\ggt$,\ defined in terms of
the simple roots $\,\a_i,\ i\in\ovl{1,{\rm rank} \,\ggt}\,$ of
$\,\ggt\,$ and its longest root $\,\theta$;
\item[(D.ii)] the $\cGk$-bi-brane maps, defined component-wise by
the embedding $\,\iota_{Q^\p_\sfk}\vert_{\xcC_\la}\equiv\imath_\la:
\xcC_\la\emb\txG\,$ of the conjugacy class $\,\xcC_\la\,$ in the
group manifold, and the constant map $\,\bullet:Q_\sfk^\p\to\{
\bullet\}$;
\item[(D.iii)] the curvature, also defined component-wise as
\qq\label{eq:WZW-brane-curv}
\om_\sfk^\p\vert_{\xcC_\la}=\tfrac{\sfk}{8\pi}\,\imath_\la^*
\tr_\ggt\left(\tht_L\wedge\tfrac{\id_\ggt+\Ad_\cdot}{\id_\ggt-
\Ad_\cdot}\,\tht_L\right)=:\om^\p_{\sfk,\la}\,;
\qqq
\item[(D.iv)] the $\cGk$-bi-brane 1-isomorphism given on each
component $\,\xcC_\la\,$ of $\,Q^\p_\sfk\,$ by the corresponding
trivialisation
\qq\nn
\Phi^\p_\sfk\vert_{\xcC_\la}=:\Phi^\p_{\sfk,\la}\ :\ \imath_\la^*
\cGk\xrightarrow{\cong}I_{\om^\p_{\sfk,\la}}
\qqq
of the restricted gerbe $\,\cGk$.
\eit
\eit
The DGC obtained by plugging the above data into \Reqref{eq:DGC} is
the familiar statement
\qq\nn
(J_L-J_R)\vert_\ell=0
\qqq
of maximal symmetry of the boundary defect, the symmetry being
determined by a single copy of the level-$\sfk$ Ka\v c--Moody
algebra $\,\ggtk\,$ embedded diagonally in the current-symmetry
algebra $\,\ggtk^L\oplus\ggtk^R\,$ of the bulk theory.

It is vital to note that -- as was argued for $\,\txG={\rm SU}(N)\,$
in \Rxcite{Sec.\,8.1}{Gawedzki:2002se} and for an arbitrary compact
simple 1-connected Lie group $\,\txG\,$ in
\Rxcite{Sec.\,5.1}{Gawedzki:2004tu} -- the stable isomorphisms
$\,\Phi^\p_{\sfk,\la}\,$ exist for $\,\la\in\faff{\ggt}\,$
exclusively, and so they single out a subset of conjugacy classes in
$\,\txG\,$ which coincides with the set of world-volumes of stable
(untwisted) maximally symmetric D-branes of the WZW model at level
$\,\sfk$,\ cf., e.g., \Rcite{Felder:1999ka}, which -- in turn -- are
in a one-to-one correspondence with the (untwisted) maximally
symmetric boundary states of the associated BCFT (ib.). \medskip

\noindent \emph{\textbf{The non-boundary $\cGk$-bi-brane.}} The next
type of maximally symmetric WZW defects that we want to discuss are
those implementing jumps by elements of the target Lie group in the
sense that the limiting values attained at a point $\,p\,$ on the
defect circle $\,\ell\,$ by the one-sided local extensions
$\,(g_{|1},g_{|2})\,$ of the embedding map $\,g:\Si\setminus\G\to
\txG\,$ to the defect line $\,\ell$,\ described in Definition
\ref{def:net-field}, are two generically distinct points in the
group manifold. A special class of such defects -- the central-jump
defects at which $\,g_{|2}=z\cdot g_{|1}\,$ for $\,z\,$ from the
centre $\,Z(\txG)\,$ of $\,\txG\,$ -- were considered at length in
\Rcite{Runkel:2008gr}. The more general jump defects, with -- as
above -- the jump given by $\,g_{|1}^{-1}\cdot g_{|2}\in\txG$,\ were
first studied in \Rcite{Fuchs:2007fw}, where the notion of a
bi-brane was introduced. They shall be expanded upon in
\Rcite{Runkel:2010}. In the conventions of the latter paper, the
string background $\,\Bgt_\sfk= (\cM_\sfk,\cB_\sfk,\cdot)\,$ for
these jump defects consists of
\bit
\item[(TB)] the target $\,\cM_\sfk=(\txG,\txg_\sfk,\cGk)\,$ of the
defect-free model;
\item[(B)] the $\cGk$-bi-brane $\,\cB_\sfk=(Q_\sfk,d_1,d_0,\om_\sfk
,\Phi_\sfk)$,\ with
\bit
\item[(B.i)] the world-volume
\qq\nn
Q_\sfk=\txG\x Q^\p_\sfk\,;
\qqq
\item[(B.ii)] the $\cGk$-bi-brane maps, defined explicitly as
\qq\nn
d_0(g,h_\la)=g\cdot h_\la\,,\qquad\qquad d_1(g,h_\la)=g\,;
\qqq
\item[(B.iii)] the curvature, defined component-wise as
\qq\nn
\om_\sfk\vert_{\txG\x\xcC_\la}=-\pr_2^*\om^\p_{\sfk,\la}+\rho_\sfk=:
\om_{\sfk,\la}\,,\qquad\qquad\rho_\sfk=\tfrac{\sfk}{4\pi}\,\tr_\ggt
\bigl(\pr_1^*\tht_L\wedge\pr_2^*\tht_R\bigr)
\qqq
in terms of the canonical projections $\,\pr_\a:\txG\x\txG\to\txG,\
\a\in\{1,2\}$;
\item[(B.iv)] the $\cGk$-bi-brane 1-isomorphism with a
component-wise definition
\qq\nn
\Phi_\sfk\vert_{\txG\x\xcC_\la}=:\Phi_{\sfk,\la}\ &:&\ d_1^*\cGk
\equiv d_1^*\cGk\ox I_{\pr_2^*\om^\p_{\sfk,\la}}\ox I_{-\pr_2^*
\om^\p_{\sfk,\la}}\xrightarrow{\quad\id_{d_1^*\cGk}\ox\Phi^{\p\,
\vee}_{\sfk,\la}\ox\id_{ I_{-\pr_2^*\om^\p_{\sfk,\la}}}\quad}
\pr_1^*\cGk\ox\pr_2^*\cGk\ox I_{-\pr_2^*\om^\p_{\sfk,\la}}\cr\cr
&&\hspace{5.2cm}\xrightarrow{\quad\cM_\sfk\ox\id_{ I_{-\pr_2^*
\om^\p_{\sfk,\la}}}\quad}d_0^*\cGk\ox I_{\om_{\sfk,\la}}\,,
\qqq
invoking the 1-isomorphism
\qq\nn
\cM_\sfk\ :\ \pr_1^*\cGk\ox\pr_2^*\cGk\xrightarrow{\cong}\txm^*
\cGk\ox I_{\rho_\sfk}\,,\qquad\qquad\txm\ :\ \txG\x\txG\to\txG\ :\
(g,h)\mapsto g\cdot h
\qqq
of the multiplicative structure on $\,\cGk$,\ as introduced in
\Rcite{Carey:2004xt} and developed in
Refs.\,\cite{Waldorf:2008mult,Gawedzki:2009jj}.
\eit
\eit
The DGC is, once more, the statement of maximal symmetry,
\qq\nn
(J^1_L-J^2_L)\vert_\ell=0=(J^1_R-J^2_R)\vert_\ell\,,
\qqq
the symmetry being determined by the full bi-chiral level-$\sfk$
Ka\v c--Moody algebra $\,\ggtk^L\oplus\ggtk^R\,$ of the bulk theory.
The chiral currents $\,J^\a_L,J^\a_R\,$ are defined as previously
but using the respective one-sided local extensions $\,g_{|\a}\,$ of
the patch component of the $\si$-model field. Owing to the form of
the energy-momentum tensor, given by a sum of terms quadratic in the
chiral currents, the continuity of the latter across the defect line
ensures the topologicality of the defect associated to
$\,\cB_\sfk$,\ as defined in \Rcite{Runkel:2008gr}.

\brem\label{rem:mult-str} It ought to be emphasised that the
existence and uniqueness (up to a 2-isomorphism) of the
1-isomorphism $\,\cM_\sfk\,$ on a compact simple 1-connected Lie
group $\,\txG\,$ implies, via a simple topological argument (cf.\
\Rcite{Runkel:2010}), that the maximally symmetric WZW
$\cGk$-bi-brane $\,\cB_\sfk\,$ has precisely as many connected
components (labelled by weights $\,\la\in\faff{\ggt}$) as its
boundary analogon $\,\cB_\sfk^\p$. \erem \brem The world-volume
$\,Q_\sfk\,$ of $\,\cB_\sfk\,$ is $\txG\x\txG$-equivariantly
isomorphic to the disjoint union of the bi-conjugacy classes
\qq
\xcB_{(t_\la,e)}=\bigl\{\ \bigl(x\cdot t_\la\cdot y^{-1},x\cdot y^{-
1}\bigr) \quad\big\vert\quad x,y\in\txG \ \}\,,
\label{eq:biconj-def}
\qqq
of \Rcite{Fuchs:2007fw} for the pairs of group elements $\,(t_\la,e)
\in\xcC_\la\x\{e\}\subset\txG\x\txG$. \erem
\eeg

\section{The canonical structure and pre-quantisation of the
$\si$-model} \label{sec:can}

Having written out the $\si$-model action functional of interest in
\Reqref{eq:sigma}, we may now analyse the symplectic structure on
its state space, a task best completed within the framework of
covariant classical field theory (or first-order formalism) of
Refs.\,\cite{Gawedzki:1972ms,Kijowski:1973gi,Kijowski:1974mp,Kijowski:1976ze,Szczyrba:1976,Kijowski:1979dj},
cf.\ also \Rcite{Gotay:2006cft} for an exhaustive exposition of the
modern approach and a comprehensive list of references. This
formalism enables us to interpret the (inter-)bi-brane data in terms
of the canonical structure of the underlying two-dimensional field
theory, whereupon a clear-cut field-theoretic statement can be made
in regard to the relation between defects and dualities of the
$\si$-model, in the spirit of, e.g., \Rcite{Frohlich:2006ch}. We
begin by briefly reviewing those elements of the general formalism
that are instrumental in the subsequent analysis of the physical
system of interest.
\subsection{Elements of the covariant formalism}\label{sub:can-gen}
Let us first introduce some basic notions.
\bedef\label{def:Cartan}
Let $\,\pi_\cF:\cF\to\xcM\,$ be a fibre bundle over a
(pseudo-)riemannian base $\,(\xcM,\txg)\,$ of dimension
$\,\dim\,\xcM=:d$,\ and let $\,\sfJ^1\cF\to\xcM\,$ be the first-jet
bundle of $\,\cF$,\ with local coordinates
$\,(x^\mu,\phi^A,\xi^B_\nu),\ \mu,\nu\in\ovl{1,d} ,\
A,B\in\ovl{1,N}$,\ where $\,N\,$ is the dimension of the typical
fibre of $\,\cF$.\ Consider an $\cF$-field theory $\,\xcF\,$ on
$\,\xcM$,\ i.e.\ a theory of continuously differentiable sections
$\,(\phi^A)^{A\in\ovl{1,N}}\,$ of the bundle $\,\cF\,$ (termed the
\textbf{covariant configuration bundle of $\cF$-field theory
$\,\xcF\,$} in this context), determined by the principle of least
action applied to the action functional
\qq\label{eq:F-theory}
S_\xcF[\phi^A]=\int_\xcM\,\ceL_\xcF(x^\mu,\phi^A,\xi^B_\nu)
\vert_{\xi^B_\nu=\p_\nu\phi^B}\,\sfd^d x\,,
\qqq
in which $\,\sfd^d x=\sfd x^1\wedge\sfd x^2\wedge\ldots\wedge\sfd
x^d\in\G(\wedge^d\sfT^*\xcM)\,$ is the volume form in local
coordinates $\,x^\mu\,$ on $\,\xcM$,\ $\,\p_\mu=\frac{\p\ }{\p
x^\mu}\,$ are the associated partial derivatives, and the map
$\,\ceL_\xcF\,$ on $\,\sfJ^1\cF\,$ with values in the space of
scalar densities (of weight $0$) on $\,\xcM\,$ is termed the
\textbf{lagrangean (density) of $\cF$-field theory $\,\xcF\,$} and
considered \textbf{regular} iff the matrix functional
$\,\frac{\d^2\ceL_\xcF}{\d\xi^A_\mu\d \xi^B_\nu}\,$ (with a
multi-index $\,{}^\mu_A$) is invertible on sections of $\,\cF\,$
that extremise $\,S_\xcF$.\ The \textbf{Cartan form $\,\Th_\xcF\,$
of $\cF$-field theory $\,\xcF\,$} is the $d$-form on $\,\sfJ^1\cF\,$
given by the formula
\qq\nn
\Th_\xcF(x^\mu,\phi^A,\xi^B_\nu)=\left(\ceL-\xi^C_\la\,\tfrac{\d
\ceL}{\d\xi^C_\la}\right)(x^\mu,\phi^A,\xi^B_\nu)\,\sfd^d x+
\tfrac{\d\ceL}{\d\xi^C_\la}(x^\mu,\phi^A,\xi^B_\nu)\,\d\phi^C
\wedge(\p_\la\con\sfd^d x)\,.
\qqq
Above, and in what follows, we use the symbol $\,\d\,$ to
distinguish differentiation in the direction of the fibre of
$\,\sfJ^1\cF$,\ termed \textbf{$\cF$-vertical},\ from that along the
base $\,\xcM$,\ e.g., $\,\d\phi^A\,$ and $\,\frac{\d\ \ }{\d
\xi^A_\mu}\,$ vs.\ $\,\sfd x^\mu\,$ and $\,\frac{\p\ \ }{\p x^\mu}$.
\exdef \noindent The significance of the Cartan form rests on the
following
\berop\cite{Gawedzki:1990jc}\label{prop:Cartan}
Let $\,\xcF\,$ be an $\cF$-field theory on a (pseudo-)riemannian
manifold $\,(\xcM,\txg)$,\ determined by a regular lagrangean
density $\,\ceL_\xcF$,\ with a covariant configuration bundle
$\,\pi_\cF:\cF \to\xcM$,\ the attendant first-jet bundle
$\,\sfJ^1\cF\to\xcM$,\ and the Cartan form $\,\Th_\xcF\,$ on the
latter. Then,
\bit
\item[i)] the principle of least action applied to the functional
\qq\nn
S_{\Th_\xcF}[\Psi]=\int_\xcM\,\Psi^*\Th_\xcF\,,\qquad\qquad\Psi\in
\G(\sfJ^1\cF)
\qqq
yields, as the Euler--Lagrange equations, the field equations of
$\,\xcF$,\ that is the Euler--Lagrange equations of the action
functional \eqref{eq:F-theory}, and -- for $\,\xcM\,$ with a
non-empty boundary -- also the boundary conditions of $\,\xcF$;\ the
equations follow from the condition
\qq\label{eq:EuLagAl}
0=\xcV\con\d S_{\Th_\xcF}[\Psi_{\rm cl}]\,,
\qqq
to be satisfied by the extremal (or \emph{classical}) sections
$\,\Psi_{\rm cl}\,$ of $\,\sfJ^1\cF\,$ for an arbitrary
$\cF$-vertical vector field $\,\xcV$,\ i.e.\ for $\,\xcV\in\G(\sfT
\sfJ^1\cF)\cap\ker\,\pi_{\cF\,*}=:\G\bigl((\sfT J^1\cF)^{\perp_\cF}
\bigr)\,$ (in particular, the boundary conditions of $\,\xcF\,$ are
implied by the vanishing of the boundary term on the right-hand side
of \Reqref{eq:EuLagAl});
\item[ii)] $\Th_\xcF\,$ canonically determines a closed 2-form
$\,\Om_\xcF\,$ on the space $\,\sfP_\xcF\,$ of extremal sections of
$\,\sfJ^1\cF$.
\eit
\eerop
\noindent A complete proof can be extracted from the original paper.
However, the proof being constructive in nature, it appears useful
to give at least an idea thereof -- after \Rcite{Gawedzki:2007fo} --
to prepare the reader for the subsequent considerations.
\bit
\item[Ad i)] The key point is to note that the Euler--Lagrange
equations obtained for the distinguished choice $\,\xcV=\tfrac{\d\
}{\d\xi^A_\mu}\,$ of a $\cF$-vertical vector field on
$\,\sfJ^1\cF\,$ read
\qq\label{eq:eom-xipar}
\tfrac{\d^2\ceL_\xcF}{\d\xi^C_\kappa\d\xi^D_\la}(x^\mu,\phi^A,
\xi^B_\nu)\,(\xi^D_\la-\p_\la\phi^D)=0\,,
\qqq
and so, assuming regularity of $\,\ceL_\xcF$,\ we conclude that the
equality
\qq\label{eq:xi-as-pX}
\xi^A_\mu=\p_\mu\phi^A
\qqq
holds true on $\,\sfP_\xcF$.\ The Euler--Lagrange equations of the
action functional \eqref{eq:F-theory} then follow straightforwardly.
\item[Ad ii)] Assume $\,\p\xcM=\emptyset$.\ Pick up a pair
$\,C_\a,\ \a\in\{1,2\}\,$ of Cauchy hypersurfaces in $\,\xcM\,$ and
cut out a region $\,\xcM_{1,2}\subset\xcM\,$ such that $\,\p
\xcM_{1,2}=C_1\sqcup(-C_2)$,\ where the minus in front of $\,C_2\,$
represents the reversal of the orientation on $\,C_2\,$ induced from
that on $\,\xcM$,\ and such that the two hypersurfaces can be
homotopically transformed into one another across $\,\xcM_{1,2}$.\
Write
\qq\nn
S_{1,2}[\Psi_{\rm cl}]:=\int_{\xcM_{1,2}}\,\bigl(\Psi_{\rm cl}
\vert_{\xcM_{1,2}}\bigr)^*\Th_\xcF\,.
\qqq
Using \Reqref{eq:EuLagAl}, we find
\qq\nn
\d S_{1,2}[\Psi_{\rm cl}]=\int_{C_1}\,(\Psi_{\rm cl}
\vert_{C_1})^*\Th_\xcF-\int_{C_2}\,(\Psi_{\rm cl}
\vert_{C_2})^*\Th_\xcF\,,
\qqq
and therefore conclude that the 2-form
\qq\label{eq:presympF}
\Om_\xcF[\Psi_{\rm cl}]:=\int_\xcC\,(\Psi_{\rm cl}\vert_\xcC)^*
\d\Th_\xcF\,,
\qqq
written for an \emph{arbitrary} Cauchy hypersurface $\,\xcC$,\ is
manifestly closed (and independent of the choice of $\,\xcC$) and
hence defines a presymplectic form on $\,\sfP_\xcF$.\ The proof
proceeds analogously for $\,\p\xcM\neq\emptyset$,\ and the analysis
below (for $\,\xcM=\Si\,$ with domain walls) is readily seen to
cover that case. The sole difference is the appearance of a more
complicated expression
\qq\nn
\d S_{1,2}[\Psi_{\rm cl}]=\Xi_{C_1}[\Psi_{\rm cl}\vert_{C_1}]-
\Xi_{C_2}[\Psi_{\rm cl}\vert_{C_2}]\,,
\qqq
with the two functional 1-form contributions once more localised on
the two Cauchy hypersurfaces.
\eit
\brem\label{rem:Mars-Wein} Vector fields that span the kernel of the
presymplectic form $\,\Om_\xcF\,$ are identified with generators of
infinitesimal gauge transformations of $\,\xcF$.\ Upon performing
the standard symplectic reduction on the state space $\,\sfP_\xcF\,$
\wrt the characteristic distribution $\,K_\xcF\,$ of $\,\Om_\xcF\,$
(assumed reducible) and subsequently restricting $\,\Om_\xcF\,$ to
the space $\,\ovl{\ovl\sfP}_\xcF=\sfP_\xcF// K_\xcF\,$ of leaves of
this distribution, we ultimately obtain a canonical form
$\,\ovl{\ovl\Om}_\xcF\,$ on the physical (reduced) state space
$\,\ovl{\ovl\sfP}_\xcF$,\ alongside a Poisson bracket
\qq
\{\xcO_1,\xcO_2\}_\xcF[\Psi_{\rm cl}]=\ovl{\ovl\Om}_\xcF[\Psi_{\rm
cl}]( X_{\xcO_1},X_{\xcO_2})
\qqq
of \textbf{hamiltonian functions $\,\xcO_i$},\ i.e.\ functionals on
the reduced state space which generate \textbf{hamiltonian vector
fields} associated with $\,\xcO_i\,$ as per
\qq
\d\xcO_i=-\xcX_{\xcO_i}\con\,\ovl{\ovl\Om}_\xcF[ \Psi_{\rm cl}]\,.
\qqq
Here, $\,X_{\xcO_i}\,$ are vectors tangent to $\,\sfP_\xcF \,$ at
the state $\,\Psi_{\rm cl}$.\ They are defined by the corresponding
$\cF$-vertical vector fields $\,\xcX_{\xcO_i}\,$ tangent to
$\,\sfJ^1\cF\,$ and satisfying the linearised variant of the field
equations of $\,\xcF$.\ In the case of a space-time with a non-empty
boundary (or domain walls), the $\,\xcX_{\xcO_i}\,$ are additionally
required to obey the linearised version of the boundary (resp.\
domain-wall gluing) conditions of $\,\xcF$.\erem

The reconstruction of the symplectic form on the state space of the
$\cF$-field theory $\,\xcF\,$ is the first step towards a geometric
quantisation of the latter as it induces, iff $\,\frac{1}{2\pi}\,
\Om_\xcF\,$ has integral periods over 2-cycles of $\,\sfP_\xcF$,\ a
circle bundle over $\,\sfP_\xcF\,$ whose space of sections, when
suitably polarised, can be identified with the Hilbert space of
$\,\xcF$,\ cf., e.g., \Rcite{Woodhouse:1992de}. In the present
paper, we do not address the question of the choice of the
polarisation, and so we content ourselves with the following
\bedef\label{def:prequantise}
Let $\,\xcF\,$ be an $\cF$-field theory on a (pseudo-)riemannian
manifold $\,(\xcM,\txg)\,$ with a covariant configuration bundle
$\,\pi_\cF: \cF\to\xcM$,\ and let $\,(\sfP_\xcF,\Om_\xcF)\,$ be the
symplectic space of extremal sections of the first-jet bundle
$\,\sfJ^1\cF\,$ of $\,\cF$,\ equipped with a symplectic form
$\,\Om_\xcF\,$ of Proposition \ref{prop:Cartan}. The
\textbf{pre-quantum bundle $\,\pi_{\ceL_\xcF}:\ceL_\xcF\to
\sfP_\xcF\,$ of $\cF$-field theory $\,\xcF\,$} is a circle bundle
over $\,\sfP_\xcF\,$ with connection $\,\nabla_{\ceL_\xcF}\,$ of
curvature
\qq\nn
\curv(\nabla_{\ceL_\xcF})=\pi_{\ceL_\xcF}^*\Om_\xcF\,.
\qqq
Fix a choice $\,\cO_{\sfP_\xcF}=\{\cO^{\sfP_\xcF}_i\}_{i\in
\xcI_{\sfP_\xcF}}\,$ of an open cover of $\,\sfP_\xcF\,$ and a local
presentation, in the sense of Definition \ref{def:loco}, of the
pre-quantum bundle in terms of its \v Cech--Deligne data $\,
\ceL_\xcF\xrightarrow{\rm loc.}(\theta_{\xcF\,i},\g_{\xcF\,ij})\in
\cA^{2,1}(\cO_{\sfP_\xcF})\,$ associated with $\,\cO_{\sfP_\xcF}\,$
and subject to the cohomological constraints
\qq\nn
D_{(1)}(\theta_{\xcF\,i},\g_{\xcF\,ij})=(\Om_\xcF
\vert_{\cO^{\sfP_\xcF}_i},0,1)\,.
\qqq
A \textbf{pre-quantisation of $\cF$-field theory $\,\xcF$},
understood in the sense of, e.g., \Rcite{Woodhouse:1992de}, is an
assignment, to every smooth function $\,h\in C^\infty(\sfP_\xcF,\bR
)\,$ and to the associated \textbf{(global) hamiltonian vector field
$\,\xcX_h\,$} on $\,\sfP_\xcF$,\ determined by the relation
\qq\nn
\xcX_h\con\Om_\xcF=-\d h\,,
\qqq
of a collection $\,\widehat\cO_h:=(\widehat h_i)_{i\in\xcI_\xcF}\,$
of local linear operators
\qq\label{eq:pre-ham-gen}
\widehat h_i:=-\sfi\,\pLie{\xcX_h}-\xcX_h\con\theta_{\xcF\,i}+h
\vert_{\cO^{\sfP_\xcF}_i}
\qqq
on the space $\,\G(\ceL_\xcF)\,$ of sections of the pre-quantum
bundle, the latter being regarded as the \textbf{pre-quantisation
Hilbert space}. The collection $\,\widehat\cO_h\,$ shall be termed
the \textbf{pre-quantum hamiltonian for $\,h$}.\ By the very
construction, the commutator of a pair $\,\widehat\cO_{h_\a},\ \a\in
\{1,2\}\,$ of pre-quantum hamiltonians takes the canonical form
\qq\label{eq:comm-preq-ham}
[\,\widehat\cO_{h_1}\,,\,\widehat\cO_{h_2}\,]=-\sfi\,\widehat
\cO_{\{\,h_1\,,\,h_2\,\}_{\Om_\xcF}}\,.
\qqq
\exdef

\subsection{The covariant formalism for the $\si$-model}

Specialisation of the above general discussion to the non-linear
$\si$-model of \Reqref{eq:sigma} prerequires a number of
modifications, which -- while preserving the basic conceptual
framework -- serve to adapt the tools introduced to the setting in
hand, in which forms on the space-time $\,\Si\,$ are replaced by
locally smooth forms associated with a given triangulation
$\,\triangle(\Si)$,\ and in which the space-time itself is split
into domains, supporting the respective phases of the
two-dimensional field theory. As for the latter point, the reader is
advised to acquaint herself or himself, by way of a warm-up, with
the treatment of the world-sheet with a non-empty boundary in
\Rcite{Gawedzki:2001rm}.

The first modification consists in replacing the covariant
configuration bundle $\,\cF\,$ with
\bedef\label{def:cov-conf-bdle-si}
The \textbf{covariant configuration bundles $\,\pi_{\cF_\si}:\cF_\si
\to\Si\,$ of the non-linear $\si$-model for network-field
configurations $\,(X\,\vert\,\G)\,$ in string background $\,\Bgt\,$
on world-sheet $\,(\Si,\g)\,$ with defect quiver $\,\G\,$} are given
by a disjoint sum of fibre bundles over the respective components of
the disjoint union of elements of $\,\Pgt_\Si,\Egt_\G\,$ and
$\,\Vgt_\G\,$ with restrictions
\qq\nn
\cF_\si\vert_{\wp\in\Pgt_\Si}:=\wp\x M\to\wp\,,\qquad\qquad\cF_\si
\vert_{\ell\in\Egt_\G}:=\ell\x Q\to\ell\,,\qquad\qquad\cF_\si
\vert_{\jmath\in\Vgt_\G}:=\jmath\x T_{n_\jmath}\to\jmath\,.
\qqq
The associated first-jet bundles, $\,\sfJ^1\cF_\si\to\Si$,\ admit
local coordinates
\bit
\item $\bigl(\si^a,X^\mu,\xi^\nu_b)\,$ over $\,\wp\in\Pgt_\Si$,\
where $\,\si^a\,$ are local coordinates on the patch $\,\wp$,\ and
$\,X^\mu\,$ are local coordinates on $\,M$;
\item $\,(\varphi,X^A,\xi^B_\varphi)\,$ over $\,\ell\in\Egt_\G$,\
where $\,\varphi\,$ is a local coordinate on the defect line
$\,\ell$,\ and $\,X^A\,$ are local coordinates on $\,Q$;
\item $(\si_\jmath,X^i)\,$ over $\,\jmath\in\Vgt_\G$,\ where
$\,\si_\jmath\,$ are the coordinates of the defect junction
$\,\jmath\,$ within $\,\Si$,\ and $\,X^i\,$ are local coordinates on
$\,T_{n_\jmath}\,$ (the first-jet extension is trivial over the
point $\,\jmath$).
\eit
\exdef \noindent The action functional \eqref{eq:sigma} being
defined in terms of local expressions sourced by plaquettes of a
triangulation $\,\triangle(\Si)\,$ of the world-sheet (and their
lower-dimensional submanifolds), the Cartan form naturally splits
into a sum over terms supported by the particular plaquettes. Below,
we define the Cartan form as an object glued up from these
contribution. Our point of departure is the local term $\,\ceL_p(\si
,X,\p X)\,$ of the lagrangean density for $\,S_\si\,$ coming from
the plaquette $\,p\in\triangle(\Si)$.\ In the minkowskian gauge
$\,\g=\eta\,$ of the intrinsic world-sheet metric, it is given by
the formula\footnote{In the formula, we employ a shorthand notation
$\,e\subset p\cap\G\,$ to denote those edges of the plaquette
$\,p\,$ which lie on a defect line (and hence belong to the
triangulation $\,\triangle(\G\setminus\Vgt_\G)$), and similarly for
$\,\jmath\in p\cap\Vgt_\G$,\ the latter denoting the defect
junctions among the vertices of the triangulation of the plaquette
$\,p$.}
\qq\nn
\ceL_p(\si,X,\xi)\,\sfd^2\si&=&\tfrac{1}{2}\,\txg_{\mu\nu}(X)\,
\star_\eta\bigl(\xi^\mu\wedge\star_\eta\xi^\nu\bigr)\,\sfd^2\si-
B_{i_p,\mu\nu}(X)\,\star_\eta\bigl(\xi^\mu\wedge\xi^\nu\bigr)\,
\sfd^2\si\cr\cr
&&+\sum_{e\subset p}\,\left(A_{i_p i_e,\mu}(X)\,\xi^\mu\wedge\d_e-
\sfi\,\sum_{v\in e}\,\d_v\,\log g_{i_p i_e i_v}^{\vep_{pev}}(X)\,
\sfd^2\si\right)\cr\cr
&&+\sum_{e\subset p\cap\G}\,\left(P_{i_e,A}(X)\,\xi^A\wedge\d_e-
\sfi\,\sum_{v\in e}\,\d_v\,\log K_{i_e i_v}^{-\vep_{ev}}(X)\,\sfd^2
\si\right)\cr\cr
&&-\sfi\,\sum_{\jmath\in p\cap\Vgt_\G}\,\d_\jmath\,\log f_{n_\jmath
,i_\jmath}(X)\,\sfd^2\si\,,
\qqq
with
\qq\nn
\tfrac{\d^2\ceL_p}{\d\xi^\mu_a\d\xi^\nu_b}(\si,X,\xi)=-\bigl(
\txg_{\mu\nu}\,\eta^{ab}-2B_{i_p,\mu\nu}\,\vep^{ab}\bigr)(X)=:-L^{a
b}_{i_p,\mu\nu}(X)\,,
\qqq
all written in terms of local coordinates $\,\si^a,\ a\in\{1,2\}\,$
on $\,p$,\ with $\,\sfd^2\si=\sfd\si^1\wedge\sfd\si^2$,\ alongside
objects $\,\xi=\xi_a\,\exd\si^a$,\ and the Dirac distributions
$\,\d_x\equiv\d^{(2)}(\si-\si_x)\,$ on $\,\Si$,\ as well as the
singular (Dirac-type) currents $\,\d_e\,$ supported over $\,e\subset
p$,\ with the defining property
\qq\nn
\forall_{\tht\in\Om^1(p)}\ :\ \int_p\,\tht\wedge\d_e=\int_e\,
\iota_e^*\tht\,,
\qqq
where $\,\iota_e:e\emb p\,$ is the embedding map. In the minkowskian
gauge, we have
\qq\nn
\star_\eta
1=\sfd^2\si\,,\qquad\star_\eta\sfd\si^1=-\sfd\si^2\,,\qquad
\star_\eta\sfd\si^2=-\sfd\si^1\,,\qquad\star_\eta\sfd^2\si=-1\,.
\qqq
This yields
\bedef\label{def:Cart-form-si}
Let $\,\Bgt=(\cM,\cB,\cJ)\,$ be a string background of Definition
\ref{def:bckgrnd}. The \textbf{Cartan form of the non-linear
$\si$-model for network-field configurations $\,(X\,\vert\,\G)\,$ in
string background $\,\Bgt\,$ on world-sheet $\,(\Si,\g)\,$ with
defect quiver $\,\G\,$} is a 2-form $\,\Th_\si\,$ on the first-jet
bundles $\,\sfJ^1\cF_\si\,$ of the covariant configuration bundles
$\,\cF_\si\,$ of the $\si$-model, given in terms of its restrictions
$\,\Th_\si\vert_p=:\Th_p\,$ to patches $\,p\in\triangle(\Si)\,$ of a
triangulation $\,\triangle(\Si)\,$ of $\,\Si\,$ subordinate to
$\,\cO_\xcM,\ \xcM\in\{M,Q,T\}\,$ \wrt $\,(X\,\vert\,\G)\,$ that
take the form
\qq\nn
\Th_p(\si,X,\xi)&=&\tfrac{1}{2}\,\txg_{\mu\nu}(X)\,\xi^\mu\wedge
\star_\eta\xi^\nu-\txg_{\mu \nu}(X)\,\d
X^\mu\wedge\star_\eta\xi^\nu\cr\cr
&&-B_{i_p,\mu\nu}(X)\,\xi^\mu\wedge\xi^\nu+2B_{i_p,\mu\nu}(X)\,\d
X^\mu\wedge\xi^\nu\cr\cr
&&+\sum_{e\subset p}\,\bigl(A_{i_p i_e}(X)\wedge\d_e-\sfi\,\sum_{v
\in e}\,\d_v\,\log g_{i_p i_e i_v}^{\vep_{pev}}(X)\,\sfd^2\si\bigr)
\cr\cr
&&+\sum_{e\subset p\cap_\G}\,\bigl(P_{i_e}(X)\,\wedge\d_e-\sfi\,
\sum_{v\in e}\,\d_v\,\log K_{i_e i_v}^{-\vep_{ev}}(X)\,\sfd^2\si
\bigr)\cr\cr
&&-\sfi\,\sum_{\jmath\in p\cap\Vgt_\G}\,\d_\jmath\,\log f_{n_\jmath,
i_\jmath}(X)\,\sfd^2\si\,.
\qqq
\exdef \brem Consider a generic $\cF_\si$-vertical vector field
$\,\xcV\,$ on $\,\sfJ^1\cF_\si\,$ with restrictions
\qq\nn
\xcV\vert_{\Pgt}=V^\mu\,\tfrac{\d\ \ }{\d X^\mu}+V^\mu_a\,\tfrac{\d\
}{\d \xi^\mu_a}\,,\qquad\qquad\xcV\vert_{\Egt_\G}=V^A\,\tfrac{\d\ \
}{\d X^A}+V_\varphi^A\,\tfrac{\d\ }{\d\xi^A_\varphi}\,,\qquad
\qquad\xcV\vert_{\Vgt_\G}=V^i\,\tfrac{\d\ \ }{\d X^i}\,,
\qqq
where the various components are constrained as per
\qq\label{eq:tangojet-constr}
V^A\,\tfrac{\p\iota_\a^\mu}{\p X^A}=V^\mu\circ\iota_\a\,,\qquad
\qquad V^i\,\tfrac{\p\pi_n^{k,k+1\ A}}{\p X^i}=V^A\circ\pi_n^{k,k+
1}\,.
\qqq
The requirement that $\,\xcV\,$ obey the linearised version of
\Reqref{eq:eom-xipar} is tantamount to the imposition of the
relation
\qq\nn
V^\mu_a=\p_a V^\mu\,.
\qqq
Hence, a vector field tangent to the space of extremal sections at a
section $\,\Psi_{\si,{\rm cl}}\,$ is necessarily of the form
\qq\label{eq:tan-phasp}
\xcV\vert_{\Pgt}=V^\mu\,\tfrac{\d\ \ }{\d X^\mu}+\p_a
V^\mu\,\tfrac{\d\ }{\d \xi^\mu_a}\,,
\qqq
where the various components are related as in
\Reqref{eq:tangojet-constr}, and where the $\,V^\mu\,$ satisfy the
linearised version of \Reqref{eq:field-eqs}.\erem

We have
\berop\label{prop:Cart-si-def}
Let $\,\Th_\si\,$ be the Cartan form of the non-linear $\si$-model
for network-field configurations $\,(X\,\vert\,\G)\,$ in string
background $\,\Bgt\,$ on world-sheet $\,(\Si,\g)\,$ with defect
quiver $\,\G$,\ explicited in Definition \ref{def:Cart-form-si}.
Given a section $\,\Psi_\si\in\G(\sfJ^1\cF_\si)\,$ of the first-jet
bundles $\,\sfJ^1 \cF_\si\to\Si\,$ of the covariant configuration
bundles for the $\si$-model, write
\qq\nn
S_{\Th_\si}[\Psi_\si]:=\int_\Si\,\Psi^*\Th_\si\,.
\qqq
The principle of least action applied to the functional
$\,S_{\Th_\si}\,$ as per
\qq\label{eq:var-STHsi}
\xcV\con\d S_{\Th_\si}[\Psi_{\si,{\rm cl}}]=0\,,
\qqq
with $\,\xcV\in\G(\sfT\sfJ^1\cF_\si)^{\perp_{\cF_\si}}\,$ an
arbitrary $\cF_\si$-vertical vector field on $\,\sfJ^1\cF_\si$,\
yields the field equations \eqref{eq:field-eqs} alongside the Defect
Gluing Condition \eqref{eq:DGC} for classical sections $\,\Psi_{\si
,{\rm cl}}\,$ of the $\si$-model.
\eerop
\noindent As the proof of the proposition is rather technical, it
has been relegated to Appendix \ref{app:Cart-form-si}.\medskip

The Cartan form $\,\Th_\si\,$ enables us to study the canonical
structure of the classical $\si$-model and provides non-trivial
insights into its quantum r\'egime, all that through the definition
of a (pre-)symplectic form on the space of states $\,\Psi_{\si,{\rm
cl}}(\si)=\bigl(\si^a,X^I(\si),\p_b X^J(\si)\bigr)\,$ of the model
(here, $\,I\,$ and $\,J\,$ are multi-indices taking values in the
index sets associated with coordinates on $\,M\sqcup Q\sqcup T$).
The latter space admits a natural parameterisation in terms of
initial data of an extremal section $\,\Psi_{\si,{\rm cl}}\,$
localised on a Cauchy hypersurface in $\,\Si\,$ -- a space-like
contour $\,\xcC\,$ in the case in hand. As argued in Remark
\ref{rem:states}, there are two qualitatively different species of a
classical state in the presence of a defect quiver in the
world-sheet: the untwisted state and the twisted state. Accordingly,
we have
\bedef\label{def:untw-phspace}
Let $\,\Bgt\,$ be a string background with target space $\,M$.\ The
\textbf{untwisted state space $\,\sfP_{\si,\emptyset}\,$ of the
non-linear $\si$-model for network-field configurations
$\,(X\,\vert\,\G)\,$ in string background $\,\Bgt\,$ on world-sheet
$\,(\Si,\g)\,$ with defect quiver $\,\G\,$} is given by the
cotangent bundle over the free-loop space $\,\sfL M=
C^\infty(\bS^1,M)\,$ of the target space $\,M\,$ of the $\si$-model,
\qq\nn
\sfP_{\si,\emptyset}=\sfT^*\sfL M\,.
\qqq
It has local coordinates $\,(X^\mu,\sfp_\nu)$,\ where $\,X:\bS^1\to
M\,$ is a smooth loop in $\,M\,$ and $\,\sfp=\sfp_\mu\,\d X^\mu\,$
is a normal covector field on $\,X^\mu$.\exdef \noindent The twisted
counterpart is introduced in
\bedef\label{def:tw-phspace}
Let $\,\Bgt\,$ be a string background with target $\,\cM=(M,\txg,\cG
)\,$ and bi-brane $\,\cB=\bigl(Q,\iota_\a,\om,\Phi\ \vert\ \a\in\{1,
2\}\bigr)$,\ and let $\,\G\,$ be a defect quiver embedded in a
world-sheet $\,(\Si,\g)\,$ in such a manner that there exists a
closed space-like curve $\,\xcC\cong\bS^1\subset\Si\,$ that
intersects $\,I\in\bN_{>0}\,$ defect lines $\,\ell_k\in\Egt_\G,\
k\in\ovl{1, I}\,$ of $\,\G\,$ at the respective points $\,\si_k\,$
so that the tangent vectors $\,\widehat t_k\,$ of the defect lines
$\,\ell_k\,$ are all time-like or anti-time-like at the $\,\si_k$,\
with $\,X_* \widehat t_k=:V_k\in\sfT_{q_k}Q$.\ Write $\,\vep_k=+
1\,$ if $\,\widehat t_k\,$ is time-like, and $\,\vep_k=-1\,$ if
$\,\widehat t_k\,$ is anti-time-like at $\,\si_k$.\ Fix a collection
of points $\,\{P_k\}_{k\in\ovl{1,I}}\in\bS^1$,\ write
$\,\bS^1_{\{P_k\}}:= \bS^1\setminus\{P_k\}_{k\in\ovl{1,I}}\,$ and
define the space of smooth maps
\qq\nn
\sfL_{Q|\{(P_k,\vep_k)\}}M=\{\ (X,q_k\ \vert\ k\in\ovl{1,I})\in
C^\infty(\bS^1_{\{P_k\}},M)\x Q^{\x I} \quad\vert\quad \lim_{\ep\to
0^+}\,X\bigl(P_k+(-1)^{\a+1}\,\vep_k\,\ep\bigr)= \iota_\a(q_k) \
\}\,.
\qqq
Denote as $\,\widehat\tau_\a(P_k):=-\vep_k\,\lim_{\ep\to 0^+}\,X_*
\widehat t\bigl(P_k+(-1)^{\a+1}\,\vep_k\,\ep\bigr)\,$ the
(one-sided) limiting values of the pushforward of the tangent vector
field $\,\widehat t(\cdot)\,$ on $\,\bS^1_{\{P_k\}}\,$ along $\,X$,\
and write $\,(\iota_1^{+1},\iota_2^{+1}):=(\iota_1,\iota_2)\,$ and
$\,( \iota_1^{-1},\iota_2^{-1}):=(\iota_2,\iota_1)$.\ The
\textbf{$k$-twisted state space $\,\sfP_{\si,\cB|\{(P_k,\vep_k)
\}}\,$ of the non-linear $\si$-model for network-field
configurations $\,(X\,\vert\,\G)\,$ in string background $\,\Bgt\,$
on world-sheet $\,(\Si,\g)\,$ with defect quiver $\,\G\,$} is
naturally identified with the space
\qq\nn
\sfP_{\si,\cB|\{(P_k,\vep_k)\}}=\Bigg\{\ (X,\sfp=\sfp_\mu\,\d X^\mu
,q_k,V_k\ \vert\ k\in\ovl{1,I})\in\sfT^*C^\infty(\bS^1_{\{P_k\}},M)
\x\sfT Q^{\x I} \quad\bigg\vert\quad \cr\cr\cr \land\quad \left\{
\barr{l} \lim_{\ep\to 0^+}\sfp\bigl(P_k+(-1)^{\a+1}\,\ep\bigr)=\txg
\bigl(\iota_\a^{\vep_k}(q_k)\bigr)\bigl(\vep_k\,\iota_{\a\,*}^{\vep_k}
V_k,\cdot\bigr)\cr\cr
\txg\bigl(\iota_1(q_k)\bigr)\bigl(\widehat\tau_1(P_k),\iota_{1 \,*}(
\cdot)\bigr)-\txg\bigl(\iota_2(q_k)\bigr)\bigl(\widehat\tau_2(P_k),
\iota_{2\,*}(\cdot)\bigr)=V_k\con\om(q_k)\earr \right. \ \Bigg\}\,.
\qqq
The space $\,\sfP_{\si,\cB|\{(P_k,\vep_k)\}}\,$ shall be described
in terms of its local coordinates $\,(X^\mu,\sfp_\nu,q_k,V_k\ \vert\
k\in\ovl{1,I})$. \exdef \noindent We may now formulate the following
fundamental statements:
\berop\label{prop:sympl-form-si-untw}
Let $\,\Bgt\,$ be a string background with target $\,\cM=(M,\txg,\cG
)$,\ and let $\,\sfP_{\si,\emptyset}\,$ be the untwisted state space
of the non-linear $\si$-model for network-field configurations
$\,(X\,\vert\,\G)\,$ in string background $\,\Bgt\,$ on world-sheet
$\,(\Si,\g)\,$ with defect quiver $\,\G$.\ Denote by
\qq\nn
\ev_M\ :\ \sfL M\x\bS^1\to M
\qqq
the canonical evaluation map. The Cartan form $\,\Th_\si\,$ of the
$\si$-model from Definition \ref{def:Cart-form-si} canonically
defines a closed 2-form on $\,\sfP_{\si,\emptyset}$,\ given by the
formula
\qq\label{eq:Omsi-untw}
\Om_{\si,\emptyset}=\d\theta_{\sfT^*\sfL M}+\pi_{\sfT^*\sfL M}^*
\int_{\bS^1}\,\ev_M^* \txH\,,
\qqq
in which
\qq\label{eq:can-1-cot}
\theta_{\sfT^*\sfL M}[(X,\sfp)]=\int_{\bS^1}\,\Vol(\bS^1)\wedge\sfp
\,,\qquad\qquad\sfp=\sfp_\mu\,\d X^\mu
\qqq
is the canonical 1-form on the total space of the cotangent bundle
$\,\pi_{\sfT^*\sfL M}:\sfT^*\sfL M\to\sfL M$,\ written using the
volume form $\,\Vol(\bS^1)\,$ on $\,\bS^1$,\ and $\,\txH=\curv(\cG
)$.\ The 2-form is to be evaluated on an arbitrary classical section
$\,\Psi_{\si,{\rm cl}}\in\G( \sfJ^1\cF_\si)\,$ of the first-jet
bundles of the covariant configuration bundles $\,\cF_\si\,$ of the
$\si$-model. The section (state) is represented by its Cauchy data
$\,\bigl(X^\mu,\sfp_\nu\bigr)\in \sfP_{\si,\emptyset}\,$ localised
on an arbitrary untwisted Cauchy contour $\,\xcC\cong\bS^1$.
\eerop
\noindent A proof of the proposition is given in Appendix
\ref{app:sympl-form-si-untw}.
\berop\label{prop:sympl-form-si-tw}
Let $\,\Bgt\,$ be a string background with target $\,\cM=(M,\txg,\cG
)\,$ and $\cG$-bi-brane $\,\cB=(Q,\iota_\a,\om,\Phi\ \vert\
\a\in\{1,2\})$,\ and let $\,\sfP_{\si,\cB|\{(P_k,\vep_k)\}}\,$ be
the $k$-twisted state space of the non-linear $\si$-model for
network-field configurations $\,( X\,\vert\,\G)\,$ in string
background $\,\Bgt\,$ on world-sheet $\,(\Si,\g)\,$ with defect
quiver $\,\G$.\ Write $\,\bS^1_{\{P_k\}}=\bS^1\setminus\{P_k\}_{k
\in\ovl{1,I}}\,$ and denote by
\qq\nn
\ev_{M,\{P_k\}}\ :\ C^\infty(\bS^1_{\{P_k\}},M)\x\bS^1_{\{P_k\}}\to
M
\qqq
the canonical evaluation map, and by $\,\pr_{\sfT^*C^\infty
(\bS^1_{\{P_k\}},M)}:\sfP_{\si,\cB|\{(P_k,\vep_k)\}}\to\sfT^*
C^\infty(\bS^1_{\{P_k\}},M)\,$ and $\,\pr_{Q,k}:\sfP_{\si,\cB|\{(
P_k,\vep_k)\}}\to Q\,$ the canonical projections, the latter having
the $k$-th cartesian factor as the codomain. The Cartan form
$\,\Th_\si\,$ of the $\si$-model from Definition
\ref{def:Cart-form-si} canonically defines a closed 2-form on
$\,\sfP_{\si,\cB|\{(P_k,\vep_k)\}}$,\ given by the formula
\qq
\Om_{\si,\cB|\{(P_k,\vep_k)\}}=\pr_{\sfT^*C^\infty(\bS^1_{\{P_k\}},
M)}^*\bigl(\d\theta_{\sfT^*C^\infty(\bS^1_{\{P_k\}},M)}+
\pi_{\sfT^*C^\infty(\bS^1_{\{P_k\}},
M)}^*\int_{\bS^1_{\{P_k\}}}\,\ev_{M,\{P_k\}}^*\txH\bigr)+\sum_{k=1}^I\,
\vep_k\,\pr_{Q,k}^*\om\,,\cr\cr\label{eq:Omsi-tw}
\qqq
in which
\qq\label{eq:can-1-cot-tw}
\theta_{\sfT^*C^\infty(\bS^1_{\{P_k\}},M)}[(X,\sfp)]=
\int_{\bS^1_{\{P_k\}}}\,\Vol(\bS^1_{\{P_k\}})\wedge\sfp\,,\qquad
\qquad\sfp=\sfp_\mu\,\d X^\mu
\qqq
is the canonical 1-form on the total space of the cotangent bundle
$\,\pi_{\sfT^*C^\infty(\bS^1_{\{P_k\}},
M)}:\sfT^*C^\infty(\bS^1_{\{P_k\}},M)\to
C^\infty(\bS^1_{\{P_k\}},M)$,\ written using the volume form
$\,\Vol(\bS^1_{\{P_k\}})\,$ on $\,\bS^1_{\{P_k\}}$.\ The 2-form is
to be evaluated on an arbitrary classical section $\,\Psi_{\si,{\rm
cl}}\in\G(\sfJ^1\cF_\si)\,$ of the first-jet bundles of the
covariant configuration bundles $\,\cF_\si\,$ of the $\si$-model.
The section (state) is represented by its Cauchy data
$\,(X^\mu,\sfp_\nu,q_k,V_k\ \vert\ k\in\ovl{1,I})
\in\sfP_{\si,\cB|\{(P_k,\vep_k)\}}\,$ localised on an arbitrary
twisted Cauchy contour $\,\xcC\cong\bS^1_{\{P_k\}}$,\ as in
Definition \ref{def:tw-phspace}.
\eerop
\noindent A proof of the proposition is given in Appendix
\ref{app:sympl-form-si-tw}.\medskip

In order to give an explicit description of the pre-quantum bundles
for the two types of the state space of the $\si$-model, we should
first recall the necessary facts about the (Fr\'echet) manifold
$\,\sfL M$,\ as defined in \Rcite{Hamilton:1982}. We use, after
\Rcite{Gawedzki:1987ak}, the straightforward
\berop\label{prop:cover-untw}
Let $\,\sfL M=C^\infty(\bS^1,M)\,$ be the free-loop space of a
manifold $\,M$,\ the latter coming with a choice
$\,\cO_M=\{\cO^M_i\}_{i\in\xcI_\xcM}\,$ of an open cover. Consider
the non-empty open sets\footnote{The free-loop space $\,\sfL M\,$ is
equipped with the compact-open topology.}
\qq\nn
\cO_\igt=\{\ X\in\sfL M \quad\vert\quad \forall_{e,v\in\triangle(
\bS^1)}\ :\ X(e)\subset\cO^M_{i_e}\quad\land\quad X(v)\in
\cO^M_{i_v} \ \}\,,
\qqq
with the index $\,\igt\,$ given by a pair $\,\bigl(\triangle(\bS^1)
,\phi\bigr)\,$ consisting of a choice $\,\triangle(\bS^1)\,$ of the
triangulation of the unit circle, with its edges $\,e\,$ and
vertices $\,v$,\ and a choice $\,\phi:\triangle(\bS^1)\to\xcI_M:f
\mapsto i_f\,$ of the assignment of indices of $\,\cO_M\,$ to
elements of $\,\triangle(\bS^1)$.\ By varying these two choices
arbitrarily, whereby an index set $\,\xcI_{\cO_{\sfL M}}\,$ is
formed, all of $\,\sfL M\,$ is covered, thus yielding an
\textbf{open cover $\,\cO_{\sfL M}=\{\cO_\igt \}_{\xcI_{\sfL M}}\,$
of free-loop space $\,\sfL M$}.
\eerop
\noindent Similarly,
\berop\label{prop:cover-tw}
Let $\,\Bgt\,$ be a string background with target $\,\cM=(M,\txg,\cG
)\,$ and $\cG$-bi-brane $\,\cB=(Q,\iota_\a,\om,\Phi\ \vert\
\a\in\{1,2 \})$.\ Given a collection $\,\{P_k\}_{k\in \ovl{1,I}}\,$
of $\,I\in \bN_{>0}\,$ points on the unit circle $\,\bS^1$,\ and a
collection $\,\{\vep_k\}_{k\in\ovl{1,I}}\,$ of $I$ elements of
$\,\{-1,+1\}$,\ let $\,\sfL_{Q|\{(P_k,\vep_k)\}}M\,$ be the space
introduced in Definition \ref{def:tw-phspace}. Fix a choice
$\,\cO_M=\{\cO^M_i \}_{i\in\xcI_\xcM}\,$ of an open cover of $\,M$,\
and a choice $\,\cO_Q=\{\cO^Q_i\}_{i \in\xcI_Q}\,$ of an open cover
of $\,Q$,\ for which there exist \v Cech-extended $\cG$-bi-brane
maps $\,(\iota_\a,\phi_\a),\ \a\in\{1,2\}$,\ as described in
Definition \ref{def:loco}. Consider the non-empty open sets
\qq\nn
\cO_{\{(P_k,\vep_k)\}\,\igt}=\left\{\ (X,q_k\ \vert\ k\in\ovl{1,I})
\in\sfL_{Q|\{P_k,\vep_k\}}M \quad\bigg\vert\quad X(e)\subset
\cO^M_{i_e}\ \land\ X(v)\in\cO^M_{i_v}\ \land\ \left\{ \barr{l} q_k
\in\cO^Q_{i^{1,2}_{P_k}}\cr \iota_\a(q_k)\in\cO^M_{\phi_\a(i^{1,
2}_{P_k})} \earr \right. \ \right\}
\qqq
with the index $\,\igt\,$ given by a triple $\,\bigl(\triangle_{\{
P_k\}}(\bS^1),\phi,\phi^{1,2}\bigr)\,$ consisting of a choice
$\,\triangle_{\{P_k\}}(\bS^1)\,$ of the triangulation of the unit
circle, with its edges $\,e\,$ and vertices $\,v\,$ of which $I$ are
fixed at the $\,P_k,\ k\in\ovl{1,I}$,\ and choices $\,\phi:\bigl(
\triangle_{\{P_k\}}(\bS^1)\setminus\{P_k\}_{k\in\ovl{1,I}}\bigr)
\to\xcI_M:f\mapsto i_f\,$ and $\,\phi^{1,2}:\{P_k\}_{k\in\ovl{1,I}}
\to\xcI_Q:P_k\mapsto i^{1,2}_{P_k}\,$ of the assignment of indices
of the open covers $\,\cO_M\,$ and $\,\cO_Q\,$ to elements of
$\,\triangle_{\{P_k\}}(\bS^1)$.\ By varying these choices
arbitrarily, whereby an index set $\,\xcI_{\cO_{\sfL_{Q|\{P_k
,\vep_k\}}M}}\,$ is formed, all of $\,\sfL_{Q|\{(P_k,\vep_k)\}} M\,$
is covered, thus yielding an \textbf{open cover $\,\cO_{\sfL_{Q|\{
P_k,\vep_k\}}M}=\{\cO_{\{(P_k,\vep_k)\}\,\igt}\}_{\igt\in
\xcI_{\sfL_{Q|\{P_k,\vep_k\}}M}}\,$ of space $\,\sfL_{Q|\{(
P_k,\vep_k)\}}M$}.
\eerop

\brem\label{rem:overlaps} It is straightforward to describe
intersections of elements of the open cover $\,\cO_{\sfL M}$.\ In so
doing, we follow \Rcite{Gawedzki:1987ak} once more. Given a pair
$\,\cO_{\igt^n},\ n\in\{1,2\}\,$ with the respective triangulations
$\,\triangle_n(\bS^1)\,$ (consisting of edges $\,e_n\,$ and vertices
$\,v_n$) and index assignments $\,(e_n,v_n)\mapsto(
i^n_{e_n},i^n_{v_n})$,\ we consider the triangulation $\,\ovl
\triangle(\bS^1)\,$ obtained by intersecting $\,\triangle_1(\bS^1
)\,$ with $\,\triangle_2(\bS^1)$,\ by which we mean that the edges
$\,\ovl e\,$ of $\,\ovl\triangle(\bS^1)\,$ are the non-empty
intersections of the edges of the $\,\triangle_n(\bS^1)$,\ and its
vertices $\,\ovl v\,$ are taken from the set-theoretic sum of the
two vertex sets. As previously, the incoming (resp.\ outgoing) edge
of $\,\ovl\triangle(\bS^1)\,$ at the vertex $\,\ovl v\,$ is denoted
by $\,\ovl e_+(\ovl v)\,$ (resp.\ $\,\ovl e_-(\ovl v)$). A non-empty
double intersection $\,\cO_{\igt^1}\cap\cO_{\igt^2}=:\cO_{\igt^1
\igt^n}\,$ is then labelled by the triangulation $\,\ovl\triangle(
\bS^1)$,\ taken together with the indexing convention such that
$\,i^n_{\ovl e}\,$ is the \v Cech index assigned -- via $\,\igt^n\,$
-- to the edge of $\,\triangle_n (\bS^1)\,$ containing $\,\ovl e
\in\ovl\triangle(\bS^1)$,\ and $\,i^n_{\ovl v}\,$ is the \v Cech
index assigned -- via $\,\igt^n\,$ -- to $\,\ovl v\,$ if $\,\ovl
v\in\triangle_n(\bS^1)$,\ or the \v Cech index assigned -- via
$\,\igt^n\,$ -- to the edge of $\,\triangle_n(\bS^1)\,$ containing
$\,\ovl v\,$ otherwise. Analogous remarks apply to
$\,\cO_{\sfL_{Q|\{(P_k,\vep_k)\}} M}\,$.\erem

\noindent We have the fundamental result:
\bethe\cite{Gawedzki:1987ak}\label{thm:trans-untw}
Let $\,\xcM\,$ be a manifold with a gerbe $\,\cG\,$ of curvature
$\,\curv(\cG)=:\txH\,$ over it, and denote by
\qq\nn
\ev_\xcM\ :\ \sfL\xcM\x\bS^1\to\xcM
\qqq
the canonical evaluation map for the free-loop space $\,\sfL\xcM=
C^\infty(\bS^1,M)\,$ of $\,\xcM$.\ The gerbe $\,\cG\,$ canonically
defines a circle bundle $\,\pi_{\ceL_\cG}:\ceL_\cG\to\sfL M\,$ with
connection $\,\nabla_{\ceL_\cG}\,$ of curvature
\qq\nn
\curv(\nabla_{\ceL_\cG})=\int_{\bS^1}\,\ev_\xcM^*\txH\,,
\qqq
to be termed the \textbf{transgression bundle}. The assignment
$\,\cG \to\ceL_\cG\,$ yields a cohomology map\linebreak
$\,\bH^2\bigl(\xcM,\cD(2
)^\bullet\bigr)\to\bH^1\bigl(\sfL\xcM,\cD(1)^\bullet\bigr)$,\ termed
the \textbf{transgression map}. \ethe \noindent A constructive proof
can be found in the original paper. As the underlying idea shall
subsequently be extended to $\,\sfL_{Q|\{(P_k,\vep_k)\}}M$,\ we
review it below. The transgression bundle $\,\ceL_{\cG}\to\sfL
\xcM\,$ can be defined in terms of its local data $\,(E_\igt,G_{\igt
\jgt})\in\cA^{2,1}(\cO_{\sfL\xcM})\,$ associated with the open cover
$\,\cO_{\sfL\xcM}\,$ from Proposition \ref{prop:cover-untw} and
determined by the local data of $\,\cG$,\ as written out in
Definition \ref{def:loco}. Here, the connection 1-forms $\,E_\igt\,$
are the 1-forms on $\,\cO_\igt\ni X\,$ given by the formul\ae
\qq\nn
E_\igt[X]=-\sum_{e\in\triangle(\bS^1)}\,\int_e\,X_e^*B_{i_e}-
\sum_{v\in\triangle(\bS^1)}\,X^*A_{i_{e_+(v)}i_{e_-(v)}}(v)\,,
\qqq
where $\,e_+(v)\,$ and $\,e_-(v)\,$ denote the incoming and the
outgoing edge meeting at $\,v$,\ respectively, and where $\,X_e=
X\vert_e$.\ The transition functions $\,G_{\igt\jgt}\,$ are the
$\uj$-valued functionals on $\,\cO_{\igt\jgt}\ni X\,$ defined as
\qq\nn
G_{\igt\jgt}[X]=\prod_{\ovl e\in\ovl\triangle(\bS^1)}\,\ee^{-\sfi\,
\int_{\ovl e}\,X_{\ovl e}^*A_{i_{\ovl e}j_{\ovl e}}}\,\prod_{\ovl v
\in\ovl\triangle(\bS^1)}\,X^*\bigl(g_{i_{\ovl e_+(\ovl v)}i_{\ovl
e_-(\ovl v)}j_{\ovl e_+(\ovl v)}}\cdot g^{-1}_{j_{\ovl e_+(\ovl v)}
j_{\ovl e_-(\ovl v)}i_{\ovl e_-(\ovl v)}}\bigr)(\ovl v)
\qqq
in terms of edges $\,\ovl e\,$ and vertices $\,\ovl v\,$ of the
triangulation $\,\ovl\triangle(\bS^1)\,$ from Remark
\ref{rem:overlaps}, and satisfying the standard cohomological
identities
\qq\nn
E_\jgt-E_\igt=\sfi\,\d\log G_{\igt\jgt}\,,\qquad\qquad G_{\jgt\kgt}
\cdot G_{\igt\kgt}^{-1}\cdot G_{\igt\jgt}=1\,.
\qqq
Under gauge transformations \eqref{eq:gauge-trans-gerbe} of the
local data of $\,\cG$,\ the local symplectic potentials undergo
induced gauge transformations
\qq\label{eq:E-gauge}
E_\igt\mapsto E_\igt-\sfi\,\d\log H_\igt\,,
\qqq
where
\qq\nn
H_\igt[X]=\prod_{e\in\triangle(\bS^1)}\,\ee^{\sfi\,\int_e\,X_e^*
\Pi_{i_e}}\,\prod_{v\in\triangle(\bS^1)}\,X^*\chi_{i_{e_+(v)}i_{e_-
(v)}}^{-1}(v)\,.
\qqq

The physical significance of the last proposition can be phrased as
\becor\label{cor:preqb-untw}
Let $\,\Bgt\,$ be a string background with target $\,\cM=(M,\txg,\cG
)$,\ and let $\,(\sfP_{\si,\emptyset},\Om_{\si,\emptyset})\,$ be the
untwisted state space of the non-linear $\si$-model for
network-field configurations $\,(X\,\vert\,\G)\,$ in string
background $\,\Bgt\,$ on world-sheet $\,(\Si,\g)\,$ with defect
quiver $\,\G$.\ Denote by $\,\pi_{\sfT^*\sfL M}:\sfT^*\sfL M\to\sfL
M\,$ the canonical map from the total space of the cotangent bundle
over the free-loop space $\,\sfL M=C^\infty(\bS^1,M)\,$ of $\,M\,$
onto its base. The pre-quantum bundle $\,\pi_{\ceL_{\si,\emptyset}}:
\ceL_{\si,\emptyset}\to\sfP_{\si,\emptyset}\,$ for the untwisted
sector of the $\si$-model is the circle bundle
\qq\nn
\ceL_{\si,\emptyset}:=\pi_{\sfT^*\sfL M}^*\ceL_\cG\ox\bigl(\sfT^*
\sfL M\x\bS^1\bigr)\to\sfT^*\sfL M\cong\sfP_{\si,\emptyset}
\qqq
given by the tensor product of the pullback, along $\,\pi_{\sfT^*
\sfL M}$,\ of the transgression bundle $\,\ceL_\cG\,$ of Theorem
\ref{thm:trans-untw} and of the trivial circle bundle $\,\sfT^*\sfL
M\x\bS^1\to\sfT^*\sfL M\,$ with a global connection 1-form equal to
the canonical 1-form $\,\theta_{\sfT^*\sfL M}\,$ on $\,\sfT^*\sfL
M$,\ explicited in Proposition \ref{prop:sympl-form-si-untw}. In
particular, given the open cover $\,\cO_{\sfL M}=\{\cO_\igt\}_{\igt
\in\xcI_{\sfL M}}\,$ of $\,\sfL M\,$ defined in Proposition
\ref{prop:cover-untw}, local data of $\,\ceL_{\si,\emptyset}\,$
associated with the induced open cover $\,\cO_{\sfP_{\si,
\emptyset}}=\{\cO^*_\igt\}_{\igt\in \xcI_{\sfL M}},\ \cO^*_\igt:=
\pi_{\sfT^*\sfL M}^{-1}(\cO_\igt)\,$ of $\,\sfP_{\si,\emptyset}\,$
can be expressed in terms of the local data
$\,(E_\igt,G_{\igt\jgt})\,$ of the bundle $\,\ceL_\cG\,$ from
Theorem \ref{thm:trans-untw} as
\qq\label{eq:loc-dat-preq-untw}
\theta_{\si,\emptyset\,\igt}=\theta_{\sfT^*\sfL
M}\vert_{\cO^*_\igt}+\pi_{\sfT^*\sfL
M}^*E_\igt\,,\qquad\qquad\g_{\si,\emptyset\,\igt\jgt}=\pi_{\sfT^*
\sfL M}^*G_{\igt\jgt}\,.
\qqq
\ecor
\medskip
\noindent Gaw\c{e}dzki's construction is readily verified to
generalise to the twisted case.
\bethe\label{thm:trans-tw}
Let $\,\Bgt\,$ be a string background with target $\,\cM=(M,\txg,\cG
)\,$ and $\cG$-bi-brane $\,\cB=(Q,\iota_\a,\om,\Phi\ \vert\ \a\in\{1
,2\})$,\ and let $\,\sfL_{Q|\{(P_k,\vep_k)\}}M\,$ be the space
introduced in Definition \ref{def:tw-phspace}. Write $\,\bS^1_{\{P_k
\}}=\bS^1\setminus\{P_k\}_{k\in\ovl{1,I}}\,$ and denote by
\qq\nn
\ev_{M,\{P_k\}}\ :\ C^\infty(\bS^1_{\{P_k\}},M)\x\bS^1_{\{P_k\}}\to
M
\qqq
the canonical evaluation map. The pair $\,(\cG,\cB)\,$ canonically
defines a circle bundle $\,\pi_{\ceL_{(\cG,\cB)|\{(P_k,\vep_k)\}}}:
\ceL_{(\cG,\cB)|\{(P_k,\vep_k)\}}\to\sfL_{Q|\{(P_k,\vep_k)\}}M\,$
with connection $\,\nabla_{\ceL_{(\cG,\cB)|\{(P_k,\vep_k)\}}}\,$ of
curvature
\qq\nn
\curv(\nabla_{\ceL_{(\cG,\cB)|\{(P_k,\vep_k)\}}})=\pi_{\ceL_{(\cG,
\cB)|\{(P_k,\vep_k)\}}}^*\bigl(\pr_{C^\infty(\bS^1_{\{P_k\}},M)}^*
\int_{\bS^1_{\{P_k\}}}\,\ev_{M,\{P_k\}}^*\txH+\sum_{k=1}^I\,\vep_k
\,\pr_{Q,k}^*\om\bigr)\,,
\qqq
written in terms of the canonical projections $\,\pr_{C^\infty(
\bS^1_{\{P_k\}},M)}:\sfL_{Q|\{(P_k, \vep_k)\}}M\to C^\infty(
\bS^1_{\{P_k\}},M)\,$ and $\,\pr_{Q,k}:\sfL_{Q|\{(P_k,\vep_k)\}}M
\to Q$,\ the latter having the $k$-th cartesian factor as the
codomain. Under the assignment
$\,(\cG,\cB)\to\ceL_{(\cG,\cB)|\{(P_k,\vep_k)\}}$,\
(gauge-)equivalence classes of pairs $\,(\cG,\cB)$,\ as described in
Definition \ref{def:loco}, are mapped to isomorphism classes of
bundles with connection. \ethe
\beroof
We define $\,\ceL_{(\cG,\cB)|\{(P_k,\vep_k)\}}\,$ explicitly in
terms of its local data $\,(E_{\{(P_k,\vep_k)\}\,\igt},G_{\{(P_k,
\vep_k)\}\,\igt\jgt})\in\cA^{2,1}(\cO_{\sfL_{Q|\{(P_k,\vep_k)\}}M})
\,$ associated with the open cover $\,\cO_{\sfL_{Q|\{(P_k,\vep_k)\}}
M}\,$ from Proposition \ref{prop:cover-tw} and determined by local
data of $\,(\cG,\cB)$,\ as written out in Definition \ref{def:loco}.
Write $\,(X^\mu,q_k\ \vert\ k\in\ovl{1,I})\equiv(X,\{q_k\})$.\ It is
easy to check that the objects
\qq\nn
E_{\{(P_k,\vep_k)\}\,\igt}[(X,\{q_k\})]&=&-\sum_{e\in\triangle_{\{
P_k\}}(\bS^1)}\,\int_e\,X_e^*B_{i_e}-\sum_{v\in\triangle_{\{P_k\}}(
\bS^1)\setminus\{P_k\}_{k\in\ovl{1,I}}}\,X^*A_{i_{e_+(v)}i_{e_-(v
)}}(v)\cr\cr
&&+\sum_{k=1}^I\,\vep_k\,\bigl(\iota_1^*A_{i_{e_{\widetilde\si_k}(
P_k)}\phi_1(i^{1,2}_{P_k})}-\iota_2^*A_{i_{e_{\si_k}(P_k)}\phi_2(
i^{1,2}_{P_k})}+P_{i^{1,2}_{P_k}}\bigr)(q_k)\,,\cr\cr\cr
G_{\{(P_k,\vep_k)\}\,\igt\jgt}[(X,\{q_k\})]&=&\prod_{\ovl e\in\ovl
\triangle_{\{P_k\}}(\bS^1)}\,\ee^{-\sfi\,\int_{\ovl e}\,X_{\ovl e}^*
A_{i_{\ovl e}j_{\ovl e}}}\cr\cr
&&\cdot\prod_{\ovl v\in\ovl\triangle_{\{P_k\}}(\bS^1)\setminus\{P_k
\}_{k\in\ovl{1,I}}}\,X^*\bigl(g_{i_{\ovl e_+(\ovl v)}i_{\ovl e_-(
\ovl v)}j_{\ovl e_+(\ovl v)}}\cdot g^{-1}_{j_{\ovl e_+(\ovl v)}
j_{\ovl e_-(\ovl v)}i_{\ovl e_-(\ovl v)}}\bigr)(\ovl v)\cr\cr
&&\cdot\prod_{k=1}^I\,\bigl[\iota_1^*\bigl(g_{i_{\ovl e_{\widetilde
\si_k}(P_k)}j_{\ovl e_{\widetilde\si_k}(P_k)}\phi_1(i^{1,2}_{P_k})}
\cdot g^{-1}_{j_{\ovl e_{\widetilde\si_k}(P_k)}\phi_1(i^{1,2}_{P_k})
\phi_1(j^{1,2}_{P_k})}\bigr)\cr\cr
&&\hspace{.7cm}\cdot\iota_2^*\bigl(g^{-1}_{i_{\ovl e_{\si_k}(P_k)}
j_{\ovl e_{\si_k}(P_k)}\phi_2(i^{1,2}_{P_k})}\cdot g_{j_{\ovl
e_{\si_k}(P_k)}\phi_2(i^{1,2}_{P_k})\phi_2(j^{1,2}_{P_k})}\bigr)
\cdot K_{i^{1,2}_{P_k}j^{1,2}_{P_k}}\bigr]^{\vep_k}(q_k)\,,
\qqq
written in terms of $\,(\si_k,\widetilde\si_k)=(+,-)\,$ if $\,\vep_k
=+1$,\ and $\,(\si_k,\widetilde\si_k)=(-,+)\,$ otherwise, obey the
required cohomological constraints. Similarly, one verifies through
inspection that under a gauge transformation
\eqref{eq:gauge-trans-gerbe} of the local data of $\,\cG$,\
accompanied by the $\cG$-twisted gauge transformation
\eqref{eq:gauge-trans-bi} of the local data of the $\cG$-bi-brane
1-isomorphism, the pair $\,(E_{\{(P_k,\vep_k)\}\,\igt},G_{\{(P_k,
\vep_k)\}\,\igt\jgt})\,$ undergoes induced gauge transformation
\qq\label{eq:tw-preq-si-gauge}
(E_{\{(P_k,\vep_k)\}\,\igt},G_{\{(P_k,\vep_k)\}\,\igt\jgt})\mapsto
\bigr(E_{\{(P_k,\vep_k)\}\,\igt},G_{\{(P_k,\vep_k)\}\,\igt\jgt})+
D_{(0)}(H_{\{(P_k,\vep_k)\}\,\igt})\,,
\qqq
with
\qq\nn
H_{\{(P_k,\vep_k)\}\,\igt}[(X,\{q_k\})]&=&\prod_{e\in\triangle_{\{
P_k\}}(\bS^1)}\,\ee^{\sfi\,\int_e\,X_e^*\Pi_{i_e}}\,\prod_{v\in
\triangle_{\{P_k\}}(\bS^1)\setminus\{P_k\}_{k\in\ovl{1,I}}}\,X^*
\chi_{i_{e_+(v)}i_{e_-(v)}}^{-1}(v)\cr\cr
&&\cdot\prod_{k=1}^I\,\bigl(\iota_1^*\chi_{i_{e_{\widetilde\si_k}(
P_k)}\phi_1(i^{1,2}_{P_k})}\cdot\iota_2^*\chi^{-1}_{i_{e_{\si_k}(
P_k)}\phi_2(i^{1,2}_{P_k})}\cdot W^{-1}_{i^{1,2}_{P_k}}
\bigr)^{\vep_k}(q_k)\,.
\qqq
\eroof \noindent The physical content of the above result is
summarised in
\becor\label{cor:preqb-tw}
Let $\,\Bgt\,$ be a string background with target $\,\cM=(M,\txg,\cG
)\,$ and $\cG$-bi-brane $\,\cB=(Q,\iota_\a,\om,\Phi\ \vert\ \a\in\{1
,2\})$,\ and let $\,(\sfP_{\si,\cB|\{(P_k,\vep_k)\}},\Om_{\si,\cB|\{
(P_k,\vep_k)\}})\,$ be the $k$-twisted state space of the non-linear
$\si$-model for network-field configurations $\,(X\,\vert\,\G)\,$ in
string background $\,\Bgt\,$ on world-sheet $\,(\Si,\g)\,$ with
defect quiver $\,\G$.\ Denote by $\,\pr_{\sfL_{Q|\{(P_k,\vep_k)\}}
M}:\sfP_{\si,\cB|\{(P_k,\vep_k)\}}\to\sfL_{Q|\{(P_k,\vep_k)\}}M\,$
the canonical projection from the $k$-twisted state space to the
space $\,\sfL_{Q|\{(P_k,\vep_k)\}}M\,$ from Definition
\ref{def:tw-phspace}. The pre-quantum bundle $\,\pi_{\ceL_{\si,\cB|
\{(P_k,\vep_k)\}}}:\ceL_{\si,\cB|\{(P_k,\vep_k)\}}\to\sfP_{\si,\cB|
\{(P_k,\vep_k)\}}\,$ for the twisted sector of the $\si$-model is
the circle bundle
\qq\nn
\ceL_{\si,\cB|\{(P_k,\vep_k)\}}:=\pr_{\sfL_{Q|\{(P_k,\vep_k)\}}M}^*
\ceL_{(\cG,\cB)|\{(P_k,\vep_k)\}}\ox\bigl(\sfP_{\si,\cB|\{(P_k,
\vep_k)\}}\x\bS^1\bigr)\to \sfP_{\si,\cB|\{(P_k,\vep_k)\}}
\qqq
given by the tensor product of the pullback, along $\,\pr_{\sfL_{Q|
\{(P_k,\vep_k)\}}M}$,\ of the bundle $\,\ceL_{(\cG,\cB)|\{(P_k,
\vep_k)\}}\,$ of Theorem \ref{thm:trans-tw}, and of the trivial
circle bundle $\,\sfP_{\si,\cB|\{(P_k,\vep_k)\}}\x\bS^1\to
\sfP_{\si,\cB|\{(P_k,\vep_k)\}}\,$ with a global connection 1-form
equal to the pullback, along the canonical projection $\,\pr_{\sfT^*
C^\infty(\bS^1_{\{P_k\}},M)}:\sfP_{\si,\cB|\{(P_k,\vep_k)\}}\to
\sfT^*C^\infty(\bS^1_{\{P_k\}},M)$,\ of the canonical 1-form
$\,\theta_{\sfT^*C^\infty(\bS^1_{\{P_k\}},M)}\,$ on $\,\sfT^*
C^\infty(\bS^1_{\{P_k\}},M)$,\ explicited in Proposition
\ref{prop:sympl-form-si-tw}. In particular, given the open cover
$\,\cO_{\sfL_{Q\vert\{(P_k,\vep_k)\}}M}=\{\cO_{\{(P_k,\vep_k)\}\,
\igt}\}_{\igt\in\xcI_{\sfL_{Q\vert\{(P_k,\vep_k)\}}M}}\,$ of
\linebreak $\,\sfL_{Q\vert\{(P_k,\vep_k)\}}M\,$ defined in
Proposition \ref{prop:cover-tw}, local data of
$\,\ceL_{\si,\cB|\{(P_k,\vep_k) \}}\,$ associated with the induced
open cover $\,\cO_{\sfP_{\si,\cB|
\{(P_k,\vep_k)\}}}=\{\pr_{\sfL_{Q|\{(P_k,\vep_k)\}}M}^{-1}(\cO_{\{(
P_k,\vep_k)\}\,\igt})\}_{\igt\in\xcI_{\sfL_{Q\vert\{(P_k,\vep_k)\}}
M}}\,$ of $\,\sfP_{\si,\cB|\{(P_k,\vep_k)\}}\,$ can be expressed in
terms of the local data $\,(E_{\{(P_k,\vep_k)\}\,\igt},G_{\{(P_k,
\vep_k)\}\,\igt\jgt})\,$ of the bundle $\,\ceL_{(\cG,\cB)\vert\{(P_k
,\vep_k)\}}\,$ from Theorem \ref{thm:trans-tw} as
\qq
\theta_{\si,\cB\vert\{(P_k,\vep_k)\}\,\igt}&=&\pr_{\sfT^*C^\infty(
\bS^1_{\{P_k\}},M)}^*\theta_{\sfT^*C^\infty(\bS^1_{\{P_k\}},M)}+
\pr_{\sfL_{Q|\{(P_k,\vep_k)\}}M}^*E_\igt\,,\cr
\label{eq:loc-dat-preq-tw}\\
\g_{\si,\cB\vert\{(p_k,\vep_k)\}\,\igt\jgt}&=&\pr_{\sfL_{Q|\{(P_k,
\vep_k)\}}M}^*G_{\igt\jgt}\,.\nonumber
\qqq
\ecor
\bigskip

Prior to passing to the subsequent sections, in which we exploit the
knowledge, gained heretofore, of the canonical and pre-quantum
structure on the space of states of the $\si$-model in the presence
of defects, we pause to briefly discuss a simple application of our
results, of particular relevance to the study of the concept of
`emergent geometry of string theory'. \brem\label{rem:NCG}
\emph{\textbf{The non-commutative geometry of the bi-brane
world-volume.}} The presence of the defect-line contributions
\qq\nn
\om(q_k)=\om_{AB}(q_k)\,\d X^A\wedge\d X^B\,,\qquad q_k\in Q
\qqq
in \Reqref{eq:Omsi-tw} is a clear-cut indication that the
quantisation of the defect $\si$-model yields a non-commu\-ta\-tive
deformation of the algebra of functions on the bi-brane
world-volume, the latter being generated by the coordinate functions
$\,X^A$.\ This is a bi-brane variant of the long-known phenomenon of
the (gerbe-induced) non-commutativity of the D-brane geometry in the
so-called `stringy r\'egime', first discussed\footnote{For an
earlier account of the phenomenon, exhibiting its rich mathematical
structure in the \emph{closed}-string sector, cf.\
\Rcite{Frohlich:1993es} and Refs.\
\cite{Frohlich:1995mr,Frohlich:1996zc,Frohlich:1998zm}. In the more
recent \Rcite{Recknagel:2006hp}, the gerbe-related description of
the open string in the WZW model was worked out along the lines of
these earlier papers.} in \Rcite{Douglas:1997fm}. The actual form of
the non-commutativity of the quantum position operators depends
strongly on the choice of the quantisation scheme, as indicated in
\Rcite{Seiberg:1999vs}. However, under certain circumstances, one
can get some insight into the matter already on the (semi)classical
level. Indeed, assume that the string background $\,\cM\,$ comes
with a small dimensionless parameter $\,\ep\,$ (derived, e.g., from
a common length scale for $\,M\,$ and $\,Q$), and that there exists
a geometric r\'egime, to be referred to as a \textbf{decoupling
r\'egime} (e.g., a vicinity of a distinguished point in the large
target space)
\qq\nn
X^\mu,X^A=O(\ep^{d_X})\,,\qquad d_X\in\bN
\qqq
in which the target-space metric behaves as
\qq\nn
\txg=O(\ep^{d_\txg})\,,\qquad d_\txg\in\bN\,.
\qqq
The condition of a vanishing Weyl anomaly, mentioned in Remark
\ref{rem:Weyl-anom}, then fixes the scaling behaviour of the gerbe
curvature to be
\qq\nn
\txH=O(\ep^{d_\txH})\,,\qquad d_\txH\geq d_\txg
\qqq
(consistently with the field equations \eqref{eq:field-eqs}), and so
the sum of the first two terms in $\,\Om_{\si,\cB|\{(P_k,\vep_k
)\}}\,$ scales as $\,O(\ep^{d_\txg})$.\ The bi-brane curvature, on
the other hand, need not decrease at the same rate since the
relation
\qq\label{eq:curv-constr}
\iota_1^*\txH-\iota_2^*\txH=\sfd\om
\qqq
that follows from Eqs.\,\eqref{eq:DG-is-H}-\eqref{eq:DPhi-is} and
determines the scaling behaviour of the defect-line terms in the
symplectic structure contains the difference of the pullbacks along
the two bi-brane maps, which may affect the value of the critical
exponent $\,d_\om\,$ in
\qq\nn
\om=O(\ep^{d_\om})\,.
\qqq
Thus, whenever $\,d_\txg-d_\om>0$,\ we may, in the decoupling
r\'egime described, approximate the symplectic structure as
\qq\nn
\Om_{\si,\cB|\{(q_k,\vep_k)\}}=\sum_{k=1}^I\,\vep_k\,\pr_{Q,k}^*\om
\bigl(1+O\bigl(\ep^{d_\txg-d_\om}\bigr)\bigr)\,.
\qqq
From now onwards, we restrict our attention to the case of $\,k=1$,\
with $\,P_1=P,\ q_1=q\,$ and $\,\vep_1=+1$.

If the bi-brane curvature is invertible in the decoupling r\'egime,
with the inverse defining a\linebreak Poisson bivector
\qq\nn
\Pi=\Pi^{AB}\,\p_A\wedge\p_B\,,\qquad\qquad\Pi^{AB}=-\tfrac{1}{4}\,
\bigl(\om^{-1}\bigr)^{AB}\,,
\qqq
we find natural defect observables $\,X^A(P)\equiv X^A\,$
represented by the hamiltonian vector fields
\qq\nn
\xcX_{X^A}=2\Pi^{AB}(X)\,\tfrac{\d \ \ }{\d X^B} \,.
\qqq
Canonical quantisation of their Poisson bracket
\qq\nn
\{X^A,X^B\}_{\Om_{\si,\cB|\{(P,+1)\}}}=2\Pi^{AB}(X)
\qqq
gives a non-commutative algebra of stringy coordinates on $\,Q$,\ as
claimed. By the usual argument, the closedness of $\,\om\,$ ensures
the vanishing of the jacobiator of the Poisson bracket thus defined,
necessary for the associativity of the non-commutative deformation
of the algebra of functions on the bi-brane.\medskip

By way of illustration of the general phenomenon, we treat in some
detail the case of the maximally symmetric WZW $\cGk$-bi-branes from
Example \ref{eg:WZW-def}. The WZW model for a compact Lie group
$\,\txG\,$ comes with a natural parameter $\,\epk=\frac{1}{\sfk}\,$
that sets -- through the Cartan--Killing metric $\,\txg_\sfk\,$ (and
in conjunction with the string tension, which we suppressed in the
present notation) -- the characteristic length scale of the string
background and plays the r\^ole of Planck's constant $\,\hbar\,$ and
a parameter of the non-commutative deformation of the algebra of
functions on $\,\txG\,$ (or a submanifold thereof) determined by the
operator content of the quantised $\si$-model, cf., e.g.,
Refs.\,\cite{Frohlich:1993es} and \cite{Recknagel:2006hp}. Here,
very large but finite values of $\,\sfk\,$ give access to a
semiclassical approximation of the quantum geometry of the WZW
string\footnote{For a proposal of an algebraic description of that
geometry in the quantum r\'egime, consistent with the quantum-group
symmetries of the rational conformal field theory of the WZW model,
cf.\
Refs.\,\cite{Pawelczyk:2001zi,Pawelczyk:2002kd,Pawelczyk:2003nb,Pawelczyk:2005ye}.}.
More specifically, let us write elements of $\,\txG$,\ and hence
also fields of the WZW model, in terms of the canonical (Riemann
normal) coordinates $\,X^A=\epk\,\widetilde X_A\,$ on the group
manifold, suitably rescaled, to wit,
\qq\nn
g=\ee^{\epk\,\widetilde X^A\,t_A}\,,
\qqq
and subsequently pass to the decoupling r\'egime
\qq\nn
\widetilde X^A=O(1)\,,\qquad\qquad\epk\ll 1
\qqq
of \Rcite{Alekseev:1999bs}. Having thus restricted our analysis to
world-sheets embedded in an immediate vicinity of the group unit in
a large group manifold, we readily establish the equalities
\qq\nn
d_X=1\,,\qquad\qquad d_{\txg_\sfk}=1\,,\quad\qquad d_{\txH_\sfk}=2
\,,
\qqq
and -- for $\,\la\in\faff{\ggt}\,$ small and for values of the
$\cGk$-bi-brane field $\,(g,h_\la)\,$ restricted to a small
neighbourhood of $\,(e,e)\in\txG\x\txG\,$ --
\qq\nn
\om^\p_{\sfk,\la}=O(1)\,,\qquad\qquad\om_{\sfk,\la}=-\pr_2^*
\om^\p_{\sfk,\la}+O(\epk)\,.
\qqq
This follows straightforwardly from the relations
\qq\nn
\theta_L=O(\epk)\,,\qquad\qquad\id_\ggt-\Ad_g=-\epk\,
\ad_{\widetilde X}+O(\epk^2)\,.
\qqq
Consequently, the deformation of the commutative algebra of
functions on both the boundary and the non-boundary maximally
symmetric WZW $\cGk$-bi-brane, as encoded by the decoupling limit of
the respective symplectic forms, is determined by the properties of
the 2-form
\qq\nn
\om_{\sfk,\la}^\p(\widetilde X)=-\tfrac{1}{8\pi}\,\tr_\ggt\bigl(
t_A\,\ad^{-1}_{\widetilde X}t_B\bigr)\,\d\widetilde X^A\wedge\d
\widetilde X^B+O(\epk)\,,\qquad\widetilde X=\widetilde X^A\,t_A
\qqq
which coincides with the Kirillov--Kostant--Souriau symplectic form
on the coadjoint orbit $\,\xcO_\la\cong\txG/\txG_\la\,$ (for
$\,\txG_\la\,$ the $\Ad_\cdot$-stabiliser of $\,\la\,$ in $\,\txG$)
of Refs.\,\cite{Kostant:1970,Souriau:1970,Kirillov:1975}.
Equivalently, the deformation is characterised by the associated
Poisson bivector
\qq\nn
\Pi^\p_{\sfk,\la}(\widetilde X)=4\pi\,f_{ABC}\,\widetilde X^A\,
\tfrac{\d\ \ }{\d\widetilde X^B}\wedge\tfrac{\d\ \ }{\d\widetilde
X^C}+O(\epk)\,.
\qqq
Taking the latter as the germ of a deformation quantisation of the
smooth geometry of conjugacy classes $\,\xcC_\la\,$ close to the
group unit leads to the emergence of the so-called fuzzy conjugacy
classes, analogous to the fuzzy sphere of
Refs.\,\cite{Hoppe:1982PhD,Madore:1992}. These are precisely the
non-commutative geometries emerging from perturbative calculations
of the quantised WZW model. They were first explored in the present
context in
Refs.\,\cite{Alekseev:1999bs,Alekseev:2000fd,Alekseev:2000wg}, cf.\
also \Rcite{Recknagel:2006hp} for a gerbe-related discussion based
on the spectral data of the supersymmetric extension of the boundary
WZW model. \erem

\section{State-space isotropics from defects, and $\si$-model
dualities}\label{sec:def-as-iso}

Now that we have developed a symplectic formalism for the
description of the two-dimensional field theory in hand, we may
apply it to study defects. Thus, motivated by the discussion,
presented in
Refs.\,\cite{Frohlich:2004ef,Frohlich:2006ch,Bachas:2008jd}, of the
r\^ole that defects play in mediating dualities of the underlying
two-dimensional field theory, and also by the study of chosen
examples of dualities in the gerbe-theoretic context in
Refs.\,\cite{Schweigert:2007wd,Runkel:2008gr,Sarkissian:2008dq}, we
seek to establish an appropriate rigorous result within the
canonical framework, to wit, we want to describe the symplectic
relations within the untwisted state space $\,(\sfP_{\si,\emptyset}=
\sfT^*\sfL M,\Om_{\si,\emptyset})\,$ of the $\si$-model inhabiting
two adjacent patches $\,\wp_\a,\ \a\in\{1,2\}\,$ separated by a
space-like connected component $\,\ell\cong\bS^1\,$ of the defect
quiver $\,\G\,$ that are induced by the intermediary $\cG$-bi-brane
structure $\,\cB$.\ To this end, using independence of $\,\Om_{\si,
\emptyset}\,$ over each of the two patches of the choice of the
Cauchy contour used to define it, we push the respective Cauchy
contours, $\,C_1\,$ and $\,C_2$,\ to $\,\ell$.\ Following this
simple prescription, we find
\berop\label{prop:DGC-as-iso}
Let $\,\Bgt\,$ be a string background with target $\,\cM=(M,\txg,\cG
)\,$ and $\cG$-bi-brane $\,\cB=(Q,\iota_\a,\om,\Phi\ \vert\ \a\in\{
1,2\})$,\ and let $\,(\sfP_{\si,\emptyset}=\sfT^*\sfL
M,\Om_{\si,\emptyset})\,$ be the untwisted state space of the
non-linear $\si$-model for network-field configurations
$\,(X\,\vert\,\G)\,$ in string background $\,\Bgt\,$ on world-sheet
$\,(\Si,\g)\,$ with defect quiver $\,\G$.\ Consider the symplectic
structure on the product space $\,\sfP_{\si,\emptyset}\x\sfP_{\si,
\emptyset}\equiv\sfP_{\si,\emptyset}^{\x 2}\,$ determined by the
`difference' symplectic form
\qq\nn
\Om_{\si,\emptyset}^-=\pr_1^*\Om_{\si,\emptyset}-\pr_2^*\Om_{\si,
\emptyset}\,.
\qqq
The $\cG$-bi-brane $\,\cB\,$ together with the Defect Gluing
Condition \eqref{eq:DGC} canonically defines an isotropic
submanifold in $\,(\sfP_{\si,\emptyset}^{\x 2},\Om_{\si,
\emptyset}^-)$,\ given by
\qq\nn
\Igt_\si(\cB)&=&\{\ (\psi_1,\psi_2)\in\sfP_\si^{\x 2}\,,\quad
\psi_\a=( X_\a,\sfp_\a)\,,\ \a\in\{1,2\} \quad\vert\quad (X_1,X_2)
\in(\iota_1 \x\iota_2)(\sfL Q)\cr\cr
&&\hspace{.5cm}\land\quad\exists_{X\in(\iota_1\x\iota_2)^{-1}\{(X_1,
X_2)\}}\ :\ {\rm DGC}_\cB (\psi_1,\psi_2,X)=0 \ \}
\qqq
in terms of Cauchy data $\,\psi_\a$.\ The latter subspace is a
fibration over the free-loop space $\,\sfL Q$,\ and we shall
identify it with the corresponding subspace in $\,\sfP_{\si,
\emptyset}^{\x 2}\x\sfL Q\,$ in what follows.
\eerop
\beroof
Take a pair of states $\,(\psi_1,\psi_2)\in\Igt_\si(\cB)\,$ with
$\,(X_1,X_2)=\bigl(\iota_1(X),\iota_2(X)\bigr)$,\ satisfying the DGC
\eqref{eq:DGC}, i.e.
\qq\label{eq:DGC-coord}
\sfp_{2\,\mu}\,\xcW^\mu_2-\sfp_{1\,\mu}\,\xcW^\mu_1=-2(X_*\widehat t
)^A\,\om_{AB}\,\xcW^B\,,
\qqq
for $\,\xcW=\xcW^A\,\tfrac{\d\ \ }{\d X^A}\,$ an arbitrary vector
field from $\,\G(\sfT Q)\,$ restricted to $\,X$,\ the latter being
modelled on $\,\bS^1\,$ with a normalised tangent vector field
$\,\widehat t$,\ and for $\,\xcW_\a=\iota_{\a\,*}\xcW$.\ We want to
consider the distinguished subspace $\,\Tgt\Igt_\si(\cB)\,$ of
$\,\G(\sfT\sfP_\si^{\x 2})\,$ over $\,\Igt_\si(\cB)\,$ spanned by
vector fields
\qq\nn
\widetilde\xcV=\widetilde\xcV_1\oplus\widetilde\xcV_2\,,\qquad
\qquad\widetilde\xcV_\a=\xcV_\a^\mu\,\tfrac{\d\ \ }{\d X_\a^\mu}+
\xcP_{\a\,\mu}\,\tfrac{\d\ \ }{\d\sfp_{\a\,\mu}}
\qqq
describing tangential deformations $\,\xcV_\a\equiv\xcV_\a^\mu\,
\frac{\d\ \ }{\d X_\a^\mu}=\iota_{\a\,*}(\xcV^A\,\frac{\d\ \ }{\d
X^A})\,$ of the $\,X_\a \,$ induced by deformations $\,\xcV=\xcV^A\,
\frac{\d\ \ }{\d X^A}\,$ of the defect loop $\,X$,\ and augmented by
deformations $\,\xcP_{\a\, \mu}\,\frac{\d\ \ }{\d\sfp_{\a\,\mu}}\,$
of the normal covector fields satisfying a linearised version of
\Reqref{eq:DGC-coord},
\qq
&&\sum_{\a=1,2}\,(-1)^\a\,\bigl[\xcP_{\a\,\mu}\,\xcW_\a^\mu+
\sfp_{\a\,\mu}\,\xcV^A\,\bigl(\xcW^B\,\p_A\p_B\iota_\a^\mu+\p_A
\xcW^B\,\p_B\iota_\a^\mu\bigr)\bigr]\cr\cr
&=&-2(X_*\widehat t)^A\,\bigl(\xcV^B\,\bigl(\p_B\om_{AC}\,\xcW^C+
\om_{AC}\,\p_B\xcW^C\bigr)+\om_{BC}\,\xcW^C\,\p_A\xcV^B\bigr)\,,
\label{eq:lin-DGC}
\qqq
where $\,\p_A=\frac{\p\ \ }{\p X^A}\,$ and where all fields are
implicitly functions on $\,\bS^1$.\ Clearly, $\,\xcV\,$ is a vector
field on $\,Q\,$ and the pair $\,\xcW_\a,\ \a\in\{1,2\}\,$ can be
completed to another admissible deformation of the $\,X_\a$,\
\qq\nn
\widetilde\xcW=\widetilde\xcW_1\oplus\widetilde\xcW_2\,,\qquad
\qquad\widetilde\xcW_\a=\xcW_\a^\mu\,\tfrac{\d\ \ }{\d X_\a^\mu}+
\xcQ_{\a\,\mu}\,\tfrac{\d\ \ }{\d\sfp_{\a\,\mu}}\,,
\qqq
by the addition of deformations $\,\xcQ_{\a\,\mu}\,\frac{\d\ \ }{\d
\sfp_{\a\,\mu}}\,$ of the normal covector fields satisfying an
analogon of \Reqref{eq:lin-DGC}. Upon subtracting the resulting
equation from \Reqref{eq:lin-DGC} and using \Reqref{eq:DGC} in
conjunction with the closedness of $\,\G(\sfT Q)\,$ under the Lie
bracket, we obtain the relation
\qq\label{eq:lin-DGC-asym}\qquad\qquad
\sum_{\a=1,2}\,(-1)^\a\,\bigl(\xcQ_{\a\,\mu}\,\xcV_\a^\mu-\xcP_{\a
\,\mu}\,\xcW_\a^\mu\bigr)=-\xcW\con\xcV\con X_*\widehat t\con\sfd
\om+X_*\widehat t\con\sfd\bigl(\xcW\con\xcV\con\om\bigr)\,.
\qqq
We may now evaluate $\,\Om_{\si,\emptyset}^-$,\ taken at an extremal
section $\,(\Psi_\si^1,\Psi_\si^2)\,$ determined by the Cauchy data
$\,(\psi_1,\psi_2)\in\Igt_\si(\cB;X)$,\ on a pair $\,(\widetilde V,
\widetilde W)\,$ of vectors obtained by evaluating the pair $\,(
\widetilde\xcV,\widetilde\xcW)\,$ of vector fields from $\,\Tgt
\Igt_\si(\cB)\,$ at $\,(\psi_1,\psi_2)$.\ Putting
Eqs.\,\eqref{eq:Omsi-untw} and \eqref{eq:lin-DGC-asym} together and
employing \Reqref{eq:curv-constr}, we readily verify the desired
identity
\qq\nn
\Om_{\si,\emptyset}^-[(\psi_1,\psi_2)](\widetilde V,\widetilde
W)=0\,.
\qqq
It leads us to conclude that the defect defines an isotropic
subspace $\,\Tgt\Igt_\si(\cB)\,$ within $\,\G(\sfT \sfP_\si^{\x
2}\vert_{\Igt_\si(\cB)})\,$ over the distinguished submanifold
$\,\Igt_\si(\cB)$. \eroof\medskip

Whenever the subspace $\,\Igt_\si(\cB)\,$ with an isotropic tangent
$\,\Tgt\Igt_\si(\cB)\,$ is actually a graph, the $\cG$-bi-brane
defines a symplectomorphism of the untwisted state space\footnote{A
symplectomorphism can be viewed as a maximal isotropic in $\,\sfT
\sfP_\si^{\x 2}$, cf., e.g., \Rcite{Woodhouse:1992de}.}, which can
be understood as an identification between states, chosen
arbitrarily, incident on the defect carrying the data of the
$\cG$-bi-brane $\,\cB\,$ from one side of the defect line and those
emerging from it on the other side. However, it is to be kept in
mind that states on either side of the defect line carry charges of
the symmetries of the untwisted $\si$-model, notably, the energy
and, in the case of extended (internal) symmetry, additional charges
to which the symmetry currents couple. Thus, for the defect to
describe a duality\footnote{Note that the introduction of the
bi-brane, whose world-volume is a priori not related to the target
space, allows for a unified treatment of \emph{symmetries} which do
not leave a connected component of the target space (and solely put
in correspondence extremal sections that map the world-sheet into
different regions thereof) and proper \emph{dualities} which act
between different connected components of the target space,
oftentimes of inequivalent topology. A notable example of the latter
type is the T-duality between principal torus bundles, cf.\ Example
\ref{ex:Tdual}.} of the untwisted theory, we should demand that the
charges of the two states identified with one another at the defect
match, or, equivalently, that the corresponding symmetry currents be
continuous at the defect. Whereas, in concrete examples, one may
wish to impose weaker correspondence constraints, allowing for a
partial breakdown of some internal symmetries, the gauge symmetry of
the $\si$-model, that is the conformal symmetry, should always be
preserved. As was shown in \Rcite{Runkel:2008gr}, one linear
combination of the conformal currents, to wit, the one that
generates diffeomorphisms preserving the defect line,
$\,T_{++}-T_{--},\ T_{\pm\pm}=\txg_{\mu \nu}(X)\,\p_\pm
X^\mu\,\p_\pm X^\nu\,$ (in the adapted coordinates), is
automatically preserved at the defect by virtue of the DGC -- this
is the content of Theorem \ref{thm:conf-def}. The last property
identifies the defects considered as \textbf{conformal} in the sense
of \Rcite{Oshikawa:1996dj}. For a space-like defect line, which is
what we have been considering in the present section, it is the
other linear combination,
\qq\label{eq:ham-dens-si}
T_{++}+T_{--}\equiv\tfrac{1}{2}\,\bigl(\bigl(\txg^{-1}\bigr)^{\mu
\nu}(X)\,\sfp_\mu(X)\,\sfp_\nu(X)+\txg_{\mu\nu}(X)\,(X_*\widehat
t)^\mu\,(X_*\widehat t)^\nu\bigr)=\ceH_\si\,,
\qqq
that gives the hamiltonian density $\,\ceH_\si\,$ of the
$\si$-model. It generates conformal transformations on the
world-sheet which deform the defect line, and hence the continuity
of $\,\ceH_\si\,$ at the latter is related to the extendibility of
the defect, as made precise in
\bedef
Let $\,\Bgt\,$ be a string background with target $\,\cM=(M,\txg,\cG
)\,$ and $\cG$-bi-brane $\,\cB=(Q,\iota_\a,\om,\Phi\ \vert\ \a\in\{1
,2\})$,\ and let $\,(X\,\vert\,\G)\,$ be a network-field
configuration in string background $\,\Bgt\,$ on world-sheet
$\,(\Si,\g)\,$ with defect quiver $\,\G$.\ Denote by $\,U\,$ a
tubular neighbourhood of an edge $\,\ell\,$ of $\,\G\,$ within
$\,\Si$,\ with the property $\,U\cap\G=\ell$.\ The neighbourhood
$\,U\,$ is split by the oriented line $\,\ell\,$ into subsets
$\,U_\a,\ \a\in\{1,2\}\,$ as in Definition \ref{def:net-field}. An
\textbf{extension $\,\widehat X\,$ of network-field configuration
$\,(X\,\vert\,\G)\,$ on neighbourhood $\,U\,$ of defect line
$\,\ell\,$} is a map $\,\widehat X:U\to Q$,\ such that
\qq\nn
\widehat X\vert_\ell=X\,,\qquad\qquad\iota_\a\circ\widehat
X\vert_{U_\a}=X\vert_{U_\a}\,,\qquad\qquad
\qqq
and such that the relation
\qq\nn
\iota_1^*\txg\bigl(\widehat X(p)\bigr)\bigl(\widehat X_*\widehat
u^\perp,\cdot\bigr)-\iota_2^*\txg\bigl(\widehat X(p)\bigr)\bigl(
\widehat X_*\widehat u^\perp,\cdot\bigr)-\widehat X_*\widehat u\con
\om\bigl(\widehat X(p)\bigr)=0\,,\qquad\qquad\widehat u^\perp=
\g^{-1}\bigl(\widehat u\con\Vol(\Si,\g),\cdot\bigr)
\qqq
is satisfied at any point $\,p\in U\,$ and for an arbitrary vector
$\,\widehat u\in\sfT_p\Si$.\ A defect $\,\ell\,$ of a network field
configuration that admits an extension on a neighbourhood of
$\,\ell\,$ shall be termed \textbf{extendible}.\exdef \noindent The
less restrictive condition of continuity of the conformal current
gives rise to
\bedef\label{def:def-top}
Let $\,\Bgt\,$ be a string background with target $\,\cM=(M,\txg,\cG
)$,\ and let $\,\G\,$ be a defect quiver embedded in a world-sheet
$\,\Si$.\ Consider the non-linear $\si$-model for network-field
configurations $\,(X\,\vert\,\G)\,$ in string background $\,\Bgt\,$
on world-sheet $\,(\Si,\g)\,$ with defect quiver $\,\G$.\ Choose a
local coordinate system $\,\{\si^a \}^{a\in\{1,2\}}\,$ in the
neighbourhood of a point in $\,\G$.\ The defect $\,\G\,$ is called
\textbf{topological} iff the conformal current $\,T$,\ with local
components
\qq\nn
T^{ab}=\tfrac{2}{\sqrt{\det\,\g}}\,\tfrac{\d S_\si}{\d\g_{ab}}\,,
\qqq
is continuous across $\,\G$. \exdef \brem The notion of
topologicality can be regarded as a classical counterpart of the
quantum concept introduced in \Rcite{Petkova:2000ip}. \erem
Extendible defects have the desired property of topologicality, as
stated in
\bethe\cite[Sec.\,2.9]{Runkel:2008gr}
The non-linear $\si$-model of Definition \ref{def:sigmod} for
network-field configurations $\,(\G ,X)\,$ in string background
$\,\Bgt\,$ on world-sheet $\,(\Si,\g)\,$ with defect quiver $\,\G\,$
composed of extendible defects is invariant with respect to
arbitrary (gauge) transformations
\qq\nn
&X\mapsto X\circ D\,,\qquad\qquad\g\mapsto D^*\g\,,\qquad D\in
\Diff^+(\Si)\,,&\cr\cr
&\g\mapsto\ee^{2w}\cdot\g\,,\qquad\ee^{2w}\in{\rm Weyl}(\g)&
\qqq
from the semidirect product $\,\Diff^+(\Si)\lx{\rm Weyl}(\g)\,$ of
the group $\,\Diff^+(\Si)\,$ of (orientation-preserving)
diffeomorphisms of $\,\Si\,$ with the group $\,{\rm Weyl}(\g)\,$ of
Weyl rescalings of the metric $\,\g$.\ All components of the
conformal current are continuous across the defect lines of $\,\G$,\
that is the defect is \textbf{topological} in the sense of
Definition \ref{def:def-top}. \ethe \brem From the point of view of
the categorial quantisation of the $\si$-model in the presence of
defects, as discussed, e.g., in \Rcite{Runkel:2008gr}, it is natural
to expect a topological defect to be deformable, and hence
necessarily extendible. However, the statement in the quantum theory
is usually formulated in terms of correlation functions assigned to
(decorated) world-sheets with embedded defect quivers, and so -- in
the present setting -- the issue of finding its proper classical
counterpart is obscured by aspects of the quantisation procedure
such as the choice of the renormalisation scheme that affects the
field-theoretic functionals entering the DGC, cf., e.g.,
Refs.\,\cite{Bachas:2004sy,Alekseev:2007in,Bachas:2009mc} for an
illustration. As we are not addressing here the issue of
quantisation beyond the construction of the pre-quantum bundle, we
shall restrict ourselves to topological rather than extendible
defects in what follows. \erem \noindent Thus, it is amidst
topological defects that we should look for those that describe
dualities of the untwisted $\si$-model. Before we do that, however,
let us make the very notion of duality precise, using the various
field-theoretic constructs introduced hitherto.
\bedef\label{def:pqsymm}
Let $\,\Bgt\,$ be a string background with target $\,\cM=(M,\txg,\cG
)$,\ and let $\,(\sfP_{\si,\emptyset}=\sfT^*\sfL M,\Om_{\si,
\emptyset})\,$ be the untwisted state space of the non-linear
$\si$-model for network-field configurations $\,(X\,\vert\,\G )\,$
in string background $\,\Bgt\,$ on world-sheet $\,(\Si,\g)\,$ with
defect quiver $\,\G$.\ Furthermore, let $\,\pi_{\ceL_{\si,
\emptyset}}:\ceL_{\si,\emptyset}\to\sfP_{\si,\emptyset}\,$ be the
pre-quantum bundle for the untwisted sector of the $\si$-model,
constructed in Corollary \ref{cor:preqb-untw}. A \textbf{pre-quantum
duality of the untwisted sector of the non-linear $\si$-model for
network-field configurations $\,(X\,\vert\,\G)\,$ in string
background $\,\Bgt\,$ on world-sheet $\,(\Si,\g)\,$ with defect
quiver $\,\G\,$} is a pair $\,(\Igt_\si, \Dgt_\si)\,$ which consists
of
\bit
\item a graph $\,\Igt_\si\subset\sfP_{\si,\emptyset}^{\x 2}$,\
isotropic \wrt the `difference' symplectic form $\,\Om_{\si,
\emptyset}^-\,$ of Proposition \ref{prop:DGC-as-iso}, and having the
property that the difference
\qq\label{eq:Hsi-}
\ceH_\si^-=\pr_1^*\ceH_\si-\pr_2^*\ceH_\si
\qqq
of the pullbacks, along the canonical projections $\,\pr_\a:
\sfP_{\si,\emptyset}\x\sfP_{\si,\emptyset}\to \sfP_{\si,\emptyset},\
\a\in\{1,2\}\,$,\ of the hamiltonian density $\,\ceH_\si\,$ of the
$\si$-model, as given in \Reqref{eq:ham-dens-si}, vanishes
identically on restriction to $\,\Igt_\si$;
\item a bundle isomorphism
\qq\label{eq:Dsi}
\Dgt_\si\ :\ \pr_1^*\ceL_{\si,\emptyset}\vert_{\Igt_\si}
\xrightarrow{\cong}\pr_2^*\ceL_{\si,\emptyset}\vert_{\Igt_\si}\,,
\qqq
between the restrictions to $\,\Igt_\si\,$ of the pullbacks of
$\,\ceL_{\si,\emptyset}\,$ along the canonical projections
$\,\pr_\a$.
\eit\exdef
\brem Pre-quantum dualities of the $\si$-model which are consistent
with a given choice of the polarisation of the pre-quantum bundle
defining the Hilbert space of the theory give rise to bona fide
dualities of the quantised $\si$-model. \erem \noindent The first
relation between conformal defects and $\si$-model dualities is
established in the following
\bethe\label{thm:def-dual}
Let $\,\Bgt\,$ be a string background with target $\,\cM=(M,\txg,\cG
)\,$ and $\cG$-bi-brane $\,\cB=\bigl(Q,\iota_\a,\om,\Phi\ \vert\ \a
\in\{ 1,2\}\bigr)$,\ and consider the non-linear $\si$-model for
network-field configurations $\,(X\,\vert\,\G)\,$ in string
background $\,\Bgt\,$ on world-sheet $\,(\Si,\g)\,$ with defect
quiver $\,\G$.\ The $\cG$-bi-brane $\,\cB\,$ together with the
Defect Gluing Condition \eqref{eq:DGC} canonically defines a
pre-quantum duality of the untwisted sector of the $\si$-model iff
the following conditions are satisfied:
\bit
\item[i)] both loop-space maps
\qq\nn
\widetilde\iota_\a\ :\ \sfL Q\to\sfL M\ :\ X\mapsto\iota_\a\circ X
\,,\quad\a\in\{1,2\}\,,
\qqq
induced by the $\cG$-bi-brane maps $\,\iota_\a\,$ (and hence also
the latter), are surjective submersions onto connected components of
$\,\sfL M$;
\item[ii)] let $\,s_{\igt_\a,\a}:\cO_{\igt_\a}\to\sfL Q\,$ denote
smooth local sections of $\,\widetilde\iota_\a\,$ over elements
$\,\cO_{\igt_\a}\,$ of an open cover $\,\cO_{\sfL M}=\{\cO_\igt
\}_{\igt\in\xcI_{\sfL M}}\,$ of $\,\sfL M$,\ e.g., the one from
Proposition \ref{prop:cover-untw} induced from a sufficiently fine
open cover of $\,M$;\ given a pair $\,s_{\igt_\a,\a}^n,\ n\in\{1,
2\}\,$ of such sections satisfying $\,\widetilde\iota_\a\circ
s_{\igt_\a,\a}^1=\widetilde\iota_\a\circ s_{\igt_\a,\a}^2$,\ the
relations
\qq\label{eq:iotasI-iotasJ-tan}
\widetilde\iota_{3-\a}\circ s_{\igt_\a,\a}^1=\widetilde\iota_{3-\a}
\circ s_{\igt_\a,\a}^2\,,\qquad\qquad\widetilde\iota_{3-\a\,*}\circ
s_{\igt_\a,\a\,*}^1=\widetilde\iota_{3-\a\,*}\circ s_{\igt_\a,
\a\,*}^2
\qqq
hold true;
\item[iii)] let $\,s_{i_\a,\a}:\cO_{i_\a}\to Q\,$ denote smooth
local sections of $\,\iota_\a\,$ over elements $\,\cO_{i_\a}\,$ of
an open cover $\,\cO_M=\{\cO_i\}_{i\in\xcI_M}\,$ of $\,M\,$
compatible, in an obvious manner, with the $\,s_{\igt_\a,\a}\,$
introduced previously; given a pair $\,s_{i_\a,\a}^n,\ n\in\{1,2
\}\,$ of such sections associated with a pair $\,s_{\igt_\a,\a}^n,\
n\in\{1,2\}\,$ as above, the relations
\qq\label{eq:sIom-sJom}
s^{1\,*}_{i_\a,\a}\om=s^{2\,*}_{i_\a,\a}\om
\qqq
obtain;
\item[iv)] for arbitrary $\,((X_1,\sfp_1),(X_2,\sfp_2))\in
\cO_{\igt_1}\x\cO_{\igt_2}\subset\Igt_\si\,$ and, in the notation of
the preceding points, for any $\,(s_{\igt_1,1},s_{\igt_2,2})\,$ such
that $\,s_{\igt_1,1}(X_1)=s_{\igt_2,2}(X_2)$,\ the following
identity is satisfied:
\qq\nn
\bigl(\txg^{-1}\bigr)^{\mu\nu}\bigl(\widetilde\iota_2\circ s_{i_1,
1}(X_1)\bigr)\,\tfrac{\p s_{\igt_2,2}^A}{\p X_1^\mu}\,\tfrac{\p
s_{\igt_2,2}^B}{\p X_1^\nu}\,\bigl[\sfp_{1\,\rho}\,\tfrac{\p
\iota_1^\rho}{\p X^A}+2\om_{AC}\bigl(s_{\igt_1,1}(X_1)\bigr)\,
\tfrac{\p s_{\igt_1,1}^C}{\p X_1^\rho}\,(X_{1\,*}\widehat t)^\rho
\bigr]\cr\cr \cdot\bigl[\sfp_{1\,\si}\,\tfrac{\p\iota_1^\si}{\p X^B}
+2\om_{BD}\bigl(s_{\igt_1,1}(X_1)\bigr)\,\tfrac{\p s_{\igt_1,
1}^D}{\p X_1^\si}\,(X_{1\,*}\widehat t)^\si\bigr]+\txg_{\mu\nu}
\bigl(\widetilde\iota_2\circ s_{i_1,1}(X_1)\bigr)\,\tfrac{\p
\iota_2^\mu}{\p X^A}\,\tfrac{\p\iota_2^\nu}{\p X^B}\,\tfrac{\p
s_{\igt_1,1}^A}{\p X_1^\rho}\,\tfrac{\p s_{\igt_1,1}^B}{\p X_1^\si}
\,(X_{1\,*}\widehat t)^\rho\,(X_{1\,*}\widehat t)^\si\cr\cr
=\bigl(\txg^{-1}\bigr)^{\mu\nu}(X_1)\,\sfp_{1\,\mu}\,\sfp_{1\,\nu}+
\txg_{\mu\nu}(X_1)\,(X_{1\,*}\widehat t)^\mu\,(X_{1\,*}\widehat t
)^\nu\,.
\qqq
\eit
\ethe
\beroof
Recall that, in virtue of Proposition \ref{prop:DGC-as-iso},
$\,\cB\,$ canonically defines an isotropic submanifold $\,\Igt_\si(
\cB)\subset\sfP_{\si,\emptyset}^{\x 2}$.\ Choose good open covers
$\,\cO_M=\{\cO^M_i\}_{i\in\xcI_M}\,$ and $\,\cO_Q=\{\cO^Q_i\}_{i\in
\xcI_Q}\,$ such that there exist \v Cech extensions $\,(\iota_\a,
\phi_\a)\,$ of the $\cG$-bi-brane maps and an open cover of
$\,\Igt_\si(\cB)\,$ is induced, as in Proposition
\ref{prop:cover-untw}, with triangulations of both loops in $\,(
\psi_1,\psi_2)\in\Igt_\si(\cB)\,$ coming from a triangulation of the
parent loop $\,X\in\sfL Q$.\ Thus, in particular, at each point
$\,(\psi_1,\psi_2)\in\Igt_\si(\cB)$,\ we have a common triangulation
$\,\triangle(\bS^1)$,\ with edges $\,e\,$ and vertices $\,v$,\ of
the unit circle parameterising $\,X\,$ and $\,X_\a=\iota_\a\circ
X$,\ and, for each element $\,f \in\triangle(\bS^1)$,\ a triple of
indices $\,(i_f^1,i_f^2,i_f^{1,2})\in\xcI_M\x\xcI_M\x\xcI_Q$,\
related as per
\qq\label{eq:comm-triang-index}
i_f^\a=\phi_\a(i_f^{1,2})\,.
\qqq
Next, fix a local presentation of $\,\Bgt\,$ associated with this
choice of covers as in Definition \ref{def:loco}. It is then a
matter of a simple calculation to verify that the local data
$\,(\theta_{\si,\emptyset\,\igt},\g_{\si,\emptyset\,\igt\jgt})\,$ of
the pre-quantum bundle $\,\ceL_{\si,\emptyset}\,$ associated -- as
in Corollary \ref{cor:preqb-untw} -- with the open cover of a
cartesian factor in $\,\Igt_\si(\cB)\,$ (induced as in Proposition
\ref{prop:cover-untw}) satisfy the identities
\qq
\pr_2^*\theta_{\si,\emptyset\,\igt^2}-\pr_1^*\theta_{\si,\emptyset
\,\igt^1}&=&-\sfi\,\sfd\log f_{\si,\cB\,(\igt^1,\igt^2)}\,,
\label{eq:f12-as-iso}\\\cr
\pr_2^*\g_{\si,\emptyset\,\igt^2\jgt^2}&=&f_{\si,\cB\,(\igt^1,
\igt^2)}\cdot\pr_1^*\g_{\si,\emptyset\,\igt^1\jgt^1}\cdot f_{\si,
\cB\,(\jgt^1,\jgt^2)}^{-1}\,,
\qqq
written in terms of the canonical projections $\,\pr_\a:\Igt_\si(\cB
)\to\sfP_{\si,\emptyset},\ \a\in\{1,2\}\,$ and of the $\uj$-valued
functionals
\qq\label{eq:duality-iso-bib}\qquad\qquad
f_{\si,\cB\,(\igt^1,\igt^2)}[(\psi_1,\psi_2)]=\prod_{e\in\triangle(
\bS^1)}\,\ee^{\sfi\,\int_e\,X_e^*P_{i^{1,2}_e}}\cdot\prod_{v\in
\triangle(\bS^1)}\,X^*K^{-1}_{i_{e_+(v)}^{1,2}i_{e_-(v)}^{1,2}}(v)
\qqq
on $\,\cO^*_{\igt^1}\x\cO^*_{\igt^2}\subset\Igt_\si(\cB)$,\ where
$\,\cO^*_{\igt^\a}=\pi_{\sfT^*\sfL M}^{-1}\bigl(\cO_{\igt^\a}
\bigr)$.\ Hence, the $\,f_{\si,\cB\,\igt^{1,2}}\,$ can be identified
with local data of an isomorphism $\,\Dgt_\si(\cB)\,$ from
Definition \ref{def:pqsymm}. It remains to establish the conditions
under which the isotropic submanifold $\,\Igt_\si(\cB)\subset
\sfP_{\si,\emptyset}^{\x 2}\,$ becomes a graph.

For this to be the case, it is necessary that the two maps
$\,\widetilde\iota_\a\,$ be surjective so that any loop in $\,M\,$
can be descended from a loop in $\,Q$.\void{More specifically, it
may happen that the target space $\,M\,$ is a disjoint sum of
several manifolds, with the patch embedding map $\,X_{|\a}\,$
sending $\,\wp_\a\,$ into a separate connected component $\,M_\a\,$
of $\,M\,$ by continuity. In this case, we should demand that
$\,\iota_\a\,$ covers all of $\,\sfL M_\a\,$ as we vary the parent
loop $\,X\vert_\ell$.} Having thus established a correspondence,
fibred over $\,Q$,\ between loops in either cartesian factor of
$\,\sfP_{\si,\emptyset}^{\x 2}$,\ or -- in the world-sheet picture
-- on either side of the defect line, we still have to require that
upon choosing a specific parent loop $\,X\in\sfL Q\,$ and thus
picking up a pair $\,(X_1,X_2)\,$ of loops from $\,\sfL M\,$ and
putting them in correspondence, and upon determining either of the
loop momenta, $\,\sfp_1\,$ or $\,\sfp_2$,\ the other loop momentum
is already fixed uniquely by the DGC. Inspection of the latter,
\qq\nn
\sfp_1\circ\widetilde\iota_{1\,*}-\sfp_2\circ\widetilde\iota_{2\,*}
-X_*\widehat t\con\om=0\,,
\qqq
reveals that for this to hold, also the tangent maps $\,\widetilde
\iota_{\a\,*}\,$ must admit local right inverses. That is,
altogether, the $\,\widetilde\iota_\a\,$ should be surjective
submersions, with smooth local sections $\,s_{\igt_\a,\a}:
\cO_{\igt_\a}\to\sfL Q\,$ satisfying the identities
\qq\nn
\widetilde\iota_\a\circ s_{\igt_\a,\a}=\id_{\cO_{\igt_\a}}\,,\qquad
\qquad\widetilde\iota_{\a\,*}\circ s_{\igt_\a,\a\,*}=\id_{\G(\sfT\sfL
M\vert_{\cO_{\igt_\a}})}\,.
\qqq
Indeed, for $\,X_\a\in\cO_{\igt_\a},\ X\in(\iota_1\x\iota_2)^{-1}\{(
X_1,X_2)\}\,$ and a pair of sections $\,(s_{\igt_1,1},s_{\igt_2,2}
)\,$ such that
\qq\label{eq:X-as-sec}
X=s_{\igt_\a,\a}(X_\a)\,,
\qqq
there arise functional relations
\qq\label{eq:Xal-of-Xal}
X_2(X_1)=\widetilde\iota_2\circ s_{\igt_1,1}(X_1)\,,\qquad\qquad
X_1(X_2)=\widetilde\iota_1\circ s_{\igt_2,2}(X_2)
\qqq
between the loop coordinates, alongside the functional relations
\qq
\sfp_2(\sfp_1,X_1)&=&\sfp_1\circ\widetilde\iota_{1\,*}\circ
s_{\igt_2,2\,*}- \bigl(X_*\widehat t\con\om(X)\bigr)\circ s_{\igt_2,
2\,*}\,,\cr\label{eq:pal-of-pal}&&\\
\sfp_1(\sfp_2,X_2)&=&\sfp_2\circ\widetilde\iota_{2\,*}\circ
s_{\igt_1,1\,*}+\bigl(X_*\widehat t\con\om(X)\bigr)\circ s_{\igt_1,
1\,*}\nonumber
\qqq
between the loop momentum coordinates on $\,\Igt_\si(\cB)$.\ The
dependence of the loop momentum $\,\sfp_\a\,$ on the loop coordinate
$\,X_{3-\a}\,$ is given by \Reqref{eq:X-as-sec}. The above are
statements valid at every point along the loop, and the pushforward
operators $\,s_{\igt_\a,\a\,*}\,$ are to be understood as
characterising the local sections of the $\,\widetilde\iota_\a\,$
that enter the definition of the $\,s_{\igt_\a,\a}$.\ The relations
define a graph iff they agree for any two choices $\,s_{\igt_\a,
\a}^n,\ n\in\{1,2\}\,$ of sections (i.e.\ for any two choices
$\,X^n\in\sfL Q,\ n\in\{1,2\}\,$ of the parent loop) corresponding
to a given pair $\,X_\a\,$ of loops in $\,M$.\ This is tantamount to
imposing conditions \eqref{eq:iotasI-iotasJ-tan} (which ensure that
arbitrary \emph{curves} of loops in $\,\sfL M\,$ are mapped into one
another in a unique manner), together with
\qq\nn
\bigl(X^1_*\widehat t\con\om(X^1)\bigr)\circ s^1_{\igt_\a,\a\,*}=
\bigl(X^2_*\widehat t\con\om(X^2)\bigr)\circ s^2_{\igt_\a,\a\,*}\,,
\qqq
or -- equivalently --
\qq\nn
X_{\a\,*}\widehat t\con\bigl(s^{1\,*}_{\igt_\a,\a}\om-s^{2\,
*}_{\igt_\a,\a}\om\bigr)(X_\a)=0\,.
\qqq
The corresponding local statement, at a given point $\,X_\a(\varphi)
\in\cO^M_{i_\a}\,$ along the loop in $\,M$,\ yields
\Reqref{eq:sIom-sJom} by virtue of the arbitrariness of the vector
$\,X_{\a\,*}\widehat t$.

Finally, the identity from point iv) is a simple rewrite of the
condition $\,\ceH_\si^-\vert_{\Igt_\si}\equiv 0\,$ taking into
account the relations \eqref{eq:X-as-sec}-\eqref{eq:pal-of-pal}.
\eroof \brem It deserves to be noted that the requirement that the
state correspondence engendered by the defect be independent of the
choice of the local section of the surjective submersion
$\,\widetilde\iota_\a:\sfL Q\to\sfL M\,$ is automatically satisfied
in the (physically) most natural setting, which is that of $\,Q\,$
being a submanifold within $\,M\x M\,$ projecting surjectively on
both cartesian factors. \erem \brem Our discussion indicates that
surjective submersions play a prominent r\^ole in the canonical
description of dualities of the $\si$-model on world-sheets with
defect quivers. This is to be compared with the categorial treatment
of gerbes and gerbe bi-modules in \Rcite{Fuchs:2009si} which also
appears to distinguish maps of this kind, albeit in a more formal
manner. \erem\medskip

Once the conditions for the defect to describe a duality of the
untwisted sector of the $\si$-model have been established, it is
tempting to reverse the question and enquire as to the necessary
conditions for a duality to define a bi-brane that can subsequently
be put over a defect line. General as it stands, the question falls
beyond the compass of the present paper. We may, nonetheless, try to
draw useful insights from the study of a wide class of dualities for
which there exists a concise explicit description in terms of
canonical transformations on the state space of the untwisted sector
of the $\si$-model, determined by generating functionals of a
restricted `linear' form, to be described below. Dualities of this
type, including abelian and non-abelian dualities, as well as the
Poisson--Lie T-duality of the WZW model, were examined in a series
of papers by Alvarez, Refs.\,\cite{Alvarez:2000bh,Alvarez:2000bi},
from which we borrow some of our conventions and a number of
observations.

The chief idea of the approach advertised above consists in
explicitly enforcing the isotropy of the space $\,\Tgt\Igt_\si\,$ of
sections of the tangent bundle of a graph $\,\Igt_\si\,$ in
$\,\sfP_\si^{\x 2}\,$ representing the duality by trivialising the
symplectic potential of $\,\Om^-_{\si,\emptyset}\,$ with the help of
a generating functional of a canonical transformation $\,\psi_1
\mapsto\psi_2\,$ determined by the graph $\,\Igt_\si\ni(\psi_1,
\psi_2),\ \psi_\a=(X_\a,\sfp_\a),\ \a\in\{1,2\}$.\ In so doing, the
generating functional is chosen such that the transformation between
the two sets of variables: $\,(X_{1 \,*} \widehat t,\sfp_1)\,$ and
$\,(X_{2\,*}\widehat t,\sfp_2)\,$ induced by the canonical
transformation is invertible and preserves the hamiltonian density
\eqref{eq:ham-dens-si}. The latter condition, in conjunction with
the distinguished form of the hamiltonian density (a sum of terms
quadratic in $\,\sfp\,$ and $\,X_*\widehat t$,\ respectively), was
used in \Rcite{Alvarez:2000bh} to restrict the choice of the
generating functional, for a specific trivialisation of
$\,\Om^-_{\si, \emptyset}$,\ to those depending linearly on the
$\,X_{\a*}\widehat t\,$ and further constrained by the requirement
of orthogonality with respect to the metric
$\,\bigl(\txg,\txg^{-1}\bigr)\,$ entering the definition of the
hamiltonian density.

An obvious problem with the above description of a canonical
transformation lies with the lack of a global definition of the
symplectic potential of $\,\Om^-_{\si,\emptyset}\,$ in general, a
simple variation on the theme of the lack of a global definition of
the topological term in the $\si$-model action functional, only
transferred one degree lower in cohomology and -- simultaneously --
from the target space to its free-loop space. Below, we resolve this
problem by considering the full structure of the pre-quantum bundle
over the state space of the untwisted sector of the $\si$-model, in
a local presentation suggested by Corollary \ref{cor:preqb-untw} in
conjunction with Proposition \ref{prop:cover-untw}. Moreover, we
extend the analysis to a larger class of trivialisations of
$\,\Om^-_{\si,\emptyset}$,\ thereby gaining access to a canonical
description of geometric symmetries of the $\si$-model.

With view towards organising the discussion of our results, we begin
by providing a precise description of the class of dualities to be
considered in the sequel.
\bedef\label{def:dualiTN}
Let $\,\Bgt\,$ be a string background with target $\,\cM=(M,\txg,\cG
)$,\ and let $\,(\Igt_\si,\Dgt_\si)\,$ be a pre-quantum duality of
the untwisted sector of the non-linear $\si$-model for network-field
configurations $\,(X\,\vert\,\G)\,$ in string background $\,\Bgt\,$
on world-sheet $\,(\Si,\g)\,$ with defect quiver $\,\G$.\ Assume
that $\,(\Igt_\si,\Dgt_\si )\,$ is determined by a generating
functional $\,\xcF_\si\,$ of a canonical transformation $\,\xcD_\si:
\sfP_{\si,\emptyset}\to\sfP_{\si,\emptyset}$,\ given as a collection
of smooth real-valued functionals $\,\Phi_{\si\,\igt}\,$ on elements
of an open cover $\,\cO_{\Igt_\si}=\{\cO^{\Igt_\si}_\igt\}_{\igt\in
\xcI_{\Igt_\si}}\,$ of $\,\Igt_\si$,\ i.e.\ $\,\Igt_\si\,$ is the
graph of $\,\xcD_\si\,$ and the local data of $\,\xcF_\si\,$ yield a
local presentation of the bundle isomorphism $\,\Dgt_\si$.\ Assume
further that the $\,\Phi_{\si\,\igt}\,$ depend \emph{at most
linearly} on the variables $\,(X_{\a\,*}\widehat t,\sfp_\a),\ \a\in
\{1,2\}$.\ Fix $\,\cO_{\Igt_\si}\,$ to be the open cover induced
from the open covers $\,\cO_{\sfL M}\,$ of the free-loop space
$\,\sfL M=C^\infty(\bS^1, M)\,$ of the target space $\,M\,$ from
Proposition \ref{prop:cover-untw} on the cartesian factors of
$\,\Igt_\si$,\ coming from a sufficiently fine good open cover
$\,\cO_M\,$ of $\,M$,\ so that elements of $\,\cO_{\Igt_\si}\,$ are
of the form
\qq\nn
\cO^{\Igt_\si}_{(\igt^1,\igt^2)}=\cO^*_{\igt^1}\x\cO^*_{\igt^2}
\qqq
for $\,\cO^*_{\igt^\a}\,$ as defined in Corollary
\ref{cor:preqb-untw}. We call $\,(\Igt_\si,\Dgt_\si)\,$ a
\textbf{pre-quantum duality of type $T$ of the untwisted sector of
the non-linear $\si$-model for network-field configurations
$\,(X\,\vert\,\G)\,$ in string background $\,\Bgt\,$ on world-sheet
$\,(\Si,\g)\,$ with defect quiver $\,\G\,$} iff its local data can
be put in the form
\qq
\Phi_{\si\,(\igt_1,\igt_2)}[(\psi_1,\psi_2)]=\sum_{e\in\triangle
(\bS^1)}\,\int_e\,(X_{1\,e},X_{2\,e})^*P_{(i^1_e,i^2_e)}+\sfi\,
\sum_{v\in\triangle(\bS^1)}\,(X_1,X_2)^*\log K_{(i^1_{e_+(v)},
i^2_{e_+(v)})(i^1_{e_-(v)},i^2_{e_-(v)})}(v)\cr\cr
\label{eq:Phi-om-loc}
\qqq
for $\,(\psi_1,\psi_2)\in\cO^{\Igt_\si}_{(\igt^1,\igt^2)}\,$ with
$\,\psi_\a=(X_\a,\sfp_\a)$,\ some smooth 1-forms $\,P_{(i^1,i^2
)}\,$ on $\,\cO^M_{i^1}\x\cO^M_{i^2}\,$ and some smooth $\uj$-valued
maps $\,K_{(i^1,i^2)(j^1,j^2)}=K^{-1}_{(j^1,j^2)(i^1,i^2)}\,$ on
$\,\cO^M_{i^1 j^1}\x\cO^M_{i^2 j^2}$.\ The data are required to
satisfy the identities
\qq
\pr_1^*\bigl(\theta_{\sfT^*\sfL M}+\pi_{\sfT^*\sfL M}^*E_{\igt^1}
\bigr)-\pr_2^*\bigl(\theta_{\sfT^*\sfL M}+\pi_{\sfT^*\sfL M}^*
E_{\igt^2}\bigr)&=&-\sfi\,\d\log f_{\si\,(\igt_1,\igt_2)}\,,\cr
\label{eq:dual-preq-triv}&&\\ \pr_1^* \pi_{\sfT^*\sfL M}^*G_{\igt^1
\jgt^1}\cdot\pr_2^*\pi_{\sfT^*\sfL M}^*G_{\igt^2\jgt^2}^{-1}&=&
f_{\si\,(\igt_1,\igt_2)}\cdot f_{\si\,(\jgt_1,\jgt_2)}^{-1}\,,
\nonumber
\qqq
written in terms of the smooth $\uj$-valued functionals
\qq\label{eq:Phifa}
f_{\si\,(\igt_1,\igt_2)}=\ee^{-\sfi\,\Phi_{\si\,(\igt_1,\igt_2)}}
\qqq
and of the canonical projections $\,\pr_\a:\sfP_{\si,\emptyset}\x
\sfP_{\si,\emptyset}\to\sfP_{\si,\emptyset},\ \a\in\{1,2\}$,\ the
canonical 1-form $\,\theta_{\sfT^*\sfL M}\,$ on the total space of
the cotangent bundle $\,\pi_{\sfT^*\sfL M}:\sfT^*\sfL M\to\sfL M\,$
from Proposition \ref{prop:sympl-form-si-untw} and the local data
$\,(E_\igt,G_{\igt\jgt})\,$ of the transgression bundle
$\,\ceL_\cG\to\sfL M\,$ from Theorem \ref{thm:trans-untw}.

Analogously, we call $\,(\Igt_\si,\Dgt_\si)\,$ a \textbf{pre-quantum
duality of type $N$ of the untwisted sector of the non-linear
$\si$-model for network-field configurations $\,(X\,\vert\,\G)\,$ in
string background $\,\Bgt\,$ on world-sheet $\,(\Si,\g )\,$ with
defect quiver $\,\G\,$} iff, in the same notation, its local data
can be put in the form
\qq\label{eq:Phi-unipot}
\Phi_{\si\,(\igt_1,\igt_2)}[(\psi_1,\psi_2)]=-\int_{\bS^1}\,\Vol(
\bS^1)\,\sfp_{2\,\mu}\,F^\mu(X_1)+W_{(\igt^1,\igt^2)}[(\psi_1,
\psi_2)]
\qqq
for an arbitrary smooth map
\qq\nn
F\ :\ M\to M\,,
\qqq
and for smooth real-valued functionals $\,W_{(\igt^1,\igt^2 )}\,$ on
the $\,\cO^{\Igt_\si}_{(\igt^1,\igt^2)}$,\ of the form
\qq
W_{(\igt_1,\igt_2)}[(\psi_1,\psi_2)]=\sum_{e\in\triangle(\bS^1)}\,
\int_e\,(X_{1\,e},X_{2\,e})^*P_{(i^1_e,i^2_e)}+\sfi\,\sum_{v\in
\triangle(\bS^1)}\,(X_1,X_2)^*\log K_{(i^1_{e_+(v)},i^2_{e_+(v)})(
i^1_{e_-(v)},i^2_{e_-(v)})}(v)\,,\cr\cr\label{eq:Wi1i2}
\qqq
with $\,P_{(i^1,i^2)}\in\Om^1(\cO^M_{i^1}\x\cO^M_{i^2})\,$ and
$\,K_{(i^1,i^2)(j^1,j^2)}=K^{-1}_{(j^1,j^2)(i^1,i^2)}\in
\uj_{\cO^M_{i^1 j^1}\x\cO^M_{i^2 j^2}}$.\ (It is understood that
there is no dependence on the $\,\sfp_\a\,$ in $\,W_{(\igt^1,\igt^2
)}$.) Here, the identities to be satisfied by $\,f_{\si\,(\igt_1,
\igt_2)}\,$ as in \Reqref{eq:Phifa} read
\qq
\pr_1^*\bigl(\theta_{\sfT^*\sfL M}+\pi_{\sfT^*\sfL M}^*E_{\igt^1}
\bigr)-\pr_2^*\bigl(\theta_{\sfT^*\sfL M}^*+\pi_{\sfT^*\sfL M}^*
E_{\igt^2}\bigr)&=&-\sfi\,\d\log f_{\si\,(\igt_1,\igt_2)}\,,\cr
\label{eq:dual-preq-triv-bis}&&\\
\pr_1^*\pi_{\sfT^*\sfL M}^*G_{\igt^1\jgt^1}\cdot\pr_2^*\pi_{\sfT^*
\sfL M}^*G_{\igt^2\jgt^2}^{-1}&=&f_{\si \,(\igt_1,\igt_2)}\cdot
f_{\si\,(\jgt_1,\jgt_2)}^{-1}\,,\nonumber
\qqq
where
\qq\nn
\theta_{\sfT^*\sfL M}^*[(X,\sfp)]=-\int_{\bS^1}\,\Vol(\bS^1)\wedge
X^\mu\,\d\sfp_\mu\,.
\qqq
\exdef \brem The form of the edge terms in the definition
\eqref{eq:Phi-om-loc} of the local data of the generating functional
of the duality of type $T$ is dictated by the requirement that the
ensuing canonical transformation induce a linear map $\,(X_{1\,
*}\widehat t,\sfp_1)\mapsto(X_{2\,*}\widehat t,\sfp_2)$,\ as
discussed earlier in this section and in Alvarez's papers, and the
vertex corrections are perfectly consistent with this requirement in
the local description of the generating functional.

The expression \eqref{eq:Phi-unipot} defining the generating
functional of the duality of type $N$, on the other hand, should be
regarded as a natural local deformation of the global generating
functional
\qq\nn
\Phi_\id[(\psi_1,\psi_2)]=-\int_{\bS^1}\,\Vol(\bS^1)\,\sfp_{2\,\mu}
\,X_1^\mu\,,
\qqq
readily verified to yield the identity canonical transformation on
$\,\sfP_{\si,\emptyset}$.\ Thus, unlike dualities of type $T$,
dualities of type $N$ are continuously deformable to the trivial
duality (i.e.\ to the identity symplectomorphism).\erem

We are now ready to present our findings which can be summarised as
follows
\bethe\label{thm:duali-T-bib} Let $\,\Bgt\,$ be a string background
with target $\,\cM=(M,\txg,\cG)$.\ Consider the non-linear
$\si$-model for network-field configurations $\,(X\,\vert\,\G)\,$ in
string background $\,\Bgt\,$ on world-sheet $\,(\Si,\g)\,$ with
defect quiver $\,\G$.\ For every duality $\,(\Igt_\si,\Dgt_\si)\,$
of type $T$ of the $\si$-model, there exists a topological defect
with a $\cG$-bi-brane $\,\cB_{\Dgt_\si}=\bigl(Q,\iota_\a,\om,\Phi\
\vert\ \a\in\{1,2\} \bigr)\,$ over it with the following properties:
\bit
\item[i)] the world-volume $\,Q\,$ is a submanifold of the cartesian
square $\,M\x M\,$ of the target space $\,M$;
\item[ii)] the $\cG$-bi-brane maps are given by the canonical
projections $\,\iota_\a=\pr_\a:Q\to M,\ \a\in\{1,2\}$;
\item[iii)] $Q\,$ carries a symplectic form
\qq\nn
\Om_{\Dgt_\si}(X_1,X_2):=\om^{1\wedge 2}_{\mu\nu}(X_1,X_2)\,\sfd
X_1^\mu\wedge\sfd X_2^\nu\,,\qquad\qquad(X_1,X_2)\in Q
\qqq
defined by the curvature (no summation over the repeated indices)
\qq\label{eq:curv-dualB-wedge}
\om(X_1,X_2)=\sum_{0<i\leq j<3}\,(-1)^{i+1}\,\om_{i\wedge j}\,,
\qquad\qquad \om_{i\wedge j}=\om^{i\wedge j}_{\mu\nu}(X_1,X_2)\,\sfd
X_i^\mu\wedge\sfd X_j^\nu
\qqq
of $\,\cB_{\Dgt_\si}$,\ the latter being given in terms of globally
smooth maps $\,\om^{i\wedge j}_{\mu\nu}\in C^\infty(Q,\bR)$;
\item[iv)] the \textbf{duality background}
\qq\nn
\Bgt_{\Dgt_\si}=(\cM,\cB_{\Dgt_\si},\cdot)
\qqq
satisfies the \textbf{duality-background constraints}
\qq\label{eq:dualiT-back-constr}
\txE_2=-\om_{1\wedge 2}\circ\txE_1^{-1}\circ\om_{1\wedge 2}\,,
\qqq
written in terms of the \textbf{background operators}
\qq\nn
\txE_\a=:\txg_\a+\om_{\a\wedge\a}\ :\ \G(\sfT Q)\to\G(\sfT^*Q)\ :\
\xcV\mapsto\txE_\a(\xcV,\cdot)\,,\qquad\qquad\txg_\a=\pr_\a^*\txg
\,,\qquad\a\in\{1,2\}\,.
\qqq
\eit
\ethe
\beroof
Let us adopt the notation of Definition \ref{def:dualiTN}. Using the
identity
\qq\nn
\d\int_e\,X_e^*\eta=-\int_e\,X_\ell^*\d\eta+X^*\eta\vert_{\p e}\,,
\qqq
valid for an arbitrary edge $\,e\in\triangle(\bS^1)\,$ and for any
$\,\eta\in\Om^1\bigl(X(e)\bigr)$,\ we readily extract from the first
of identities \eqref{eq:dual-preq-triv} the relations
\qq
\sfp_1-\sfp_2=(X_{1\,*}\widehat t,X_{2\,*}\widehat t)\con\bigl(
\pr_1^*B_{i^1}-\pr_2^*B_{i^2}+\sfd P_{(i^1,i^2)}\bigr)(X_1,X_2)\,,
\label{eq:p-diff-symm}\\\cr \pr_1^*A_{i^1 j^1}-\pr_2^*A_{i^2 j^2}+
P_{(j^1,j^2)}-P_{(i^1,i^2)}-\sfi\,\sfd\log K_{(i^1,i^2)(j^1,j^2)}=0
\,, \label{eq:can-trans-cond-loc}
\qqq
implied by the requirement that both the edge term and the vertex
term of the identity vanish independently. The relations are to be
satisfied on the submanifold $\,Q\subset M\x M\,$ obtained by taking
the set of all pairs of points in $\,M\,$ intersected by pairs of
loops from $\,(\pi_{\sfT^*\sfL M},\pi_{\sfT^*\sfL M})(\Igt_\si)$.\
The manifold $\,Q\,$ canonically projects \emph{onto} $\,M$.\ Taking
the exterior derivative of both sides of
\Reqref{eq:can-trans-cond-loc} and, subsequently, using
\Reqref{eq:DG-is-H}, we obtain the equality
\qq\nn
\pr_1^*B_{j^1}-\pr_2^*B_{j^2}+\sfd P_{(j^1,j^2)}=\pr_1^*B_{i^1}-
\pr_2^*B_{i^2}+\sfd P_{(i^1,i^2)}\,,
\qqq
from which we infer the existence of a globally defined 2-form
\qq\label{eq:om-def}
\om:=\pr_1^*B_{i^1}-\pr_2^*B_{i^2}+\sfd P_{(i^1,i^2)}\in\G(\wedge^2
\sfT^*Q)\,.
\qqq
This is in keeping with \Reqref{eq:p-diff-symm} as the latter
requires that the expression on the right-hand side be a smooth
1-form. We also note that it yields a relation
\qq
K_{(j^1,j^2)(k^1,k^2)}\cdot K_{(i^1,i^2)(k^1,k^2)}^{-1}\cdot K_{(i^1
,i^2)(j^1,j^2)}\cdot\pr_2^*g_{i^2 j^2 k^2}\cdot\pr_1^*g_{i^1 j^1
k^1}^{-1}=:C_{(i^1,i^2)(j^1,j^2)(k^1,k^2)}\,,\cr\cr\label{eq:KC-gg-C}
\qqq
in which $\,(C_{(i^1,i^2)(j^1,j^2)(k^1,k^2)})\,$ is a locally
constant $\uj$-valued \v Cech 2-cochain on $\,Q$.\ Clearly,
\qq\label{eq:Cobstr-cech}
\bigl(\vd^{(2)}C\bigr)_{(i^1,i^2)(j^1,j^2)(k^1,k^2)(l^1,l^2)}=1\,,
\qqq
and the class $\,[(C_{(i^1,i^2 )(j^1,j^2)(k^1,k^2)})]\in\vH^2\bigl(Q
,\uj\bigr)\,$ is readily seen to define the obstruction to the
existence of a $\cG$-bi-brane $\,(Q,\om,\pr_1,\pr_2,\Phi)\,$ with
1-isomorphism $\,\Phi:\pr_1^*\cG\xrightarrow{\cong}\pr_2^*\cG\ox
I_\om\,$ with local data $\,(P_{(i^1,i^2)},K_{(i^1,i^2)(j^1,j^2)}
)$.\ Indeed, Eqs.\,\eqref{eq:can-trans-cond-loc}, \eqref{eq:om-def}
and \eqref{eq:KC-gg-C} can be rewritten concisely in the familiar
form
\qq\nn
\pr_1^*(B_{i^1},A_{i^1 j^1},g_{i^1 j^1 k^1})+D_{(1)}(P_{(i^1,i^2)},
K_{(i^1,i^2)(j^1,j^2)})+(0,0,C_{(i^1,i^2)(j^1,j^2)(k^1,k^2)})\cr\cr
=\pr_2^*(B_{i^2},A_{i^2 j^2},g_{i^2 j^2 k^2})+(\om\vert_{\cO^Q_{(i_1
,i^2)}},0,0)\,,
\qqq
and it is immediately clear that a pair of 2-cochains $\,(C_{(i^1,
i^2)(j^1,j^2)(k^1,k^2)})\,$ and $\,(C_{(i^1,i^2)(j^1,j^2)(k^1,k^2)}
')\,$ cohomologous as per $\,(C_{(i^1,i^2)(j^1,j^2)(k^1,k^2)}')=(
C_{(i^1, i^2)(j^1,j^2)(k^1,k^2)})\cdot\vd^{(1)}c\,$ for some
(locally constant) 1-cochain $\,c\,$ corresponds to a pair of
1-cochains $\,( K_{(i^1,i^2)(j^1,j^2)})\,$ and
$\,(K_{(i^1,i^2)(j^1,j^2)}')\,$ related by the shift
$\,(K_{(i^1,i^2)(j^1,j^2)}')=(K_{(i^1,i^2)(j^1 ,j^2)})\cdot
c^{-1}$.\

Finally, \Reqref{eq:p-diff-symm} rewrites as
\qq\label{eq:DGC-from-dualiT}
\sfp_1-\sfp_2-(X_{1\,*}\widehat t,X_{2\,*}\widehat t)\con\om(X_1,X_2
)=0\,,
\qqq
and so we recover the complete description of a conformal defect up
to the obstruction\linebreak
$\,[(C_{(i^1,i^2)(j^1,j^2)(k^1,k^2)})]$.\ The latter is removed on
taking into account the second of identities
\eqref{eq:dual-preq-triv}. Indeed, using
Eqs.\,\eqref{eq:can-trans-cond-loc} and \eqref{eq:KC-gg-C}, we
readily cast the above relation in the compact form
\qq\nn
\prod_{\ovl v\in\triangle(\bS^1)}\,(X_1,X_2)^*\left(\tfrac{C_{(
i^1_{\ovl e_+(\ovl v)},i^2_{\ovl e_+(\ovl v)})(i^1_{\ovl e_-(\ovl v
)},i^2_{\ovl e_-(\ovl v)})(j^1_{\ovl e_+(\ovl v)},j^2_{\ovl e_+(\ovl
v )})}}{C_{(j^1_{\ovl e_+(\ovl v)},j^2_{\ovl e_+(\ovl v)})(j^1_{\ovl
e_-(\ovl v)},j^2_{\ovl e_-(\ovl v)})(i^1_{\ovl e_-(\ovl v)},
i^2_{\ovl e_-(\ovl v)})}}\right)(\ovl v)=1\,,
\qqq
which -- in view of the arbitrariness of $\,(X_1,X_2)(\ovl v)\,$ and
of the triangulation used -- requires
\qq\label{eq:C-is-C}
C_{(i^1,i^2)(j^1,j^2)(k^1,k^2)}=C_{(i^1,i^2)(j^1,j^2)(l^1,l^2)}
\qqq
for any pair of quadruples $\,(i^\a,j^\a,k^\a,l^\a)\in\xcI_M^4,\ \a
\in\{1,2\}\,$ such that $\,\cO^M_{i^\a j^\a k^\a l^\a}\neq
\emptyset$.\ Hence,
\qq\nn
C_{(i^1,i^2)(j^1,j^2)(k^1,k^2)}=C_{(i^1,i^2)(j^1,j^2)(j^1,j^2)}=:
\widetilde C_{(i^1,i^2)(j^1,j^2)}\,,
\qqq
and the newly defined maps $\,\widetilde C_{(i^1,i^2)(j^1,j^2)}$,\
with values in the set $\,\{-1,1\}$,\ form a locally constant
2-cochain -- in particular,
\qq\nn
\widetilde C_{(j^1,j^2)(i^1,i^2)}=\widetilde C_{(i^1,i^2)(j^1,j^2
)}^{-1}\,.
\qqq
Using Eqs.\,\eqref{eq:Cobstr-cech} and \eqref{eq:C-is-C}, we then
find
\qq\nn
C_{(i^1,i^2)(j^1,j^2)(k^1,k^2)}&=&C_{(j^1,j^2)(k^1,k^2)(l^1,l^2)}
\cdot C_{(i^1,i^2)(k^1,k^2)(l^1,l^2)}^{-1}\cdot C_{(i^1,i^2)(j^1,j^2
)(l^1,l^2)}\cr\cr
&=&\widetilde C_{(j^1,j^2)(k^1,k^2)}\cdot\widetilde C_{(i^1 ,i^2)(
k^1,k^2)}^{-1}\cdot\widetilde C_{(i^1,i^2)(j^1,j^2)}\,.
\qqq
Clearly, the \v Cech cohomology class of this 2-cochain is trivial
and it can be absorbed into a redefinition of the local data of the
1-isomorphism $\,\Phi$,\ cf.\ \Reqref{eq:KC-gg-C}. This leaves us
with statements iii) and iv) of the theorem to demonstrate.

The remainder of the proof uses solely elementary analysis of
canonical transformations defined in terms of generating functions,
cf., e.g., \Rxcite{Sec.\,6.5}{Marsden:1994}. Thus, upon recalling
that the space $\,\Igt_\si\,$ is -- by assumption -- diffeomorphic
to $\,\sfT^*\sfL M$,\ we can choose the loop variables $\,(X_1,X_2
)\,$ as independent local coordinates on $\,\Igt_\si$,\ which has
the following two consequences: First of all, \Reqref{eq:om-def}
yields three independent relations:
\qq
\om_{1\wedge 1}&=&\pr_1^*B_{i^1}+[\sfd P_{(i^1,i^2)}]_{1\wedge 1}\,,
\label{eq:om-1wedge1}\\\cr \om_{2\wedge 2}&=&\pr_2^*B_{i^2}-[\sfd
P_{(i^1,i^2)}]_{2\wedge 2}\,,\label{eq:om-2wedge2}\\\cr \om_{1\wedge
2}&=&[\sfd P_{(i^1,i^2)}]_{1\wedge 2}\,, \label{eq:om-1wedge2}
\qqq
written in terms of the components $\,\om_{i\wedge j}\,$ of
$\,\om\,$ from \Reqref{eq:curv-dualB-wedge} and those of $\,\sfd
P_{(i^1,i^2)}$,\ defined analogously. Secondly, we may extract from
\Reqref{eq:DGC-from-dualiT} a pair of coupled equations
\qq\nn
\left(\barr{cc} -\om_{1\wedge 1} & \id_{\G(\sfT^*\sfL M)}\cr\cr
-\om_{1\wedge 2} & 0 \earr\right)\, \left(\barr{c} X_{1\,*}\widehat
t\cr\cr \sfp_1 \earr\right)=\left(\barr{cc} \om_{1\wedge 2} & 0
\cr\cr -\om_{2\wedge 2} & \id_{\G(\sfT^*\sfL M)} \earr\right)\,
\left(\barr{c} X_{2\,*}\widehat t \cr\cr \sfp_2 \earr\right)\,,
\qqq
to be understood as representing the action of linear operators on
sections of $\,\sfT\sfL Q\oplus\sfT^*\sfL Q$,\ with 2-form fields
acting on vector fields through contraction, i.e.\ as per $\,\om_{i
\wedge j}\lact X_{\a\,*}\widehat t:=X_{\a\,*}\widehat t\con\om_{i
\wedge j}$.\ Clearly, for the transformation between the two pairs
$\,\bigl(X_{\a \,*}\widehat t,\sfp_\a\bigr),\ \a\in\{1,2\}\,$ thus
defined to be invertible, we have to demand that $\,\om_{1 \wedge
2}$,\ regarded as a map from $\,\G(\sfT Q)\,$ to $\,\G(\sfT^* Q)$,\
possess an inverse,
\qq\nn
\om_{1\wedge 2}^{-1}=\tfrac{1}{4}\,\bigl(\bigl(\om^{1\wedge 2}
\bigr)^{-1}\bigr)^{\mu\nu}\,\p_\mu\wedge\p_\nu\,,\qquad\qquad\bigl(
\bigl(\om^{1\wedge 2}\bigr)^{-1}\bigr)^{\la\mu}\,\om^{1\wedge
2}_{\mu\nu}=\d^\la_{\ \nu}\,,
\qqq
acting on 1-forms as $\,\om_{1\wedge 2}^{-1}\lact(\eta_\mu\,\sfd
X^\mu):=\frac{1}{2}\,\bigl(\bigl(\om^{1\wedge 2}\bigr)^{-1}
\bigr)^{\mu\nu}\,\eta_\mu\,\p_\nu$.\ This proves statement iii) of
the theorem. Having ensured the invertibility of $\,\om_{1\wedge
2}$,\ we may express $\,\bigl(X_{\a \,*}\widehat t,\sfp_\a\bigr)\,$
through $\,\bigl(X_{3-\a\,*}\widehat t,\sfp_{3-\a}\bigr)$.\
Demanding that $\,\ceH^-_\si\,$ of \Reqref{eq:Hsi-} vanish
identically on $\,\Igt_\si\,$ then produces a relation
\qq\label{eq:MgMg-om}
\txM_\om^{\rm T}\circ\widehat\txg_1\circ\txM_\om=\widehat\txg_2\,,
\qqq
written in terms of the operators
\qq\nn
\widehat\txg_\a=\left(\barr{cc} \txg_\a & 0 \cr\cr 0 & \txg^{-1}_\a
\earr\right)\,,\qquad\a\in\{1,2\}
\qqq
and
\qq\nn
\txM_\om=\left(\barr{cc} \om_{1\wedge 2}^{-1}\circ\om_{2\wedge 2} &
-\om_{1\wedge 2}^{-1}\cr\cr \om_{1\wedge 2}+\om_{1\wedge 1}\circ
\om_{1\wedge 2}^{-1}\circ\om_{2\wedge 2} & -\om_{1\wedge 1}\circ
\om_{1\wedge 2}^{-1} \earr\right)\,,
\qqq
and of the transpose of the latter,
\qq\nn
\txM_\om^{\rm T}=\left(\barr{cc} \om_{2\wedge 2}\circ\om_{1\wedge
2}^{-1} & -\om_{1\wedge 2}-\om_{2\wedge 2}\circ\om_{1\wedge 2}^{-1}
\circ\om_{1\wedge 1}\cr\cr \om_{1\wedge 2}^{-1} & -\om_{1\wedge
2}^{-1}\circ\om_{1\wedge 1} \earr\right)\,.
\qqq
We shall next demonstrate that \Reqref{eq:MgMg-om} is equivalent to
the duality-background constraints \eqref{eq:dualiT-back-constr}. To
this end, we first note that the latter actually encodes a pair of
independent relations for the symmetric and antisymmetric component
of the background operator $\,\txE_2$,\ respectively. Explicitly,
\qq\nn
\txg_2&=&-\tfrac{1}{2}\,\om_{1\wedge 2}\circ\bigl(\txE_1^{-1}+
\txE_1^{-1\,{\rm T}}\bigr)\circ\om_{1\wedge 2}\equiv-\tfrac{1}{2}\,
\om_{1\wedge 2}\circ\bigl[(\txg_1+\om_{1\wedge 1})^{-1}\circ(\txg_1
-\om_{1\wedge 1})\circ(\txg_1-\om_{1\wedge 1})^{-1}\cr\cr
&&+(\txg_1+\om_{1\wedge 1})^{-1}\circ(\txg_1+\om_{1\wedge 1})\circ(
\txg_1-\om_{1\wedge 1})^{-1}\bigr]\circ\om_{1\wedge 2}=-\om_{1
\wedge 2}\circ(\txg_1+\om_{1\wedge 1})^{-1}\circ\txg_1\circ(\txg_1-
\om_{1\wedge 1})^{-1}\circ\om_{1\wedge 2}\cr\cr
&=&-\om_{1\wedge 2}\circ\bigl[(\txg_1-\om_{1\wedge 1})\circ
\txg_1^{-1}\circ(\txg_1+\om_{1\wedge 1})\bigr]^{-1}\circ\om_{1
\wedge 2}\equiv-\om_{1\wedge 2}\circ(\txg_1-\om_{1\wedge 1}\circ
\txg_1^{-1}\circ\om_{1\wedge 1})^{-1}\circ\om_{1\wedge 2}\,,
\qqq
and, analogously,
\qq\nn
\om_{2\wedge 2}=\om_{1\wedge 2}\circ\txg_1^{-1}\circ\om_{1\wedge 1}
\circ(\txg_1-\om_{1\wedge 1}\circ \txg_1^{-1}\circ\om_{1\wedge 1}
)^{-1}\circ\om_{1\wedge 2}\,.
\qqq
The above are to be compared with the independent relations
determined by the continuity constraint \eqref{eq:MgMg-om}. These
are easily found to be
\qq\nn
\txg_2^{-1}&=&\om_{1\wedge 2}^{-1}\circ(\om_{1\wedge 1}\circ
\txg_1^{-1}\circ\om_{1\wedge 1}-\txg_1)\circ\om_{1\wedge 2}^{-1}\,,
\cr\cr
\txg_2&=&\om_{2\wedge 2}\circ\om_{1\wedge 2}^{-1}\circ\txg_1\circ
\om_{1\wedge 2}^{-1}\circ\om_{2\wedge 2}-\bigl(\om_{2\wedge 2}\circ
\om_{1\wedge 2}^{-1}\circ\om_{1\wedge 1}+\om_{1\wedge 2}\bigr)\circ
\txg_1^{-1}\circ\bigl(\om_{1\wedge 2}+\om_{1\wedge 1}\circ\om_{1
\wedge 2}^{-1}\circ\om_{2\wedge 2}\bigr)\,,\cr\cr 0&=&\om_{1\wedge
2}^{-1}\circ\txg_1\circ\om_{1\wedge 2}^{-1}\circ\om_{2\wedge 2}-
\om_{1\wedge 2}^{-1}\circ\om_{1\wedge 1}\circ\txg_1^{-1}\circ\bigl(
\om_{1\wedge 2}+\om_{1\wedge 1}\circ\om_{1\wedge 2}^{-1}\circ\om_{2
\wedge 2}\bigr)\,.
\qqq
The remaining relation is a transpose of the bottom one. Evidently,
the top one is an inverse of the symmetric component of
\Reqref{eq:dualiT-back-constr}, and so we are left with the other
two to examine.

Upon using the bottom relation in the middle one, we reduce the
latter to the form
\qq\label{eq:rem-g2}
\txg_2=-\om_{1\wedge 2}\circ\txg_1^{-1}\circ\bigl(\om_{1\wedge 2}+
\om_{1\wedge 1}\circ\om_{1\wedge 2}^{-1}\circ\om_{2\wedge 2}\bigr)
\,,
\qqq
which can be combined with (the inverse of) the top one and
subsequently substituted back into the bottom relation to yield
\qq\nn
\om_{2\wedge 2}&=&\om_{1\wedge 2}\circ\txg_1^{-1}\circ\om_{1\wedge
1}\circ\txg_1^{-1}\circ\bigl(\om_{1\wedge 2}+\om_{1\wedge 1}\circ
\om_{1\wedge 2}^{-1}\circ\om_{2\wedge 2}\bigr)=-\om_{1\wedge 2}\circ
\txg_1^{-1}\circ\om_{1\wedge 1}\circ\om_{1\wedge 2}^{-1}\circ\txg_2
\cr\cr
&=&-\om_{1\wedge 2}\circ\txg_1^{-1}\circ\om_{1\wedge 1}\circ\bigl(
\om_{1\wedge 1}\circ\txg_1^{-1}\circ\om_{1\wedge 1}-\txg_1\bigr)^{-
1}\circ\om_{1\wedge 2}\,,
\qqq
which is the desired form of the antisymmetric component of
\Reqref{eq:dualiT-back-constr}. At this stage, it remains to verify
that the two components found hitherto ensure that the remaining
relation \eqref{eq:rem-g2} is satisfied identically. With $\,\om_{2
\wedge 2}\,$ as above, its right-hand side takes the form
\qq\nn
\txg_2=-\om_{1\wedge 2} \circ\txg_1^{-1}\circ\bigl[\om_{1\wedge 2}-
\om_{1\wedge 1}\circ\txg_1^{-1}\circ\om_{1\wedge 1}\circ\bigl(
\om_{1\wedge 1}\circ\txg_1^{-1}\circ\om_{1\wedge 1}-\txg_1\bigr)^{-
1}\circ\om_{1\wedge 2}\bigr]\,,
\qqq
and so we must show the identity
\qq\nn
-\om_{1\wedge 2}\circ \txg_1^{-1}\circ\bigl[\om_{1\wedge 2}-\om_{1
\wedge 1}\circ\txg_1^{-1}\circ\om_{1\wedge 1}\circ\bigl(\om_{1\wedge
1}\circ\txg_1^{-1}\circ\om_{1\wedge 1}-\txg_1\bigr)^{-1}\circ\om_{1
\wedge 2}\bigr]\cr\cr =\om_{1\wedge 2}\circ(\om_{1\wedge 1}\circ
\txg_1^{-1}\circ\om_{1\wedge 1}-\txg_1)^{-1}\circ\om_{1\wedge 2}\,,
\qqq
which follows straightforwardly upon regrouping its terms.\eroof
\medskip

A similar result can be established for dualities of type $N$,\
namely,
\bethe\label{thm:duali-N-bib}
Let $\,\Bgt\,$ be a string background with target $\,\cM=(M,\txg,\cG
)$.\ Consider the non-linear $\si$-model for network-field
configurations $\,(X\,\vert\,\G)\,$ in string background $\,\Bgt\,$
on world-sheet $\,(\Si,\g)\,$ with defect quiver $\,\G$.\ To every
duality $\,(\Igt_\si, \Dgt_\si)\,$ of type $N$ of the $\si$-model,
there is associated a topological defect with a $\cG$-bi-brane
$\,\cB_{\Dgt_\si}=\bigl(Q, \iota_\a,\om,\Phi\ \vert \
\a\in\{1,2\}\bigr)\,$ over it with the following properties:
\bit
\item[i)] the world-volume $\,Q\,$ is a submanifold $\,Q=(\id_M\x F)
(M)\subset M\x M\,$ of the cartesian square $\,M\x M\,$ of the
target space $\,M$;
\item[ii)] $F\,$ is an isometry of the metric manifold $\,(M,\txg)$;
\item[iii)] the $\cG$-bi-brane maps are given by the canonical
projections $\,\iota_\a=\pr_\a:Q\to M,\ \a\in\{1,2\}$;
\item[iv)] the curvature $\,\om\,$ vanishes identically;
\item[v)] the pullback of the 1-isomorphism $\,\Phi\,$ along
the isomorphism $\,\id_M\x F:M\xrightarrow{\cong}Q\,$ is of the form
\qq\label{eq:iso-for-sympl-nom}
(\id_M\x F)^*\Phi\ :\ \cG\xrightarrow{\cong}F^*\cG\,.
\qqq
\eit
\ethe
\beroof
We use the notation of Definition \ref{def:dualiTN} and -- reasoning
as in the proof of Theorem \ref{thm:duali-T-bib} -- choose
$\,(X_1,\sfp_2)\,$ as independent local coordinates on
$\,\Igt_\si$,\ which leads to the relation
\qq\label{eq:XFX}
X_2=F[X_1]\,,
\qqq
extracted from the first of identities \eqref{eq:dual-preq-triv}.
Clearly, for the generating functional \eqref{eq:Phi-unipot} to
define a duality of the $\si$-model, $\,F\,$ has to induce an
invertible map on $\,M$.\ The remaining relation encoded by the
first of identities \eqref{eq:dual-preq-triv} reads
\qq\label{eq:sep-sympl-id}
\int_{\bS^1}\,\Vol(\bS^1)\wedge\bigl(\sfp_1-\widehat F_*\lact\sfp_2
\bigr)=E_{\igt_2}[X_2]-E_{\igt_1}[X_1]-\d W_{\igt_1\igt_2}[(\psi_1,
\psi_2)]\,,
\qqq
where we introduced the operator $\,\widehat F_*=\frac{\d F^\mu}{\d
X_1^\nu}\,\frac{\d \ }{\d X_2^\mu}\ox\d X_1^\nu$,\ acting on
$\,\sfp_2=\sfp_{2\,\mu}\,\d X_2^\mu\,$ through contraction,
\qq\nn
\widehat F_*\lact\d X_2^\mu:=\frac{\d F^\mu}{\d X_1^\nu}\,\d
X_1^\nu\,.
\qqq
The left-hand side of \Reqref{eq:sep-sympl-id} being globally
defined, so must be its right-hand side, hence
\qq\label{eq:bigO}
E_{\igt_2}[X_2]-E_{\igt_1}[X_1]-\d W_{\igt_1\igt_2}[(\psi_1,\psi_2)
]=:O[X_1]\,,
\qqq
for some $\,O\in\G(\wedge^1\sfT^*\sfL M)\,$ induced by a global
2-form on $\,M\,$ as per
\qq\nn
O=\int_{\bS^1}\,\ev_M^*\om\,,
\qqq
where $\,\ev_M:\sfL M\x\bS^1\to M\,$ is the canonical evaluation
map. Here, we made explicit use of relation \eqref{eq:XFX} to
express the combination of local objects on the left-hand side of
\Reqref{eq:bigO} as a functional of the independent variable
$\,X_1\,$ exclusively. Substituting formula \eqref{eq:bigO} back
into \Reqref{eq:sep-sympl-id}, we now establish a linear
transformation between the pairs $\,(X_{1\,*}\widehat t,\sfp_1)\,$
and $\,(X_{2\,*}\widehat t,\sfp_2)\,$ which reads
\qq\nn
\left(\barr{cc} \widehat F_*^{\rm T} & 0 \cr\cr -\om & \id_{\G(
\sfT^*\sfL M)} \earr\right)\,\left(\barr{c} X_{1\,*}\widehat t
\cr\cr \sfp_1 \earr\right)=\left(\barr{cc} \id_{\G(\sfT\sfL M)}
& 0 \cr\cr 0 & \widehat F_* \earr\right)\,\left(\barr{c} X_{2\,*}
\widehat t \cr\cr \sfp_2 \earr\right)\,,
\qqq
with  $\,\widehat F_*^{\rm T}=\frac{\d F^\mu}{\d X_1^\nu}\,\d
X_1^\nu\ox\frac{\d \ }{\d X_2^\mu}\,$ acting on $\,X_{1\,*}\widehat
t\,$ via contraction,
\qq\nn
\widehat F_*^{\rm T}\lact X_{1\,*}\widehat t=(X_{1\,*}\widehat
t)^\nu\,\frac{\d F^\mu}{\d X_1^\nu}\,\frac{\d \ }{\d X_2^\mu}\,.
\qqq
The invertibility of the transformation thus defined necessitates
the existence of an inverse of the tangent map $\,F_*$,\ which
identifies $\,F\,$ as a ($C^1$-)diffeomorphism of $\,M$.\ Demanding,
furthermore, that the transformation preserve the hamiltonian
density yields the constraints
\qq\label{eq:Ndual-constr}
\om=0\,,\qquad\qquad F^*\txg=\txg\,,
\qqq
and so $\,F\,$ is a ($C^1$-)isometry of $\,(M,\txg)$.

Finally, taking into account the assumed form of the functionals
$\,W_{\igt_1\igt_2}$,\ we readily establish -- reasoning along the
same lines as in the proof of Theorem \ref{thm:duali-T-bib} -- that
local data $\,(P_{(i_1,i_2)},K_{(i_1,i_2)(j_1,j_2)})\,$ define a
1-isomorphism
\qq\nn
\Phi\ :\ \pr_1^*\cG\xrightarrow{\cong}\pr_2^*\cG
\qqq
over the manifold $\,(\id_M\x F)(M)\cong M$,\ with a pullback along
$\,\id_M\x F\,$ as claimed in the thesis of the theorem.\eroof
\medskip

Prior to passing to the discussion of the canonical interpretation
of defect junctions, we pause to present a couple of examples that
give some flesh to the abstract constructions of the present
section.\bigskip

\beg\textbf{Duality of type $T$ from the T-duality defect.}
\label{ex:Tdual}\\[-8pt]

\noindent An important example of a proper duality that
(generically) involves a non-trivial change of the topology of the
connected component of the target space is provided by T-duality,
generalising the duality between the $\si$-model with target space
$\,\bS^1_R$,\ i.e.\ a circle of radius $\,R$,\ and that with target
space $\,\bS^1_{\frac{1}{R}}$,\ i.e.\ a circle of the (T-)dual
radius $\,\frac{1}{R}\,$ (in certain natural units). In the latter
case, translational charges of the string are interchanged with the
winding charges under the duality. A local description of the
algebraic relations between the various components of the background
established by the duality was first worked out in
Refs.\,\cite{Buscher:1987qj,Buscher:1987sk}, whence they are called
\textbf{the Buscher rules}, cf.\ also \Rcite{Giveon:1994fu} for a
review of the early studies of the subject, and
Refs.\,\cite{Alvarez:2000bh,Alvarez:2000bi} for an analysis carried
out in the canonical framework. Global issues were attacked in
\Rcite{Alvarez:1993qi} and, more recently, in
Refs.\,\cite{Bouwknegt:2003vb}, where the important concept of a
correspondence space was introduced and the topological transitions
accompanying T-dualisation in the presence of non-trivial
backgrounds were studied in a systematic manner (cf.\ also
\Rcite{Belov:2007qj} for an attempt at a full-fledged
gerbe-theoretic formulation). The duality was also studied in the
context of the lagrangean description of the string in the presence
of world-sheet defects in \Rcite{Sarkissian:2008dq}.

The string background $\,\Bgt_T=(\cM_T,\cB_T,\cdot)\,$ for the
T-duality defect that we want to consider here consists of
\bit
\item[(TT)] the target $\,\cM_T=(M_T,\txg_T,\cG_T)\,$ with the
target space
\qq\nn
M_T=\bT^n_1\sqcup\bT^n_2
\qqq
given by the disjoint union of a pair of $n$-dimensional tori, with
the metric $\,\txg_T\,$ of constant restrictions
\qq\nn
\txg_T\vert_{\bT^n_\a}=\txg_\a\,,
\qqq
and the gerbe $\,\cG_T\,$ of trivial restrictions
\qq\nn
\cG_T\vert_{\bT^n_\a}= I_{\txB_\a}
\qqq
with constant curvings $\,\txB_\a\in\G(\wedge^2\sfT^*\bT^n_\a)$;
\item[(T)] the $\cG_T$-bi-brane $\,\cB_T=(Q_T,\pr_1,\pr_2,\om_T,\Phi_T
)$,\ with
\bit
\item[(T.i)] the world-volume
\qq\nn
Q_T=\bT^n_1\x\bT^n_2\subset M_T\x M_T\,;
\qqq
\item[(T.ii)] the $\cG_T$-bi-brane maps, given by the canonical
projections
\qq\nn
\iota_\a=\pr_\a\ :\ \bT^n_1\x\bT^n_2\to\bT^n_\a\subset M_T\,;
\qqq
\item[(T.iii)] the closed curvature $\,\om_T$,\ with components
\qq\nn
\om_{T\,\a\wedge\a}=\pr_\a^*\txB_\a\,,\qquad\qquad\om_{T\,1\wedge 2}
=\txF_{\rm P}\,,
\qqq
given in terms of the curvings $\,\txB_\a\,$ and of the curvature
2-form $\,\pi_{P_{\bT^n_1\x \bT^n_2}}^*\txF_{\rm P}=\curv(
\nabla_{P_{\bT^n_1\x\bT^n_2}})\,$ of a connection
$\,\nabla_{P_{\bT^n_1\x\bT^n_2}}\,$ on the Poincar\'e bundle
$\,\pi_{P_{\bT^n_1\x\bT^n_2}}:P_{\bT^n_1\x\bT^n_2}\to\bT^n_1\x
\bT^n_2\,$ over the double torus $\,\bT^n_1\x\bT^n_2$;
\item[(T.iv)] the $\cG_T$-bi-brane 1-isomorphism
\qq\nn
\Phi_T\ :\ I_{\pr_1^*\txB_1}\xrightarrow{\cong} I_{\pr_2^*\txB_2
+\om_T}\,,
\qqq
induced (e.g., on the level of the local data) by the Poincar\'e
bundle $\,P_{\bT^n_1\x\bT^n_2}$.
\eit
\eit
Given these background data, the DGC \eqref{eq:DGC} produces the
compact formul\ae
\qq\nn
\pi_2=-X_{1\,*}\widehat t\con\txF_{\rm P}\,,\qquad\qquad\pi_1=X_{2
\,*}\widehat t\con\txF_{\rm P}
\qqq
defining an isotropic graph $\,\Igt_T\subset\sfP_{\si,
\emptyset}^{\x 2}\,$ and written here in terms of the
\emph{canonical} momentum fields
\qq\nn
\pi_\a=\txp_\a-X_{\a\,*}\widehat t\con\txB_\a\,.
\qqq
In local angle coordinates $\,\theta^\mu_\a,\ \mu\in\ovl{1,n}\,$ on
$\,\bT^n_\a$,\ we have a simple expression for the curvature 2-form
of the Poincar\'e bundle:
\qq\nn
\txF_{\rm P}=\tfrac{1}{2\pi}\,\d_{\mu\nu}\,\sfd\theta_1^\mu\wedge
\sfd\theta_2^\nu\,.
\qqq
This form ensures the required symplecticity of $\,(Q,\txF_{\rm P}
)$.\ The duality-background constraints
\eqref{eq:dualiT-back-constr}, on the other hand, are identical with
the Buscher rules of Refs.\,\cite{Buscher:1987qj,Buscher:1987sk},
relating components of the T-dual pairs $\,(\txg_\a,\txB_\a)\,$ as
per
\qq\nn
\txg_2=-\txF_{\rm P}\circ(\txg_1-\txB_1\circ\txg_1^{-1}\circ\txB_1
)^{-1}\circ\txF_{\rm P}\,,\qquad\qquad\txB_2=-\txF_{\rm P}\circ
\txg_1^{-1}\circ\txB_1\circ\txF_{\rm P}^{-1}\circ\txg_2\,.
\qqq
\eeg\medskip

\beg\textbf{Duality of type $N$ from the central-jump WZW defect.}
\label{ex:ZGjump}\\[-8pt]

\noindent An example of a geometric duality engendered by an
extendible defect associated with an isometry of the target is
provided by the central-jump WZW defect -- a subdefect of the
non-boundary maximally symmetric WZW defect at which the
discontinuity of the $\txG$-valued lagrangean field $\,g:\Si\to
\txG\,$ of the $\si$-model is constrained to take values in the
disjoint union of the distinguished point-like conjugacy classes
$\,\cC_{\la_z}=\{z\}\,$ of elements $\,z\in Z(\txG)\,$ of the centre
$\,Z(\txG)\,$ of the target Lie group $\,\txG$,
\qq\nn
g_{|2}=z\cdot g_{|1}\,.
\qqq
The world-volume of the associated $\cGk$-bi-brane, equipped with a
$Z(\txG)$-invariant (Cartan--Killing) metric and of a vanishing
curvature, all in conformity with \Reqref{eq:Ndual-constr}, is
identified with $\,\txG\x Z(\txG)$,\ and the (pullback)
$\cGk$-bi-brane 1-isomorphisms of \Reqref{eq:iso-for-sympl-nom},
\qq\nn
\cA_{\sfk,z}\ :\ \cGk\xrightarrow{\cong}\bigl(z^{-1}\bigr)^*\cGk\,,
\qqq
one for each element of $\,Z(\txG)$,\ form part of the data of the
$Z(\txG)$-equivariant structure on $\,\cGk\,$ constructed explicitly
in \Rcite{Gawedzki:2003pm}. The extendibility of the defect was
verified in \Rcite{Runkel:2008gr}, where the defect data were
subsequently shown to encode a piece of the Moore--Seiberg data of
the WZW model, to wit, the fusing matrix restricted to the
simple-current sector of the quantised CFT.
\eeg

\section{Fusion of states through defect junctions}\label{sec:fusion}

Hereunder, we continue to unravel, in the canonical framework
adopted in the present paper, the physical contents of the gluing
conditions satisfied by the $\si$-model field and components of the
string background at the defect quiver, this time focusing on the
DJI
\qq\label{eq:DJI}
\D_{T_{n_\jmath}}\om=0\,,
\qqq
to be imposed at any defect junction $\,\jmath\in\Vgt_\G\,$ of
valence $\,n_\jmath$.\ From the point of view of the underlying
gerbe theory, the identity expresses a consistency condition for the
trivialising 2-isomorphism $\,\varphi_{n_\jmath}\,$ assigned to
$\,\jmath$,\ cf.\ \Reqref{eq:Dphin-is}. Much in the same fashion as
the DGC \eqref{eq:DGC} constrains propagation of states in the
world-sheet with an embedded defect quiver by determining which
states of the untwisted sector of the theory are transmitted through
the defect line, the DJI turns out to be associated intimately with
the natural geometric splitting-joining interactions of the string
in that it restricts the spectrum of states emerging from a
collision taking place at the defect quiver with defect
junctions\footnote{Throughout the present section, one ought to keep
in mind the contents of the clarifying footnote
\footref{foot:mink-vs-eukl}.}. Thus, in particular, it will be
shown, in the companion paper \cite{Suszek:2010b}, to define an
intertwiner for a representation of the symmetry algebra of the
$\si$-model on the space of multi-string states associated with an
interaction vertex decorated with a defect junction, cf.\ the recent
findings of Refs.\,\cite{Runkel:2009su,Runkel:2010} to this effect.
This result can be regarded as a straightforward completion of the
chain of results: the old one, reported in \Rcite{Gawedzki:1987ak},
which shows that the $\si$-model gerbe transgresses to a circle
bundle over the configuration space of the untwisted sector of the
theory and thus defines a pre-quantum bundle of the theory, and the
novel one, presented in the previous section, which demonstrates
that the bi-brane, considered together with the attendant DGC, on
one hand transgresses to an isomorphism of the pre-quantum bundle
over an isotropic submanifold in the space of two-string states, and
on the other hand canonically defines a pre-quantum bundle of the
twisted sector of the theory.

In order to illustrate our point and -- in so doing -- introduce
convenient means of description, let us consider the following
(simplest possible)\bigskip

\beg\textbf{The splitting-joining interaction in the absence of
defects.}\label{ex:fusion-triv}\\[-8pt]

\begin{figure}[hbt]~\\[5pt]

$$
 \raisebox{-50pt}{\begin{picture}(50,50)
  \put(-79,-4){\scalebox{0.25}{\includegraphics{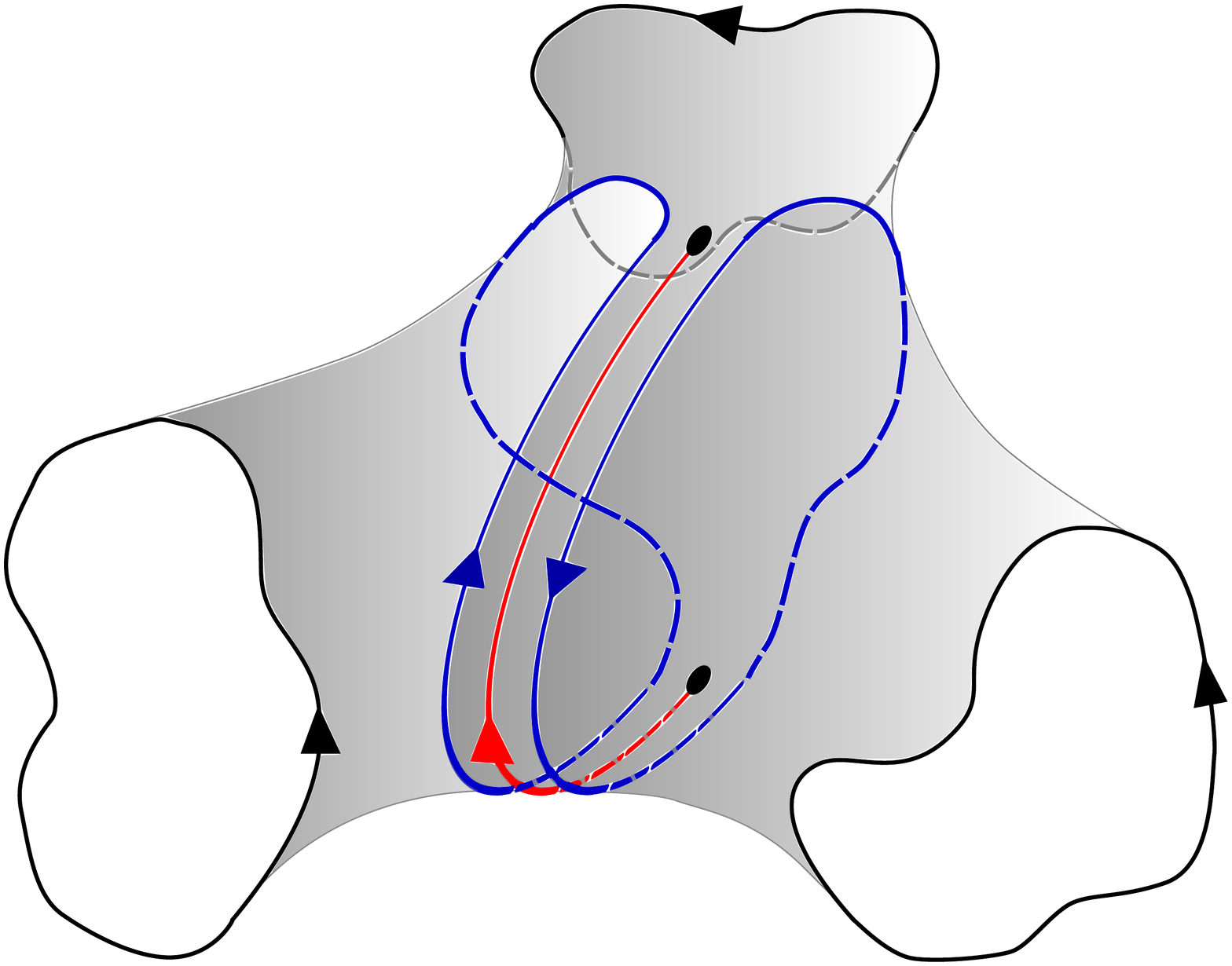}}}
  \end{picture}
  \put(0,0){
     \setlength{\unitlength}{.60pt}\put(-28,-16){
     \put(-30,0)      { (a) }
     \put(-35,40)     { $\ell$  }
     \put(-76,105)    { $\psi_1$ }
     \put(-17,105)    { $\psi_2$ }
     \put(-100,170)   { $\Si$   }
           }\setlength{\unitlength}{1pt}}}
 \hspace{6cm}
 \raisebox{-50pt}{\begin{picture}(50,50)
  \put(-79,-4){\scalebox{0.25}{\includegraphics{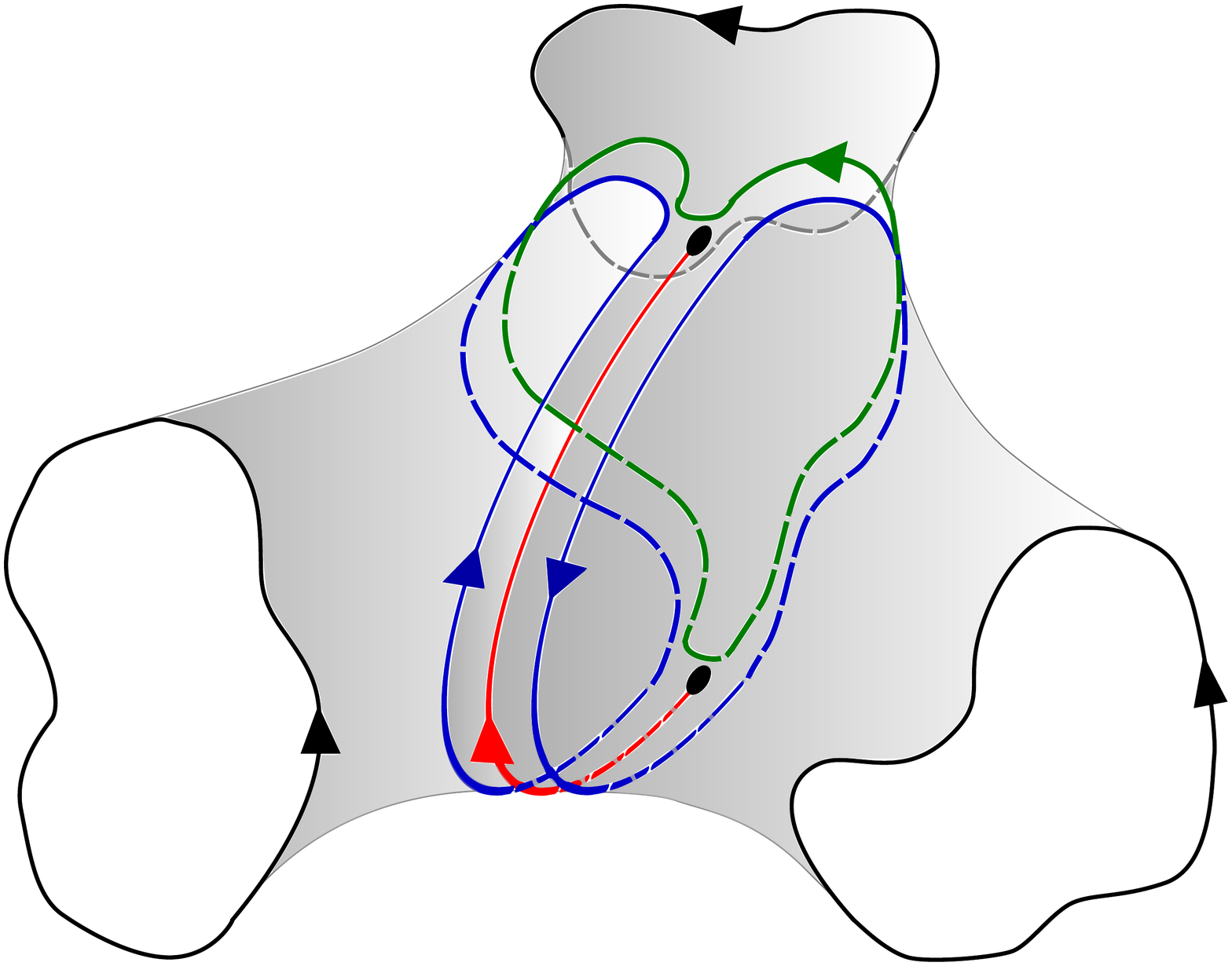}}}
  \end{picture}
  \put(0,0){
     \setlength{\unitlength}{.60pt}\put(-28,-16){
     \put(-30,0)      { (b) }
     \put(-35,40)     { $\ell$  }
     \put(-76,105)    { $\psi_1$ }
     \put(-17,105)    { $\psi_2$ }
     \put(8,215)      { $\psi_3$ }
     \put(-100,170)   { $\Si$   }
           }\setlength{\unitlength}{1pt}}}
$$

\caption{A canonical description of the splitting-joining
interaction. (a) Fusion of the states $\,\psi_1\,$ and $\,\psi_2\,$
along a half-loop $\,\ell\subset\Si,\ \ell\cong\sfI$.\ (b) The
interaction $\,\igt_{\si,(\circledast\cB_{\rm triv}:\cJ_{\rm triv}:
\cB_{\rm triv})}\,$ sends a two-string state $\,(\psi_1,\psi_2)\,$
from the $\cB_{\rm triv}$-fusion subspace into an emergent state
$\,\psi_3\,$ across the loose half-loops.} \label{fig:pants}
\end{figure}

\noindent Let $\,\sfI=[0,\pi]\,$ denote the closed $\pi$-unit
interval, and write
\qq\label{eq:prev-shift-id}\qquad\qquad
\vsi_1=\id_{\bS^1}\,,\qquad\qquad\vsi_2\ :\ \varphi\mapsto 2\pi-
\varphi\,,\qquad\qquad\tau\ :\ \varphi\mapsto\varphi+\pi\,,\qquad
\varphi\in\bS^1
\qqq
for the identity map, the standard parity-reversal map and the
$\pi$-shift map on the unit circle, respectively. We shall think of
$\,\sfI\,$ as a submanifold of the unit circle $\,\bS^1$,\ and so,
in particular, $\,\vsi_2(\sfI)=-[\pi,2\pi]\,$ (the minus denotes the
orientation reversal) and $\,\tau(\sfI)=[\pi,2\pi]$.\ We then take
the cartesian product $\,\sfP_{\si,\emptyset}^{\x 2}\,$ of two
copies of the untwisted state space $\,\sfP_{\si,\emptyset}=\sfT^*
\sfL M$,\ and, for an arbitrarily chosen free open path $\,Y_{1,2}
\in C^\infty(\sfI,M)\equiv\sfI M\,$ in $\,M$,\ define a subspace
\qq\label{eq:fusion-sub-comp-triv}\qquad\qquad
\sfP_{\si,\emptyset}^{\circledast(\cB_{\rm triv};Y_{1,2})}=\left\{\
(\psi_1,\psi_2)\in\sfP_{\si,\emptyset}^{\x 2}\,,\quad \psi_\a=(X_\a,
\sfp_\a)\,,\ \a\in\{1,2\} \quad\middle\vert \quad \left\{ \barr{l}
X_\a\vert_{\vsi_\a(\sfI)}=Y_{1,2} \cr
\sfp_1\vert_\sfI=\sfp_2\vert_{\vsi_2(\sfI)} \earr \right. \
\right\}\,,
\qqq
where $\,\cB_{\rm triv}\,$ stands for the trivial $\cG$-bi-brane
from Example \ref{eg:triv-def}. The gluing condition for the loop
momenta of the two states can be thought of as a trivial instance of
the DGC \eqref{eq:DGC} imposed along the half-loop
interval\footnote{The direction of $\,\sfp_2\,$ is determined,
according to our original conventions, by the orientation of the
reversed half-loop $\,\vsi_2(\sfI)$.}. Clearly, elements of
$\,\sfP_{\si,\emptyset}^{\circledast(\cB_{\rm triv};Y_{1,2})}\,$ are
generic states assigned to the two incoming legs of the standard
stringy `pair-of-pants' diagram with the contour $\,\ell\cong\sfI$,\
which carries no extra string-background data, fixed (arbitrarily)
within the world-sheet $\,\Si\,$ as in \Rfig{fig:pants}. Upon
varying the half-loop $\,Y_{1,2}$,\ we obtain a subspace
\qq\label{eq:fusion-sub-triv}
\sfP_{\si,\emptyset}^{\circledast\cB_{\rm triv}}=\bigcup_{Y_{1,2}
\in\sfI M}\,\sfP_{\si,\emptyset}^{\circledast(\cB_{\rm triv};Y_{1,2}
)}\subset\sfP_{\si,\emptyset}^{\x 2}\,.
\qqq
in the space of untwisted two-string states, which, for the reason
just named and also for other reasons that shall become clear
shortly when we come to discuss less trivial examples, we choose to
call the \textbf{$\cB_{\rm triv}$-fusion subspace of the untwisted
string}.

We may next consider a mapping
\qq\label{eq:int-map-triv}
\igt_{\si,(\circledast\cB_{\rm triv}:\cJ_{\rm triv}:\cB_{\rm triv})}
\ :\ \sfP_{\si,\emptyset}^{\circledast\cB_{\rm triv}}\to\sfP_{\si,
\emptyset}\,,
\qqq
labelled by the trivial inter-bi-brane of Example \ref{eg:triv-def},
which assigns to a pair of states $\,(\psi_1,\psi_2)\,$ a third
state $\,\psi_3\,$ with the loop embedding field satisfying a pair
of `half-loop' gluing conditions
\qq\label{eq:X-glue-int-triv}
X_2\vert_\sfI=X_3\vert_\sfI\,,\qquad\qquad X_1\vert_{\tau(\sfI)}=
X_3\vert_{\tau(\sfI)}\,,
\qqq
and with the loop momentum field constrained analogously as per
\qq\label{eq:p-glue-int-triv}
\sfp_2\vert_\sfI=\sfp_3\vert_\sfI\,,\qquad\qquad\sfp_1\vert_{\tau(
\sfI)}=\sfp_3\vert_{\tau(\sfI)}\,,
\qqq
across -- in the simple case in hand -- the distinguished defect
$\,(\cB_{\rm triv};X_3)$.\ The conditions identify $\,\psi_3\,$ as a
generic state to be placed around the `waist' in the `pair-of-pants'
diagram of \Rfig{fig:pants}. Accordingly, we call
$\,\igt_{\si,(\circledast\cB_{\rm triv}:\cJ_{\rm triv}: \cB_{\rm
triv})}\,$ the \textbf{$2\to 1$ cross-$(\cB_{\rm triv}, \cJ_{\rm
triv})$ interaction of the untwisted string}.
Due to the trivial character of the gluing conditions, the
interaction $\,\igt_{\si,(\circledast\cB_{\rm triv}:\cJ_{\rm triv}:
\cB_{\rm triv})}\,$ is manifestly surjective and many-to-one (pairs
of loops differing by the choice of the differentiable extension of
the given `loose' half-loop embedding field and an extension of the
attendant momentum field to the fused half-loop all map to the same
loop in $\,\sfP_{\si,\emptyset}$). It leads us to
\bedef\label{def:cross-int-sub-triv}
Let $\,\Bgt\,$ be a string background with target $\,\cM=(M,\txg,\cG
)$,\ and let $\,(\sfP_{\si,\emptyset},\Om_{\si,\emptyset})\,$ be the
untwisted state space of the non-linear $\si$-model for
network-field configurations $\,(X\,\vert\,\G)\,$ in string
background $\,\Bgt\,$ on world-sheet $\,(\Si,\g)\,$ with defect
quiver $\,\G$.\ Furthermore, let $\,\sfP_{\si,
\emptyset}^{\circledast\cB_{\rm triv}}\,$ be the $\cB_{\rm
triv}$-fusion subspace in $\,\sfP_{\si,\emptyset}^{\x 2}\equiv
\sfP_{\si,\emptyset}\x\sfP_{\si,\emptyset}\,$ given in
Eqs.\,\eqref{eq:fusion-sub-comp-triv}-\eqref{eq:fusion-sub-triv},
and $\,\igt_{\si,(\circledast\cB_{\rm triv}:\cJ_{\rm triv}:\cB_{\rm
triv})}\,$ the $2\to 1$ cross-$(\cB_{\rm triv},\cJ_{\rm triv})$
interaction defined by
Eqs.\,\eqref{eq:int-map-triv}-\eqref{eq:p-glue-int-triv}. The
\textbf{$2\to 1$ cross-$(\cB_{\rm triv},\cJ_{\rm triv})$ interaction
subspace of the untwisted string} is the space
\qq\nn
\Igt_\si(\circledast\cB_{\rm triv}:\cJ_{\rm triv}:\cB_{\rm triv})&=&
\bigl\{\ (\psi_1,\psi_2,\psi_3)\in\sfP_{\si,\emptyset}^{\circledast
\cB_{\rm triv}}\x\sfP_{\si,\emptyset} \quad\vert\quad \psi_3=
\igt_{\si,(\circledast\cB_{\rm triv}:\cJ_{\rm triv}:\cB_{\rm triv})}
(\psi_1,\psi_2) \ \bigr\}\,.
\qqq
\exdef It is physically pertinent to enquire as to the distinctive
features of $\,\Igt_\si(\circledast\cB_{\rm triv}:\cJ_{\rm triv}:
\cB_{\rm triv})$,\ the latter viewed as a subspace in a symplectic
space. These could then be interpreted as a canonical manifestation
of the basic interaction process in the string theory in hand. The
answer is contained in the following
\berop\label{prop:int-sub-triv}
Let $\,\Bgt\,$ be a string background with target $\,\cM=(M,\txg,\cG
)$,\ and let $\,(\sfP_{\si,\emptyset},\Om_{\si,\emptyset})\,$ be the
untwisted state space of the non-linear $\si$-model for
network-field configurations $\,(X\,\vert\,\G)\,$ in string
background $\,\Bgt\,$ on world-sheet $\,(\Si,\g)\,$ with defect
quiver $\,\G$.\ Consider the symplectic manifold $\,(\sfP_{\si,
\emptyset}^{\x 3},\Om_{\si,\emptyset}^{+-})\,$ defined as
\qq\nn
\sfP_{\si,\emptyset}^{\x 3}:=\sfP_{\si,\emptyset}\x\sfP_{\si,
\emptyset}\x\sfP_{\si,\emptyset}\,,\qquad\qquad\Om_{\si,
\emptyset}^{+-}:=\pr_1^*\Om_{\si,\emptyset}+\pr_2^*\Om_{\si,
\emptyset}-\pr_3^*\Om_{\si,\emptyset}
\qqq
in terms of the canonical projections $\,\pr_\a:\sfP_{\si,
\emptyset}^{\x 3}\to\sfP_{\si,\emptyset}$.\ Furthermore, let
$\,\pi_{\ceL_{\si,\emptyset}}:\ceL_{\si,\emptyset}\to\sfP_{\si,
\emptyset}\,$ be the pre-quantum bundle for the untwisted sector of
the $\si$-model. Then, the $2\to 1$ cross-$(\cB_{\rm triv},\cJ_{\rm
triv})$ interaction subspace $\,\Igt_\si(\circledast\cB_{\rm triv}:
\cJ_{\rm triv}:\cB_{\rm triv} )\,$ is an isotropic submanifold in
$\,(\sfP_{\si,\emptyset}^{\x 3},\Om_{\si, \emptyset}^{+-})\,$ and
there exists a canonical bundle isomorphism
\qq\nn
\Jgt_{\si,(\circledast\cB_{\rm triv}:\cJ_{\rm triv}:\cB_{\rm triv}
)}\ :\ \bigl(\pr_1^*\ceL_{\si,\emptyset}\ox\pr_2^*\ceL_{\si,
\emptyset}\bigr)\vert_{\Igt_\si(\circledast\cB_{\rm triv}: \cJ_{\rm
triv}:\cB_{\rm triv})}\xrightarrow{\cong}\pr_3^*\ceL_{\si,\emptyset}
\vert_{\Igt_\si(\circledast\cB_{\rm triv}: \cJ_{\rm triv}:\cB_{\rm
triv})}
\qqq
between the restrictions to $\,\Igt_\si(\circledast\cB_{\rm triv}:
\cJ_{\rm triv}:\cB_{\rm triv})\,$ of the (tensor) pullback bundles.
\eerop
\noindent A proof of the proposition can be obtained through
specialisation of the proof of Theorem \ref{thm:cross-def-int-untw}
upon setting $\,\cB\equiv\cB_{\rm triv}\,$ and $\,\cJ\equiv\cJ_{\rm
triv}$,\ the latter two being as in Example \ref{eg:triv-def}.

It is owing to the purely geometric nature of the field theory under
consideration that we obtain a simple yet structured representation
of the interaction in the canonical description, which -- as is
obvious from the hitherto discussion -- generalises
straightforwardly to higher-rank fusion and interaction subspaces.
\eeg\medskip

We are now ready to go directly to the main point of interest of
this section, which is a canonical interpretation of the DJI for
world-sheets decorated with non-trivial defect quivers. In order to
describe these in a fashion suggested by the above example, we shall
have to modify our construction of the fusion subspace and that of
the cross-defect interaction suitably.

\subsection{Interactions in the untwisted sector}

\begin{figure}[hbt]~\\[5pt]

$$
 \raisebox{-50pt}{\begin{picture}(50,50)
  \put(-79,-4){\scalebox{0.25}{\includegraphics{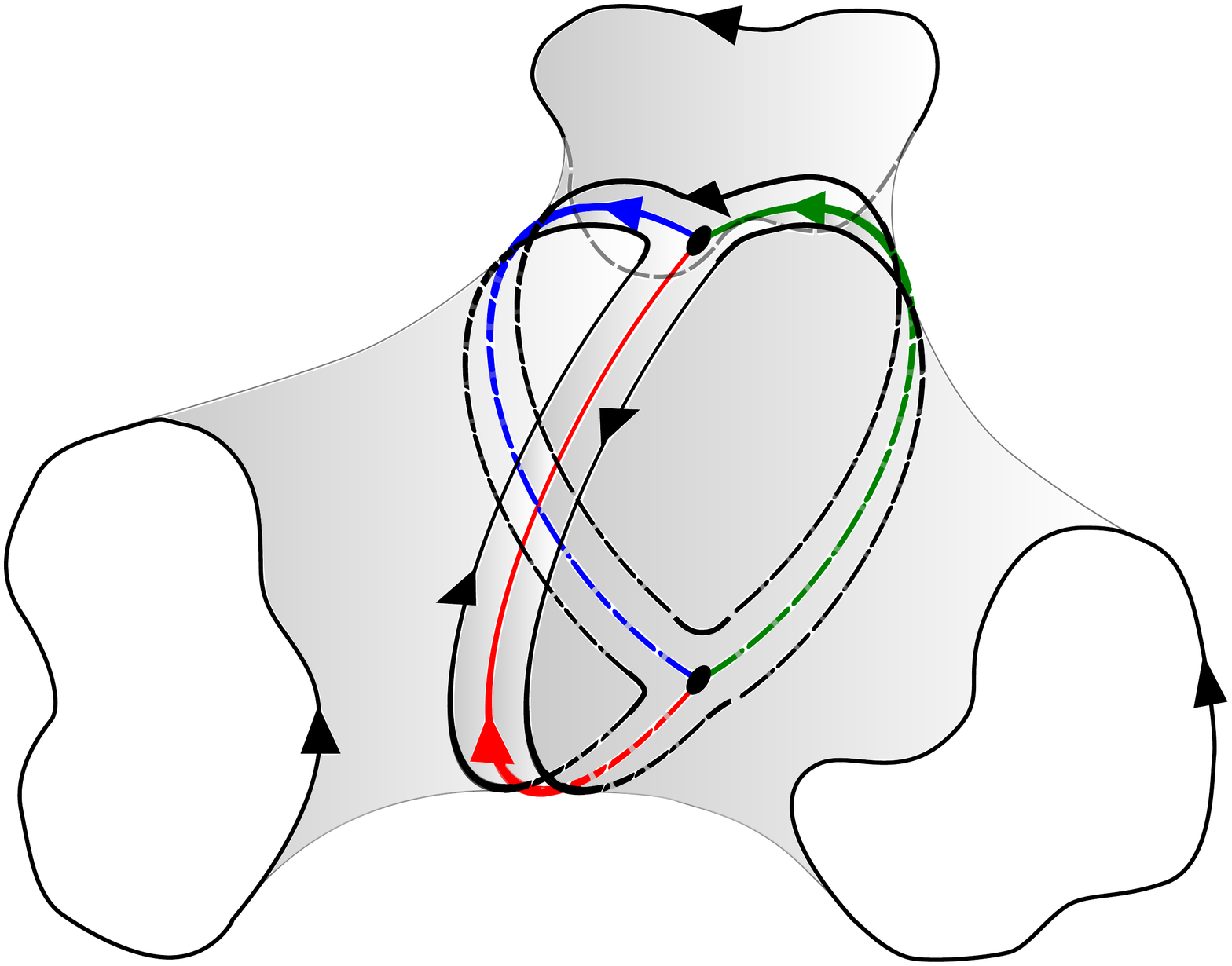}}}
  \end{picture}
  \put(0,0){
     \setlength{\unitlength}{.60pt}\put(-28,-16){
     \put(-35,40)     { $\ell_{1,2}$  }
     \put(-78,103)    { $\psi_1$ }
     \put(-2,148)     { $\psi_2$ }
     \put(1,215)      { $\psi_3$ }
     \put(63,177)     { $\ell_{2,3}$ }
     \put(-65,192)    { $\ell_{3,1}$ }
     \put(-120,160)   { $\wp_1$ }
     \put(105,133)    { $\wp_2$   }
     \put(70,240)     { $\wp_3$ }
     \put(10,180)     { $\jmath$ }
     \put(8,73)       { $\jmath^\vee$ }
           }\setlength{\unitlength}{1pt}}}
$$

\caption{The splitting-joining interaction mediated by defects,
crossing at a pair of defect junctions $\,\jmath\,$ and
$\,\jmath^\vee$.\ Fusion of the states $\,\psi_1\,$ and $\,\psi_2\,$
along the defect $\,(\cB;Y_{1,2})\,$ produces an emergent state
$\,\psi_3\,$ via the $2\to 1$ cross-$(\cB,\cJ)$ interaction.}
\label{fig:fusion}
\end{figure}

We begin with the canonical analysis of the splitting-joining
interaction of untwisted states across a non-trivial defect
(sub-)quiver. While reading the formal definitions and mathematical
expressions ~appearing in this part of the section, it is good to
keep in mind the physical situation depicted in \Rfig{fig:fusion}
that is being modelled by them.
\bedef\label{def:int-sub-untw}
Let $\,\Bgt\,$ be a string background with target $\,\cM=(M,\txg,\cG
)$,\ $\cG$-bi-brane $\,\cB=\bigl(Q,\iota_\a,\om,\Phi\ \vert\ \a\in
\{1,2\}\bigr)\,$ and $(\cG,\cB)$-inter-bi-brane $\,\cJ=\bigl(T_n,
\bigl(\vep^{k,k+1}_n,\pi^{k,k+1}_n \ \vert\ k\in\ovl{1,n}\bigr),
\varphi_n\ \vert\ n\in\bN_{\geq 3}\bigr)$,\ and let $\,(\sfP_{\si,
\emptyset},\Om_{\si,\emptyset})\,$ be the untwisted state space of
the non-linear $\si$-model for network-field configurations
$\,(X\,\vert\,\G)\,$ in string background $\,\Bgt\,$ on world-sheet
$\,(\Si,\g)\,$ with defect quiver $\,\G$.\ For $\,\sfI=[0,\pi]$,\
denote by $\,\sfI Q=C^\infty(\sfI,Q)\,$ the free open-path space of
$\,Q$.\ The \textbf{$\cB$-fusion subspace of the untwisted string}
is the subset of $\,\sfP_{\si,\emptyset}^{\x
2}=\sfP_{\si,\emptyset}\x \sfP_{\si,\emptyset}\,$ given by the
formula
\qq
\sfP_{\si,\emptyset}^{\circledast\cB}&=&\{\ (\psi_1,\psi_2)\in
\sfP_\si^{\x 2}\,,\quad \psi_\a=(X_\a,\sfp_\a)\,,\ \a\in\{1,2\}
\quad\vert\quad (X_1\vert_\sfI,X_2\vert_{\vsi_2(\sfI)})\in(\iota_1
\x\iota_2)(\sfI Q)\cr\cr
&&\hspace{.5cm}\land\quad\exists_{Y_{1,2}\in(\iota_1\x\iota_2)^{-1}
\{(X_1,X_2)\}}\ :\ \tx{DGC}_\cB(\psi_1\vert_\sfI,\psi_2
\vert_{\vsi_2(\sfI)},Y_{1,2})=0 \ \}\,.\label{eq:fus-prod-untw}
\qqq
It is a fibration over the free-path space $\,\sfI Q$,\ and we shall
identify it with the corresponding subspace in $\,\sfP_{\si,
\emptyset}^{\x 2}\x \sfI Q\,$ in what follows. Consider a map
$\,\igt^{2\to 1}_{\si,(\circledast\cB:\cJ:\cB)}$,\ to be termed the
\textbf{$2\to 1$ cross-$(\cB,\cJ)$ interaction of the untwisted
string}, which assigns to pairs of states from
$\,\sfP_{\si,\emptyset}^{\circledast\cB}\,$ subsets of $\,\sfP_{\si,
\emptyset}\,$ such that a pair $\,(\psi_1, \psi_2)\,$ fused along a
free open path $\,Y_{1,2}\in\sfI Q\,$ is mapped to the set of all
those states $\,\psi_3=(X_3,\sfp_3)\,$ that satisfy the relations
\qq\qquad\qquad
X_2\vert_\sfI=\iota_1\circ Y_{2,3}\,,\qquad X_3\vert_\sfI=\iota_2
\circ Y_{2,3}\,,\qquad X_1\vert_{\tau(\sfI)}=\iota_1\circ Y_{1,3}
\,,\qquad X_3\vert_{\tau(\sfI)}=\iota_2\circ Y_{1,3}\,,
\label{eq:nfix-half}\\\cr Y_{I,J}\vert_{\p\sfI}=\pi_3^{I,J}\circ Z
\,,\qquad(I,J)\in\{(1,2),(2,3),(1,3)\}\,,\\\cr
\tx{DGC}_\cB(\psi_2\vert_\sfI,\psi_3\vert_\sfI,Y_{2,3})=0\,,\qquad
\qquad\tx{DGC}_\cB(\psi_1\vert_{\tau(\sfI)},\psi_2\vert_{\tau(\sfI
)},Y_{1,3})=0\label{eq:loop-nfix-half}
\qqq
for $\,\pi_3^{1,3}\equiv\pi_3^{3,1}$,\ some free open paths $\,Y_{1,
3},Y_{2,3}\in\sfI Q\,$ and a map $\,Z:\p\sfI\to T_{3,++-}\cup T_{3,-
-+}\,$ from the set $\,\{0,\pi\}\,$ into the components $\,T_{3,++
-}\,$ and $\,T_{3,--+}\,$ of $\,T_3\subset T\,$ corresponding to the
values $\,\vep_3^{1,2}=\pm 1= \vep_3^{2,3}=-\vep_3^{3,1}\,$ of the
orientation maps, with $\,Z(0)\in T_{3,--+}\,$ and $\,Z(\pi)\in T_{3
,++-}$.\ The \textbf{$2\to 1$ cross-$(\cB,\cJ)$ interaction subspace
of the untwisted string} is then the subset of
$\,\sfP_{\si,\emptyset}^{\x 3}=
\sfP_{\si,\emptyset}\x\sfP_{\si,\emptyset}\x\sfP_{\si,\emptyset}\,$
given by the formula
\qq\nn
\Igt_\si(\circledast\cB:\cJ:\cB)=\{\ (\psi_1,\psi_2,\psi_3)\in
\sfP_{\si,\emptyset}^{\circledast\cB}\x\sfP_{\si,\emptyset}
\quad\vert\quad \psi_3\in\igt^{2\to 1}_{\si,(\circledast\cB:\cJ:\cB
)}(\psi_1,\psi_2) \ \}\,.
\qqq
Once again, the latter subspace is a fibration over the cartesian
cube $\,\sfI Q^3$,\ and we shall identify it with the corresponding
subspace in $\,\sfP_\si^{\x 3}\x\sfI Q^{\x 3}\,$ in what follows.
\exdef \noindent We have
\bethe\label{thm:cross-def-int-untw}
Let $\,\Bgt=(\cM,\cB,\cJ)\,$ be a string background, and let
$\,(\sfP_{\si,\emptyset},\Om_{\si,\emptyset})\,$ be the untwisted
state space of the non-linear $\si$-model for network-field
configurations $\,(X\,\vert\,\G)\,$ in string background $\,\Bgt\,$
on world-sheet $\,(\Si,\g)\,$ with defect quiver $\,\G$,\ with the
pre-quantum bundle for the untwisted sector of the $\si$-model over
it, $\,\pi_{\ceL_{\si,\emptyset}}:\ceL_{\si,\emptyset}\to\sfP_{\si,
\emptyset}$.\ Furthermore, let $\,(\sfP_{\si,\emptyset}^{\x 3},
\Om_{\si,\emptyset}^{+-})\,$ be the symplectic manifold defined in
Proposition \ref{prop:int-sub-triv}. Then, the following statements
hold true:
\bit
\item[i)] the $2\to 1$ cross-$\cB$ interaction subspace of the
untwisted string, $\,\Igt_\si(\circledast\cB:\cJ:\cB)$,\ constructed
in Definition \ref{def:int-sub-untw}, is an isotropic submanifold of
$\,(\sfP_{\si,\emptyset}^{\x 3},\Om_{\si, \emptyset}^{+-})$;
\item[ii)] the background $\,\Bgt\,$ canonically
induces a bundle isomorphism
\qq\nn
\Jgt_{\si,(\circledast\cB:\cJ:\cB)}\ :\ \bigl(\pr_1^*\ceL_{\si,
\emptyset}\ox\pr_2^*\ceL_{\si,\emptyset}\bigr)\vert_{\Igt_\si(
\circledast\cB:\cJ:\cB)}\xrightarrow{\cong}\pr_3^*\ceL_{\si,
\emptyset}\vert_{\Igt_\si(\circledast\cB:\cJ:\cB)}
\qqq
between the restrictions to $\,\Igt_\si(\circledast\cB:\cJ:\cB)\,$
of the (tensor) pullback bundles.
\eit
\ethe \noindent A proof of the theorem is given in Appendix
\ref{app:cross-def-int-untw}. \brem\label{rem:duality-scheme} The
relation between defects and dualities of the $\si$-model worked out
in the previous section distinguished those bi-branes whose maps
$\,\iota_\a:Q\to M\,$ are \emph{surjective submersions}. The
canonical analysis of the splitting-joining interaction of the
untwisted string immediately leads to similar conclusions for the
inter-bi-brane. Indeed, it is clear that for a given interaction
vertex of \Rfig{fig:fusion} to allow the appearance of arbitrary
outgoing and incoming states, i.e.\ for the interaction subspace to
project onto each cartesian component
$\,\sfP_{\si,\emptyset}\subset\sfP_{\si, \emptyset}^{\x n}$,\ the
inter-bi-brane maps $\,\pi_n^{k,k+1}:T_n \to Q\,$ should all be
surjective. Taking into account the additional requirement of
topologicality of the defect, we note that -- at least in the case
of extendible defects -- the inter-bi-brane maps should, moreover,
be submersions, so that, once more, surjective submersions become
singled out. These are particularly interesting in the case of
inter-bi-branes admitting \textbf{induction}, as introduced in
\Rxcite{Sec.\,2.8}{Runkel:2008gr}. The latter is motivated by the
physical observation that a defect junction $\,\jmath\,$ of valence
$\,n_\jmath>3$,\ represented by a defect-field insertion in the
underlying CFT, can be regarded as a product of a stepwise limiting
procedure in which a collection of `elementary' 3-valent vertices
are merged by sending the lengths of the interconnecting defect
lines to zero, whereby the associated defect fields of the CFT may
have to be renormalised in order to remove the ensuing divergencies.
In what follows, we briefly recall the idea of induction in
restriction to the component of the background obtained by fixing
the values of the orientation maps $\,\vep_n^{k,k+1}\,$ to be all
$\,+1\,$ except for $\,\vep_n^{n,1}=-1\,$ for all $\,n\in \bN_{\geq
3}\,$ and taking the inter-bi-brane 2-isomorphisms $\,\varphi_n\,$
restricted to the corresponding submanifolds
$\,T_{n,++\cdots+-}\subset T_n\,$ (the remaining components of the
background are left unrestricted). As the very construction of the
$\si$-model in string background $\,\Bgt\,$ clearly indicates, there
are no additional geometric insights to gain from considering the
more general case. Indeed, as long as we are concerned with a
\emph{single} defect junction (which is, in particular, all we need
to determine the associated field-space data), we are at liberty to
choose an arbitrary relative-orientation pattern for the defect
lines converging at that junction.

\begin{figure}[hbt]~\\[5pt]

$$
 \raisebox{-50pt}{\begin{picture}(50,50)
  \put(-70,-4){\scalebox{0.25}{\includegraphics{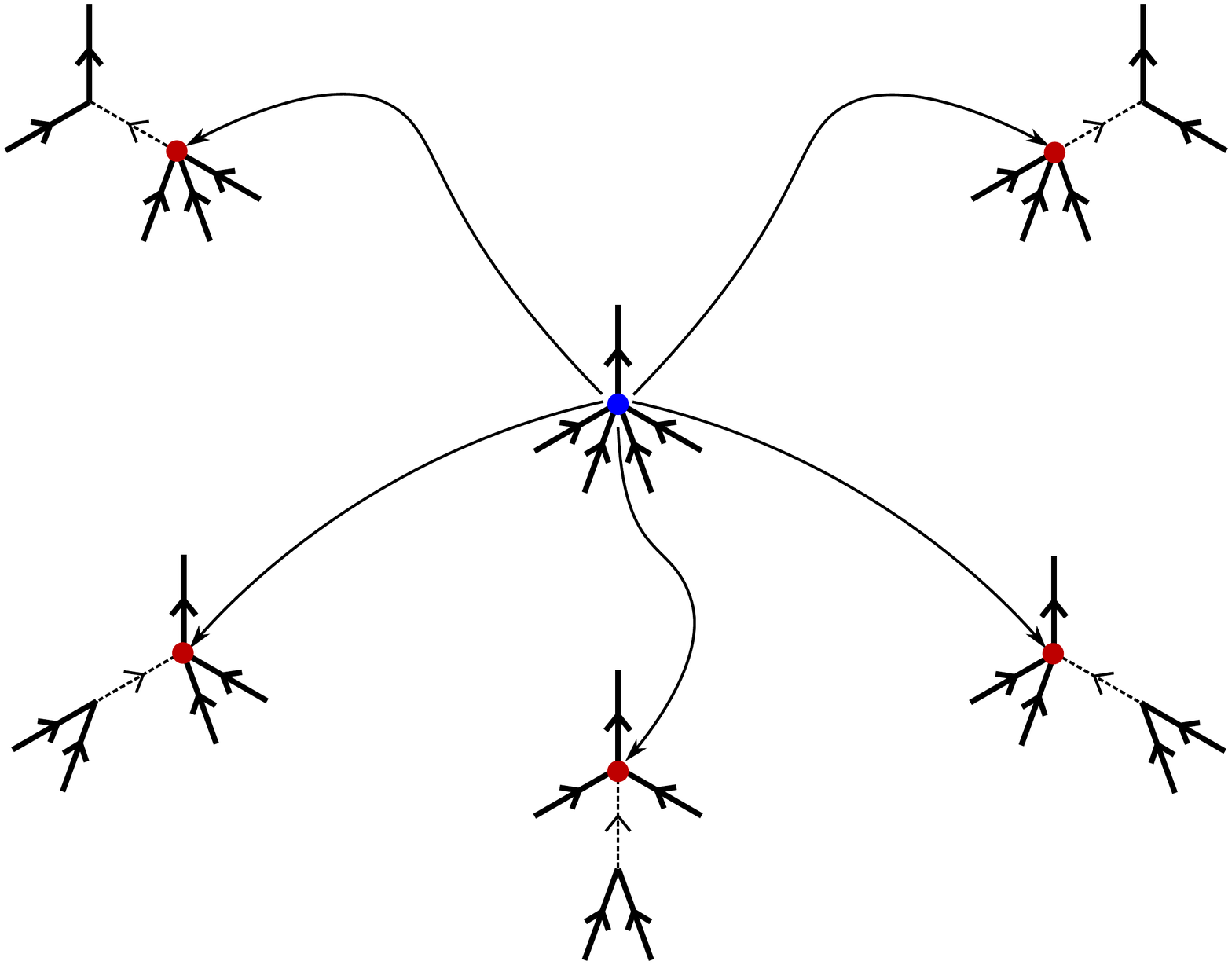}}}
  \end{picture}
  \put(0,0){
     \setlength{\unitlength}{.60pt}\put(-28,-16){
     \put(-80,235)    { $d^{(4)}_4$  }
     \put(-80,137)    { $d^{(4)}_3$  }
     \put(17,100)     { $d^{(4)}_2$  }
     \put(55,137)     { $d^{(4)}_1$  }
     \put(55,235)     { $d^{(4)}_0$  }
           }\setlength{\unitlength}{1pt}}}
$$

\caption{Simplicial moves on a defect junction of valence 5.}
\label{fig:induction}
\end{figure}

Denote the inter-bi-brane maps $\,\pi_3^{k,k+1},\ k\in\{1,2,3\}\,$
as
\qq\nn
\pi_3^{1,2}=d^{(2)}_0\,,\qquad\qquad\pi_3^{2,3}=d^{(2)}_2\,,\qquad
\qquad\pi_3^{3,1}=d^{(2)}_1\,.
\qqq
The background $\,\Bgt\,$ shall be termed a \textbf{string
background with induction} iff, for each $\,n\geq 3$,\ there exist
smooth maps
\qq\nn
d^{(n)}_i\ :\ T_{n+1,++\cdots+-}\to T_{n,++\cdots+-}\,,\qquad
i\in\ovl{0,n}
\qqq
satisfying the identities
\qq\label{eq:simpl-id-dd}
d^{(n-1)}_i\circ d^{(n)}_j=d^{(n-1)}_{j-1}\circ d^{(n)}_i\qquad
\tx{for}\quad i<j\,,
\qqq
and such that the inter-bi-brane 2-isomorphisms $\,\varphi_n\,$ are
induced from $\,\varphi_3\,$ in a natural manner illustrated in
\Rxcite{Sec.\,2.8}{Runkel:2008gr} on an explicit example amenable to
a straightforward generalisation  (which we leave out here for the
sake of conciseness). Identities \eqref{eq:simpl-id-dd} arise as
simple consistency conditions to be imposed on the limiting values
attained by the $\si$-model field at the defect junctions of an
embedded defect quiver as we decompose the defect quiver at these
junctions into clusters of defect junctions of lower valence bridged
by intermediate defect lines, prior to passing to the limit of the
vanishing length of the intermediate defect lines, cf.\
\Rfig{fig:induction}.
On the other hand, it is tempting to view them as the simplicial
identities obeyed by the face maps of a simplicial space composed by
the family of manifolds $\,\{M,Q,T_{3,++-},T_{4,+++-},\ldots\}$.\
That this is the proper manner of thinking of the string background
with induction can be seen as follows: First of all, the bi-brane
maps provide a natural completion of the family $\,(d^{(n)}_i\
\vert\ i\in\ovl{0,n}\,,\ n\in\bN_{\geq 2})\,$ of smooth maps
interrelated as per \Reqref{eq:simpl-id-dd}, which can readily be
seen upon setting
\qq\nn
d^{(1)}_0:=\iota_1\,,\qquad\qquad d^{(1)}_1:=\iota_2
\qqq
and recalling relations \eqref{eq:proto-simpl}. Furthermore, it is
natural, from the point of view of the associated $\si$-model, to
incorporate the trivial defect into the formal definition of the
string background by allowing degenerate defect quivers in which
some defect lines carry the (trivial) data of the trivial defect.
Indeed, we should always be able to insert a circular trivial defect
into the world-sheet, or attach a trivial-defect line to a given
defect junction, between any two of its defect lines, whereby the
valence of the defect junction is increased by $1$. It is easy to
see, going through similar consistency checks of the limiting values
of the $\si$-model fields as those used in the derivation of
\Reqref{eq:simpl-id-dd}, that this can be formalised as a
requirement of the existence of distinguished sections
\qq
s^{(n-1)}_i\ :\ T_{n,++\cdots+-}\to T_{n+1,++\cdots+-}\,,\qquad i\in
\ovl{0,n-1}\,,\cr\label{eq:deg-sec}\\ s^{(1)}_j\ :\ Q\to
T_{3,++-}\,,\qquad j\in\{0,1 \}\,,\qquad\qquad\qquad s^{(0)}_0\ :\
M\to Q\nonumber
\qqq
of the respective surjective submersions, satisfying the identities
\qq
s_i^{(n+1)}\circ s_j^{(n)}=s_{j+1}^{(n+1)}\circ s_i^{(n)}\qquad
\tx{if}\quad i\leq j\,,\label{eq:simpl-id-ss}\\\cr d_i^{(n+1)}\circ
s_j^{(n)}=\left\{\barr{lcl} s_{j-1}^{(n-1)}\circ d_i^{(n)}\quad &
\tx{if} & i<j\cr\cr \id_{T_{n+1,++\cdots+-}}\quad & \tx{if} & i=j
\quad\lor\quad i=j+1\cr\cr s_j^{(n-1)}\circ d_{i-1}^{(n)}\quad &
\tx{if} & i>j+1 \earr\right..\label{eq:simpl-id-ds}
\qqq
Here, $\,s^{(0)}_0\,$ puts the image, \wrt $\,X:\Si\setminus\G\to
M$,\ of an arbitrary point from the interior of a world-sheet patch
on the world-volume of the trivial defect. Similarly, $\,s^{(1
)}_j\,$ expresses the possibility of viewing the image, \wrt $\,X:
\G\setminus\Egt_\G\to Q$,\ of a point from the interior of a defect
line embedded in $\,\Si\,$ as the image of the degenerate 3-valent
defect junction, with the trivial-defect line joining the original
one from the side of $\,U_1\,$ (for $\,j=1$) or $\,U_2\,$ (for $\,j=
0$) in the notation of Definition \ref{def:net-field}. Finally, the
maps $\,s^{(n-1)}_i\,$ represent the process of increasing the
valence of a given defect junction through attachment of a
trivial-defect line between the neighbouring defect lines $\,\ell_{n
-i-1,n-i}\,$ and $\,\ell_{n-i,n-i+1}\,$ that converge at this
junction (with the usual convention $\,\ell_{0,1}\equiv\ell_{n,1}
\equiv\ell_{n,n+1}$).

Altogether, Eqs.\,\eqref{eq:simpl-id-dd}-\eqref{eq:simpl-id-ds}
reproduce the full set of simplicial identities for the face maps
$\,d^{(n)}_i\,$ and degeneracy maps $\,s^{(n)}_i\,$ of a simplicial
space
\qq\nn
\alxydim{@C=1.cm@R=.05cm}{\cdots \ar@<1ex>[r]^{d^{(4)}_i}
\ar@<.5ex>[r] \ar@<0ex>[r] \ar@<-.5ex>[r] \ar@<-1ex>[r] & T_4
\ar@<.75ex>[r]^{d^{(3)}_i} \ar@<.25ex>[r] \ar@<-.25ex>[r]
\ar@<-.75ex>[r] & T_3 \ar@<.5ex>[r]^{d^{(2)}_i} \ar@<0.ex>[r]
\ar@<-.5ex>[r] & Q \ar@<.5ex>[r]^{d^{(1)}_i} \ar@<-.5ex>[r]
& M}
\qqq
A \textbf{simplicial string background $\,\Bgt=(\cM,\cB,\cJ)$},\
i.e.\ a string background with induction, equipped with the family
\eqref{eq:deg-sec} of sections, can be regarded as a straightforward
generalisation of a simplicial background describing a proper
(geometric) symmetry of the $\si$-model with target $\,\cM=(M,\txg,
\cG)\,$ endowed with structure of a $\txK$-space for some group
$\,\txK$,\ acting on $\,(M,\txg)\,$ by isometries
\qq\nn
\ell\ :\ \txK\x M\to M\ :\ (g,m)\mapsto\ell_g(m)=:g.m\,,\qquad\qquad
\ell_g^*\txg=\txg
\qqq
that lift to the gerbe $\,\cG\,$ in the sense made precise in
\Rcite{Gawedzki:2010rn}. The relevant simplicial space is given by
the nerve
\qq\nn
\alxydim{@C=1.cm@R=.05cm}{\sfN(\txK\lx M)^\bullet\quad :
\hspace{-.5cm} & \cdots \ar@<1ex>[r]^{\Gup d^{(4)}_i} \ar@<.5ex>[r]
\ar@<0ex>[r] \ar@<-.5ex>[r] \ar@<-1ex>[r] & \txK^3\x M
\ar@<.75ex>[r]^{\Gup d^{(3)}_i} \ar@<.25ex>[r] \ar@<-.25ex>[r]
\ar@<-.75ex>[r] & \txK^2\x M \ar@<.5ex>[r]^{\Gup d^{(2)}_i}
\ar@<0.ex>[r] \ar@<-.5ex>[r] & \txK\x M \ar@<.5ex>[r]^{\Gup
d^{(1)}_i} \ar@<-.5ex>[r] & M}\,,
\qqq
of the action groupoid
\qq\nn
\alxydim{@C=2.cm}{\txK\lx M\quad : \hspace{-2cm} & \txK\x M
\ar@<.5ex>[r]^{\pr_2=:\Gup d^{(1)}_0} \ar@<-.5ex>[r]_{\ell=:\Gup
d^{(1)}_1} & M}\,,
\qqq
written in terms of the action $\,\ell\,$ and of the canonical
projection $\,\pr_2$.\ The action groupoid is understood as the
small category with object and morphism sets
\qq\nn
\obj\,(\txK\lx M)=M\,,\qquad\qquad\morf\,(\txK\lx M)=\txK\x M\,,
\qqq
with the identity morphism ($e\,$ is the group unit)
\qq\nn
\id_m=(e,m)\,,
\qqq
and with source and target maps
\qq\nn
s(g,m)=m\,,\qquad\qquad t(g,m)=g.m\,,
\qqq
which, altogether, lead to a natural identification of the spaces of
$n$-tuples of composable morphisms (i.e.\ the remaining members of
the family $\,\sfN(\txK\lx M)^\bullet\,$ of spaces) with the
respective product spaces $\,\txK^n\x M$.\ The (inter-)bi-brane data
compose a $\txK$-equivariant structure on $\,\cG\,$ as in
\Rcite{Gawedzki:2010rn}. This structure was shown to be a
prerequisite of the gauging of the internal (rigid) $\txK$-symmetry
of the $\si$-model (the latter being obtained as a lift of the
geometric action $\,\ell\,$ to the phase space of the $\si$-model)
and it is readily proven necessary for the gauged $\si$-model to
descend to the coset $\,M/\txK$.\ Reasoning by analogy, we
conjecture that the existence of a simplicial string background
associated with a given $\si$-model duality is a necessary
ingredient of a consistent formulation of string theory on a
`quotient' background of a $\si$-model descended from the original
one upon `gauging' the duality group, whenever such a group and the
attendant `quotient' can be defined. Here, the submersive
surjectivity of the face maps is necessary to establish duality
equivalences on the entire phase space and to ensure that all states
are transmitted by any defect quiver carrying the duality data. More
specifically, the bi-brane face maps $\,d^{(1)}_i:Q\to M\,$ encode
an element-wise presentation of the set of duality transformations
on the space of states, the 3-valent inter-bi-brane face maps
$\,d^{(2)}_i:T_3\to Q\,$ render the presentation distributive with
respect to the group operation on the set of dualities, and the
requirement that the 4-valent inter-bi-brane structure induced from
the 3-valent one by means of the face maps $\,d^{(3)}_i:T_4\to
T_3\,$ be independent of the choice of the defining simplicial move
enforce the associativity of the presentation. Finally, the
existence of an induced inter-bi-brane structure on the component
world-volumes $\,T_n\,$ of valence $\,n\geq 5$,\ independent of the
choice of defining simplicial moves, guarantees that the associative
presentation of the duality group carries over to arbitrary
interaction schemes (i.e.\ to arbitrary defect quivers). As the
treatment of coset $\si$-models in \Rcite{Gawedzki:2010rn} suggests,
there are, generically, extra constraints to be imposed on the thus
obtained `duality-equivariant' structure for the $\si$-model to
descend to the duality quotient. A motivating explicit instantiation
of this idea (if also far from being understood rigorously to date)
is the notion of a T-fold, advanced in \Rcite{Hull:2004in}, in which
the target is described in terms of local charts (carrying local
metric and gerbe data) patched together using T-duality
transformations. Let us also point out that the above
\textbf{duality scheme} bears a deep affinity with the
\textbf{categorial descent scheme} discussed in
\Rcite{Fuchs:2009si}. We hope to return to these issues in the near
future.\erem

\subsection{Interactions in the twisted sector}

\begin{figure}[hbt]~\\[5pt]

$$
 \raisebox{-50pt}{\begin{picture}(50,50)
  \put(-79,-4){\scalebox{0.25}{\includegraphics{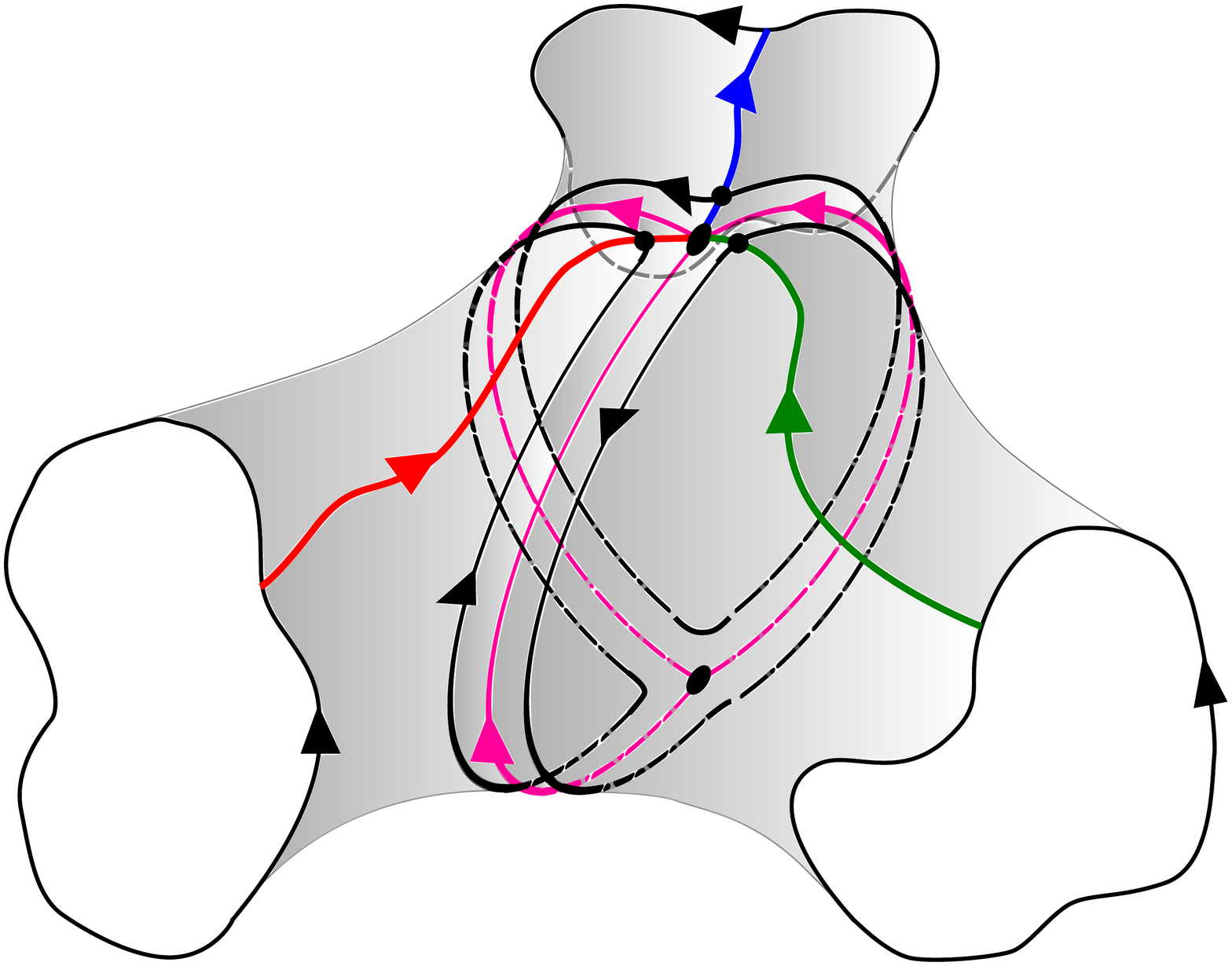}}}
  \end{picture}
  \put(0,0){
     \setlength{\unitlength}{.60pt}\put(-28,-16){
     \put(-35,40)     { $\ell$  }
     \put(-90,105)    { $\psi_{\ell_{1,2}}$ }
     \put(-5,145)     { $\psi_{\ell_{2,3}}$ }
     \put(23,213)     { $\psi_{\ell_{3,1}}$ }
     \put(-95,140)    { $\ell_{1,2}$ }
     \put(55,112)     { $\ell_{2,3}$ }
     \put(-10,220)    { $\ell_{3,1}$ }
     \put(63,177)     { $\ell_R$ }
     \put(-65,192)    { $\ell_L$ }
     \put(10,180)     { $\jmath$ }
           }\setlength{\unitlength}{1pt}}}
$$

\caption{The splitting-joining interaction of a triple of 1-twisted
states. The Cauchy contours representing the states are drawn in
black.} \label{fig:tw-fusion}
\end{figure}

Another class of string processes in which the inter-bi-brane and
the associated DJI are naturally expected to transgress to the
canonical description is the splitting-joining interaction of
twisted states. When assembling the necessary formal ingredients, we
are guided by the depiction of the corresponding network-field
configuration on the world-sheet, the simplest of its kind,
presented in \Rfig{fig:tw-fusion}. Thus, we consider three defect
lines, $\,\ell_{1,2},\ell_{2,3}\,$ and $\,\ell_{3,1}$,\ converging
at a defect junction $\,\jmath$.\ To each defect line, we attach the
corresponding Cauchy contour $\,C_{I,J},\
(I,J)\in\{(1,2),(2,3),(3,1)\}$,\ oriented as the ones drawn in the
figure and crossing the respective defect lines transversally, each
at a single point. The contours are next pushed towards one another
in such a manner that they overlap pairwise along open arcs, all
crossing at a pair of points in $\,\Si$,\ the defect junction
$\,\jmath\,$ being one of them.

Our first task consists in identifying the various subspaces within
the cartesian square and the cartesian cube of the 1-twisted state
space of the $\si$-model, of relevance to the problem in hand.
\bedef\label{def:int-sub-tw}
Let $\,\Bgt=(\cM,\cB,\cJ)\,$ be a string background as in Definition
\ref{def:bckgrnd}. Fix $\,P\in\bS^1\,$ and $\,\vep\in\{-1,+1\}$,\
and let $\,(\sfP_{\si,\cB|(P,\vep)},\Om_{\si,\cB|(P,\vep)})\,$ be
the 1-twisted state space of the non-linear $\si$-model for
network-field configurations $\,(X\,\vert\,\G)\,$ in string
background $\,\Bgt\,$ on world-sheet $\,(\Si,\g)\,$ with defect
quiver $\,\G$.\ The \textbf{$\cB_{\rm triv}$-fusion subspace of the
1-twisted string} is a subspace in
\qq\label{eq:Ptw2}
\sfP_{\si,\cB|(\vep_1,\vep_2)}^{\x 2}:=\sfP_{\si,\cB|(\pi,\vep_1)}\x
\sfP_{\si,\cB|(\pi,\vep_2)}
\qqq
given by the formula
\qq
\sfP_{\si,\cB|(\vep_1,\vep_2)}^{\circledast\cB_{\rm triv}}&=&\{\ (
\psi_1,\psi_2)\in\sfP_{\si,\cB|(\vep_1,\vep_2 )}^{\x 2}\,,\quad
\psi_\a=(X,\sfp,q_\a,V_\a)\,,\ \a\in\{1,2\}\cr\cr
&&\hspace{.5cm}\vert\quad X_1\vert_\sfI=X_2\vert_{\vsi_2(\sfI)}\quad
\land\quad\sfp_1\vert_\sfI=\sfp_2\vert_{\vsi_2(\sfI)} \ \}\,.
\label{eq:fus-prod-tw}
\qqq
Clearly, the specific choice $\,P_1=\pi=P_2\,$ of the intersection
point is immaterial to the outcome of our analysis. Consider a map
$\,\igt^{2\to 1}_{\si,(\circledast\cB_{\rm triv}:\cJ:\cB_{\rm triv}
)}$,\ to be termed the \textbf{$2\to 1$ cross-$(\cB_{\rm triv},\cJ)$
interaction of the untwisted string}, which assigns to pairs of
states from $\,\sfP_{\si,\cB|(\vep_1,\vep_2)}^{\circledast\cB_{\rm
triv}}\,$ subsets of $\,\sfP_{\si,\cB|(\pi,\vep_3)}\,$ such that a
pair $\,(\psi_1,\psi_2)\,$ is mapped to the set of all those states
$\,\psi_3=(X,\sfp,q_3,V_3)\,$ that satisfy the relations
\qq\label{eq:glue-inter-tw}
\barr{cc} X_2\vert_\sfI=X_3\vert_\sfI\,,\qquad & X_1\vert_{\tau(\sfI
)}=X_3\vert_{\tau(\sfI)}\,,\cr\cr
\sfp_2\vert_\sfI=\sfp_3\vert_\sfI\,,\qquad & \sfp_1\vert_{\tau(\sfI
)}=\sfp_3\vert_{\tau(\sfI )}\,, \earr
\qqq
augmented with the constraints
\qq\label{eq:glue-inter-junct}
q_k=\pi_3^{k,k+1}(t_3)\,,\qquad k\in\{1,2,3\}\,,
\qqq
the latter being expressed in terms of a fixed point $\,t_3\in T_{3,
++-}\subset T\,$ from the component $\,T_{3,++-}\,$ of $\,T_3\subset
T\,$ corresponding to the values $\,\vep_3^{1,2}=+1=\vep_3^{2,3}=-
\vep_3^{3,1}\,$ of the orientation maps. The \textbf{$2\to 1$
cross-$(\cB_{\rm triv},\cJ )$ interaction subspace of the 1-twisted
string} is then the subset of the space
\qq\nn
\sfP_{\si,\cB|(\vep_1,\vep_2,\vep_3)}^{+-}:=\sfP_{\si,\cB|(\pi,
\vep_1)}\x\sfP_{\si,\cB|(\pi,\vep_2)}\x\sfP_{\si ,\cB|(\pi,\vep_3)}
\qqq
given by the formula
\qq\nn
\Igt_\si(\circledast\cB_{\rm triv}:\cJ:\cB_{\rm triv})^{\cB|(\vep_1
,\vep_2,\vep_3)}=\{\ (\psi_1,\psi_2,\psi_3)\in\sfP_{\si,\cB|(
\vep_1,\vep_2)}^{\circledast\cB_{\rm triv}}\x\sfP_{\si,\cB|(\pi,
\vep_3)}\quad\vert\quad \psi_3\in\igt^{2\to 1}_{\si,(\circledast
\cB_{\rm triv}:\cJ:\cB_{\rm triv})}(\psi_1,\psi_2)\ \}\,.
\qqq
\exdef \noindent We are now ready to give
\bethe\label{thm:cross-def-int-tw}
Let $\,\Bgt=(\cM,\cB,\cJ)\,$ be a string background as in Definition
\ref{def:bckgrnd}, and let
$\,(\sfP_{\si,\cB|(\pi,\vep)},\linebreak\Om_{\si,\cB|(\pi,\vep)}),\
\pi\in\bS^1\,$ be the 1-twisted state space of the non-linear
$\si$-model for network-field configurations $\,(X\,\vert\,\G)\,$ in
string background $\,\Bgt\,$ on world-sheet $\,(\Si,\g)\,$ with
defect quiver $\,\G$,\ with the pre-quantum bundle for the 1-twisted
sector of the $\si$-model over it,
$\,\pi_{\ceL_{\si,\cB|(\pi,\vep)}}:\ceL_{\si,\cB|(\pi,\vep)}\to
\sfP_{\si,\cB|(\pi,\vep)}$.\ Endow the space
\qq\nn
\sfP_{\si,\cB|(\vep_1,\vep_2,\vep_3)}^{\x 3}:=\sfP_{\si,\cB|(\pi,
\vep_1)}\x\sfP_{\si,\cB|(\pi,\vep_2)}\x\sfP_{\si,\cB|(\pi,\vep_3)}
\qqq
with the symplectic structure defined by the 2-form
\qq\nn
\Om^{+-}_{\si,\cB|(\vep_1,\vep_2,\vep_3)}:=\pr_1^*\Om_{\si,\cB|(\pi,
\vep_1)}+\pr_2^*\Om_{\si,\cB|(\pi,\vep_2)}-\pr_3^*\Om_{\si,\cB|(\pi,
\vep_3)}
\qqq
in terms of the canonical projections $\,\pr_k:\sfP_{\si,\cB|(\vep_1
,\vep_2,\vep_3)}^{\x 3}\to\sfP_{\si,\cB|(\pi,\vep_k)},\ k\in\{1,2,3
\}$.\ and consider the pullback circle bundle
\qq\nn
\ceL_{\si,\cB|(\vep_1,\vep_2,\vep_3)}^{+-}:=\to\sfP_{\si,\cB|(\vep_1,\vep_2,
\vep_3)}^{\x 3}\,.
\qqq
Then, the following statements hold true:
\bit
\item[i)] the $2\to 1$ cross-$(\cB_{\rm triv},\cJ)$ interaction
subspace $\,\Igt_\si(\circledast\cB_{\rm triv}:\cJ:\cB_{\rm
triv})^{\cB|(\vep_1,\vep_2,\vep_3)}$ of the 1-twisted string is an
isotropic submanifold in the symplectic manifold $\,(\sfP_{\si,\cB|(
\vep_1,\vep_2,\vep_3)}^{\x 3},\Om^{+-}_{\si,\cB|(\vep_1,\vep_2,
\vep_3)})$;
\item[ii)] the background $\,\Bgt\,$ canonically
induces a trivialisation
\qq\nn
\Jgt_{\si,(\circledast\cB_{\rm triv}:\cJ:\cB_{\rm triv})}^{\cB|(
\vep_1,\vep_2,\vep_3)}\ :\ \bigl(\pr_1^*\ceL_{\si,\cB|(\pi,\vep_1)}
\ox\pr_2^*\ceL_{\si,\cB|(\pi,\vep_2)}\bigr)\vert_{{\Igt_\si(
\circledast\cB_{\rm triv}:\cJ:\cB_{\rm triv})^{\cB|(\vep_1,\vep_2,
\vep_3)}}}\xrightarrow{\cong}\pr_3^*\ceL_{\si,\cB|(\pi,\vep_3)}
\vert_{\Igt_\si(\circledast\cB_{\rm triv}:\cJ:\cB_{\rm triv})^{\cB
|(\vep_1,\vep_2,\vep_3)}}
\qqq
between the restrictions to $\,\Igt_\si( \circledast\cB_{\rm triv}:
\cJ:\cB_{\rm triv})^{\cB|(\vep_1,\vep_2,\vep_3)}$ of the (tensor)
pullback bundles.
\eit
\ethe \noindent A proof of the theorem is given in Appendix
\ref{app:cross-def-int-tw}. \brem A generic interaction process on a
world-sheet with an arbitrary embedded defect quiver is a
combination of the two `pure' types considered in detail above, with
higher-rank fusion and interaction subspaces involved. Our
conclusions are readily seen to generalise to arbitrary such
processes. \erem\bigskip

\beg\textbf{The WZW fusion ring and the maximally symmetric
inter-bi-brane.}\label{ex:Verlinde}\\[-8pt]

\noindent The correspondence between inter-bi-brane 2-isomorphisms
and interaction subspaces in the space of states of the untwisted
string may, in fact, carry over to the quantised theory, as
demonstrated in Refs.\,\cite{Runkel:2009su,Runkel:2010}, where the
inter-bi-brane for the non-boundary maximally symmetric WZW bi-brane
of Example \ref{eg:WZW-def} was reconstructed following the general
scheme laid out in \Rcite{Runkel:2008gr}.

The point of departure of the reconstruction scheme proposed in
\Rcite{Runkel:2010} is the simplicial $\txG\x\txG$-space given by
the nerve
\qq\nn
\alxydim{@C=1.cm@R=.05cm}{\sfN(\txG\lx\txG)^\bullet\quad :
\hspace{-.5cm} & \cdots \ar@<1ex>[r]^{\Gup d^{(4)}_i} \ar@<.5ex>[r]
\ar@<0ex>[r] \ar@<-.5ex>[r] \ar@<-1ex>[r] & \txG^4
\ar@<.75ex>[r]^{\Gup d^{(3)}_i} \ar@<.25ex>[r] \ar@<-.25ex>[r]
\ar@<-.75ex>[r] & \txG^3 \ar@<.5ex>[r]^{\Gup d^{(2)}_i}
\ar@<0.ex>[r] \ar@<-.5ex>[r] & \txG^2 \ar@<.5ex>[r]^{\Gup d^{(1)}_i}
\ar@<-.5ex>[r] & \txG}\,,
\qqq
of the action groupoid
\qq\nn
\alxydim{@C=2.cm}{\txG\lx\txG\quad : \hspace{-2cm} & \txG^2
\ar@<.5ex>[r]^{\pr_2=:\Gup d^{(1)}_0} \ar@<-.5ex>[r]_{\varrho=:\Gup
d^{(1)}_1} & \txG}\,,
\qqq
defined in terms of the \emph{right} regular action
\qq\nn
\varrho\ :\ \txG\x\txG\to\txG\ :\ (g,h)\mapsto g\cdot h=:\varrho_h(g
)\,.
\qqq
Note that it is the first factor in each component $\,\txG^n\,$ of
the nerve that plays the r\^ole of the $\txG$-space from Remark
\ref{rem:duality-scheme}. Apart from that, the construction follows
that of the nerve of the group (viewed as a small category) first
presented in \Rcite{Segal:1968}. The component inter-bi-brane
world-volumes $\,T_n,\ n\in\bN_{\geq 3}\,$ were assumed to be
composed of full orbits under the $\txG\x\txG$-action on
$\,\txG^n\,$ intertwined,  by the manifestly
$\txG\x\txG$-equivariant face maps of $\,\sfN(\txG\lx\txG
)^\bullet$,\ with the standard action of $\,\txG\x\txG\,$ on
$\,\txG\equiv\sfN(\txG\lx\txG)^{(0)}\,$ by left and right regular
translations. They were then shown to split into disjoint unions of
such orbits,
\qq\label{eq:IBB-WZW}
T_{n,++\cdots-}=\txG\x\widetilde\bigsqcup_{\la_1,\la_2,\ldots,\la_n
\in\faff{\ggt}}\,\cT_{\la_1,\la_2,\ldots,\la_{n-1}}^{\hspace{35pt}
\la_n}\,,
\qqq
e.g.
\qq\label{eq:IBB-WZW-3}
T_{n,++-}=\txG\x\widetilde\bigsqcup_{\la,\mu,\nu\in\faff{\ggt}}\,
\cT_{\la,\mu}^{\ \ \nu}\,,
\qqq
with
\qq
\cT_{\la,\mu}^{\ \ \nu}=\bigsqcup_{[w]\in\cS_\la\backslash\txG/
\cS_\mu}\,\left\{\ \bigl(\Ad_x\ee_\la,\Ad_{x\cdot w}\ee_\mu\bigr)
\quad\middle\vert\quad x\in\txG\quad\land\quad\ee_\la\cdot\Ad_w
\ee_\mu\in\xcC_\nu\ \right\}\,,
\qqq
where $\,\cS_\la\,$ is the $\Ad_\bullet$-stabiliser of the Cartan
element $\,\ee_\la\,$ (cf.\ \Rcite{Runkel:2010} for a general
definition). The proof of the splitting of the inter-bi-brane
world-volume into a disjoint union of diagonal $\Ad_\bullet$-orbits,
each multiplied with the `reference' factor $\,\txG$,\ rests upon
the identification of the sign-weighted sum of pullbacks, along the
inter-bi-brane maps, of the bi-brane world-volume $\,\om_\sfk\,$
appearing in the DJI \eqref{eq:DJI} (to be satisfied necessarily by
all vector fields tangent to the inter-bi-brane world-volume) as a
pre-symplectic form on partially symplectically reduced, in a manner
detailed in \Rcite{Alekseev:1993rj}, space of classical field
configurations of the level-$\sfk$ Chern--Simons theory with gauge
group $\,\txG_{\bR \x\bC P^1}\,$ on $\,\bR\x\bC P^1\,$ (with
$\,\bR\,$ playing the r\^ole of the time axis) in the presence of
$n-1$ vertical time-like Wilson lines of holonomies fixed to lie in
the respective conjugacy classes $\,\xcC_{\la_i},\
i\in\ovl{1,n-1}$,\ and a single vertical anti-time-like Wilson line
of a holonomy constrained to lie in $\,\xcC_{\la_n}$.\ The
quantisation of the weight labels, all taken from the discrete set
$\,\faff{\ggt}\,$ of \Reqref{eq:faffggt}, expresses the requirement
that there exist component 1-isomorphisms $\,\Phi_{\sfk,\la_i},\
i\in\ovl{1,n}\,$ entering the definition of the inter-bi-brane
2-isomorphism, cf.\ Diagram \eqref{diag:2iso}. Finally, the set of
admissible components of the inter-bi-brane world-volume is
restricted, as indicated by the tilde over the symbol of the
disjoint union over $\,\faff{\ggt}\,$ in Eqs.\,\eqref{eq:IBB-WZW}
and \eqref{eq:IBB-WZW-3}, to those which support a
\emph{non-vanishing} inter-bi-brane 2-isomorphism (conjectured to
correspond to the non-vanishing Verlinde fusion coefficients). The
existence of the latter is topologically obstructed on a generic
space $\,\txG\x\cT_{\la_1,\la_2,\ldots,\la_{n-1}}^{\hspace{1.2cm}
\la_n}\,$ owing to, in particular, the non-simple connectedness of
that space, cf.\ Proposition \ref{prop:2iso-class-1iso}.

In the presence of the multiplicative structure on $\,\cGk$,\
mentioned in Example \ref{eg:WZW-def}, cf.\ also Remark
\ref{rem:mult-str}, the question of existence of the inter-bi-brane
2-isomorphism over $\,\txG\x\cT_{\la_1,\la_2,\ldots,\la_{n-
1}}^{\hspace{1.2cm}\la_n}\,$ reduces to the same question for a
so-called fusion 2-isomorphism over $\,\cT_{\la_1,\la_2,\ldots,
\la_{n-1}}^{\hspace{35pt}\la_n}$.\ Thus, for instance, in the case
of the elementary inter-bi-brane $\,T_{3,++-}$,\ it boils down to
constructing a 2-isomorphism
\qq\nn
\alxydim{@C=4em@R=3em}{\bigl(\pr_1^*\cGk\ox\pr_2^*\cGk\bigr)\big
\vert_{\cT_{\la_1,\la_2}^{\quad\la_3}} \ar[d]_{\pr_1^*\Phi^\p_{\sfk
,\la_1}\ox\pr_2^*\Phi^\p_{\sfk,\la_2}} \ar[r]^{\cM_\sfk
\vert_{\cT_{\la_1,\la_2}^{\quad\la_3}}} & \bigl(\txm^*\cGk\ox
 I_{\rho_\sfk}\bigr)\big\vert_{\cT_{\la_1,\la_2}^{\quad\la_3}}
\ar[d]^{\txm^*\Phi^\p_{\sfk,\la_3}\ox\id_{ I_{\rho_\sfk}}} \\
 I_{\pr_1^*\om^\p_{\sfk,\la_1}+\pr_2^*\om^\p_{\sfk,\la_2}}\big
\vert_{\cT_{\la_1,\la_2}^{\quad\la_3}} \ar@{=>}[ur]|{\varphi_{\la_1,
\la_2}^{\quad\la_3}} \ar@{=}[r] &  I_{\txm^*\om^\p_{\sfk,\la_3}+
\rho_\sfk}\big\vert_{\cT_{\la_1,\la_2}^{\quad\la_3}}}
\qqq
whose definition involves, beside the 1-isomorphism $\,\cM_\sfk\,$
of the multiplicative structure, the component 1-isomorphisms
$\,\Phi^\p_{\sfk,\la_i},\ i\in\{1,2,3\}\,$ of the \emph{boundary}
maximally symmetric WZW bi-brane described in Example
\ref{eg:WZW-def}. The task in hand was explicitly carried out for
$\,\txG=\sug\,$ in \Rcite{Runkel:2009su}, whereby it was shown that
the inter-bi-brane 2-isomorphism exists iff the Verlinde fusion
coefficient $\,N_{\la_1,\la_2}^{\quad\la_3}\,$ for the triple of
chiral sectors of the quantised WZW model labelled by $\,(\la_1,
\la_2,\la_3)\,$ does not vanish, in accord with the prediction of
the categorial quantisation of the WZW model elaborated in
Refs.\,\cite{Fuchs:2004xi,Frohlich:2006ch}. This result constitutes
a quantum variant of the pre-quantum result discussed in the present
section.
\eeg
\brem Incidentally, the last example indicates the possibility of
adding further structure to the correspondence between the
inter-bi-brane and the interaction subspace, valid whenever the
corresponding defect perserves some of the symmetry of the untwisted
sector of the $\si$-model. The structure in question is that of an
intertwiner of the representations of the symmetry algebra
associated with the states undergoing the splitting-joining
interaction mediated by the defect quiver. Making this observation
rigorous calls for a detailed study of the issue of symmetry
transmission across the defect, which shall be addressed in the
framework of generalised geometry in the companion paper
\cite{Suszek:2010b}. \erem

\section{Conclusions and outlook}\label{sec:out}

At the focus of our interest in the present paper lay the
state-space interpretation of conformal defects in the classical and
(pre-)quantum formulation of the two-dimensional (bosonic)
non-linear $\si$-model on an oriented multi-phase world-sheet. The
latter theory is defined in terms of cohomological structures,
termed the gerbe $\,\cG$,\ the $\cG$-bi-brane $\,\cB\,$ and the
$(\cG,\cB)$-inter-bi-brane $\,\cJ$,\ coming from the 2-category
$\,\bgrb^\nabla(M\sqcup Q\sqcup T)\,$ of bundle gerbes (with
connection) over the codomain $\,M\sqcup Q\sqcup T\,$ of the
$\si$-fields fields. The issue of interest was addressed in the
canonical framework of description of the state space of the theory,
reconstructed in the classical r\'egime using the techniques of
covariant classical field theory (Propositions
\ref{prop:sympl-form-si-untw} and \ref{prop:sympl-form-si-tw}) and
subsequently extended to the quantum r\'egime by means of
transgression maps (Theorems \ref{thm:trans-untw} and
\ref{thm:trans-tw}), derived along the lines of the long-known
explicit construction for the closed string with a mono-phase
world-sheet. The transgression maps were employed to induce a
pre-quantum bundle over the state space of the $\si$-model from
gerbe and bi-brane data, in both the untwisted sector and the
twisted sector of that space (Corollaries \ref{cor:preqb-untw} and
\ref{cor:preqb-tw}). In the presence of the pre-quantum bundles, the
notion of a pre-quantum duality of the $\si$-model was formalised,
whereupon a precise correspondence was established between
pre-quantum dualities and (non-intersecting) topological world-sheet
defects. The correspondence associates a duality to a topological
defect with bi-brane maps given by surjective submersions satisfying
some additional technical conditions (Theorem \ref{thm:def-dual}),
and -- conversely -- identifies the bi-brane encoded by the data of
a duality of one of the two distinguished types: type $T$ (Theorem
\ref{thm:duali-T-bib}) and type $N$ (Theorem \ref{thm:duali-N-bib}).
From a further extension of the canonical framework to the
interacting multi-string state space, both classical and
(pre-)quantum, an intuitive picture was shown to emerge of the
defect junctions and the attendant geometric data from
$\,\bgrb^\nabla(M\sqcup Q\sqcup T)\,$ acting as intertwiners between
representations of the symmetry algebra realised on the spaces of
states of the string in interaction, be it untwisted or twisted
(Theorems \ref{thm:cross-def-int-untw} and
\ref{thm:cross-def-int-tw}). In the minimal scenario, the symmetry
algebra in question is the Virasoro algebra of the conformal group
in two dimensions. The case in which this algebra is extended by
some Ka\v c--Moody algebra induced from distinguished isometries of
the (pseudo-)riemannian target space of the $\si$-model is
elaborated in \Rcite{Suszek:2010b}. As a by-product of our analysis
of conformal defects, the notion of a simplicial string background
was introduced (Remark \ref{rem:duality-scheme}) and a conjectural
statement was made with regard to its r\^ole in defining string
theory on (generically non-geometric) `duality quotients',
generalising the concept of a T-fold in a manner dictated by a
duality scheme. The scheme is largely motivated by the idea of
categorial descent for the 2-category of bundle gerbes and the
gerbe-theoretic construction of the gauged $\si$-model.\medskip

There are two main inferences that can be drawn from our findings
recapitulated above: The first is the manifest naturality, i.e.\
completeness and minimality of the full-blown 2-categorial structure
$\,\bgrb^\nabla(M\sqcup Q\sqcup T)\,$ viewed as a scheme of
description of the two-dimensional dynamics (including the purely
geometric interactions) of the field theory in hand, and of its
modifications obtained through gauging and orientifolding.
Additional evidence in favour of the said naturality can be
extracted from the analysis of the algebraic structure on the set of
continuous internal symmetries of the multi-phase $\si$-model. This
issue is examined at length in the companion paper
\cite{Suszek:2010b}, whereby the concept of generalised geometry
twisted by the background $\,\Bgt\,$ is seen to arise, encompassing
the familiar construction of
Refs.\,\cite{Hitchin:2004ut,Gualtieri:2003dx} as a special case. The
second basic inference is the fundamental r\^ole of world-sheet
defects (and so also of the attendant cohomological structures over
$\,M\sqcup Q\sqcup T$) in probing the `topography' of the moduli
space of two-dimensional (bosonic) non-linear $\si$-models via the
associated dualities. Taken in conjunction, the two offer insights
into the very deep structure underlying the lagrangean formulation
of (critical) string theory, and that from the level of readily
tractable geometric constructs from the smooth category. This alone
provides strong motivation for further work in the directions
suggested by the hitherto results.

An outstanding problem from this last category is the precise
relation between the 2-category $\,\bgrb^\nabla(M\sqcup Q\sqcup
T)$,\ considered together with the transgressed structures over the
state space of the $\si$-model, and elements of the categorial
quantisation scheme thereof (in the sense of Segal, cp.\
\Rcite{Segal:1987sk}). There is ample and highly non-trivial
evidence indicating that certain topologically protected (and
quantised) results of the gerbe-theoretic approach, such as, e.g.,
the existence and uniqueness results for the $\si$-model in a given
string background, its quotients and orientifolds, as well as the
cohomological data describing the fusion of topological defects
(e.g., the conditions of existence of inter-bi-brane 2-isomorphisms,
admitting a straightforward interpretation in terms of spaces of
conformal blocks, cp.\ \Rcite{Runkel:2010}, and the recoupling
coefficients for simple associator moves on topological defect
quivers embedded in the world-sheet, related to the fusing matrices
of Moore and Seiberg, cp.\ \Rcite{Runkel:2008gr}), carry over
\emph{unaltered} to the quantum r\'egime, as defined rigorously by
the categorial quantisation (or by the operator-algebraic
quantisation, for that matter). It therefore seems apposite to
enquire whether the cohomological structure of the gerbe and its
2-categorial descendants is sufficiently rigid to encode (still
more) essential non-perturbative data of the quantised $\si$-model,
also for string backgrounds which -- unlike the previously examined
cases of canonical gerbes on compact Lie groups and their maximally
symmetric (inter-)bi-branes -- are devoid of a rich symmetry that
could independently constrain the quatisation procedure. It stands
to reason that our understanding in this matter can be furthered by
a search for a direct `holographic' relation between the 2-category
$\,\bgrb^\nabla(M\sqcup Q\sqcup T)\,$ for the (rational)
two-dimensional $\si$-model and the higher categorial structure
behind a three-dimensional Topological Field Theory (TFT) that
defines the categorial quantisation scheme of the $\si$-model in a
manner detailed in Refs.\,\cite{Felder:1999cv,Felder:1999mq}. A link
between the two structures was established in the largely tractable
WZW setting (in which the relevant TFT is the Chern--Simons theory
with the gauge group given by the target Lie group of the WZW model)
in \Rcite{Carey:2004xt} but even in this highly symmetric example an
exhaustive analysis of the relation between the two-dimensional CFT
on a multi-phase world-sheet and the corresponding three-dimensional
TFT coupled to a collection of intersecting Wilson lines is lacking
to date.

In an attempt at gauging the actual extent to which gerbe theory is
an intrinsic element of a CFT description of string theory, one
could be even more audacious and explore string backgrounds away
from criticality (at which the Weyl anomaly vanishes), implicitly
chosen as the basis (or completion) of the $\si$-model discussion.
One possible way of grappling with the issue in hand might be the
concept of a generalised Ricci flow, winning an ever increasing
popularity of late. Here, the hope would be that the ideas of
Refs.\,\cite{Streets:2007PhD,Young:2008PhD}, originally applied to
(principal) fibre bundles, could be successfully adapted to handle
gerbes over (pseudo-)riemannian bases. A novel alternative for this
line of development appears to be offered by the study of the
so-called String structures of \Rcite{Killingback:1986rd}, cp.\
\Rcite{Waldorf:2009uf} for a modern treatment in the
higher-categorial language.

Another important question merely touched upon by hitherto
gerbe-theoretic analyses carried out with reference to world-sheet
defects takes its origin in the concept of a non-geometric string
background. While the general notion of a simplicial string
background forwarded in the present paper seems to be perfectly
tailored to describe the latter, specific conditions for the
existence of a `duality quotient' and extra constraints prerequisite
for gauging a `duality group' should be worked out, and explicit
examples of those intricate stringy (non-)geometries should be
found. An obvious point of departure for these general
considerations is an in-depth understanding of the cohomological
duality between principal torus bundles with gerbes, or T-duality.
Already this outwardly (physically) well-studied subject offers
interesting conceptual challenges such as, e.g., the geometric
description of the procedure of descending the gauged $\si$-model
from the correspondence space (i.e.\ from the intermediary
bi-toroidal fibration linking the dual backgrounds) to the T-dual
toroidal fibration via elimination of the gauge field and symplectic
reduction. Its peculiarity consists in that it mixes various
tensorial objects defining the string background, as accounted for,
e.g., by the Buscher rules of Example \ref{ex:Tdual} (or, more
generally, by the duality-background constraints
\eqref{eq:dualiT-back-constr}). This prompts to conceive a
significant departure from the established mode of description of
geometric constructs such as the metric structure, with its global
tensorial representation by the metric field, and the gerbe, with
its local differential-geometric presentation. Such a departure,
capable of incorporating also the dilaton field, should lead to the
emergence of a unified geometric treatment of the various components
of the full multiplet of massless closed-string excitations. It is
worth pointing out that a possible first step towards such a unified
treatment is Hitchin's construction, advanced in
\Rcite{Hitchin:2005in}, of a generalised metric on the generalised
tangent bundle with a torsion-full metric connection, combining, in
a most natural manner, the metric tensor and local gerbe data.

Last but not least, given the prominent r\^ole played by
$\,\bgrb^\nabla(M\sqcup Q\sqcup T)\,$ in the geometric quantisation
of the $\si$-model, it is tempting to investigate the conditions of
compatibility of a choice of polarisation of the pre-quantised
theory with structures carried by the conformal defect, and the
ensuing constraints on the admissible \emph{quantum} dualities. The
passage from the classical to the quantum r\'egime is bound to
result in a renormalisation of the functional relations determining
the bahaviour of the $\si$-model fields at the defect (cp., e.g.,
Refs.\,\cite{Bachas:2004sy,Alekseev:2007in}), and it would be
desirable to attain a good understanding of these quantum effects.
\medskip

Thus, altogether, it seems fair to conclude the present paper with
the constatation that the hitherto incursions into the physics of
conformal defects of the two-dimensional non-linear $\si$-model,
aided substantially and organised neatly by gerbe theory, present us
with a panoply of string-theoretic and mathematical problems,
including those of a truly fundamental nature, of which but a small
proportion have been elucidated so far. We are hoping to return to
some of them in a future publication.

\appendix

\section{A proof of Proposition \ref{prop:Cart-si-def}}
\label{app:Cart-form-si}

Take an arbitrary $\cF_\si$-vertical vector field $\,\xcV\,$ on
$\,\sfJ^1 \cF_\si\,$ with restrictions
\qq\label{eq:Votan}
\xcV\vert_\Pgt=V^\mu\,\tfrac{\d\ \ }{\d X^\mu}+V^\mu_a\,\tfrac{\d\
}{\d \xi^\mu_a}\,,\qquad\qquad\xcV\vert_{\Egt_\G}=V^A\,\tfrac{\d\ \
}{\d X^A}+V_\varphi^A\,\tfrac{\d\ }{\d \xi^A_\varphi}
\,,\qquad\qquad\xcV\vert_{\Vgt_\G}=V^i\,\tfrac{\d\ \ }{\d X^i}\,,
\qqq
with components constrained as in \Reqref{eq:tangojet-constr}, and
denote by $\,\ovl\xcV\,$ a vector field on the field space $\,\xcF=
M\sqcup Q\sqcup T\,$ with restrictions
\qq\label{eq:barVotan}
\ovl\xcV\vert_M=V^\mu\,\tfrac{\d\ \ }{\d X^\mu}\,,\qquad\qquad\ovl
\xcV\vert_Q=V^A\,\tfrac{\d\ \ }{\d X^A}\,,\qquad\qquad\ovl\xcV
\vert_T=V^i\,\tfrac{\d\ \ }{\d X^i}\,.
\qqq
Using the identity
\qq\nn
\star_\eta\sfd\si^a=\eta^{ab}\,\vep_{bc}\,\sfd\si^c
\qqq
in conjunction with the relation
\qq\nn
\d\int_e\,X_e^*\eta=-\int_e\,X_e^*\d\eta+X^*\eta\vert_{\p e}\,,
\qqq
valid for any edge $\,e\in\triangle(\Si)$,\ an arbitrary 1-form
$\,\eta\in\Om^1\bigl(X(e)\bigr)\,$ and for $\,X_e:=X \vert_e$,\ we
may express the Lie derivative of the functional $\,S_{\Th_\si}\,$
along $\,\xcV\,$ as
\qq
&&\hspace{12.7cm}\xcV\con\d S_{\Th_\si}[\Psi]\label{eq:V-con-delS-prim}\\\cr
&=&\sum_{p\in\triangle(\Si)}\,\bigg\{\int_p\,\Psi_p^*\bigl[\bigl(
\sfd^2\si\,\xi^\mu_a\,\bigl(\tfrac{1}{2}\,\xi^\nu_b\,V^\la\,\p_\la+
V^\nu_b\bigr)+\sfd\si^c\wedge\vep_{ca}\,\bigl(V^\mu\,\d\xi^\nu_b-
V^\nu_b\,\d X^\mu-\xi^\nu_b\,V^{[\la}\,\d X^{\mu]}\,\p_\la\bigr)
\bigr)\,L^{ab}_{i_p,\mu\nu}\bigr]\cr\cr
&&+\sum_{e\subset p}\,\bigg[\int_e\,\Psi_e^*\bigl(-V^\nu\,\xi^\mu_a\,
L^{ab}_{i_p,\mu\nu}\,\vep_{bc}\,\sfd\si^c+\ovl\xcV\con\d A_{i_p
i_e}\bigr)+\sum_{v\in e}
\,\vep_{pev}\,\Psi^*\bigl(\ovl\xcV\con\bigl(A_{i_p i_e}-\sfi\,\d\log
g_{i_p i_e i_v}\bigr)\bigr)(v)\bigg]\cr\cr
&&+\sum_{e\subset p\cap\G}\,\bigg[\int_e\,\Psi_e^*\bigl(\ovl\xcV\con
\d P_{i_e}\bigr)+\sum_{v\in e}\,\vep_{ev}\,\Psi^*\bigl(\ovl\xcV\con
\bigl(P_{i_e}+\sfi\,\d\log K_{i_e i_v}\bigr)\bigr)(v)\bigg]\cr\cr
&&-\sfi\,\sum_{\jmath\in p\cap\Vgt_\G}\,\Psi^*\bigl(\ovl\xcV\con\d
\log f_{n_\jmath,i_\jmath}\bigr)(\jmath)\bigg\}\nonumber
\qqq
in terms of the restrictions $\,\Psi_f:=\Psi\vert_f,\ f\in\triangle
(\Si)\,$ and the standard antisymmetriser $\,V^{[\mu}\,W^{\nu]}:=
V^\mu\,W^\nu-V^\nu\,W^\mu$.

We begin by considering the distinguished $\cF_\si$-vertical vector
field $\,\frac{\d\ }{\d\xi^\mu_a}$,\ for which
\qq\nn
0&=&\frac{\d\ }{\d\xi^\mu_a}\con\d S_{\Th_\si}[\Psi_{\si,{\rm cl}}]=
\sum_{p\in\triangle(\Si)}\,\int_p\,\Psi_{\si,{\rm cl}\ p}^*\bigl(
\bigl(\sfd^2\si\,\xi^\mu_a-\sfd\si^c\wedge\vep_{ca}\,\d X^\mu\bigr)
\,L^{ab}_{i_p,\mu\nu}\,V^\nu_b\bigr)\cr\cr
&=&\sum_{p\in\triangle(\Si)}\,\int_p\,\sfd^2\si\,\Psi_{\si,{\rm cl}\
p}^*\bigl(\bigl(\xi^\mu_a-\p_a X^\mu\bigr)\,L^{ab}_{i_p,\mu\nu}\,
V^\nu_b\bigr)\,,
\qqq
whence the classical relation
\qq\nn
\xi^\mu_a=\p_a X^\mu\,.
\qqq
Upon taking the latter into account, invoking the identity
\qq\nn
L_{i_p,\mu\nu}^{ab}=L_{i_p,\nu\mu}^{ba}
\qqq
as well as Eqs.\,\eqref{eq:DG-is-H}, \eqref{eq:DPhi-is} and
\eqref{eq:Dphin-is}, and -- finally -- denoting by $\,\Vol(e)\,$ and
$\,\widehat t\,$ the volume form and the versor tangent to
$\,e\in\Egt_\G$,\ respectively, we readily reduce
\Reqref{eq:var-STHsi} to the simpler form
\qq
&&\hspace{11.8cm}\xcV\con\d S_{\Th_\si}[\Psi_{\si,{\rm cl}}]\label{eq:V-con-delS-bis}\\
\cr
&=&\sum_{p\in\triangle(\Si)}\,\bigg\{-\int_p\sfd^2\si\,\Psi_{\si,{\rm
cl}\ p}^*\bigl[V^\mu\,\bigl(\txg_{\mu\nu}\,\eta^{ab}\,\p_a\p_b X^\nu
+\tfrac{1}{2}\,\bigl(\p_\mu L_{i_p,\rho\si}^{ab}-\p_\rho L_{i_p,\mu
\si}^{ab}-\p_\si L_{i_p,\rho\mu}^{ab}\bigr)\,\p_a X^\rho\,\p_b X^\si
\bigr)\bigr]\cr\cr
&&+\sum_{e\subset p\setminus\G}\,\bigg(\int_e\,
\Psi_{\si,{\rm cl}\ e}^*(\ovl\xcV\con B_{i_e})-\sum_{v\in e}\,
\vep_{pev}\,\Psi^*\bigl(\ovl\xcV\con A_{i_e i_v}\bigr)(v)\bigg)\cr
\cr
&&+\sum_{e\subset p\cap\G}\,\bigg[\int_e\,\Vol(e)\,\Psi_e^*\bigl(
\iota_{1\,*}\ovl\xcV\con\sfp(X_{|1})-\iota_{2\,*}\ovl\xcV\con\sfp(
X_{|2})-\ovl\xcV\con X_*\widehat t\con\bigl(\check\iota_1^*B_{i_e}-
\check\iota_2^*B_{i_e}+\d P_{i_e}\bigr)\bigr)\cr\cr
&&+\sum_{v\in e}\,
\vep_{ev}\,\Psi^*\bigl(\ovl\xcV\con\bigl(\check\iota_2^*A_{i_e i_v}-
\check\iota_1^*A_{i_e i_v}+P_{i_e}+\sfi\,\d\log K_{i_e
i_v}\bigr)\bigr)(v) \bigg]-\sfi\,\sum_{\jmath\in
p\cap\Vgt_\G}\,\Psi^*\bigl(\ovl\xcV\con\d\log
f_{n_\jmath,i_\jmath}\bigr)(\jmath)\bigg\}\cr\cr
&=&\sum_{p\in\triangle(\Si)}\,\bigg[-\int_p\sfd^2\si\,\Psi_{\si,{\rm
cl}\ p}^*\bigl[V^\mu\,\bigl(\txg_{\mu\nu}\,\eta^{ab}\,\p_a\p_b X^\nu
+\tfrac{1}{2}\,\bigl(\p_\mu L_{i_p,\rho\si}^{ab}-\p_\rho L_{i_p,\mu
\si}^{ab}-\p_\si L_{i_p,\rho\mu}^{ab}\bigr)\,\p_a X^\rho\,\p_b X^\si
\bigr)\bigr]\cr\cr
&&+\sum_{e\subset p\cap\G}\,\int_e\,\Vol(e)\,
\Psi_e^*\bigl(\iota_{1\,*}\ovl\xcV\con\sfp(X_{|1})-\iota_{2\,*}\ovl
\xcV\con\sfp(X_{|2})-\ovl\xcV\con X_*\widehat t\con\om\bigr)\cr\cr
&&+\sum_{\jmath\in p\cap\Vgt_\G}\,\Psi^*\bigl(\ovl\xcV\con\bigl(
\sum_{k=1}^{n_\jmath}\,\vep_{n_\jmath}^{k,k+1}\,\check
\pi_{n_\jmath}^{k,k+1\,*}P_{i_\jmath}-\sfi\,\d\log
f_{n_\jmath,i_\jmath} \bigr)\bigr)(\jmath)\bigg]\cr\cr
&=&\sum_{p\in\triangle(\Si)}\,\bigg[-\int_p\sfd^2\si\,\Psi_{\si,{\rm
cl}\ p}^*\bigl[V^\mu\,\bigl(\txg_{\mu\nu}\,\eta^{ab}\,\p_a\p_b X^\nu
+\tfrac{1}{2}\,\bigl(\p_\mu L_{i_p,\rho\si}^{ab}-\p_\rho L_{i_p,\mu
\si}^{ab}-\p_\si L_{i_p,\rho\mu}^{ab}\bigr)\,\p_a X^\rho\,\p_b X^\si
\bigr)\bigr]\cr\cr
&&+\sum_{e\subset p\cap\G}\,\int_e\,\Vol(e)\,
\Psi_e^*\bigl(\iota_{1\,*}\ovl\xcV\con\sfp(X_{|1})-\iota_{2\,*}\ovl
\xcV\con\sfp(X_{|2})-\ovl\xcV\con X_*\widehat t\con\om\bigr)\bigg]
\,.\nonumber
\qqq
It is now a matter of a simple check to verify that the upper
integrand coincides with the pullback of the contraction of the
field equation \eqref{eq:field-eqs} with the arbitrary vector
$\,\ovl\xcV$,\ whereas the bottom one is the pullback of the
contraction of the DGC \eqref{eq:DGC} with the same vector field.
Thus, demanding that the variation vanish for all $\,\xcV\,$ is
tantamount to imposing the field equations and the Defect Gluing
Condition of the $\si$-model given by the action functional
\eqref{eq:sigma}.\qed

\section{A proof of Proposition \ref{prop:sympl-form-si-untw}}
\label{app:sympl-form-si-untw}

\begin{figure}[hbt]~\\[5pt]

$$
\begin{picture}(50,50)
  \put(-70,-45){\scalebox{0.30}{\includegraphics{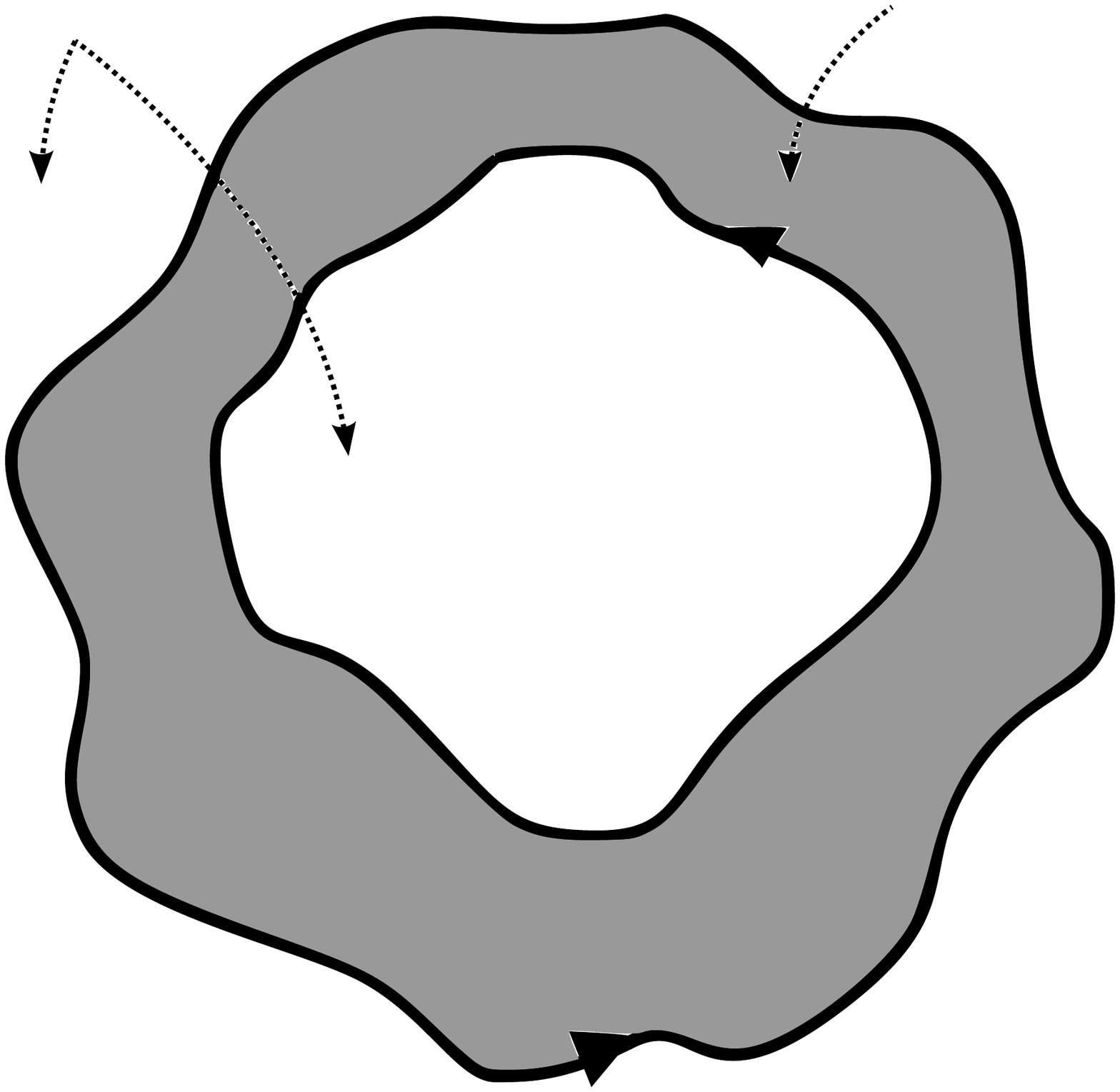}}}
  \end{picture}
  \put(0,0){
     \setlength{\unitlength}{.60pt}\put(-28,-16){
     \put(-160,267)  { $\Si\setminus\Si_{1,2}$ }
     \put(53,273)   { $\Si_{1,2}$ }
     \put(0,200)    { $C_1$ }
     \put(-20,0)    { $C_2$ }
          }\setlength{\unitlength}{1pt}}
$$

\caption{A pair of disjoint untwisted Cauchy contours $\,C_1\,$ and
$\,C_2\,$ (time runs radially, from $\,C_1\,$ towards $\,C_2$). The
intermediate region $\,\Si_{1,2}$,\ cut out from the world-sheet
$\,\Si\,$ by the two contours, does not contain any elements of the
defect quiver $\,\G$.\ The latter is embedded in the complement
$\,\Si\setminus\Si_{1,2}\,$ of an arbitrary topology. The arrows
represent the orientation of the Cauchy contours.}
\label{fig:Cauchy-untw}
\end{figure}

The proof is an adaptation of the constructive proof of Proposition
\ref{prop:Cartan} to the circumstances in hand. Thus, we consider a
region $\,\Si_{1,2}\,$ in the world-sheet $\,\Si\,$ bounded, as
$\,\p\Si_{1,2}=C_2\sqcup(-C_1)$,\ by a pair of Cauchy contours
$\,C_A,\ A\in\{1,2\}\,$ with orientation as in
\Rfig{fig:Cauchy-untw}.
Choose a triangulation $\,\triangle(\Si)\,$ of $\,\Si\,$ subordinate
to $\,\cO_\xcM,\ \xcM\in\{M,Q,T\}\,$ \wrt $\,(X\,\vert\,\G)\,$ such
that it induces a triangulation $\,\triangle(\Si_{1,2})\,$ of
$\,\Si_{1, 2}$,\ and so also triangulations $\,\triangle(C_A)\,$ of
the two Cauchy contours, the latter consisting of the respective
edges $\,e_A\,$ and vertices $\,v_A\,$ (this can always be achieved
via refinement of a given triangulation of the world-sheet). Define
\qq\nn
S_{1,2}[\Psi_{\si,{\rm cl}}]=\int_{\Si_{1,2}}\,\bigl(\Psi_{\si,{\rm
cl}}\vert_{\Si_{1,2}}\bigr)^*\Th_\si
\qqq
and use the previous result \eqref{eq:V-con-delS-prim} alongside the
first three lines of the computation \eqref{eq:V-con-delS-bis} to
write, for $\,\xcV\,$ tangent to $\,\sfP_{\si,\emptyset}$,\ and
$\,\ovl\xcV\,$ as in Eqs.\,\eqref{eq:Votan} and \eqref{eq:barVotan},
respectively,
\qq
\xcV\con\d S_{1,2}[\Psi_{\si,{\rm cl}}]&=&\sum_{A=1}^2\,(-1)^A\,
\bigg(\int_{C_A}\,\Vol(C_A)\,\bigl(\Psi_{\si,{\rm
cl}}\vert_{C_A}\bigr)^*(\xcV\con\sfp)+\sum_{e_A\in\triangle(
C_A)}\,\int_{e_A}\,\Psi_{\si,{\rm cl}\ e_A}^*(\ovl\xcV\con
B_{i_{e_A}})\cr\cr
&&\hspace{.7cm}-\sum_{v_A\in\triangle(C_A)}\,\ovl\xcV\con\bigl(
A_{i_{e_+(v_A)}i_{e_-(v_A)}}-\sfi\,\d\log g_{i_{e_+(v_A)}i_{e_-(v_A
)}i_{v_A}}\bigr)\bigl(X(v_A)\bigr)\bigg)\,,\label{eq:delS12}
\qqq
where $\,\Vol(C_A)\,$ is a volume form on $\,C_A\,$ and
$\,e_+(v_A)\,$ (resp.\ $\,e_-(v_A)$) denotes the incoming (resp.\
outgoing) edge at $\,v_A$,\ and where
\qq\nn
\int_{e_A}\,\Psi_{\si,{\rm cl}\ e_A}^*(\ovl\xcV\con B_{i_{e_A}})
\equiv\int_{e_A}\,\Vol(e_A)\,\Psi_{\si,{\rm cl}\ e_A}^*(\widehat
t_{C_A}\con\ovl\xcV\con
B_{i_{e_A}})=-\int_{e_A}\,\Vol(e_A)\,\Psi_{\si,{\rm cl}\
e_A}^*(\ovl\xcV\con \widehat t_{C_A}\con B_{i_{e_A}})
\qqq
for $\,\widehat t_{C_A}\,$ the tangent vector field on $\,C_A$.\ The
last equality enables us to derive, from \Reqref{eq:delS12}, the
sought-after expression
\qq
\d S_{1,2}[\Psi_{\si,{\rm cl}}]&=&\sum_{A=1}^2\,(-1)^A\,\bigg(
\int_{C_A}\,\Vol(C_A)\wedge\bigl(\Psi_{\si,{\rm
cl}}\vert_{C_A}\bigr)^*\sfp-\sum_{e_A\in\triangle(C_A)}\,
\int_{e_A}\,\Psi_{\si,{\rm cl}\ e_A}^*B_{i_{e_A}}\cr\cr
&&\hspace{.5cm}-\sum_{v_A\in\triangle(C_A)}\,\Psi_{\si,{\rm
cl}}^*\bigl(A_{i_{e_+( v_A)}i_{e_-(v_A)}}-\sfi\,\d\log g_{i_{e_+(
v_A)}i_{e_-(v_A)}i_{v_A}}\bigr)(v_A)\bigg)\,,\label{eq:delS12-untw}
\qqq
whence the thesis of the proposition follows straightforwardly upon
defining
\qq\nn
\Om_{\si,\emptyset}[\Psi_{\si,{\rm cl}}]&:=&\d\bigg(\int_\xcC\,\Vol
(\xcC)\wedge\bigl(\Psi_{\si,{\rm cl}}\vert_\xcC\bigr)^*\sfp-\sum_{e
\in\triangle(\xcC)}\,\int_e\,\Psi_{\si,{\rm cl}\ e}^*B_{i_e}\cr\cr
&&\hspace{.5cm}-\sum_{v\in\triangle(\xcC)}\,\Psi_{\si,{\rm cl}}^*
\bigl(A_{i_{e_+(v)} i_{e_-(v)}}-\sfi\,\d\log g_{i_{e_+(v)}i_{e_-(v
)}i_{v}}\bigr)(v)\bigg)
\qqq
(manifestly independent of the choice of the Cauchy contour
$\,\xcC$) and upon employing \Reqref{eq:DG-is-H} and the identity
\qq\label{eq:delint-B}
\d\int_e\,\Psi_{\si,{\rm cl}\ e}^*B_{i_e}=-\int_e\,\Psi_{\si,{\rm
cl}\ e}^*\d B_{i_e}+\Psi_{\si,{\rm cl}}^*B_{i_e}\vert_{\p e}\,.
\qqq
\qed

\section{A proof of Proposition \ref{prop:sympl-form-si-tw}}
\label{app:sympl-form-si-tw}

\begin{figure}[hbt]~\\[5pt]

$$
 \raisebox{-50pt}{\begin{picture}(50,50)
  \put(-67,-45){\scalebox{0.30}{\includegraphics{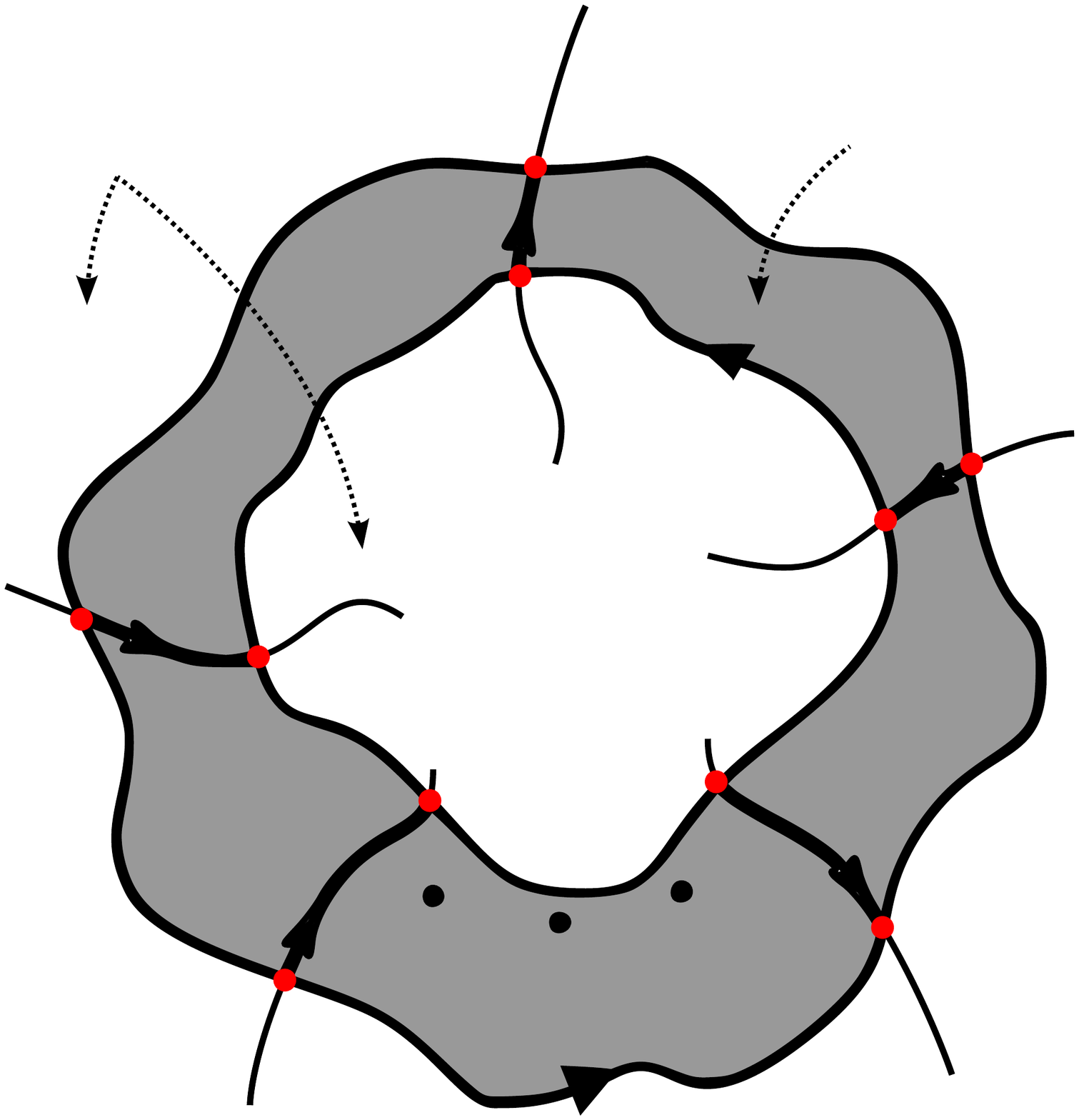}}}
  \end{picture}
  \put(0,0){
     \setlength{\unitlength}{.60pt}\put(-28,-16){
     \put(-160,267)  { $\Si\setminus\Si_{1,2}$ }
     \put(53,273)   { $\Si_{1,2}$ }
     \put(-10,315)   { $\ell_1$ }
     \put(-180,153)  { $\ell_2$ }
     \put(128,190)    { $\ell_I$ }
     \put(-105,1)    { $\ell_3$ }
     \put(78,8)      { $\ell_{I-1}$ }
     \put(-41,275)   { $v^2_1$ }
     \put(-42,210)   { $v^1_1$   }
     \put(-85,127)   { $v^1_2$   }
     \put(-163,129)   { $v^2_2$   }
     \put(-43,97)   { $v^1_3$   }
     \put(-115,37)   { $v^2_3$   }
     \put(-12,100)   { $v^1_{I-1}$   }
     \put(77,60)   { $v^2_{I-1}$   }
     \put(50,147)   { $v^1_I$   }
     \put(98,203)   { $v^2_I$   }
     \put(0,200)    { $C_1$ }
     \put(-20,0)    { $C_2$ }
          }\setlength{\unitlength}{1pt}}}
$$

\caption{A pair of disjoint twisted Cauchy contours, $\,C_1\,$ and
$\,C_2$,\ intersecting the defect quiver $\,\G\,$ at points
$\,v^A_k,\ k\in\ovl{1,I}\,$ from the respective (anti-)time-like
defect lines $\,\ell_k\,$ (time runs radially, from $\,C_1\,$
towards $\,C_2$). The intermediate region $\,\Si_{1,2}$,\ cut out
from the world-sheet $\,\Si\,$ by the two contours, is free of
defect junctions, whereas its complement $\,\Si\setminus\Si_{1,
2}\,$ contains an arbitrary sub-graph of $\,\G\,$ with $I$ free
legs, embedded in an arbitrary world-sheet topology. The arrows
represent the orientation of the various curves.}
\label{fig:Cauchy-tw}
\end{figure}

The proof develops essentially along the lines of the constructive
proof of Proposition \ref{prop:Cartan}. Here, we consider a region
$\,\Si_{1,2}\,$ in the world-sheet $\,\Si\,$ bounded, as
$\,\p\Si_{1,2}=C_2\sqcup(- C_1)$,\ by a pair of twisted Cauchy
contours $\,C_A,\ A\in\{1,2\}\,$ with orientation as in
\Rfig{fig:Cauchy-tw}, each intersecting a family of
$\,I\in\bN_{>0}\,$ defect lines $\,\ell_k,\ k\in\ovl{1,I}$.
Once again, we choose a triangulation $\,\triangle(\Si)\,$ of
$\,\Si\,$ subordinate to $\,\cO_\xcM,\ \xcM\in\{M,Q,T\}\,$ \wrt
$\,(X\,\vert\,\G)\,$ which induces a triangulation
$\,\triangle(\Si_{1,2} )\,$ of $\,\Si_{1,2}$,\ and so also
triangulations $\,\triangle( C_A)\,$ of the two Cauchy contours
(superimposed upon that of the segments of $\,\ell_k\,$ cut out by
the two Cauchy contours), the latter consisting of the respective
edges $\,e_A\,$ and vertices $\,v_A$,\ with the distinguished
vertices $\,v_k^A\,$ at intersections $\,\ell_k\cap C_A$.\ Define
\qq\nn
S_{1,2}[\Psi_{\si,{\rm cl}}]=\int_{\Si_{1,2}}\,\bigl(\Psi_{\si,{\rm
cl}}\vert_{\Si_{1,2}}\bigr)^*\Th_\si\,.
\qqq
The only difference with respect to the previous result,
\Reqref{eq:delS12-untw}, appears at the $\,v^A_k\,$ and yields
\qq\nn
\d S_{1,2}[\Psi_{\si,{\rm cl}}]&=&\sum_{A=1}^2\,(-1)^A\,\bigg(
\int_{C_A}\,\Vol(C_A)\wedge\bigl(\Psi_{\si,{\rm
cl}}\vert_{C_A}\bigr)^*\sfp-\sum_{e_A\in\triangle(C_A)}\,
\int_{e_A}\,\Psi_{\si,{\rm cl}\ e_A}^*B_{i_{e_A}}\cr\cr
&&\hspace{.5cm}-\sum_{v_A\in\triangle(C_A)\setminus\G}\,
\Psi_{\si,{\rm cl}}^*\bigl(A_{i_{e_+(v_A)}i_{e_-(v_A)}}-\sfi\,\d
\log g_{i_{e_+(v_A)}i_{e_-(v_A )}i_{v_A}}\bigr)(v_A)\cr\cr
&&\hspace{.5cm}+\sum_{k=1}^I\,\Psi_{\si,{\rm cl}}^*\bigl(
\iota_1^{\vep_k\,*}A_{i_{e_-(v^A_k)}\phi_1^{\vep_k}(i_{v^A_k})}-
\iota_2^{\vep_k\,*}A_{i_{e_+(v^A_k)}\phi_2^{\vep_k}(i_{v^A_k})}+
\vep_k\,P_{i_{v^A_k}}\bigr)(v^A_k)\bigg)\,,
\qqq
where $\,\vep_k=+1\,$ if $\,\ell_k\,$ is time-like, $\,\vep_k=-1\,$
if $\,\ell_k\,$ is anti-time-like, and where $\,(\phi_1^{+1},
\phi_2^{+1})=(\phi_1,\phi_2)\,$ and $\,(\phi_1^{-1},\phi_2^{-1})=(
\phi_2,\phi_1)\,$.\ The thesis of the proposition is demonstrated by
defining a 2-form
\qq\nn
\Om_{\si,\cB|\{(P_k,\vep_k)\}}[\Psi_{\si,{\rm cl}}]&:=&\d\bigg(
\int_\xcC\,\Vol(\xcC)\wedge\bigl(\Psi_{\si,{\rm cl}}\vert_\xcC
\bigr)^*\sfp-\sum_{e \in\triangle(\xcC)}\,\int_e\,\Psi_{\si,{\rm cl}
\ e}^*B_{i_e}\cr\cr
&&\hspace{.5cm}-\sum_{v\in\triangle(\xcC)\setminus\{P_k\}_{k\in
\ovl{1,I}}}\,\Psi_{\si,{\rm cl}}^*\bigl(A_{i_{e_+(v)}i_{e_-(v)}}-
\sfi\,\d\log g_{i_{e_+(v )}i_{e_-(v)}i_{v}}\bigr)(v)\cr\cr
&&\hspace{.5cm}+\sum_{k=1}^I\,\Psi_{\si,{\rm cl}}^*\bigl(
\iota_1^{\vep_k\,*}A_{i_{e_-(P_k)}\phi_1^{\vep_k}(i_{P_k})}-
\iota_2^{\vep_k\,*}A_{i_{e_+(P_k)}\phi_1^{\vep_k}(i_{P_k})}+\vep_k
\,P_{i_{P_k}}\bigr)(P_k)\bigg)\,,
\qqq
manifestly independent of the choice of the Cauchy contour
$\,\xcC$.\ The 2-form acquires the desired form after a simple
computation using Eqs.\,\eqref{eq:DG-is-H}, \eqref{eq:DPhi-is} and
\eqref{eq:delint-B}. \qed

\section{A proof of Theorem \ref{thm:cross-def-int-untw}}
\label{app:cross-def-int-untw}

\begin{figure}[hbt]~\\[25pt]

$$
 \raisebox{-50pt}{\begin{picture}(50,50)
  \put(-146,-7){\scalebox{0.35}{\includegraphics{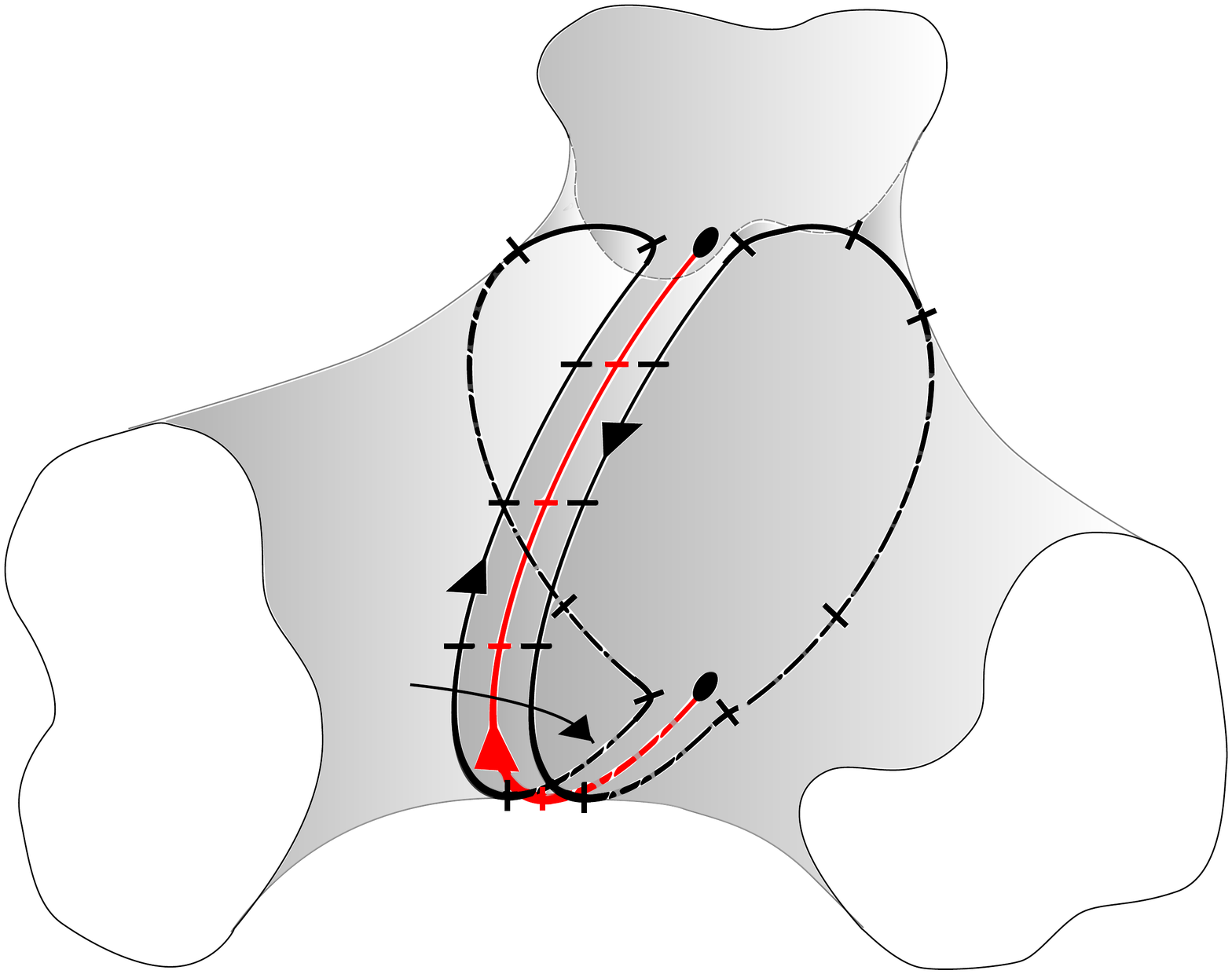}}}
  \end{picture}
  \put(0,0){
     \setlength{\unitlength}{.60pt}\put(-28,-16){
     \put(-90,53)     { $Y_{1,2}$  }
     \put(-32,269)    { $v_+$ }
     \put(42,252)     { \tiny{$i^2_{e_+(v_2)}=i^2_{e_-(v_2')}$} }
     \put(-85,247)    { $v_1$ }
     \put(15,250)     { $v_2$ }
     \put(29,228)     { $v_2'$ }
     \put(-79,82)     { $v$ }
     \put(-10,248)    { \tiny{$i^2_{v_+}$} }
     \put(-65,252)    { \tiny{$i^1_{v_+}$} }
     \put(-27,233)    { \tiny{$i^2_{e^{(1,2)}_+(v_+)}$} }
     \put(-120,238)   { \tiny{$i^1_{e^{(1,2)}_+(v_+)}$} }
     \put(-135,275)   { \tiny{$i^1_{e_-(v_+)}=i^1_{e_+(v_1)}$} }
     \put(-10,275)    { \tiny{$i^2_{e_+(v_+)}=i^2_{e_-(v_2)}$} }
     \put(-14,91)     { \tiny{$i^2_{v_-}$} }
     \put(-65,103)    { \tiny{$i^1_{v_-}$} }
     \put(-37,77)     { \tiny{$i^2_{e^{(1,2)}_-(v_-)}=i^2_{e^{(1,2)}_+(v)}$} }
     \put(-232,112)   { \tiny{$i^1_{e^{(1,2)}_-(v_-)}=i^1_{e^{(1,2)}_+(v)}$} }
     \put(5,110)      { \tiny{$i^2_{e_-(v_-)}$} }
     \put(-64,133)    { \tiny{$i^1_{e_-(v_-)}$} }
     \put(-31,118)    { $v_-$ }
     \put(-128,190)   { $X_1$ }
     \put(55,190)     { $X_2$ }
           }\setlength{\unitlength}{1pt}}}
$$

\caption{Triangulations and indexing of loops from the fusion
subspace $\,\sfP_{\si,\emptyset}^{\circledast\cB}$.\ The
triangulation of the parent half-loop $\,Y_{1,2}\in\sfI Q\,$ induces
triangulations of the half-loops under fusion and fixes a pair of
vertices $\,v_+,v_-$.\ The free half-loops are triangulated and
indexed independently of one another.}
\label{fig:fusion-triangulation}
\end{figure}

\bit
\item[Ad i)] First, through direct inspection of
Eqs.\,\eqref{eq:lin-DGC-asym} and \eqref{eq:curv-constr}, we
establish that the `sum' symplectic form
\qq\nn
\Om^+_{\si,\emptyset}=\pr_1^*\Om_{\si,\emptyset}+\pr_2^*\Om_{\si,
\emptyset}
\qqq
on $\,\sfP_{\si,\emptyset}^{\x 2}$,\ the latter space being
considered with the two canonical projections
$\,\pr_\a:\sfP_{\si,\emptyset}^{\x 2}\to \sfP_{\si,\emptyset},\
\a\in\{1,2\}$,\ restricts -- the restriction being marked by the bar
over $\,\Om_{\si,\emptyset}^+\,$ -- as
\qq
\ovl\Om_{\si,\emptyset}^+[(\psi_1,\psi_2)]&=&\int_\sfI\,\exd\varphi
\,(\Psi_{\si,{\rm cl}}^2\vert_\sfI)^*\bigl(\d\sfp_2+\p_\varphi X_2
\con\txH\bigr)+\int_{\tau(\sfI)}\,\exd\varphi\,( \Psi_{\si,{\rm
cl}}^1\vert_{\tau(\sfI)})^*\bigl(\d\sfp_1+\p_\varphi X_1\con\txH
\bigr)\cr\cr
&&+Y_{1,2}^*\om(\pi)-Y_{1,2}^*\om(0)\label{eq:Om+-restr-fus}
\qqq
to the tangent $\,\Tgt\sfP_{\si,\emptyset}^{\circledast\cB}\,$ of
the $\cB$-fusion subspace $\,\sfP_{\si,
\emptyset}^{\circledast\cB}\,$ within $\,\G\bigl(\sfT\sfP_{\si,
\emptyset}^{\x 2}\vert_{\sfP_{\si,\emptyset}^{\circledast\cB}}
\bigr)$.\ Here, $\,(\Psi_{\si,{\rm cl}}^1,\Psi_{\si,{\rm cl}}^2)\,$
is the pair of extremal sections represented by the Cauchy data
$\,(\psi_1,\psi_2)$,\ with$\,\psi_\a=(X_\a,\sfp_\a),\
\a\in\{1,2\}$.\ Given the above result, we readily check, using
\Reqref{eq:lin-DGC-asym}, that $\,\Om_{\si,\emptyset}^{+-}\,$
restricts to the subspace $\,\Tgt\Igt_\si(\circledast\cB:\cJ:\cB)\,$
within $\,\G\bigl(\sfT\sfP_{\si,\emptyset}^{\x 3}\vert_{\Igt_\si(
\circledast\cB:\cJ:\cB)}\bigr)\,$ spanned by vector fields tangent
to $\,\Igt_\si( \circledast\cB:\cJ:\cB)\,$ as
\qq\nn
\ovl\Om_{\si,\emptyset}^{+-}[(\psi_1,\psi_2,\psi_3)]=\bigl(\pi_3^{1
,2\,*}\om+\pi_3^{2,3\,*}\om-\pi_3^{3,1\,*}\om\bigr)\circ
Z\vert_0^\pi=0\,,
\qqq
which proves the first statement by virtue of the relevant DJI.
\item[Ad ii)] We begin by describing a circle bundle $\,\ceL_{\si,
\circledast\cB}\to\sfP_{\si,\emptyset}^{\circledast\cB}\,$ over the
$\cB$-fusion subspace of the untwisted string, with curvature equal
to the restriction $\,\ovl\Om_{\si,\emptyset}^+\,$ of
\Reqref{eq:Om+-restr-fus}. The bundle, given by the restriction to
$\,\sfP_{\si,\emptyset}^{\circledast\cB}\,$ of the the tensor
product
\qq\nn
\ceL_{\si,\circledast\cB}:=\bigl(\pr_1^*\ceL_{\si,\emptyset}\ox
\pr_2^*\ceL_{\si,\emptyset}\bigr)\vert_{\sfP_{\si,
\emptyset}^{\circledast\cB}}
\qqq
of the pullbacks of the pre-quantum bundle for the untwisted sector
of the $\si$-model along the canonical projections $\,\pr_\a$,\
provides -- for a choice of the polarisation -- a definition of the
untwisted two-string Hilbert space. Writing out the local data of
the bundle, which we shall need in subsequent computations,
prerequires fixing an open cover of $\,\sfP_{\si,
\emptyset}^{\circledast\cB}$.\ Similarly to the proof of Theorem
\ref{thm:def-dual}, we induce it from the open cover of the
loop-space basis $\,(\pi_{\sfT^*\sfL M},\pi_{\sfT^*\sfL M})(
\sfP_{\si,\emptyset}^{\circledast\cB})\,$ of the fusion space,
obtained by varying triangulations of the two loops in
$\,(\psi_1,\psi_2)\in \sfP_{\si,\emptyset}^{\circledast\cB}\,$ of
the following form, cf.\ \Rfig{fig:fusion-triangulation}: Along the
two half-loops $\,X_\a \vert_{\vsi_\a(\sfI)},\ \a\in\{1,2\}$,\ both
triangulations come from a triangulation of the free open path
$\,Y_{1,2}$,\ and so, for a choice $\,\triangle(\sfI)=\triangle_{1,
2}\,$ of a triangulation,\ with edges $\,e\,$ and vertices $\,v$,\
of the $\pi$-unit interval $\,\sfI\,$ parameterising $\,Y_{1,2}\,$
and $\,X_\a\vert_{\vsi_\a(\sfI)}$,\ we have an assignment of indices
$\,f\mapsto(i^1_f,i^2_f,i^{1,2}_f)\in\xcI_M\x\xcI_M\x\xcI_Q\,$ to
every element $\,f\in\triangle_{1,2}$,\ related as per
\Reqref{eq:comm-triang-index}. Note that this entails fixing a pair
of vertices $\,v_+,v_-\,$ of the triangulation corresponding to the
two boundary points of $\,\sfI$.\ The triangulation of each of the
two loops $\,X_\a\,$ is next completed by specifying an arbitrary
triangulation of the interval parameterising the free half-loop, to
wit, $\,\triangle\bigr(\tau( \sfI)\bigr)=\triangle_1\,$ for
$\,\a=1\,$ and $\,\triangle(\sfI)= \triangle_2\,$ for $\,\a=2$,\
together with the respective index assignments.

Thus, let the pair \v Cech index $\,\igt_\a\,$ of the open cover of
$\,\sfL M$,\ as described previously, encode the choice $\,(
\triangle_{1,2},\triangle_\a)\,$ of the triangulation of the two
half-loops that compose $\,X_\a$,\ together with the choice of the
indexing maps $\,\triangle_{1,2}\to\xcI_M\,$ and $\,\triangle_\a\to
\xcI_M$,\ of which the former is induced, as described, by an
indexing map $\,\triangle_{1,2}\to\xcI_Q\,$ and \v Cech extensions
$\,\phi_\a,\ \a\in\{1,2\}\,$ of the $\,\iota_\a$.\ Last, introduce
the shorthand notation $\,\unl\triangle_\a\,$ and
$\,\unl\triangle_{1,2}\,$ for the set of edges and vertices of the
triangulations $\,\triangle_\a\,$ and $\,\triangle_{1,2}$,\
respectively, with the boundary vertices $\,v_\pm\,$ removed, and
fix a local presentation of $\,\Bgt\,$ associated with the ensuing
choice of open covers of the $\cB$-fusion subspace, in conformity
with Definition \ref{def:loco}. After a tedious but otherwise
completely straightforward calculation, we find the explicit
expressions
\qq\nn
\tht_{\si,\circledast\cB\,(\igt^1,\igt^2)}[(\psi_1,\psi_2)]&=&
\int_\sfI\,\sfd\varphi\wedge\bigl(\Psi_\si^2\vert_\sfI\bigr)^*
\sfp_2+\int_{\tau(\sfI)}\,\sfd\varphi\wedge\bigl(\Psi_\si^1
\vert_{\tau(\sfI)}\bigr)^*\sfp_1\cr\cr
&&-\sum_{e\in\triangle_2}\,\int_e\,X_{2\,e}^*B_{i^2_e}-\sum_{e\in
\triangle_1}\,\int_e\,X_{1\,e}^*B_{i^1_e}\cr\cr
&&-\sum_{v\in\unl\triangle_2}\,X_2^*A_{i^2_{e_+(v)}i^2_{e_-(v)}}(v)
-\sum_{v\in\unl\triangle_1}\,X_1^*A_{i^1_{e_+(v)}i^1_{e_-(v)}}(v)
\cr\cr
&&-\bigl(X_2^*A_{i^2_{e_+(v_+)}i^2_{v_+}}-X_1^*A_{i^1_{e_-(v_+)}
i^1_{v_+}}-Y_{1,2}^*P_{i^{1,2}_{v_+}}\bigr)(v_+)\cr\cr
&&+\bigl(X_2^*A_{i^2_{e_-(v_-)}i^2_{v_-}}-X_1^*A_{i^1_{e_+(v_-)}
i^1_{v_-}}-Y_{1,2}^*P_{i^{1,2}_{v_-}}\bigr)(v_-)\cr\cr
&&+\sfi\,\d\log f^+_{\si,\circledast\cB,\igt_1\igt_2}[(\psi_1,
\psi_2)]
\qqq
and (defining the $\,\ovl{\unl\triangle}_\a\,$ in analogy with the
$\,\ovl\triangle_\a$)
\qq\nn
\g_{\si,\circledast\cB\,(\igt^1,\igt^2)(\jgt^1,\jgt^2)}[(\psi_1,
\psi_2)]&=&\prod_{\ovl e\in\ovl\triangle_2}\,\ee^{-\sfi\,\int_{\ovl
e}\,X_{2\,\ovl e}^*A_{i^2_{\ovl e}j^2_{\ovl e}}}\cdot\prod_{\ovl e
\in\ovl\triangle_1}\,\ee^{-\sfi\,\int_{\ovl e}\,X_{1\,\ovl e}^*
A_{i^1_{\ovl e}j^1_{\ovl e}}}\cr\cr
&&\cdot\prod_{\ovl v\in\ovl{\unl\triangle}_2}\,X_2^*\bigl(
g_{j^2_{\ovl e_+(\ovl v)}j^2_{\ovl e_-(\ovl v)}i^2_{\ovl e_-(\ovl v
)}}^{-1}\cdot g_{i^2_{\ovl e_+(\ovl v)}i^2_{\ovl e_-(\ovl v)}
j^2_{\ovl e_+(\ovl v)}}\bigr)(\ovl v)\cr\cr
&&\cdot\prod_{\ovl v\in\ovl{\unl\triangle}_1}\,X_1^*\bigl(
g_{j^1_{\ovl e_+(\ovl v)}j^1_{\ovl e_-(\ovl v)}i^1_{\ovl e_-(\ovl v
)}}^{-1}\cdot g_{i^1_{\ovl e_+(\ovl v)}i^1_{\ovl e_-(\ovl v)}
j^1_{\ovl e_+(\ovl v)}}\bigr)(\ovl v)\cr\cr
&&\cdot\bigl[Y_{1,2}^*K_{i^{1,2}_{v_+}j^{1,2}_{v_+}}\cdot X_1^*
\bigl(g_{j^1_{\ovl e_-(v_+)}i^1_{v_+}j^1_{v_+}}^{-1}\cdot
g_{i^1_{\ovl e_-(v_+)}j^1_{\ovl e_-(v_+)}i^1_{v_+})}\bigr)\cr\cr
&&\ \cdot X_2^*\bigl(g_{i^2_{\ovl e_+(v_+)}j^2_{\ovl e_+(v_+)}
i^2_{v_+}}^{-1}\cdot g_{j^2_{\ovl e_+(v_+)}i^2_{v_+}j^2_{v_+}}
\bigr)\bigr](v_+)\cr\cr
&&\cdot\bigl[Y_{1,2}^*K_{i^{1,2}_{v_-}j^{1,2}_{v_-}}^{-1}\cdot X_1^*
\bigl(g_{i^1_{\ovl e_+(v_-)}j^1_{\ovl e_+(v_-)}i^1_{v_-}}^{-1}\cdot
g_{j^1_{\ovl e_+(v_-)}i^1_{v_-}j^1_{v_-})}\bigr)\cr\cr
&&\ \cdot X_2^*\bigl(g_{j^2_{\ovl e_-(v_-)}i^2_{v_-}j^2_{v_-}}^{-1}
\cdot g_{i^2_{\ovl e_-(v_-)}j^2_{\ovl e_-(v_-)}i^2_{v_-}}\bigr)
\bigr](v_-)\cr\cr
&&\cdot\bigl(\bigl(f^+_{\si,\circledast\cB,\igt_1\igt_2}\bigr)^{-1}
\cdot f^+_{\si,\circledast\cB,\jgt_1\jgt_2}\bigr)[(\psi_1,\psi_2)]
\qqq
for the local connection 1-forms
\qq\nn
\tht_{\si,\circledast\cB\,(\igt^1,\igt^2)}=\bigl(\pr_1^*\tht_{\si,
\emptyset\,\igt^1}+\pr_2^*\tht_{\si,\emptyset\,\igt_2}\bigr)
\vert_{\sfP_{\si,\emptyset}^{\circledast\cB}}
\qqq
and the local transition functions
\qq\nn
\g_{\si,\circledast\cB\,(\igt^1,\igt^2)(\jgt^1,\jgt^2)}=\bigl(
\pr_1^*\g_{\si,\emptyset\,\igt^1\jgt^1}\cdot\pr_2^*\g_{\si,
\emptyset\,\igt^2\jgt^2}\bigr)\vert_{\sfP_{\si,
\emptyset}^{\circledast\cB}}
\qqq
of the bundle $\,\ceL_{\si,\circledast\cB}$,\ defined in terms of
the connection 1-forms and transition functions of $\,\ceL_{\si,
\emptyset}$,\ cf.\ \Reqref{eq:loc-dat-preq-untw}. Above, we have
introduced local functionals
\qq
f^+_{\si,\circledast\cB\,(\igt^1,\igt^2)}[(\psi_1,\psi_2)]&=&
\prod_{e^{(1,2)}\in\triangle_{1,2}}\,\ee^{\sfi\,\int_{e^{(1,2)}}\,
Y_{1,2\,e^{(1,2)}}^*P_{i^{1,2}_{e^{(1,2)}}}}\cr\cr
&&\cdot\prod_{v^{(1,2)}\in\unl\triangle_{1,2}}\,Y_{1,2}^*K^{-
1}_{i^{1,2}_{e^{(1,2)}_+(v^{(1,2)})}i^{1,2}_{e^{(1,2)}_-(v^{(1,2)}
)}}\bigl(v^{(1,2)}\bigr)\cr\cr
&&\cdot\bigl(Y_{1,2}^*K^{-1}_{i^{1,2}_{e^{(1,2)}_+(v_+)}i^{1,
2}_{v_+}}\cdot X_1^*g^{-1}_{i^1_{e^{(1,2)}_+(v_+)}i^1_{e_-(v_+)}
i^1_{v_+}}\cdot X_2^*g_{i^2_{e^{(1,2)}_+(v_+)}i^2_{e_+(v_+)}
i^2_{v_+}}\bigr)(v_+)\cr\cr
&&\cdot\bigl(Y_{1,2}^*K_{i^{1,2}_{e^{(1,2)}_-(v_-)}i^{1,2}_{v_-}}
\cdot X_1^*g_{i^1_{e^{(1,2)}_-(v_-)}i^1_{e_+(v_-)}i^1_{v_-}}\cdot
X_2^*g^{-1}_{i^2_{e^{(1,2)}_-(v_-)}i^2_{e_-(v_-)}i^2_{v_-}}\bigr)(
v_-)\,,\label{eq:loc-dat-dtwist-trf}
\qqq
using the labelling conventions detailed above, cf.\
\Rfig{fig:fusion-triangulation}, and analogous conventions for
double intersections, with the associated intersection
triangulations preserving the fixed vertices $\,v_\pm$.

In the last step, we pass to the $2\to 1$ cross-$\cB$ interaction
subspace $\,\Igt_\si(\circledast\cB:\cJ:\cB)\,$ and restrict to it
the pullback bundles $\,\pr_1^*\ceL_{\si,\emptyset}\ox\pr_2^*
\ceL_{\si,\emptyset}\,$ and $\,\pr_3^*\ceL_{\si,\emptyset}$.\ The
vanishing of the 2-form $\,\Om_{\si,\emptyset}^{+-}\,$ under the
same restriction, demonstrated in the proof of statement i) of the
present theorem, indicates that the restricted bundles may, indeed,
be isomorphic. We readily convince ourselves that this is the case
through a direct computation employing Definition
\ref{def:int-sub-untw} and
Eqs.\,\eqref{eq:nfix-half}-\eqref{eq:loop-nfix-half} which describe
the base of the bundles of interest locally. In so doing, we assume
-- as previously -- the triangulations of the various half-loops of
the untwisted states in interaction to be induced by the
triangulations of the respective parent half-loops $\,Y_{I,J}\in\sfI
Q\,$ across which the states are identified. This entails, in
particular, fixing a pair of vertices $\,v_+,v_-\,$ in each of the
triangulations, corresponding to the pair $\,\jmath,\jmath^\vee\,$
of defect junctions in \Rfig{fig:fusion}. A lengthy yet direct
computation, invoking the defining relations
\eqref{eq:DG-is-H}-\eqref{eq:Dphin-is}, yields the anticipated
identities
\qq
(\pr_1\x\pr_2)^*\tht_{\si,\circledast\cB\,(\igt^1,\igt^2)}-\pr_3^*
\tht_{\si,\emptyset\,\igt_3}&=&-\sfi\,\sfd\log f^{+-}_{\si\,(\igt^1,
\igt^2,\igt^3)}\,,\cr\label{eq:f+-dlog}&&\\
(\pr_1\x\pr_2)^*\g_{\si,\circledast\cB\,(\igt^1,\igt^2)(\jgt^1,
\jgt^2)}\cdot\pr_3^*\g_{\si,\emptyset\,\igt_3\jgt_3}^{-1}&=&\bigl(
f^{+-}_{\si\,(\jgt^1,\jgt^2,\jgt^3)}\bigr)^{-1}\cdot f^{+-}_{\si\,(
\igt^1,\igt^2,\igt^3)}\,.\nonumber
\qqq
Here, the index $\,\igt^3\,$ represents a composite triangulation of
the circle that parameterises the end-state loop $\,X_3$,\ induced
from those of the parent half-loops $\,Y_{1,3}\,$
($\triangle_{1,3}\,$ induces the part of $\,\igt_3\,$ associated
with the half-loop $\,\tau(\sfI)$) and $\,Y_{2,3}\,$
($\triangle_{2,3}\,$ induces the remaining part, associated with the
half-loop $\,\sfI$), together with the corresponding indexing maps.
The latter are subject to some obvious relations
\qq\nn
\barr{l} i^1=\phi_1\bigl(i^{1,2}\bigr)\,,\qquad\qquad i^2=\phi_2
\bigl(i^{1,2}\bigr)\qquad{\rm along}\quad Y_{1,2}\,,\cr\cr
i^1=\phi_1\bigl(i^{1,3}\bigr)\,,\qquad\qquad i^3=\phi_2\bigl(i^{1,
3}\bigr)\qquad{\rm along}\quad Y_{1,3}\,,\cr\cr
i^2=\phi_1\bigl(i^{2,3}\bigr)\,,\qquad\qquad i^3=\phi_2\bigl(i^{2,
3}\bigr)\qquad{\rm along}\quad Y_{2,3}\,,\earr
\qqq
as well as the relations
\qq\nn
i^{I,J}=\psi_3^{I,J}\bigl(i^{1,2,3}\bigr)
\qqq
satisfied at the triple junctions and written in terms of the index
maps $\,\psi_3^{I,J}:\xcI_{\cO_{T_3}}\to\xcI_{\cO_Q}\,$ covering the
$\,\pi^{I,J}_3\,$ (for $\,\psi_3^{1,3}\equiv\psi_3^{3,1}\,$ and
$\,\pi_3^{1,3}\equiv\pi_3^{3,1}$).\ For completeness, we also give
the local data of the isomorphism -- they read (the
$\,\unl\triangle_{I,J}\,$ are defined in analogy with
$\,\triangle_{1,2}$)
\qq
f^{+-}_{\si\,(\igt^1,\igt^2,\igt^3)}[(\psi_1,\psi_2,\psi_3)]&=&
\prod_{(I,J)\in\{(1,2),(2,3),(1,3)\}}\,\bigl(\prod_{e^{(I,J)}\in
\triangle_{I,J}}\,\ee^{-\sfi\,\int_{e^{(I,J)}}\,Y_{I,J\,e^{(I,J
)}}^*P_{i^{I,J}_{e^{(I,J)}}}}\cr\cr
&&\cdot\prod_{v^{(I,J)}\in\unl\triangle_{I,J}}\,Y_{I,J}^*K_{i^{I,
J}_{e^{(I,J)}_+(v^{(I,J)})}i^{I,J}_{e^{(I,J)}_-(v^{(I,J)})}}(v^{(I,
J)})\bigr)\cr\cr
&&\cdot\bigl(X_1^*g_{i^1_{e^{(1,2)}_+(v_+)}i^1_{e^{(1,3)}_-(v_+)}
i^1_{v_+}}\cdot X_2^*g_{i^2_{e^{(2,3)}_+(v_+)}i^2_{e^{(1,2)}_+(v_+
)}i^2_{v_+}}\cdot X_3^*g_{i^3_{e^{(1,3)}_-(v_+)}i^3_{e^{(2,3)}_+(v_+
)}i^3_{v_+}}\cr\cr
&&\cdot Y_{1,2}^*K_{i^{1,2}_{e^{(1,2)}_+(v_+)}i^{1,2}_{v_+}}\cdot
Y_{2,3}^*K_{i^{2,3}_{e^{(2,3)}_+(v_+)}i^{2,3}_{v_+}}\cdot Y_{1,3}^*
K^{-1}_{i^{1,3}_{e^{(1,3)}_-(v_+)}i^{1,3}_{v_+}}\cdot Z^*f^{-1}_{3,
i^{1,2,3}_{v_+}}\bigr)(v_+)\cr\cr
&&\cdot\bigl(X_1^*g_{i^1_{e^{(1,3)}_+(v_-)}i^1_{e^{(1,2)}_-(v_-)}
i^1_{v_-}}\cdot X_2^*g_{i^2_{e^{(1,2)}_-(v_-)}i^2_{e^{(2,3)}_-(v_-
)}i^2_{v_-}}\cdot X_3^*g_{i^3_{e^{(2,3)}_-(v_-)}i^3_{e^{(1,3)}_+(
v_-)}i^3_{v_-}}\cr\cr
&&\cdot Y_{1,2}^*K^{-1}_{i^{1,2}_{e^{(1,2)}_-(v_-)}i^{1,2}_{v_-}}
\cdot Y_{2,3}^*K^{-1}_{i^{2,3}_{e^{(2,3)}_-(v_-)}i^{2,3}_{v_-}}
\cdot Y_{1,3}^*K_{i^{1,3}_{e^{(1,3)}_+(v_-)}i^{1,3}_{v_-}}\cdot Z^*
f_{3,i^{1,2,3}_{v_-}}\bigr)(v_-)\,.\label{eq:f+-iso}
\qqq
This completes the proof of statement ii), and so also the proof of
the theorem.\qed
\eit

\section{A proof of Theorem \ref{thm:cross-def-int-tw}}
\label{app:cross-def-int-tw}

\bit
\item[Ad i)] Take the manifold $\,\sfP_{\si,\cB|(\vep_1,\vep_2
)}^{\x 2}\,$ of \Reqref{eq:Ptw2}, together with the `sum' symplectic
form
\qq\nn
\Om^+_{\si,\cB|(\vep_1,\vep_2)}=\pr_1^*\Om_{\si,\cB|(\pi,\vep_1)}+
\pr_2^*\Om_{\si,\cB|(\pi,\vep_2)}
\qqq
on it, the latter being expressed in terms of the canonical
projections $\,\pr_\a:\sfP_{\si,\cB|(\vep_1,\vep_2)}^{\x 2}\to
\sfP_{\si,\cB|(\pi,\vep_\a)},\ \a\in\{1,2\}$.\ The symplectic form
$\,\Om^+_{\si,\cB|( \vep_1,\vep_2)}\,$ is readily checked to
restrict -- with the restriction marked by the bar over
$\,\Om^+_{\si,\cB|(\vep_1,\vep_2 )}\,$ -- to the tangent $\,\Tgt
\sfP_{\si,\cB|(\vep_1,\vep_2)}^{\x 2}\,$ of the $\cB_{\rm
triv}$-fusion subspace $\,\sfP_{\si,\cB|(\vep_1,\vep_2)}^{\x 2}\,$
within $\,\G(\sfT\sfP_{\si,\cB|(\vep_1,\vep_2 )}^{\x 2}
\vert_{\sfP_{\si,\cB|(\vep_1,\vep_2)}^{\x 2}})\,$ as
\qq\nn
\ovl\Om^+_{\si,\cB|(\vep_1,\vep_2)}[(\psi_1,\psi_2)]\cr\cr
=\int_\sfI\,\sfd\varphi\,(\Psi^2_{\si,{\rm cl}}\vert_\sfI)^*\bigl(
\d\sfp_2+\p_\varphi X_2\con\txH\bigr)+\int_{\tau(\sfI)}\,\sfd
\varphi\,(\Psi^1_{\si,{\rm cl}}\vert_{\tau(\sfI)})^*\bigl(\d\sfp_1+
\p_\varphi X_1\con\txH\bigr)+\vep_1\,\om(q_1)+\vep_2\,\om(q_2)\,.
\qqq
Here, $\,(\Psi^1_{\si,{\rm cl}},\Psi^2_{\si,{\rm cl}})\,$ is a pair
of extremal sections represented by the Cauchy data $\,(\psi_1,
\psi_2)\,$ with $\,\psi_\a=(X_\a,\sfp_\a,q_\a,V_\a),\ \a\in\{1,2
\}$.\ Using this, we readily check, through inspection, that the
symplectic form $\,\Om^{+-}_{\si,\cB|(\vep_1,\vep_2,\vep_3)}\,$ on
$\,\sfP_{\si,\cB|(\vep_1,\vep_2,\vep_3)}^{\x 3}\,$ vanishes
identically when restricted to the subspace $\,\Tgt\Igt_\si(
\circledast\cB_{\rm triv}:\cJ:\cB_{\rm triv})^{\cB|(\vep_1,\vep_2,
\vep_3)}\subset\G\bigl(\sfT\sfP_{\si,\cB |(\vep_1,\vep_2,\vep_3
)}^{\x 3}\vert_{\Igt_\si(\circledast\cB_{\rm triv}:\cJ:\cB_{\rm
triv})^{\cB|(\vep_1,\vep_2,\vep_3)}}\bigr)\,$ tangent to $\,\Igt_\si
(\circledast\cB_{\rm triv}:\cJ:\cB_{\rm triv})^{\cB|(\vep_1,\vep_2,
\vep_3)}$,\ as per
\qq\nn
\ovl\Om^{+-}_{\si,\cB|(\vep_1,\vep_2,\vep_3)}[(\psi_1,\psi_2,\psi_3
)]=\sum_{k=1}^3\,\vep_k\,\om(q_k)=\sum_{k=1}^3\,\vep_k\,\pi_3^{k,k+
1\,*}\om(t_3)=0\,.
\qqq
This proves statement i).
\item[Ad ii)] We begin by reconstructing local data of the bundle
\qq\nn
\ceL_{\si,\circledast\cB_{\rm triv}}^{\cB|(\vep_1,\vep_2)}:=(\pr_1^*
\ceL_{\si,\cB|(\pi,\vep_1)}\ox\pr_2^*\ceL_{\si,\cB|(\pi,\vep_2)})
\vert_{\sfP_{\si,\cB|(\vep_1,\vep_2)}^{\circledast\cB_{\rm triv}}}
\,,
\qqq
produced by restricting, to the $\cB_{\rm triv}$-fusion subspace,
the tensor product of the pullbacks of the pre-quantum bundles for
the 1-twisted sector of the $\si$-model along the canonical
projections $\,\pr_\a$.\ To this end, we make a convenient choice of
an open cover of the base of the bundle, consistent with the various
half-loop identifications present in the definition of
$\,\sfP_{\si,\cB|(\vep_1,\vep_2)}^{\circledast\cB_{\rm triv}}$.\ As
the construction parallels that carried out in the proof of
statement ii) of Theorem \ref{thm:cross-def-int-untw}, we restrict
here to naming the differences. Thus, we fix a common triangulation
$\,\triangle_{1,2}\,$ of the parameterising $\pi$-unit interval
$\,\sfI\,$ and a common index assignment for the two half-loops
$\,X_\a\vert_{\vsi_\a(\sfI)},\ \a\in\{1,2 \}\,$ (with the common
vertex $\,v_+\,$ removed), the sole difference with respect to the
situation depicted in \Rfig{fig:fusion-triangulation} being --
beside the appearance of the two intersection points $\,q_\a\,$ --
the equality of the indices
\qq\nn
i^1_{f^{(1,2)}}=i^2_{f^{(1,2)}}=:i^{(1,2)}_{f^{(1,2)}}\,,\qquad
\qquad f^{(1,2)}\in\triangle_{1,2}\,,
\qqq
and the presence of two independent \v Cech indices from
$\,\xcI_Q\,$ assigned to the distinguished common vertex $\,v_+\,$
(corresponding to the boundary point $\,\pi\,$ in $\,\sfI$), to be
mapped to the defect junction $\,\jmath$,\ namely $\,i^{1\,1,
2}_{v_+}\,$ (for $\,X_1$) and $\,i^{2\,1,2}_{v_+}\,$ (for $\,X_2$).
They give rise, through the index maps $\,\phi_\a\,$ covering the
$\,\iota_\a$,\ to a triple of \v Cech indices from $\,\xcI_M$,\ to
wit,
\qq\label{eq:fus-ind-v+-tw}
\qquad\qquad i^1_{v_+}=\phi_1\bigl(i^{1\,1,2}_{v_+}\bigr)\,,\qquad
\qquad i^{(1,2)}_{v_+}=\phi_2\bigl(i^{1\,1,2}_{v_+}\bigr)=\phi_1
\bigl(i^{2\,1,2}_{v_+}\bigr)\,,\qquad\qquad i^2_{v_+}=\phi_2\bigl(
i^{2\,1,2}_{v_+}\bigr)\,.
\qqq
The other boundary vertex of the triangulation of the $\pi$-unit
interval $\,\sfI$,\ corresponding to the value of the angular
parameter $\,\varphi=0$,\ shall be denoted by $\,v_-$.\ The local
data of the bundle $\,\ceL_{\si,\circledast\cB_{\rm triv}}^{\cB|(
\vep_1,\vep_2)}\,$ associated with the ensuing open cover of its
base $\,\sfP_{\si,\cB|(\vep_1,\vep_2)}^{\circledast\cB_{\rm
triv}}\,$ take the form
\qq\nn
\tht_{\si,\circledast\cB_{\rm triv}\,(\igt^1,\igt^2)}^{\cB|(
\vep_1,\vep_2)}[(\psi_1,\psi_2)]&=&\int_\sfI\,\sfd\varphi\wedge(
\Psi^2_{\si,{\rm cl}}\vert_\sfI)^*\sfp_2+\int_{\tau(\sfI)}\,\sfd
\varphi\wedge(\Psi^1_{\si,{\rm cl}}\vert_{\tau(\sfI)})^*\sfp_1\cr
\cr
&&-\sum_{e\in\triangle_2}\,\int_e\,X_{2\,e}^*B_{i^2_e}-\sum_{e
\in\triangle_1}\,\int_e\,X_{1\,e}^*B_{i^1_e}\cr\cr
&&-\sum_{v\in\unl\triangle_2}\,X_2^*A_{i^2_{e_+(v)}i^2_{e_-(v
)}}(v)-\sum_{v\in\unl\triangle_1}\,X_1^*A_{i^1_{e_+(v)}
i^1_{e_-(v)}}(v)\cr\cr
&&-X_1^*A_{i^1_{e_+(v_-)}i^2_{e_-(v_-)}}(v_-)\cr\cr
&&+\bigl(\iota_1^{\vep_1\,*}A_{i^1_{e_-(v_+)}i^1_{v_+}}+\vep_1\,
P_{i^{1\,1,2}_{v_+}}\bigr)(q_1)-\bigl(\iota_2^{\vep_2\,*}
A_{i^2_{e_+(v_+)}i^2_{v_+}}-\vep_2\,P_{i^{2\,1,2}_{v_+}}\bigr)(q_2)
\cr\cr
&&+\sfi\,\d\log X_1^*g_{i^{(1,2)}_{e^{(1,2)}_-(v_-)}i^1_{e_+(v_-)}
i^2_{e_-(v_-)}}(v_-)\,,\cr\cr\cr \g_{\si,\circledast\cB_{\rm triv}\,
(\igt^1,\igt^2)(\jgt^1,\jgt^2)}^{\cB|(\vep_1,\vep_2)}[(\psi_1,
\psi_2)]&=&\prod_{\ovl e\in\ovl\triangle_2}\,\ee^{-\sfi\,\int_{\ovl
e}\,X_{2\,\ovl e}^*A_{i^2_{\ovl e}j^2_{\ovl e}}}\cdot\prod_{\ovl
e\in\ovl\triangle_1}\,\ee^{-\sfi\,\int_{\ovl e}\,X_{1\,\ovl
e}^*A_{i^1_{\ovl e}j^1_{\ovl e}}}\cr\cr
&&\cdot\prod_{\ovl v\in\ovl{\unl\triangle}_2}\,X_2^*\bigl(
g_{i^2_{\ovl e_+(\ovl v)}i^2_{\ovl e_-(\ovl v)}j^2_{\ovl e_+(\ovl v
)}}\cdot g^{-1}_{j^2_{\ovl e_+(\ovl v)}j^2_{\ovl e_-(\ovl v)}
i^2_{\ovl e_-(\ovl v )}}\bigr)(\ovl v)\cr\cr
&&\cdot\prod_{\ovl v\in\ovl{\unl\triangle}_1}\,
X_1^*\bigl(g_{i^1_{\ovl e_+(\ovl v)}i^1_{\ovl e_-(\ovl v)}j^1_{\ovl
e_+(\ovl v)}}\cdot g^{-1}_{j^1_{\ovl e_+(\ovl v)}j^1_{\ovl e_-(\ovl
v)}i^1_{\ovl e_-(\ovl v )}}\bigr)(\ovl v)\cr\cr
&&\cdot X_1^*\bigl(g_{i^1_{\ovl e_+(v_-)}i^2_{\ovl e_-(v_-)}
j^1_{\ovl e_+(v_-)}}\cdot g^{-1}_{j^1_{\ovl e_+(v_-)}j^2_{\ovl e_-(
v_-)}i^2_{\ovl e_-(v_-)}}\bigr)(v_-)\cr\cr
&&\cdot\bigl(\iota_1^{\vep_1\,*}\bigl(g_{i^1_{\ovl e_-(v_+)}
j^1_{\ovl e_-(v_+ )}i^1_{v_+}}\cdot g^{-1}_{j^1_{\ovl e_-(v_+)}
i^1_{v_+}j^1_{v_+}}\bigr)\cdot K^{\vep_1}_{i^{1\,1,2}_{v_+}j^{1\,1,
2}_{v_+}}\bigr)(q_1)\cr\cr
&&\ \cdot\bigl(\iota_2^{\vep_2\,*}\bigl(g^{-1}_{i^2_{\ovl e_+(v_+)}
j^2_{\ovl e_+(v_+)}i^2_{v_+}}\cdot g_{j^2_{\ovl e_+(v_+)}i^2_{v_+}
j^2_{v_+}}\bigr)\cdot K^{\vep_2}_{i^{2\,1,2}_{v_+}j^{2\,1,2}_{v_+}}
\bigr)(q_2)\cr\cr
&&\cdot X_1^*\bigl(g^{-1}_{i^{(1,2)}_{\ovl e_-^{(1,2)}(v_-)}
i^1_{\ovl e_+(v_-)}i^2_{\ovl e_-(v_-)}}\cdot g_{j^{(1,2)}_{\ovl
e_-^{(1,2)}(v_-)}j^1_{\ovl e_+( v_-)}j^2_{\ovl e_-(v_-)}}\bigr)(v_-
)\,.
\qqq
Note that by virtue of the gluing conditions assumed, we can
interchange the fields $\,X_\a,\ \a\in\{1,2\}\,$ at $\,v_-\,$ --
this removes the apparent asymmetry in the above formul\ae ~due to
the presence of $\,X_1$.

The last step in our analysis consists in explicitly establishing an
isomorphism between the pullback bundles $\,(\pr_1\x\pr_2)^*
\ceL_{\si,\circledast \cB_{\rm triv}}^{\cB|(\vep_1,\vep_2)}\,$ and
$\,\pr_3^*\ceL^\vee_{\si,\cB|(\pi,\vep_3)}\,$ over the interaction
subspace of the 1-twisted string. The demonstration of its
anticipated existence proceeds along similar lines as in the proof
of Theorem \ref{thm:cross-def-int-untw}, with additional
simplifications of the triangulations and indexing involved,
peculiar to the trivial gluing conditions imposed, namely: We choose
a common triangulation and indexing for the pairs
$\,(X_2,X_3)\vert_{\sfI-\{v_+\}}\,$ and $\,(X_1,X_3)
\vert_{\tau(\sfI)-\{v_+\}}$.\ At the distinguished common vertex
$\,v_+$,\ we now have, in addition to relations
\eqref{eq:fus-ind-v+-tw}, those obeyed by the extra indices $\,i^{1,
2,3}_{v_+}\in\xcI_{\cO_{T_3}}\,$ and $\,i^{3,1}_{v_+}\in
\xcI_{\cO_Q}$,\ to wit,
\qq\nn
i^{1\,1,2}_{v_+}=\psi_3^{1,2}\bigl(i^{1,2,3}_{v_+}\bigr)\,,\qquad
\qquad i^{2\,1,2}_{v_+}=\psi_3^{2,3}\bigl(i^{1,2,3}_{v_+}\bigr)\,,
\qquad\qquad i^{3,1}_{v_+}=\psi_3^{3,1}\bigl(i^{1,2,3}_{v_+}\bigr)
\,.
\qqq
It is now completely straightforward to verify, using
\Reqref{eq:Dphin-is}, the identities
\qq\nn
(\pr_1\x\pr_2)^*\tht_{\si,\circledast\cB_{\rm
triv}\,(\igt^1,\igt^2)}^{\cB|(\vep_1,
\vep_2)}-\pr_3^*\tht_{\si,\cB|(\pi,\vep_3)\,\igt^3}&=&-\sfi\,\sfd\log
f^{+ -}_{\si\,(\igt^1,\igt^2,\igt^3)}\,,\cr\cr
(\pr_1\x\pr_2)^*\g_{\si,\circledast\cB_{\rm triv}\,(\igt^1,
\igt^2)(\jgt^1,\jgt^2)}^{\cB|(\vep_1,\vep_2)}\cdot\pr_3^*\g_{\si,
\cB|(\pi,\vep_3)\,\igt^3\jgt^3}^{-1}&=&
\bigl(f^{+-}_{\si\,(\jgt^1,\jgt^2,\jgt^3)}\bigr)^{-1}\cdot f^{+
-}_{\si\,(\igt^1,\igt^2,\igt^3)}\,,
\qqq
expressed in terms of the local functionals
\qq\nn
f^{+-}_{\si\,(\igt^1,\igt^2,\igt^3)}[(\psi_1,\psi_2,\psi_3)]=X_1^*
g^{-1}_{i^{(1,2)}_{e^{(1,2)}_-(v_-)}i^1_{e_+(v_-)}i^2_{e_-(v_-)}}(
v_-)\cdot f^{-1}_{3,i^{1,2,3}_{v_+}}(t_3)\,.
\qqq
The latter define the isomorphism sought after. This concludes the
proof of the theorem.\qed
\eit

\bibliographystyle{amsalpha}
\bibliography{ddgg1dsf_bib}

\end{document}